\documentclass{aa}

\usepackage{natbib}
\bibpunct{(}{)}{;}{a}{}{,} 
\usepackage{graphicx}

\usepackage[maxfloats=256]{morefloats}
\usepackage{txfonts}
\begin{document} 

   \title{Protostellar Interferometric Line Survey of the Cygnus-X region (PILS-Cygnus)}

   \subtitle{The role of the external environment in setting the chemistry of protostars}
   \author{S. J. van der Walt\inst{1} \and L. E. Kristensen\inst{1} \and H. Calcutt\inst{2} \and J. K. J\o rgensen\inst{1} \and R. T. Garrod\inst{3} 
          }    
   \institute{Niels Bohr Institute, University of Copenhagen, \O ster Voldgade 5--7, 1350 Copenhagen K., Denmark\\
              \email{sarel.vanderwalt@nbi.ku.dk}
    \and
    Institute of Astronomy, Faculty of Physics, Astronomy and Informatics, Nicolaus Copernicus University, Grudziadzka 5, 87-100 Torun, Poland
    \and
    Departments of Astronomy and Chemistry, University of Virginia, Charlottesville, VA 22904 USA}

   \date{Received 14 October 2022; accepted 25 July 2023}

\abstract
   {Molecular lines are commonly detected towards protostellar sources. However, to get a better understanding of the chemistry of these sources we need unbiased molecular surveys over a wide frequency range for as many sources as possible to shed light on the origin of this chemistry, particularly any influence from the external environment.} 
   {We present results from the PILS-Cygnus survey of ten intermediate- to high-mass protostellar sources in the nearby Cygnus-X complex, through high angular resolution interferometric observations over a wide frequency range.}
   {Using the Submillimeter Array (SMA), a spectral line survey of ten sources was performed in the frequency range 329--361 GHz, with an angular resolution of $\sim$1\farcs5, or $\sim$2000 AU at a source distance of 1.3 kpc from the Sun. Spectral modelling was performed to identify molecular emission and determine column densities and excitation temperatures for each source. Emission maps were made to study the morphology of emission. Finally, emission properties were compared across the sample. }
   {We detect CH$_3$OH towards nine of the ten sources, with CH$_3$OCH$_3$ and CH$_3$OCHO towards three sources. We further detect CH$_3$CN towards four sources. Towards five sources the chemistry is spatially differentiated, meaning that different species peak at different positions and are offset from the peak continuum emission. Low levels of deuteration are detected towards four sources in HDO emission, whereas deuterated complex organic molecule emission is detected towards one source (CH$_2$DOH towards N63). The chemical properties of each source do not correlate with their position in the Cygnus-X complex, nor do the distance or direction to the nearest OB associations. However, the five sources located in the DR21 filament do appear to show less line emission compared to the five sources outside the filament.}
   {This work shows how important wide frequency coverage observations are combined with high angular resolution observations for studying the protostellar environment. Furthermore, based on the ten sources observed here, the external environment appears to only play a minor role in setting the chemical environment on these small scales ($<$ 2000 AU). }


\maketitle 

%

\section{Introduction}

The molecular emission observed in and near star-forming regions is an important tool with which the physical conditions and evolution of protostars can be inferred. The molecules that cause the emission range from the simplest and most abundant molecules that trace the large-scale structure of the envelopes surrounding the newly formed stars, to the complex organic molecules \citep[COMs; molecules with at least six atoms and one or more carbon atoms;][]{Herbst-2009} that are observed closer to the newly formed star, in the warm inner regions of the envelope \citep[`hot cores'; e.g.][]{Kurtz-2000, Cesaroni-2010, joergensen-2016}. Different molecules therefore trace different regions in and near newly formed stars \citep[][]{Tychoniec-2021} and observing as many different molecules, both simple and complex, and as many of their transitions as possible is necessary to get a better understanding of the conditions and evolution of protostars.

Simple di- and tri-atomic molecules typically form directly in the gas phase \citep{Herbst-2009}. The historical picture of the origin of COMs has been that they primarily form on dust grains at cold temperatures, and that they are then released into the gas phase at $T\sim$100K, where they can then be observed directly at sub-millimetre wavelengths in hot cores. However, it has been proposed that COMs can also be sputtered off the icy dust-mantles by shocks from jets and outflows \citep[e.g.][]{Avery_and_Chiao-1996, Joergensen-2004a,Arce-2008,Sugimura-2011,Lefloch-2017}, shocks from material accreted from the envelope onto the disc \citep[e.g.][]{Podio-2015,ArturdelaVillarmois-2018,Csengeri-2018,Csengeri-2019}, from explosive events near forming stars \citep[e.g. Orion KL;][]{Zapata-2011,Orozco-Aguilera-2017}, and from UV irradiation of the outflow cavity walls \citep[e.g.][]{Drozdovskaya-2015}. Some of these mechanisms have been invoked to interpret why COM emission is sometimes observed to be offset from the main continuum peak, where $T$ $>$ 100 K. The dominant process in releasing COMs frozen out on dust grains into the gas phase is therefore often unclear \citep[e.g.][]{Belloche-2020, vanderWalt-2021}. This understanding is necessary to properly interpret the chemistry, both of COMs and the simpler molecules. 

Furthermore, the observed chemistry may not only depend on the local conditions, but the history of the large-scale or external environment may play a role as well. The proximity to bright UV-radiation sources could influence the chemistry as it heats up the environment surrounding the newly formed stars, for example studies of photon-dominated regions (PDRs), near OB stars \citep[][]{Goicoechea-2006,van_der_Wiel-2009,Taniguchi-2021}. Moreover, most stars, particularly higher-mass stars, form in dense hubs where filaments intersect \citep[e.g.][]{Motte-2018}, and it is unclear if the close proximity between forming stars in such a dense environment has any effect on the chemistry. Therefore, whether it is the local conditions or the external environment that sets the chemical complexity of newly formed stars is still an open question. 

To use the full potential of chemistry and to address these questions, we need to observe emission from as many molecules and their transitions as possible, which requires broad frequency coverage. This has historically been expensive observationally, which resulted in surveys that had to choose narrow frequency bands covering only a select few molecules \citep[e.g.][]{Joergensen-2007,oberg2014,Stephens-2019,Belloche-2020} 
or full line surveys of a limited number of sources \citep[e.g.][]{joergensen-2016,Codella-2017}. 
This situation has changed in recent years with the introduction at the Submillimeter Array (SMA) of the SWARM\footnote{SMA Wideband Astronomical ROACH2 (second generation Reconfigurable Open Architecture Computing Hardware) Machine.} correlator, which is capable of 48 GHz frequency coverage in one sweep at the time of writing (32 GHz at the time data for this work were acquired). 

It is important to observe a large number of sources in a single cloud, covering as broad a frequency range as possible. These sources should be observed at high angular resolution so that the spatial origin of molecular emission can be localised and any ambiguities in the physical origin of emission can be removed as best as possible \citep[e.g.][]{joergensen-2016}: hot core, outflow, envelope, etc. The reason for choosing a single cloud is to, as best as possible, avoid distance uncertainties, while at the same time also ensuring that the sources observed share common initial conditions. 

To demonstrate and exemplify what can be done with observations over such a broad frequency range, data of the single source CygX-N30 (N30) were presented in the first-results paper of the Protosteller Interferometric Line Survey of the Cygnus-X region \citep[PILS-Cygnus;][]{vanderWalt-2021}. The PILS-Cygnus survey is a molecular line and continuum survey of ten sources located in the Cygnus-X star-forming region. It utilised the SWARM correlator on the SMA to take inventory of the chemical variability throughout the Cygnus-X region. The first results from the survey are that the origin of COM emission detected towards this source are from a combination of thermal heating from newly formed stars (the canonical hot core scenario) and accretion of material onto a disc-like structure. The authors further identified chemical differentiation along a linear gradient and between O-bearing species that have peak emission close to one continuum source, while N- and S-bearing species have emission peaks closer to the second source. The authors report low levels of deuteration with only HDO detected towards this source and an upper limit of D/H < 0.1\% derived for CH$_2$DOH, which they attribute to warm temperatures of formation, >30K, with inefficient deuterium fractionation. 

Various results utilising parts of the entire sample have already been published. When looking towards the entire sample, two results stand out. First, the properties of the protostellar envelopes (e.g. mass, density and temperature structure) appear to be similar to both low- and high-mass protostars in general, when scaled with luminosity \citep{Pitts-2022}. Second, the outflows from these sources also appear similar to low- and high-mass sources, again when scaled with the luminosity \citep{Skretas2022}. These two results suggest that the PILS-Cygnus sources are not special when it comes to envelope structure as well as accretion and ejection processes. However, the question remains whether the chemistry is or has been affected by the clustered environment or the proximity to the nearby bright OB association. This question can only be addressed by analysing line emission from the full data set, which is presented here.

We present here the full data set and analysis of the PILS-Cygnus programme for the first time, where ten sources were observed at a uniform frequency coverage, spatial resolution and sensitivity with the SMA. The focus of this paper is on the spectral line properties of the dataset, and thus the chemical properties of the sources. Apart from addressing the above-mentioned science questions, this paper also serves as the data-release paper for this survey. The paper is organised as follows: Section \ref{Sec:Observations} describes the observational setup and data reduction procedure, followed by Section \ref{Sec:Results} in which the results of the data analysis are presented. In Section \ref{Sec:Discussion} a discussion of the results are given, with a conclusion and summary in Section \ref{Sec:Conclusion}.


\section{Observations}
\label{Sec:Observations}

\subsection{Sample selection}

The PILS-Cygnus sample consists of ten intermediate- and high-mass sources (see Table \ref{table:Source_props1}) located in the nearby Cygnus-X complex ($\sim$1.3--1.5 kpc; Table \ref{table:Source_props1}). These sources were first selected from the catalogue of \citet{Kryukova-2014A}, who tabulated the bolometric luminosities of $\sim$1800 protostars in the Cygnus-X complex. These luminosities were based on \textit{Spitzer} IRAC and MIPS observations of the complex, and from there, the bolometric luminosity was inferred based on the methodology outlined in \citet{Kryukova-2012}. These were then compared to the catalogue of \citet{Motte-2007}, who observed a large part of the Cygnus-X complex at a wavelength of 1.2 mm with the IRAM-30m telescope. From these two catalogues, we identified the most luminous and massive sources. 

The next step in the selection process was to check for signs of current and active star formation. This was done using \textit{Herschel}-HIFI observations of the H$_2$O 2$_{02}$--1$_{11}$ transition at 988 GHz \citep{SanJose-Garcia-2015PhDT}. This particular transition is an excellent outflow and shock tracer \citep[e.g.][]{vanDishoeck-2021}, associated with embedded star formation. Most of these sources were also observed in SiO 2--1 emission \citep{Motte-2007}, and were found to be bright in this transition. This corroborates that these sources are embedded sources driving outflows. 

Finally, these sources were already observed with the SMA at 230 GHz in a survey conducted by PI: K. Qiu (proposal ID: 2016A-S021), and were found to show at least H$_2$CO emission at 218 GHz. Based on these selection criteria, ten sources were identified. Their properties are listed in Table \ref{table:Source_props1}.

%
\begin{table*}
\caption{PILS-Cygnus source coordinates and distances.}             
\label{table:Source_props1}      
\centering          
\begin{tabular}{l c c c c r}     
\hline
\hline       
    Source &  Brightest core & RA [J2000] & Dec [J2000] & $D_\odot$\tablefootmark{a}  & $D^\mathrm{proj}_\mathrm{CygX-OB2}$\tablefootmark{b} \\ 
     &  &   &   & [pc] & [pc] \\ 
\hline  
    N30 &   MM1 & 20$^{\text{h}}$ 38$^{\text{m}}$ 36$\fs$45 & 42$^{\circ}$ 37$'$ 33$\farcs$8 &   1300 $\pm $70& 43$^{+2}_{-3}$ \\

\hline  
    S26 &  - & 20$^{\text{h}}$ 29$^{\text{m}}$ 24$\fs$87 & 40$^{\circ}$ 11$'$ 19$\farcs$3 &  3330 $\pm$ 110 & - \\ 
\hline  
    N12 &  MM2 & 20$^{\text{h}}$ 36$^{\text{m}}$ 57$\fs$68 & 42$^{\circ}$ 11$'$ 30$\farcs$8 &  1400 $\pm $100 & 31 $\pm $2\\ 

\hline  
    N63 &  - & 20$^{\text{h}}$ 40$^{\text{m}}$ 05$\fs$50 & 41$^{\circ}$ 32$'$ 12$\farcs$6 &   1400 $\pm$ 100 & 42$^{+3}_{-2}$\\
\hline  
    N53 & MM2 & 20$^{\text{h}}$ 39$^{\text{m}}$ 03$\fs$13 & 42$^{\circ}$ 25$'$ 52$\farcs$5 &  $1500^{+80}_{-70}$ & 48 $\pm $2\\  

\hline  
    N54 &  - & 20$^{\text{h}}$ 39$^{\text{m}}$ 04$\fs$03 & 42$^{\circ}$ 25$'$ 41$\farcs$1 &  $1500^{+80}_{-70}$ & 48 $\pm $2 \\  

\hline  
    N51  &  - & 20$^{\text{h}}$ 39$^{\text{m}}$ 01$\fs$97 & 42$^{\circ}$ 24$'$ 59$\farcs$2 &   $1500^{+80}_{-70}$ & 48$^{+2}_{-3}$ \\  
\hline  
    N38 &  - & 20$^{\text{h}}$ 38$^{\text{m}}$ 59$\fs$26 & 42$^{\circ}$ 22$'$ 28$\farcs$6 &   $1500^{+80}_{-70}$ & 48 $\pm $2 \\  
\hline  
    N48 &  MM1 & 20$^{\text{h}}$ 39$^{\text{m}}$ 01$\fs$46 & 42$^{\circ}$ 22$'$ 05$\farcs$9 &  $1500^{+80}_{-70}$ & 48 $\pm $2\\  

\hline  
    S8 &  NE core & 20$^{\text{h}}$ 20$^{\text{m}}$ 39$\fs$29 & 39$^{\circ}$ 37$'$ 54$\farcs$2 &  $1400^{+2600}_{-1100}$ &  >80 \\  

\hline                  
\end{tabular}
\tablefoot{\tablefoottext{a}{Distances from \cite{Rygl-2012}, with distance estimates to N12, N63 and S8 from \citet{Pitts-2022}.}
\tablefoottext{b}{Projected distance to CygX-OB2 from \citet{Pitts-2022}. We note that S26 is located at a distance of 3.3 kpc from the Earth, and thus does not have a projected distance to OB2. }
}
\end{table*}

\subsection{Data acquisition}

The ten sources were observed with the Submillimeter Array (SMA) on Mauna Kea, Hawaii, from June to November, 2017 (PI: Kristensen, project ID 2017A-S028). The observations were performed both in the compact (COM) and extended (EXT) configurations of the array, with baselines ranging from 16--226 m. Both the compact and the extended configuration observations were carried out over five tracks each. The observing strategy was chosen such that each source was observed for close to an equal amount of time each track (for uniform sensitivity across the sample) while filling out as large a part of the $uv$ plane as possible (for uniform resolution across the sample). This was achieved by cycling through all ten sources, where the order of the sources was randomised for each track. Furthermore, each source was observed for 6 minutes (on) and two sources were observed together before the gain calibrator was observed. The scan time was set to 10 seconds, in order to minimise time lost for slewing between sources and calibrators. Apart from the science targets and the gain calibrator, bandpass and flux calibration observations were also carried out for each track. The detailed observing log is provided in Table \ref{table:obslog}, which also contains weather information for each track. 

The telescope receivers were tuned to a frequency range of 329--361 GHz in order for observations to be directly comparable to the ALMA PILS survey which has a frequency range of 329--363 GHz  \citep{joergensen-2016}. Thanks to the upgraded receivers and the SWARM correlator, this frequency range could be covered in a single setting by utilising both the 345- and 400-GHz receivers, where each receiver observed 8 GHz in the lower and the upper sidebands. The 345-GHz receivers covered the frequency range from 329.2--337.2 GHz range in the lower sideband, and 345.2--353.2 GHz in the upper sideband. The 400 GHz receivers covered the ranges of 337.2--345.2 GHz and 353.2--361.2 GHz in the lower and upper sidebands. Each 8-GHz band is divided into four 2-GHz chunks. The channel size is 140 kHz across the entire frequency range (0.12 km s$^{-1}$ at 345 GHz). In order to facilitate data reduction and to improve the noise level, data were rebinned by a factor of four prior to calibration, that is a channel size of 0.48 km s$^{-1}$ at 345 GHz.

\begin{table*}[h]
\caption{Observing log \citep[from][]{vanderWalt-2021}. \label{table:obslog}}
\centering          
\begin{tabular}{c c c c c c c}     
\hline
\hline 
Observing date	&	No. of antennas	&	Configuration\tablefootmark{a}	&	Bandpass	&	Flux	&	Gain & $\tau$(225 GHz)	\\ \hline
21/06/2017	&	7	&	COM	&	3c454.3	&	Titan, Neptune	&	mwc349a & 0.05--0.07	\\
22/06/2017	&	7	&	COM	&	3c273, 3c454.3	&	Titan, Neptune	&	mwc349a & 0.10	\\
27/06/2017	&	7	&	COM	&	3c454.3	&	Callisto	&	mwc349a & 0.08	\\
10/07/2017	&	7	&	COM	&	3c454.3	&	Callisto	&	mwc349a	& 0.05--0.06 \\
07/08/2017	&	6	&	COM	&	3c84	&	Titan, Uranus	&	mwc349a & 0.05	\\
20/10/2017	&	8	&	EXT	&	3c84	&	Uranus	&	mwc349a & 0.02--0.03	\\
22/10/2017	&	7	&	EXT	&	3c84	&	Uranus	&	mwc349a & 0.08--0.10	\\
08/11/2017	&	8	&	EXT	&	3c84	&	Uranus	&	mwc349a & 0.07	\\
09/11/2017	&	7	&	EXT	&	3c84	&	Uranus	&	mwc349a & 0.07	\\
10/11/2017	&	8	&	EXT	&	3c84	&	Uranus	&	mwc349a & 0.07	\\ \hline
\end{tabular}
\tablefoot{\tablefoottext{a}{COM is for the SMA compact configuration, while EXT is for the extended configuration.}
}
\end{table*}

\subsection{Calibration}

\begin{figure*}
    \centering
    \includegraphics[width=\textwidth]{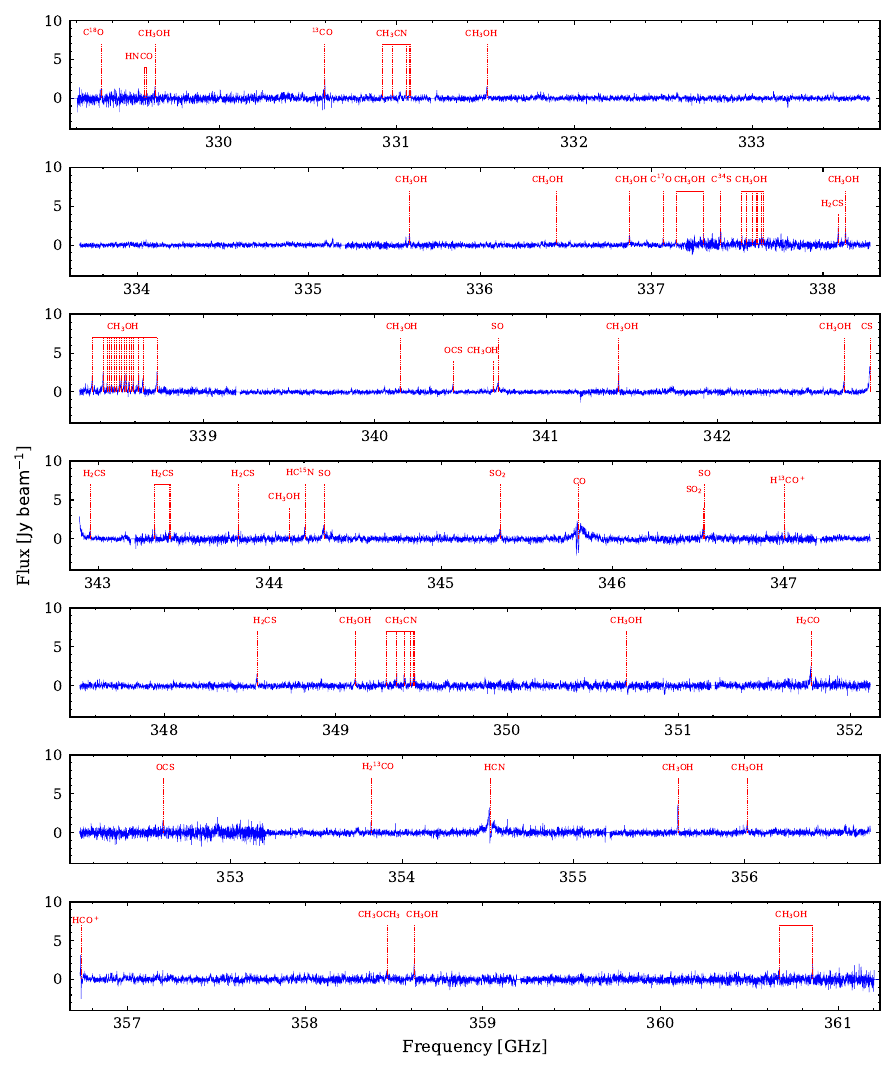}
    \caption{Spectrum towards N63, at the continuum peak. The brightest lines are marked and labelled.}
    \label{Fig:N63_spectrum}%
\end{figure*}

Data calibration was performed in CASA 4.7 \citep[Common Astronomy Software Applications;][]{McMullin-2007}. This consisted of first flagging the noisy edge channels in each 2-GHz chunk. Unfortunately, this flagging sometimes leaves small (< 0.1 GHz) gaps between spectral chunks as is evident for example around 331.2 GHz and 335.2 GHz (Fig. \ref{Fig:N63_spectrum}). The next step was flagging any channels with anomalous intensity spikes.

For the actual calibration, data from each receiver and each sideband were calibrated separately. The observations of the bandpass calibrator were first self-calibrated using a solution interval of 30 seconds to improve signal-to-noise and to correct for atmospheric effects over the 90 minutes observing time of this calibrator. The complex gains were calibrated against observations of mwc349a, which was observed for 2 minutes every 12 minutes. Finally, the flux was calibrated against observations of either Titan, Neptune, Callisto, or Uranus, depending on the time of observations (Table \ref{table:obslog}). After all calibrations were applied, the data for each source were concatenated into one measurement set prior to self calibration and imaging.

\subsection{Self calibration and imaging}

The sources observed here are all relatively strong continuum sources (peak intensities range from 0.25 -- 1.66 Jy beam$^{-1}$; Table \ref{table:Source_props2}). Thus, self calibration was attempted iteratively on each source in order to further lower the noise levels. This was achieved by first identifying the line-free channels by eye in order to isolate the continuum emission. Next, phase-only gain solutions were attempted first in solution intervals of 240 seconds, and from there going progressively down to intervals of 60 seconds. If a gain solution failed, the previous step was kept as the optimal solution. The self calibration typically improved the noise level by a factor of three, from $\sim$1.5$\times 10^{-2}$ to $\sim$5.0$\times 10^{-3}$ Jy beam$^{-1}$ for the continuum data. The self-calibration solution obtained from the continuum data was then applied to the entire cube. The continuum emission was subtracted from each chunk separately. 

The next step was to image the complex visibilities. This was done by defining a circular clean mask centred on each source, (or a central point between continuum cores), and including all continuum cores. The Briggs `robust' parameter was set to 2, corresponding to a natural weighting of the visibilities. This optimises for sensitivity at the cost of angular resolution. The resulting beam size was typically $\sim$$1\farcs5$ ($\sim$2000 AU at a source distance of 1.3 kpc) and rms noise levels of around 0.3 Jy beam$^{-1}$ in each 0.48 km s$^{-1}$ channel \citep[see][for further details]{vanderWalt-2021}. The resulting noise levels and beams are reported in Table \ref{table:Source_props2}. 

The maximum recoverable scale (MRS) of the observations is calculated using equation $MRS = 1.22 \times \lambda / D$, where $\lambda$ is the wavelength of the observations and $D$ the shortest distance between two antennae (shortest baseline) of the interferometer. With $\lambda=870\mu$m (345 GHz) and $D=16$m, an MRS of $\sim$14$''$, or $\sim$20000 AU at a distance of 1.3 kpc is obtained. This scale almost corresponds to the sizes of the envelopes \citep{Pitts-2022}. Similarly, the field of view (FoV) is $\sim$36$''\sim$50000 AU. There will therefore be some resolving-out, but this has not prevented a quantitative analysis before \citep[e.g. ][]{Joergensen-2007,Lee-2015,Stephens-2018}.

\section{Results}
\label{Sec:Results}

\subsection{Continuum emission}
\label{Sec:cont_emission}
\begin{figure*}
    \centering
    \includegraphics[width=16.2cm]{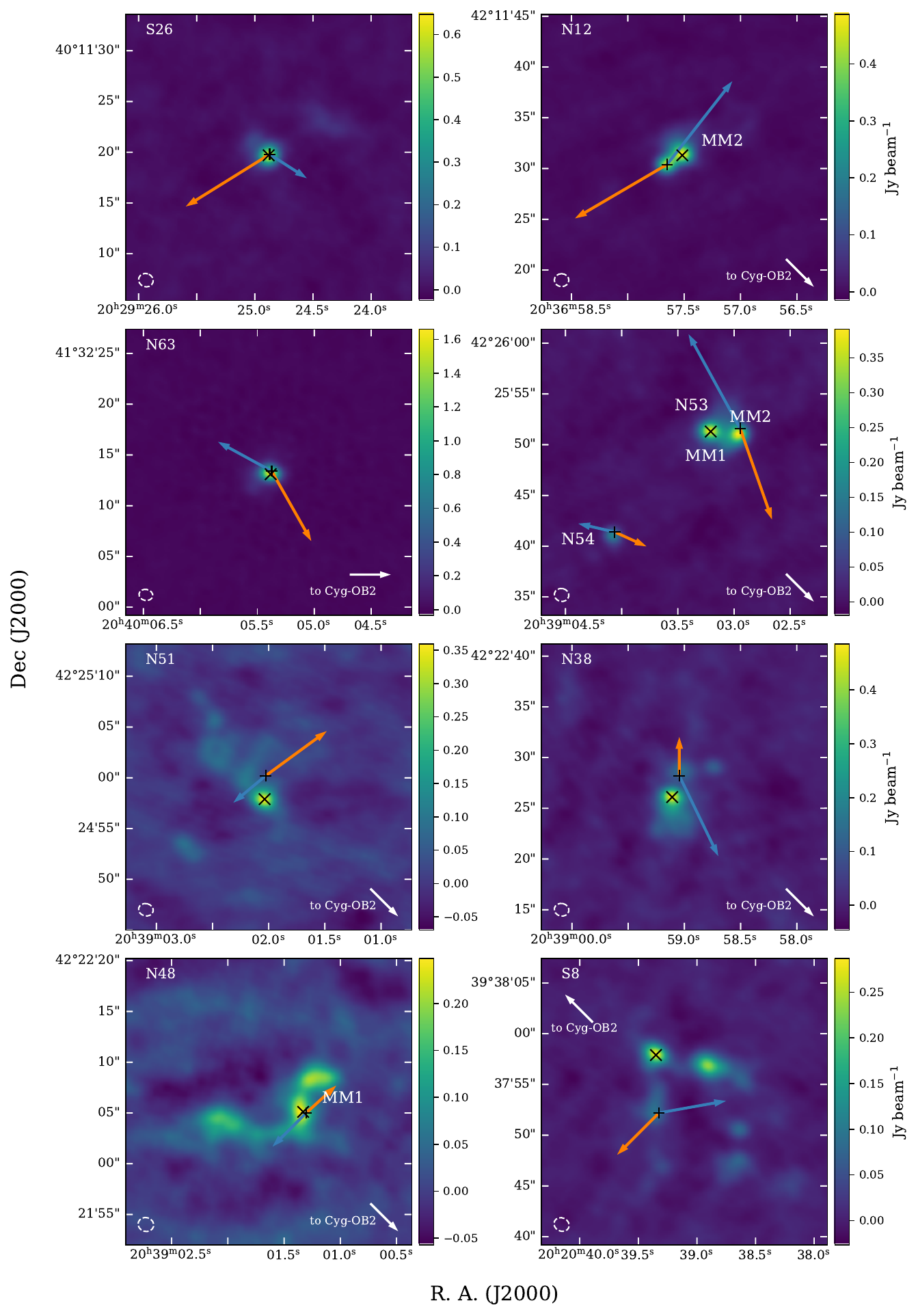}
    \caption{Continuum images for the PILS-Cygnus source sample. We note that N53 and N54 are shown in the same image. The crosses mark the brightest core and the position at which the spectrum for each source was extracted, while the plusses mark the central position between the red- and blue-shifted CO emission lobes.}
    \label{Fig:Sample_continuum}%
\end{figure*}

The PILS-Cygnus source sample is listed in Table \ref{table:Source_props1}. N30 was analysed in detail in \cite{vanderWalt-2021}, with 345 GHz ($\sim$870 $\mathrm{\mu m}$) continuum emission maps for the remaining nine sources shown in Fig. \ref{Fig:Sample_continuum}; these maps were first presented in \citet{Pitts-2022}. We note that N54 is shown together with N53, as they fall within the same field of view. 

Five of the ten sources are located in the dense DR21 ridge N53, N54, N51, N38 and N48 ($d$ = 1.5 kpc; see also Fig. \ref{Fig:Detections_number}), where several filaments intersect \citep[see e.g.][]{Motte-2007,Reipurth-2008, Pitts-2022}. N30, N12 and N63 are located to the north, west and south of DR21, respectively ($d$ = 1.3--1.4 kpc). The two remaining sources, S26 and S8, are located in the southern Cygnus-X molecular cloud, with S26 likely not being directly associated with Cygnus-X ($d$=3.3 kpc for S26, and 1.4 kpc for S8).

The first panel in Fig. \ref{Fig:Sample_continuum} shows S26, which is a single source at the $\sim$1\farcs5 resolution of the PILS-Cygnus observations. There is some extended emission to the north-east of the emission peak, as well as some extended emission $\sim$5$''$ to the north-west, independent of the emission peak. The two other northern sources, N12 and N63 are shown in the second panel and first panel of the second row, respectively. N12 shows two cores at $\sim$$1\farcs5$ resolution, with some extended emission to the northwest, while N63 is singular, but with extended emission to the south-east.

Of the DR21 sources, N53 has two components at the resolution of the PILS-Cygnus observations, with spiralling extended emission connecting the two cores. N54 is shown in the same panel as N53, with a separation of $\sim$10$''$ 
between the two sources. N51 is shown in the first panel of the third row, with N38 in the second panel. Both sources show large structure in extended emission, to the northwest of N51, and south of N38, with two faint cores north and northwest of N38. The first panel of the bottom row shows N48, with this source consisting of three elongated and connected cores, with large structure in extended emission surrounding these cores. The second panel of the bottom row shows S8, which consists of two cores, with four fainter cores to the south.

The molecular outflows may impact the observed emission, either through excitation (some molecules will more readily be excited in the dense and warm gas in outflows) or chemistry (sputtering and gas-phase reactions may lead to abundances of some molecules increasing dramatically). The sources all show outflow activity as traced by CO 3--2 emission \citep{Skretas2022}. The directions and extent of the outflows are shown in the continuum maps with red and blue arrows, for easy comparison with the observed molecular emission described below. Furthermore, a number of sources show SiO 8--7 emission, another shock and outflow tracer. The SiO emission tends to be more compact, and less extended than the CO emission \citep[$\lesssim$ a few arcsec,][]{Skretas2022}. 

Furthermore, UV radiation from the nearby OB associations may affect the chemistry observed towards these sources. For this reason, the direction of the nearest OB associations (in the plane of the sky) are also marked on the continuum maps. For the northern sources, the OB association the closest by is OB2. The typical projected distance to the sources is $\sim$ 45 pc, for an average distance of 1.4 kpc. 

Finally, the dust may act to shield the molecular emission from the protostars, if the dust is optically thick. To evaluate if the dust itself is optically thick to molecular emission, we follow the recipe outlined in, for example \citet{vanderWalt-2021}, whereby we assume a dust temperature of 30 K and a gas-to-dust mass ratio of 100. We calculate the total mass of gas and dust using the following standard relation: 
\begin{equation}
    M = \frac{S_\nu d^2}{\kappa_\mathrm{\nu}B_\nu(T)},
\end{equation} 
where $S_\nu$ is the peak intensity of the source, $d$ the distance, and $B_\nu(T)$ the Planck function at specific frequency and temperature. We used a dust opacity $\kappa_\mathrm{\nu} = 0.0175 $ cm$^2$ per gram of gas for a gas-to-dust ratio of 100 \citep[at frequency $\nu = 345$ GHz;][]{Ossenkopf-1994}. We calculate and list the resulting mass of gas and dust for each source in Table \ref{table:Source_props2} for $T$ = 30 K. An average dust temperature may be low, particularly for sources such as N30 and S26 where the 100-K radius is similar to the beam (Table \ref{table:Source_props2}). If a larger temperature is used, the mass decreases. For the specific example of a temperature of 100 K, the total mass would be lower by a factor of 4.

The optical depth can then be calculated as \citep[e.g.][]{Schoier-2002}: 
\begin{equation}
    \tau_\mathrm{\nu} = \kappa _\mathrm{\nu} \mu m_{\rm H} N_\mathrm{H_2} \ ,
\end{equation} 
where $\mu = 2.8$ is the mean molecular weight used, which also accounts for Helium \citep{kauffmann08}, $m_\mathrm{H}$ the mass of the Hydrogen atom, and $N_\mathrm{H_2}$ the H$_2$ column density \citep[see also][]{vanderWalt-2021}. The dust opacities (Table \ref{table:Source_props2}) are found to be $<$ 0.5 for all but two sources: S26 and N63. For these sources, the dust opacities are 1.02 and 0.58, respectively. Specifically for S26, the source breaks into three continuum sources when observed at higher angular resolution, but the molecular emission appears to be extended beyond these continuum peaks \citep{Suri-2021}. This suggests that although the continuum emission is moderately optically thick, it does not shield the bulk of the molecular emission. For the case of N63, the source appears to be very compact and most of the mass is contained within a radius of $\sim$ 2500 AU \citep{duarte-cabral2013}. Thus, it may very well be that if the source was observed at higher angular resolution, the dust opacity would increase. However, we do not have the data to validate this hypothesis, and we conclude that the dust may be slightly optically thick towards N63. Moreover, if the dust temperature is larger than 30 K, and the masses thus are lower, the dust opacities will also be lower and should therefore be considered as upper limits here. For the other sources, dust opacity will not play a role on the scales observed here. 

%
\begin{table*}
\caption{PILS-Cygnus continuum source properties.}             
\label{table:Source_props2}      
\centering          
\begin{tabular}{l c c c c c c r}     
\hline
\hline       
    Source & beam size & $S_{850 \mathrm{\ \mu m}}$ & rms & $M_\mathrm{beam}$ & Dust opacity ($\tau$) & $L_\mathrm{bol}$\tablefootmark{a} &  $R_\mathrm{T=100 K}$ \tablefootmark{a} \\ 
     & \arcsec$\times$\arcsec & [$\mathrm{Jy \ beam^{-1}}$] & [$\mathrm{Jy \ beam^{-1}}$] & [$M_\odot$] & & [$L_\odot$] & [AU] \\ 
\hline  
    N30-MM1 &  1.45$\times$1.31 & 1.29 & 0.014 & 7.25 & 0.31 & 2.5$^{+0.1}_{-0.3}\times 10^4$ & 2500 \\
\hline  
    S26 & 1.46$\times$1.31 &  0.60  & 0.004 & 23.98 &  1.02& 2.1$^{+0.06}_{-0.4}\times 10^5$ & 7000 \\ 
\hline  
    N12-MM2 & 1.43$\times$1.27 & 0.50 & 0.003 & 3.20 & 0.14  & 740$^{+80}_{-90}$ & 400 \\ 
\hline  
    N63 & 1.37$\times$1.10 & 1.66 & 0.005 & 10.83 & 0.58 & 380$^{+30}_{-40}$ & 300 \\
\hline  
    N53-MM2 & 1.44$\times$1.28 &  0.41 & 0.004 & 2.92 & 0.13 & 400 $\pm$ 200& 300 \\  
\hline  
    N54 & 1.42$\times$1.24 & 0.27 & 0.010 & 2.02 & 0.09 & 160 & 200 \\  
\hline  
    N51   & 1.47$\times$1.22 & 0.42 & 0.009 & 2.70 & 0.12 & 1300  $\pm$ 200 & 600 \\  
\hline  
    N38 & 1.45$\times$1.25 & 0.62 & 0.009 & 3.59 & 0.16 & 300 & 300  \\  
\hline  
    N48-MM1 & 1.56$\times$1.37 &  0.25 & 0.004 & 1.87 & 0.07 & 2000 $\pm$ 700 & 700 \\  
\hline  
    S8-NE core & 1.54$\times$1.31 &  0.30 & 0.005 & 1.89 &  0.08 & 6400$^{+200}_{-1000}$ & 1200 \\  
\hline                  
\end{tabular}
\tablefoot{\tablefoottext{a}{We use $L_\mathrm{bol}$ values from \citet{Pitts-2022}, with the values for N38 and N54 from \cite{SanJose-Garcia-2015PhDT} and \cite{Kryukova-2014A} respectively. The 100 K is the radius at which the temperature rises above 100 K  \citep[$R_\mathrm{T=100 K} \sim 2.3 \times 10^{14}(\sqrt{(L/L_\odot)}$ cm;][]{Bisschop-2007}. With a beam size of $\sim1\farcs5$, or $\sim$5000 AU for S26, and $\sim$2000 AU for all other sources, it implies that the 100 K radius is resolved for only N30 and S26.}
}
\end{table*}

\subsection{Molecular line emission}
\begin{figure}
    \centering
    \includegraphics[width=9.0cm]{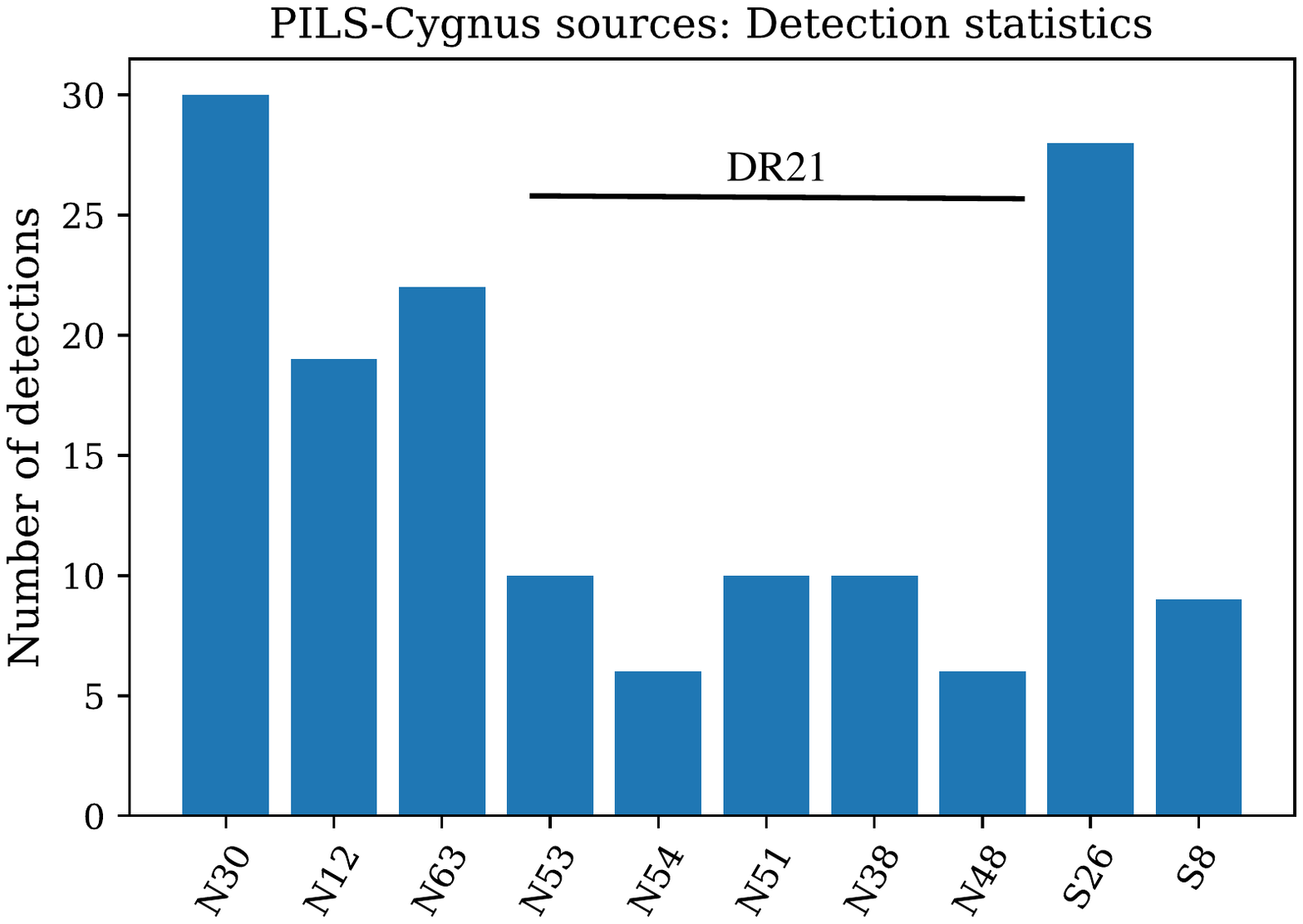}
    \caption{Number of molecules and their isotopologues detected in each source, grouped by region and with the DR21 sources labelled.}
    \label{Fig:Detections_number}%
\end{figure}

The molecular detection statistics for the source sample are shown in Table \ref{table:Detection_stats}, with the number of molecules and their isotopologues detected towards each source shown in Fig. \ref{Fig:Detections_number}. N30 and S26 have 30 and 28 molecular detections, respectively. Towards N63, 22 molecular species are detected, with 19 detected towards N12. For the DR21 sources ten  molecular species are detected towards N53, N51 and N38, and six are detected towards both N54 and N48. Nine molecular species are detected towards S8.

\begin{table*}
\caption{Detection statistics for the PILS-Cygnus sources\tablefootmark{a}.}
\label{table:Detection_stats}      
\centering          
\begin{tabular}{l c c c c c c c c c r}     
\hline
\hline 
	 &  N30  & N12  &   S26  &  N63  &  N53  & N54   &  N51	 &  N38  &  N48 &  S8  \\
\hline
   \multicolumn{11}{c}{COMs and O-bearing species} \\
\hline  
    CH$_3$OH         & oc &  se & sc & se & se & x & oe & oe & oe & oe \\
    $^{13}$CH$_3$OH  & oc &  sc & sc & se & x  & x  &  x & x  & x  & x  \\
    CH$_3$$^{18}$OH  & x  &  x  & x  & x  & x  & x  &  x & x  & x  & x  \\
    CH$_2$DOH        & x  &  x  & x  & se & x  & x  &  x & x  & x  & x  \\
    C$_2$H$_5$OH     & sc &  x  & x  & x  & x  & x  &  x & x  & x  & x  \\
    CH$_3$OCHO       & sc &  x  & sc & se & x  & x  &  x & x  & x  & x  \\
    CH$_3$OCH$_3$    & sc &  x  & sc & se & x  & x  &  x & x  & x  & x  \\
    H$_2$CO          & oe &  oe & sc & oe & se & se & oe & oe & oe & oe \\
    H$_{2}$$^{13}$CO & sc &  x  & x  & x  & x  & x  & x  & x  & x  & x  \\
    t-HCOOH          & se &  x  & sc\tablefootmark{b} & x & x & x  & x & x & x & x \\
    c-HCOOH          & x &   x  & x  & x  & x  & x  & x & x & x & x\\
    HDO    	         & oe & oe\tablefootmark{b}& sc\tablefootmark{b} & se & x & x & x & x & x & x \\
    CO               & oe &  oe & oe & oe & oe & oe & oe & oe & oe & oe \\
    $^{13}$CO        & oe &  oe & se & oe & oe & oe & oe & oe & x  & oe \\
    C$^{18}$O        & oe &  oe & se & oe & oe & x  & oe & oe & x  & oe \\
    C$^{17}$O        & oe &  x  & se & oe & x  & x  & oe & x  & x  & x  \\
    HCO$^+$          & oe &  oe & oe & oe & oe & oe & oe & oe & oe & oe \\
    SiO              & oc &  x  & sc & oe & x  & x  & x  & x  & x  & x  \\
\hline 
\multicolumn{11}{c}{S-bearing species} \\
\hline 
    SO               & sc &  sc & sc & se & se & x  & se & oe & oe & oe \\
    $^{33}$SO        & sc &  x  & sc & x  & x  & x  & x  & x  & x  & x  \\
    $^{34}$SO        & sc &  sc & sc & se & x  & x  & x  & x  & x  & x  \\
    SO$_{2}$         & sc &  sc & sc & x  & x  & x  & x  & x  & x  & x \\
    $^{33}$SO$_{2}$  & se &  x  & sc & x  & x  & x  & x  & x  & x  & x \\
    $^{34}$SO$_{2}$  & se &  x  & sc & x  & x  & x  & x  & x  & x  & x \\
    CS               & oe &  se & sc & oe & se & se & se & oe & x  & oe \\
    $^{34}$CS        & oe &  se & sc & se & se & x  & x  & oe & x  & x \\
    OCS              & oc &  se & sc & se & se & x  & x  & x  & x  & x \\
    OC$^{34}$S       & se &  se & sc & se\tablefootmark{b} & x & x & x & x & x & x \\
    H$_2$CS          & se &  se & sc & se & x  & x  & x  & oe & x  & x \\

\hline 
\multicolumn{11}{c}{N-bearing species} \\
\hline 
    CN               & oe &  oe & x  & oe & x  & x  & x  & x  & x  & x \\
    CH$_3$CN         & oc &  se & sc & se\tablefootmark{b} & x & x & x & x & x & x \\
    HNCO             & sc &  x & sc  & x  & x  & x  & x  & x  & x  & x \\
    HC$_3$N          & oc &  x & sc  & x  & x  & x  & x  & x  & x  & x \\
    HCN              & oe &  oe & se & oe & se & oe & oe & oe & oe & oe \\
    HC$^{15}$N       & oc &  oe & sc & oe & x  & x  & x  & x  & x  & x \\
\hline
\end{tabular}
\tablefoot{\tablefoottext{a}{Classification of emission is `oc', `sc', `oe' and `se', where `s' and `o' is if the emission peaks on- or off source, respectively, while `c' and `e' is whether the emission is compact or extended. `x' is for non-detections.}
\tablefoottext{b}{S/N too low to make an emission map. The emission is assumed to be compact and on source (`sc') as for most other molecules detected towards S26, and `se' for N63. For HDO towards N12 we assume emission to be off source and extended, similar to HDO emission towards N30. For N63 the S/N is too low for OC$^{34}$S and CH$_3$CN to be sure if the emission is extended or not. It is therefore assumed to be as extended as the other molecules in this source.}}
\end{table*}

Line identification was performed following the process described in \cite{vanderWalt-2021}, which follows the points laid out by \cite{Snyder-2005} and using the spectral modelling package CASSIS\footnote{Centre d’Analyse Scientifique de Spectres Instrumentaux et Synth$\Acute{\text{e}}$tiques; http://cassis.irap.omp.eu} \citep[][]{Vastel_CASSIS-2015} to construct a synthetic spectrum covering the frequency range of the PILS-Cygnus observations (329--361 GHz) using the molecular spectroscopy data from the CDMS\footnote{ Cologne Database for Molecular Spectroscopy; https://cdms.astro.uni-koeln.de/} \citep[][]{Muller-2001-CDMS,Muller-2005-CDMS,Endres-2016-CDMS} and JPL\footnote{NASA Jet Propulsion Laboratory; http://spec.jpl.nasa.gov/} \cite[][]{Pickett-1998-JPL} databases. The synthetic spectra were constructed by first adding previously identified molecules and comparing with the observed spectra to find and identify all lines with intensity above 3$\sigma$, where $\sigma$ is the noise level of the observed spectrum. 

\begin{figure*}
    \centering
    \includegraphics[width=\textwidth]{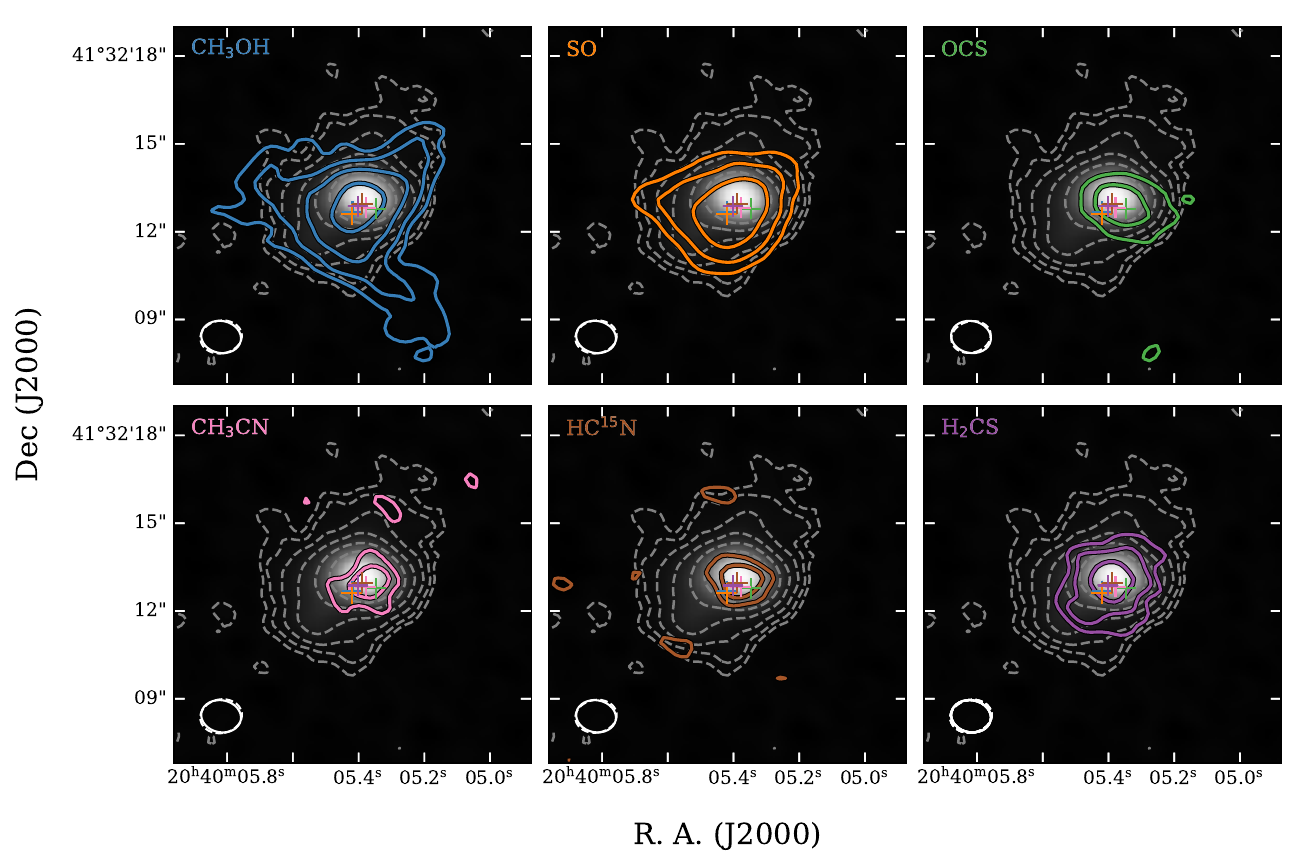}
    \caption{Molecules observed towards N63. The contour levels for both the continuum (grey dashed lines) and molecular emission (coloured) are 3, 6, 12, 24 and 48$\sigma$. The pluses are the molecular emission peak positions derived from 2D Gaussian fits (Table \ref{table:N63_Molecules_all}), and these are colour-coded according to molecule.}
    \label{Fig:N63_molecules}%
\end{figure*}

Spectra and molecular maps are shown for the source N63 in Figs. \ref{Fig:N63_spectrum} and \ref{Fig:N63_molecules}, respectively, with the molecular transitions used to produce the molecular maps listed in Table \ref{table:N63_Molecules_all}. Where more than one transition for a given molecule were detected, the lines were stacked following the methodology outlined in \cite{vanderWalt-2021}. Figures and tables for the source sample are presented in Appendix A. 
Observed line plots of the CASSIS synthetic spectral lines over-plotted on the observed lines for each source are shown in Appendix 
C, with the full line lists of observed lines above 3 $\sigma$ for each source in Appendix D.


\begin{table*}
\caption{Molecular transitions observed towards N63 used to produce the contour maps shown in Fig. \ref{Fig:N63_molecules}.}
\label{table:N63_Molecules_all}      
\centering          
\begin{tabular}{l c c c c c c r}     
\hline
\hline 
    Molecule & Frequency & Transition & $E_\mathrm{up}$ & $I_{\mathrm{peak}}$ & $I_{\mathrm{int}}$ & FWHM  & $\varv_{\text{source}}$ \\ 
     &  [GHz] & & [K] &  [$\mathrm{Jy \ beam^{-1}}$] &  [$\mathrm{Jy \ beam^{-1} \ km \ s^{-1}}$] & [$\mathrm{km \ s^{-1}}$] & [$\mathrm{km \ s^{-1}}$]\\ 
\hline
    A-CH$_3$OH & 331.5023 & $11_{1, -0} - 11_{0, +0}$ & 169.01  &    1.2 &   7.9 &    6.9 &   $-$6.0 \\
     & 335.5820 & $7_{1, +0} - 6_{1, +0}$ & 78.97  &   1.5 &        7.1 &        4.3 &       $-$4.8 \\
     & 336.8651 & $12_{1, -0} - 12_{0, +0}$ & 197.07 &    1.0 &        7.4 &        7.7 &       $-$7.2 \\
     & 338.4087 & $7_{0, +0} - 6_{0, +0}$ & 64.98 &   2.0 &       12.7 &        7.5 &       $-$4.9 \\
     & 341.4156 & $7_{1, -0} - 6_{1, -0}$ & 80.09 &    1.8 &       10.6 &        5.9 &       $-$7.2 \\
    & 349.1070 & $14_{1, -0} - 11_{0, +0}$ & 260.20 &       0.8 &        6.1 &        6.9 &       $-$4.1 \\
    & 355.6029 & $13_{0, +0} - 11_{1, +0}$ & 211.03 &       4.2 &       12.1 &        2.9 &       $-$5.8 \\
     & 356.0072 & $15_{1, -0} - 15_{0, +0}$ & 295.26 &       0.9 &        6.6 &        7.5 &       $-$7.2 \\
    E-CH$_3$OH & 338.1244 & $7_{0, 0} - 6_{0, 0}$ & 78.18 &     1.8 &        9.8 &        5.4 &       $-$5.0 \\
     & 338.3446 & $7_{-1, 0} - 6_{-1, 0}$ & 62.65 &        1.5 &       11.4 &        8.9 &      $-$4.8 \\
     & 338.7217  & $7_{2, 0} - 6_{2, 0}$ & 79.36 &       2.1 &       13.4 &        7.0 &       $-$7.3 \\
     & 360.8489 & $11_{0, 0} - 10_{1, 0}$ & 158.15 &     0.9 &        5.8 &        6.9 &      $-$7.5 \\

\hline 
     SO & 340.7141 & $8_{7} - 6_{ 6} $ &  81.25 &    1.1 &       11.4 &       12.8 &       $-$3.0 \\
    & 334.4311 & $8_{8} - 7_{7} $ &   87.48  & 1.0 &       10.8 &       17.4 &       $-$1.7 \\
\hline
   OCS & 340.4493 & $ 28 - 27 $ &  236.95 & 0.9 &        4.2 &        4.5 &       $-$4.8 \\ 
\hline
   CH$_3$CN & 330.9126 &  $18_{5} - 17_{5}$   &   329.46 &      0.6 &        3.9 &        5.0 &       $-$5.3 \\ 
    & 330.9698 &  $18_{4} - 17_{4}$   & 265.22 &  0.8 &        4.1 &        4.2 &       $-$6.0 \\ 
    & 331.0143 & $18_{3} - 17_{3}$    &   215.24 &     0.8 &        1.7 &        3.8 &       $-$1.8 \\ 
\hline    
    HC$^{15}$N & 344.2001 & $4 - 3$ &  41.30 & 0.9 &        6.6 &        6.8 &       $-$2.5 \\ 
\hline 
   H$_2$CS & 338.0832 & $ 10_{ 1, 10} - 9_{ 1, 9} $  & 102.43 & 1.7 &        8.4 &        4.9 &       $-$6.1 \\
    & 342.9464 & $ 10_{ 0, 10} -  9_{ 0, 9} $ &  90.59 & 1.1 &   6.6 &        5.8 &       $-$3.2 \\
\hline 
\end{tabular}
\end{table*}



\subsection{Emission morphology}

The emission of strong lines of each detected species was mapped and is shown in Fig.s A.9 -- A.16.
We generally identify four different types of emission morphology: (i) emission that appears unresolved and peaks on the continuum source; (ii) emission that appears unresolved but offset from the continuum source; (iii) resolved (extended) emission that peaks on the continuum source; and finally (iv) resolved emission that peaks offset from the source. Emission from the detected species is classified according to this scheme and based on a by-eye inspection of each map, with the classification shown in Table \ref{table:Detection_stats}. 

It is difficult to see a pattern in the emission morphology. The only constant is that the emission from the most simple carbon-bearing molecules (CO, HCO$^+$, HCN, and CS and their isotopologues) appears to be extended. Emission from COMs, on the other hand, appears to be both compact and extended, when detected, as for example seen towards N63 (Fig. \ref{Fig:N63_molecules}). 

\citet{vanderWalt-2021} noted that the molecular emission towards N30 was offset from the continuum source. Particularly, they found that this offset was independent of excitation, that is it was independent of the upper-level energy of the observed transition and the critical density of the transition. The same is seen for the other sources observed here: for the molecules where several bright transitions are detected, these all peak in the same place independent of excitation. This suggests that any difference in spatial extent of emission is caused by chemistry and not excitation.


\subsection{Column densities}

Column densities and excitation temperatures are derived for the observed molecules towards each source. Following the procedure described in \cite{vanderWalt-2021}, we constructed a synthetic spectrum with CASSIS, using only line emission and assuming local thermodynamic equilibrium (LTE). The best-fit spectral model for each molecule was found by computing the reduced $\chi^2$ minimum using the inbuilt regular grid function in CASSIS and covering a large parameter space. The free parameters used in the fit were source velocity ($\varv_{\text{source}} $), the full width half maximum (FWHM), the column density ($N$), and  excitation temperature ($T_\mathrm{ex}$). The FWHM and $\varv_{\text{source}}$ were estimated by fitting a Gaussian profile to a few unblended lines, which were then used in a parameter space of 4 km s$^{-1}$ around the line, in steps of 0.5 km s$^{-1}$ (approximately the channel width). Initially, a large parameter space was used for $N$ with $N_\mathrm{min}=10^{14}$ cm$^{-2}$ and $N_\mathrm{max}=10^{19}$ cm$^{-2}$ and ten logarithmically spaced steps in this range, while for $T_\mathrm{ex}$ a minimum of 90 K and a maximum of 300 K was used, also with ten steps in between these values. The parameter ranges were then reduced by running a few iterations, while keeping the steps larger than the uncertainties for each parameter. Throughout the modelling process the source sizes were assumed to be $\sim$1\farcs5, comparable to the beam size of the PILS-Cygnus observations. The derived values for N63 are shown in Table \ref{table:N63_column_densities}, with the source sample column densities presented in Appendix B
, together with the abundances relative to CH$_3$OH shown in Fig. B.1.

The uncertainties on $N$ and $T_\mathrm{ex}$ were estimated as follows: first  $N$ (or $T_\mathrm{ex}$) is fixed and then  $T_\mathrm{ex}$ (or $N$) is adjusted until the result of the change on the fit can be seen \citep[method described by][]{calcutt-2018b}. This is shown in parenthesis after the values for $N$ and $T_\mathrm{ex}$ in Tables B.1 -- B.8.

In cases where molecular lines are optically thick, isotopologues of the molecule were used and multiplied by the local ISM ratios: $^{12}$C/$^{13}$C $\sim68$ \citep[][]{Milam-2005}, $^{32}$S/$^{34}$S $\sim22$ and $^{16}$O/$^{18}$O $\sim560$ \citep[][]{Wilson&Rood-1994}. For CH$_3$CN the higher vibrational transition $v_8=1$ was used, since no isotopologues of CH$_3$CN were detected. Finally, for sources where no isotopologues of CH$_3$OH or SO were detected we give lower limits for these molecules, derived by fixing $T_\mathrm{ex}$ to typical values and fitting the synthetic spectrum to the observed lines. The lower limit reflects that the emission from these molecules is likely optically thick.

\begin{figure}
\centering
\includegraphics[width=9cm]{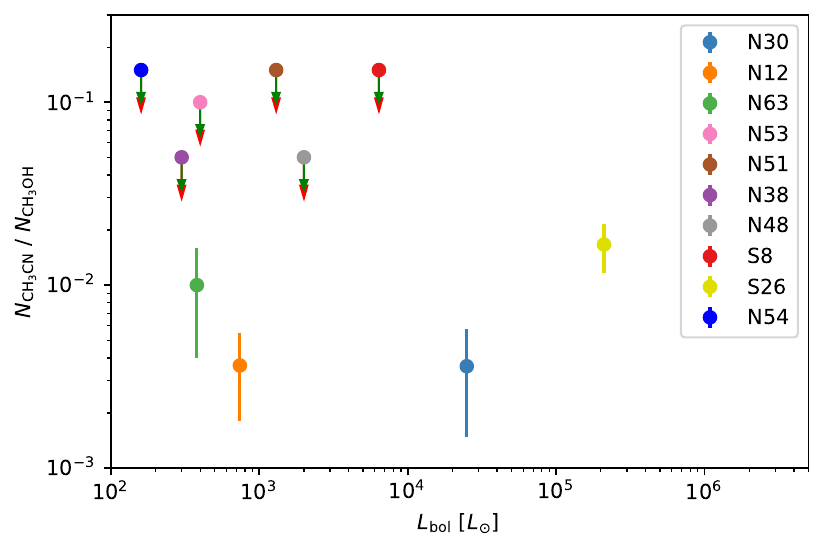}
  \caption{$N_\mathrm{CH_3CN}$/$N_\mathrm{CH_3OH}$ vs bolometric luminosity. Red arrows represent lower limits for CH$_3$OH (where no isotopologues are detected), and green arrows represent upper limits for CH$_3$CN (not detected).}
\label{Fig:CH_3CN-CH_3OH_vs_Lbol}
\end{figure}

\begin{figure}
\centering
\includegraphics[width=9cm]{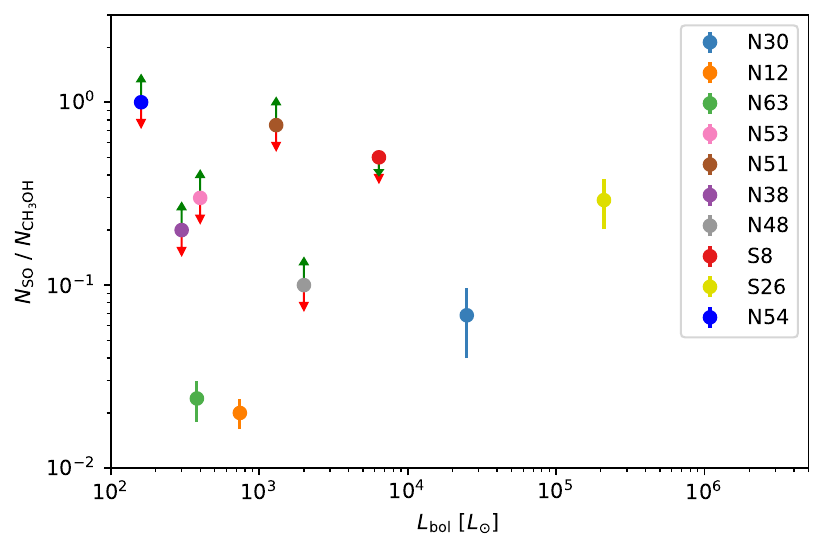}
  \caption{$N_\mathrm{SO}$/$N_\mathrm{CH_3OH}$ vs bolometric luminosity. The arrows are the same as in Fig. \ref{Fig:CH_3CN-CH_3OH_vs_Lbol}, but with the green arrows here representing upper and lower limits for SO.}
\label{Fig:SO-CH3OH_vs_Lbol}
\end{figure}

\begin{figure}
\centering
\includegraphics[width=9cm]{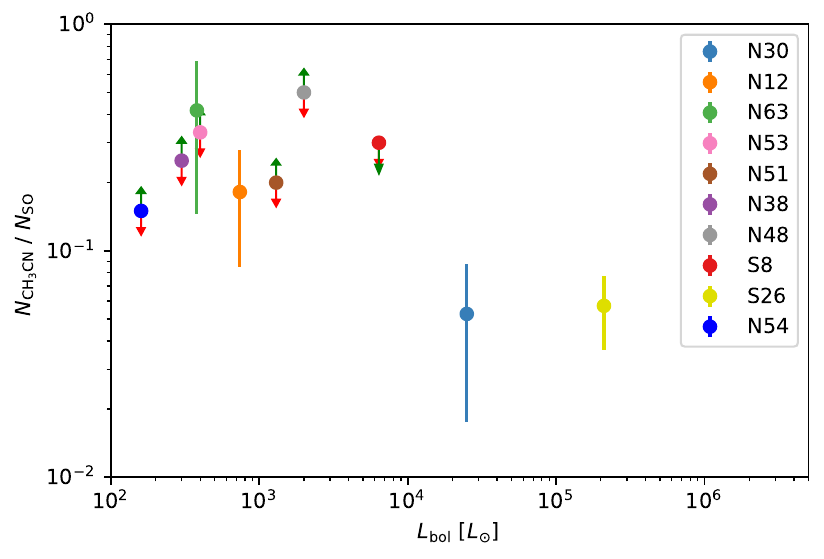}
  \caption{$N_\mathrm{CH_3CN}$/$N_\mathrm{SO}$ vs bolometric luminosity. The green arrows here represent upper and lower limits for CH$_3$CN, while red arrows represent lower limits for SO.}
\label{Fig:CH3CN-SO_vs_Lbol}
\end{figure}

The resulting column densities are represented as relative abundances in Figs. \ref{Fig:CH_3CN-CH_3OH_vs_Lbol}, \ref{Fig:SO-CH3OH_vs_Lbol} and \ref{Fig:CH3CN-SO_vs_Lbol} for three representative species. These species were chosen as a shock/outflow tracer (SO), a hot-core tracer (CH$_3$CN), and the most commonly observed COM (CH$_3$OH) \citep[e.g.][]{Tychoniec-2021}. From Fig. \ref{Fig:CH_3CN-CH_3OH_vs_Lbol} we see that the relative abundances of CH$_3$CN-to-CH$_3$OH do not exhibit a noticeable trend, whereas relative abundances of SO-to-CH$_3$OH (Fig. \ref{Fig:SO-CH3OH_vs_Lbol}) seem to generally increase with luminosity, while CH$_3$CN-to-SO (Fig. \ref{Fig:CH3CN-SO_vs_Lbol}) abundances seem to decrease with luminosity. However, with only four data-points, this result needs to be confirmed, which ideally would include more observations of sources from the same cloud.

%

\begin{table*}
\caption{Column densities, excitation temperatures, FWHM and $\varv_{\text{peak}}$ derived at the continuum peak position of N63\tablefootmark{a}. }             
\label{table:N63_column_densities}      
\centering          
\begin{tabular}{l c c c c }    
\hline  \hline 
    Molecule & Column density & $T_\mathrm{ex}$& FWHM & $\varv_{\text{peak}}$   \\ 
     & [$\mathrm{cm^{-2} }$]& [K] & [$\mathrm{km \ s^{-1} }$] & [$\mathrm{km \ s^{-1} }$]  \\
\hline
   \multicolumn{5}{c}{COMs and O-bearing species} \\
\hline   
    CH$_3$OH\tablefootmark{b}    & 5.0 (0.2) $\times 10^{17}$ &  \hspace{0.6cm} 120 (50) \hspace{0.6cm} & \hspace{1cm} 4.0 \hspace{1cm}  & \hspace{1cm} $-$6.0 \hspace{1cm}  \\
    $^{13}$CH$_3$OH\tablefootmark{c}   & 5.8 (0.3) $\times 10^{15}$ & 120 (50) & 4.0 & $-$6.0 \\
    CH$_3$$^{18}$OH & <5.0 $\times 10^{15}$ & 120 & 4.0 & $-$6.0   \\
    CH$_2$DOH  & 1.5 (0.4) $\times 10^{16}$ & 150 (40) & 4.0 & $-$5.5  \\
    C$_2$H$_5$OH     & <1.0 $\times 10^{16}$ & 120 & 4.0 & $-$6.0     \\ 
    CH$_3$OCHO        & 1.2 (0.4) $\times 10^{16}$ & 120 (40) & 4.0 & $-$5.5   \\ 
    CH$_3$OCH$_3$     & 1.4 (0.4) $\times 10^{16}$ & 120 (40) & 4.0 & $-$5.0  \\ 
    t-HCOOH            & <3.0  $\times 10^{15}$ & 120 & 4.0 & $-$6.0 \\
    c-HCOOH            & <6.0 $\times 10^{14}$ & 120 & 4.0 & $-$6.0 \\
    HDO\tablefootmark{d}         & 1.5 (1.0) $\times 10^{16}$ & 120 & 4.0 & $-$4.5   \\
\hline 
\multicolumn{5}{c}{S-bearing species} \\
\hline 
    SO\tablefootmark{b}   & 1.2 (0.3) $\times 10^{16}$ & 250 (50) & 4.5 & $-$5.5  \\
    $^{33}$SO\tablefootmark{c}   & <2.0  $\times 10^{14}$ & 250 & 4.5 & $-$5.5 \\
    $^{34}$SO   & 5.5 (1.0) $\times 10^{14}$ & 250  (50) & 4.5 & $-$5.5  \\
    SO$_{2}$    & <4.0  $\times 10^{15}$ &  250 & 4.5 & $-$5.5  \\
    $^{33}$SO$_{2}$   & <3.0 $\times 10^{15}$ & 250 & 4.5 & $-$5.5 \\
    $^{34}$SO$_{2}$   & <3.0  $\times 10^{15}$ & 250 & 4.5 & $-$5.5 \\ 
    OCS\tablefootmark{b}  & 3.3 (1.1) $\times 10^{16}$ & 100 (30) & 3.0 & $-$5.0  \\
    OC$^{34}$S   & 1.5 (0.5) $\times 10^{15}$ & 100 (30) & 3.0 & $-$5.0  \\
    H$_2$CS       &  3.0 (1.0) $\times 10^{15}$ & 150 (40)  & 5.0 & $-$5.0 \\
\hline 
\multicolumn{5}{c}{N-bearing species} \\
\hline 
    CH$_3$CN, v$_8$=1  &  5.0 (3.0)$ \times 10^{15}$ & 120 (20) & 5.0 & $-$5.0  \\
    $^{13}$CH$_3$CN\tablefootmark{d} & <1.0 $\times 10^{14}$ & 120 & 5.0 & $-$5.0  \\
    HNCO    &  <2.0 $\times 10^{15}$ &   120 & 5.0 & $-$5.0  \\
    HC$_3$N    &  <2.0 $\times 10^{14}$  & 120 & 5.0 & $-$5.0 \\
    HC$^{15}$N    &  1.0 (0.2) $\times 10^{14}$ &  220 (50) & 5.0 & $-$5.0 \\
\hline 

\end{tabular}
\tablefoot{\tablefoottext{a}{Values are derived from synthetic spectrum fitting with CASSIS, with the assumption of LTE.}
\tablefoottext{b}{Column densities for optically thick lines derived using local ISM ratios \citep[][]{Wilson-1999,Milam-2005}.}
\tablefoottext{c}{Upper limits were determined by setting $T_\mathrm{ex}$, FWHM, and $\varv_{\text{peak}}$, equal to that derived for CH$_3$OH and SO, for the O- and S-bearing species respectively.}
\tablefoottext{d}{To estimate a column density for HDO, the excitation temperature and line width of methanol was assumed.}
}

\end{table*}


\section{Discussion}
\label{Sec:Discussion}

\subsection{Cold chemistry}
\label{cold_chem}

The outer regions of the source envelope and the molecular outflows are traced by simple molecules. CO is a good tracer for the source outflows on a large scale, while its isotopologues, $^{13}$CO, C$^{18}$O and C$^{17}$O are less abundant and trace the source disc and envelope \cite[e.g.][]{Joergensen-2002}. Other simple molecules are SO, CS, CN, HCN and HCO$^+$ and they are dense gas tracers \citep{Joergensen-2004b}, tracing the source outflows, cavity walls, disc and envelope \citep[see also][for a review on which molecules traces which part of the physical environment of protostellar sources]{Tychoniec-2021}. 
At large scales, these simple molecules are detected towards all sources in the sample and the environment surrounding the source envelopes does not seem to play a role in setting their abundances.

\begin{figure}
\centering
\includegraphics[width=9cm]{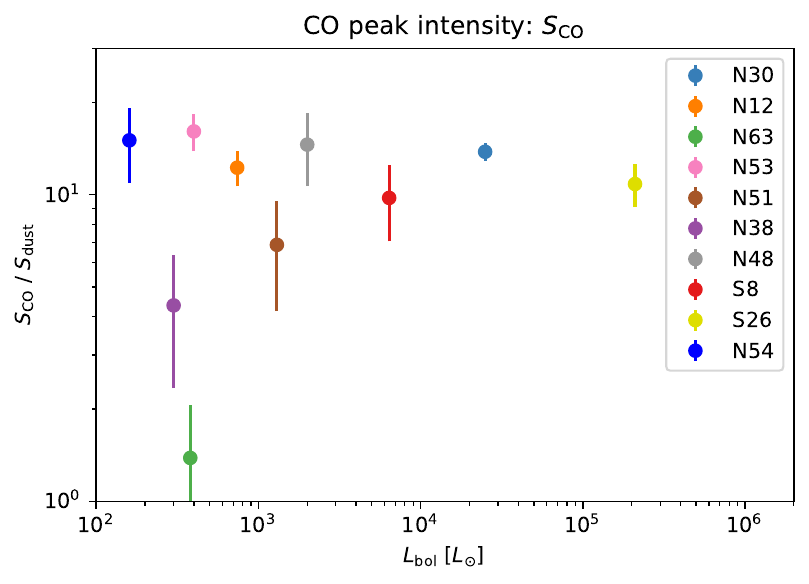}
  \caption{CO to dust ratio vs bolometric luminosity.}
\label{Fig:dust-CO_vs_Lbol}
\end{figure}

\begin{figure}
\centering
\includegraphics[width=9cm]{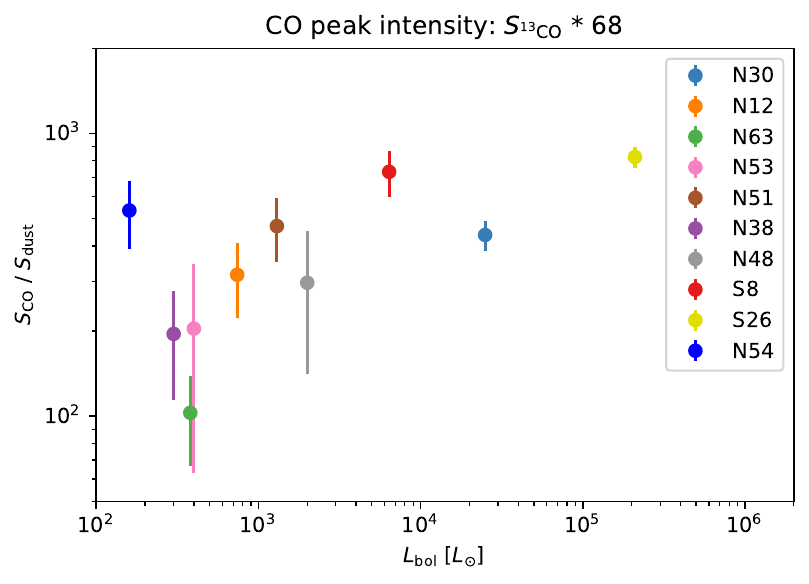}
  \caption{CO to dust ratio (derived from $^{13}$CO) vs bolometric luminosity.}
\label{Fig:dust-C13O_vs_Lbol}
\end{figure}

\begin{figure}
\centering
\includegraphics[width=9cm]{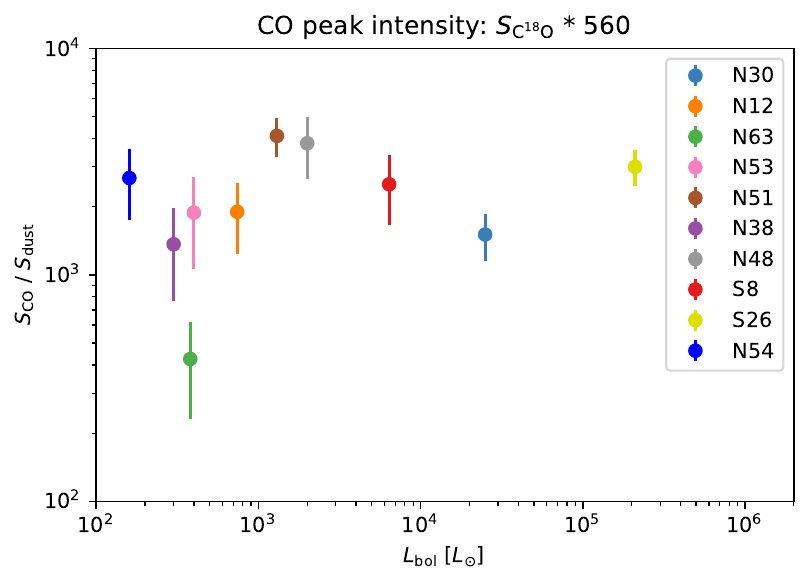}
  \caption{CO (derived from C$^{18}$O) to dust ratio vs bolometric luminosity.}
\label{Fig:dust-CO18_vs_Lbol}
\end{figure}

Figures \ref{Fig:dust-CO_vs_Lbol}, \ref{Fig:dust-C13O_vs_Lbol} and \ref{Fig:dust-CO18_vs_Lbol} show the intensity ratios of the CO, $^{13}$CO and C$^{18}$O lines to the peak continuum emission for each source. The continuum emission scales with the dust mass, for a given temperature, and thus the total mass for a given gas/dust ratio. When averaged over the area, in this case a beam, it becomes directly proportional to the total (H$_2$) column density. Similarly, under the assumption of optically thin emission, the line emission is directly proportional to the column density of the upper energy level. Adding the assumption of LTE, this can be scaled to the total column density of the molecule for a given excitation temperature. For molecules and transitions with low critical density compared to the density of the surroundings, which is the case for CO and its isotopologues, the level populations are likely to be in LTE. Thus, to a first order, this ratio reflects the CO abundance in the gas.

For the figures showing $^{13}$CO and C$^{18}$O, we have multiplied the intensities with the typical ISM ratios for $^{12}$C/$^{13}$C and $^{16}$O/$^{18}$O \citep[68 and 560, respectively,][]{Wilson-1999} to turn them into proxies for the $^{12}$CO abundance. These ratios are shown with respect to the bolometric luminosity for each source. From the figures it can be seen that the CO and $^{13}$CO lines appear to be optically thick as we would expect, since the three figures should show similar ratios if this was not the case. Therefore Fig. \ref{Fig:dust-CO18_vs_Lbol} can be taken as a more accurate representation of the CO abundance. Fig. \ref{Fig:dust-CO18_vs_Lbol} demonstrates that the CO abundance seems to be relatively constant with respect to bolometric luminosity, with values within a factor of two, with a mean (excluding N63) of 3.0$\times 10^3$ and standard deviation 1.2$\times 10^3$, whereas source luminosities vary between 10$^2$ and $\sim$2$\times 10^5$. The only exception is N63 with a low CO gas abundance (a factor of 7 lower than the mean). This source has a high peak continuum emission compared to the other sources (1.66 Jy beam$^{-1}$ compared to $\sim$ 0.4 Jy beam$^{-1}$ for the other sources), while it has a low luminosity ($\sim$3$\times 10^2$). This suggests that N63 is a colder source, but with a more massive core at the small scales that we observe. 

One caveat is that an interferometer such as the SMA, will spatially filter emission from large scales and not detect this emission. For the observations presented here, the maximum recoverable scale is $\sim$ 14$''$ or 2$\times$10$^4$ AU at the distance of Cygnus. This scale corresponds to the typical envelope size of these sources \citep{Pitts-2022}, and so is not expected to be a dominant factor. Furthermore, we are here primarily concerned with the dense inner envelope, which does appear in both dust continuum and CO isotopologue emission. In this respect, we follow previous studies \citep[e.g. ][]{Joergensen-2007, Lee-2015, Stephens-2018}.

Overall, however, it is clear that the abundance of the simplest molecules observed in this survey do not vary significantly between the different sources. This suggests that the very simple chemistry is independent of the environment and the location with respect to both the OB2 association and the DR21 filament. 

\subsection{Hot core chemistry}
\label{Hot_core_chem}
The warm inner regions (temperatures $\gtrsim$ 100K) of protostars are observed at high resolution, with COMs tracing these regions (hot cores). This is because in these warm regions the molecular ices residing on the dust grains are thermally desorbed into the gas phase \citep[e.g.][]{Herbst-2009}. As a consequence, hot core regions are often traced by molecules that formed directly on the grains and then stay on the grains, such as methanol, CH$_3$OH, until they thermally sublimate. 

CH$_3$OH is observed towards or close to all the sources in this work. CH$_3$OH is a common molecule observed towards protostars, but is also observed in outflows, as molecules frozen out on dust grains are sputtered off the grains into the gas phase \citep[e.g.][]{suutarinen2014}, as well as the surface regions of protostellar discs as material falls onto the disc and the dust is heated up to release the molecules frozen out \citep[e.g.][]{Csengeri-2019}. For the source sample in this survey it seems to be a combination of these processes that is observed \citep[see also][]{vanderWalt-2021}.  $^{13}$CH$_3$OH and CH$_3$CN are detected towards only four sources, N30, N12, S26 and N63, and CH$_3$OCH$_3$ and CH$_3$OCHO are detected towards N30, S26 and N63. The lack of COM detections towards the other sources could be because the observations are sensitivity limited towards these sources. To check this hypothesis, we calculate the relative molecular abundances $N$(CH$_3$OCHO)/ $N$(CH$_3$OH) towards N30, S26 and N63 where we detect CH$_3$OCHO and compare with N12 where this molecule is not detected. For N30, S26 and N63, a value of $\sim$ 0.02 is obtained, which, when turned into an emission spectrum, is consistent with the noise level for N12, suggesting that the observations are indeed sensitivity limited. This result is further corroborated by the 100 K radii for the sources listed in Table \ref{table:Source_props2}.  The 100 K radius for all sources, except N30 and S26, falls well within the beam size of the PILS-Cygnus observations, and any hot core emission is likely severely beam diluted.

As illustrated in the case of N30, a `hot core' is not necessarily a hot core. Towards N30 the COM emission is found to be extended and offset from the location of the continuum positions, suggesting that the emission originates from a region where thermal desorption is not the dominant desorption mechanism. Instead, it seems more plausible that the emission originates in a rotating disc-like structure \citep{vanderWalt-2021}. This type of hot-core emission that is not caused by thermal desorption of molecules is seen towards more and more sources \citep[e.g.][]{Avery_and_Chiao-1996,Joergensen-2004b,Arce-2008,Sugimura-2011,Podio-2015,Drozdovskaya-2015,Lefloch-2017,Orozco-Aguilera-2017,ArturdelaVillarmois-2018,Csengeri-2018,Csengeri-2019}. Of the sources observed here, S26 seems to be a classical hot core at the scales we observe, but at higher angular resolution it is fragmented \citep[][]{Jimenez-Serra-2012, Suri-2021}, which suggests that a more complicated morphology for the COM emission would be observed with higher angular resolution observations. N12 reveals two continuum cores at the resolution of the PILS-Cygnus observations, with COM emission peaking towards the MM2 core. However, from the emission map of CH$_3$OH in Fig. A.10
it can be seen that there is another CH$_3$OH peak to the north-west of the continuum cores, and along the CO outflow axis, suggesting this emission peak is due to sputtering caused by the outflow. This is not seen in the emission from other COMs. N63 shows only a single core at the angular resolution of the PILS-Cygnus observations, but with significant extended emission along the CO outflow direction seen in CH$_3$OH and to a lesser extent SO (Fig. \ref{Fig:N63_molecules}). Again, this suggests that at least some emission is outflow-related. 

The extent of the COM emission, when detected, does not appear to correlate with the direction to the OB2 association. This result is based on very few sources (N12, N30, and N63), and so is not enough to draw strong conclusions. However, it appears that on the small scales observed here, the COM emission is not strongly affected by the external UV radiation. On larger, envelope scales, the external environment may play a larger role.

Of the sources observed here, four harbour known compact \ion{H}{ii} regions: S26, N30, S8, and N51 \citep{Lynds-1965, Haschick-1981, Torrelles-1997, Grasdalen-1983}.  
Two of these also show COM emission (S26 and N30). In the evolutionary scheme proposed by \citet{Motte-2018}, the hot cores are the largest right before the sources develop \ion{H}{ii} regions. This suggests that these two sources, S26 and N30, are relatively evolved sources, and that protostellar evolution likely plays an important role in setting the complex chemistry. Similarly, S8 and N51 might have \ion{H}{ii} regions large enough that the hot cores are diminishing in size and not detectable to us, making them even more evolved. In this evolutionary scheme, the two other sources with COM emission, N12 and N63, would be at a stage prior to developing detectable \ion{H}{ii} regions, and the remaining sources would be at earlier evolutionary stages where the hot cores are not developed yet. We emphasise that in the absence of clear evolutionary tracers, this scenario is hypothetical.

The chemistry observed towards the four ``hot cores'' also does not seem to be affected by the external environment. When comparing the CH$_3$OH column density to the CH$_3$CN column density for the four sources (Fig. \ref{Fig:CH_3CN-CH_3OH_vs_Lbol}), the ratio is within a factor of five (with a mean of = 8.5$ \times 10^{-3}$ and standard deviation = 5.4$ \times 10^{-3}$). This again suggests, albeit based on a very small number of sources, that the external environment plays a small role in setting the chemistry. 

The value of the column-density ratio $N$(CH$_3$CN) / $N$(CH$_3$OH) ranges from 4 $\times$ 10$^{-3}$ to 1.7 $\times$ 10$^{-2}$ (Fig. \ref{Fig:CH_3CN-CH_3OH_vs_Lbol}). The lower values are similar to the results of the PILS survey of IRAS16293--2422B \citep[4.0 $\times$ 10$^{-3}$;][]{joergensen-2016} and lower than that of Sgr B2(N2) \citep[5.5 $\times$ 10$^{-2}$;][]{Belloche-2016}; the upper limits are consistent with both values and not particularly sensitive. \citet{Garrod-2022} suggest that this difference is caused by the length of the warm-up phase, where a slower warm-up phase leads to a higher column-density ratio, and vice versa. In turn, that would suggest that the warm-up phases for these protostars are more closely related to that of IRAS16293, as opposed to Sgr B2(N2). 

Interestingly, the formation paths of CH$_3$OH and CH$_3$CN may be different. It is generally accepted that CH$_3$OH forms from hydrogenation of CO frozen out on icy dust grains. CH$_3$CN, on the other hand, likely forms in the gas phase in hot cores \citep{Garrod-2022} as a second-generation molecule \citep{Herbst-2009}. Thus, to a first order, there is no a priori expectation that the two species should correlate in column density. Furthermore, since CH$_3$CN forms in dense, warm gas, the expectation is that emission is compact and peaks on the continuum position. This is indeed the case for S26. For N30, emission is extended and peaks off-source \citep{vanderWalt-2021}, whereas for N12 emission is extended and also peaks off-source. In this case, higher-angular resolution observations would be required to precisely pinpoint the origin of both CH$_3$OH and CH$_3$CN. 

To further shed light on the origin of CH$_3$OH and CH$_3$CN, their column densities are compared to that of SO. SO is often used as an outflow tracer \citep[e.g.][]{tabone-2017, ArturdelaVillarmois-2018, vanGelder-2021}, and generally traces warm gas. We do not find any signs of outflow activity in the line profiles of SO, but it is possible that this molecule traces accretion shocks onto a disc \citep{ArturdelaVillarmois-2018}, in which case the line profile would not show strong line wings or asymmetries. The column density ratio $N$(SO)/$N$(CH$_3$OH) may show a weak increase with luminosity, but with only four data-points it is not possible to draw any strong conclusions regarding trends with luminosity. Towards these sources, accretion shocks may be stronger (impacting higher-density material at higher velocities), in which case the SO abundance is expected to increase by orders of magnitude \citep{vanGelder-2021}. Furthermore, the models do not predict an enhanced CH$_3$OH abundance, and so it is expected that the ratio of $N$(SO) / $N$(CH$_3$OH) increases with increasing strength of the accretion shock. To test this scenario further, deeper observations at higher angular resolution are needed, and these should be able to constrain the SO$_2$ column density; the models make further predictions regarding the SO / SO$_2$ column density ratio, that will be used for constraining the model. Furthermore, more data-points will clearly be needed.

\subsection{Deuteration} 
We detect low levels of deuteration towards N30, N12, N63 and S26, with N12 and N63 having slightly higher levels than N30 and S26. No deuterated molecules are detected towards the remaining sources of the sample. HDO is detected towards N30, S26, N12 and N63, while CH$_2$DOH is detected towards N63, and is very tentatively present towards N12, however only a $\sim$2$\sigma$ peak can be found in this work. The D/H ratios are derived for N(CH$_2$DOH)/N(CH$_3$OH), after correcting for the three symmetries of the CH$_2$DOH molecule \cite[see e.g.][]{joergensen-2018,Manigand-2019}, towards N63 of $\sim 1\%$, with upper limits for N30 and S26 of $\sim 0.1\%$ and $\sim 0.3\%$, respectively, and with an upper limit for N12 of $\sim 1\%$. The higher D/H ratio derived for N63 is expected, since this source seems to be a cold source, as discussed in Sec. \ref{cold_chem} above \citep[see also discussion in ][]{duarte-cabral2013}. The colder the source is, the more efficient the deuteration chemistry should be \citep[e.g.][]{Ceccarelli-2007,Caselli-2019}, which leads to an increase in the deuteration fraction in molecules such as CH$_3$OH. An alternative explanation would be that it is a low-mass source, with a longer cold phase, such that even if the deuteration chemistry runs less efficiently, it has longer time to work, leading to the same result. However, given the high continuum emission peak (1.66 Jy beam$^{-1}$; Table \ref{table:Source_props2}), it suggests that this is a cold, dense and massive source (we also note in Table \ref{table:Source_props2} that N63 has a mass of >10 $M_\odot$). With a higher level of CO freeze out than the other sources, it suggests this source formed at a cooler temperature (<30K), where deuterium reactions are more efficient \cite[][]{Taquet-2014,Bogelund-2018}. More observations are necessary to confirm this result. 

N30 and S26 compare well with other high-mass sources, \citep[for example Orion KL has values around $0.2-0.8\%$ while Sgr B2 (N2) has a D/H ratio of $0.12\%$ for CH$_2$DOH; ][]{Neill-2013,Belloche-2016}. N12 and N63 have a D/H ratio slightly higher than high-mass sources, but still a factor of two or three lower than low-mass sources such as IRAS 16293--2422 B \citep[with D/H ratios of $2-3\%$;][]{joergensen-2018}. Thus, the deuteration fraction seems to be very sensitive to changes in the local environment, as expected \citep{Manigand-2019,Jensen-2021}.

\subsection{External environment}
As discussed in Sec. \ref{Sec:cont_emission} above, UV radiation from bright stars can affect the observed chemistry of newly formed stars. A strong background radiation field may heat up the environment making the formation of molecules on dust grains inefficient, since the higher temperatures will prevent the precursor molecules (e.g. CO) from freezing out onto the dust grains, and thus not participate in the further formation of COMs. 
Higher formation temperatures also result in lower deuteration, since deuterium reactions are inefficient at temperatures >30K \citep[see e.g.][]{Taquet-2014,Bogelund-2018}. 

Of the ten sources, eight are located at distances between 30--50 pc to the CygX-OB2 association. Five of these sources (N53, N54, N51, N38 and N48) are located in the DR21 ridge, and these sources show similar chemistry at the sensitivity of our observations, that is, we detect few COMs towards these sources. The three sources that surround DR21 (N30, N12 and N63) all have a higher amount of COM detections, with N63 and N12 also having a higher D/H ratio than the other sources. N12 and N63 are both closer to CygX-OB2 than the DR21 sources. The remaining two sources are S26, which is located at $\sim$ 3.3 kpc, too far from CygX-OB2 for it to have any effect, and S8 with a high uncertainty in its distance (>80 pc from CygX-OB2). 

The lack of COM detection in the DR21 ridge (and COM detections towards N30, N12 and N63) seems to suggest that the external environment does not play a noticeable role in the formation of COMs in this region and that the local environments surrounding the protostars play a bigger role. On the other hand, as discussed above, our observations are probably sensitivity limited (Sect. 4.2), implying that follow-up observations are required to confirm this speculation. The sensitivity limit was estimated using $N$(CH$_3$OH) as a reference frame.

The five sources in the DR21 ridge represent something of a conundrum. The sources have the lowest masses in the sample, inferred from the 870 $\mu$m fluxes, and the natural conclusion would be that this is consistent with low masses of the hot cores, and thus low column densities of COMs. On the other hand, the sources do not have the lowest luminosities. This means that the radii of the hot-core regions should be relatively large, and more mass should be included in the hot-core region, which should lead to higher COM column densities and detections. In this case, it appears as if the lower mass is the dominating factor since no COM emission is detected towards the sources, consistent with lower CH$_3$OH column densities towards these sources (Sect. 4.2). The relatively high luminosity compared to the mass, can likely be attributed to on-going accretion from the DR21 ridge onto these sources \citep{Pitts-2022}. This in turn suggests that the environment does play a role, in the sense that it matters whether sources form in dense molecular ridges (`hubs') or not in terms of our ability to detect COM emission. However, whether this is reflected in the chemistry or not is an open question. 

Observations of low-mass protostellar sources appear to suggest that the difference in location, specifically inside or outside a filament or ridge, plays a role in setting the chemistry. The `Astrochemical Surveys At IRAM’ \citep[ASAI;][]{Lefloch-2018} programme find that sources located in filaments appear to have a hot-core chemistry, that is the organic chemistry is dominated by C-O-bearing molecules. On the other hand, sources located outside dense filaments tend to show a chemistry dominated by C-C-bearing molecules, that is the so-called warm carbon-chain chemistry (WCCC). This result is based on observations of between four and five sources in each category as observed with a single-dish telescope. Interferometric observations reveal that on small scales, WCCC sources may be dominated by a hot-core chemistry \citep[e.g.][]{Oya-2019}, making the picture less clear. Thus, to fully address how the location, inside our outside a filament, affects the chemistry, deeper observations spanning a range of spatial scales is required for a large sample.

\subsection{Morphology}

From the observations presented here and in \citet{vanderWalt-2021} it can be seen that the origin of COMs detected in and near star-forming regions is a combination of thermal (the classical hot core scenario) and non-thermal (sputtering caused by outflows and accretion of inflowing material onto a disc structure) processes. The emission morphology of the PILS-Cygnus sources seem to point to hot cores at small scales, with COMs also detected in the surrounding envelope and outflows. We know that different molecules trace different regions of the protostar \citep[e.g.][]{Herbst-2009,Tychoniec-2021}, which make studies like PILS-Cygnus so valuable, where a large frequency range are covered at high resolution, making it possible to observe a large amount of COM transitions and in an unbiased way. This results in a more complete picture of the morphology of star-forming regions.


\section{Summary and conclusions}
\label{Sec:Conclusion}

This paper presents the source sample and results of the PILS-Cygnus survey. Ten intermediate- and high-mass sources were observed with the SMA, all located in the nearby Cygnus-X star-forming region. The observations were done at $\sim$1\farcs5 resolution ($\sim$ 2000 AU), and span the entire frequency range of 329--361 GHz. The main results of the survey can be summarised as follows: 

\begin{itemize}
    \item Continuum and line emission is detected towards all sources; the line emission is from small species, such as CO, H$_2$CO, CS, and HCO$^+$. The abundance of CO appears constant over all species, with only one exception, indicating that the chemistry of this simple molecule does not vary over the cloud. 
    \item Emission from the simplest complex organic molecule, methanol, CH$_3$OH, is detected towards nine sources, and is found to trace the hot core, the outflow, other shock activity, or a combination of these. This in turn implies that some `hot cores' are not only caused by thermal desorption but are, at least partly, due to shocks. 
    \item Emission from more complex species, for example CH$_3$OCH$_3$ and CH$_3$OCHO, is detected towards 3 sources, and typically originates in the hot core. CH$_3$CN is detected towards one additional source. Column densities have been inferred for all sources and species, and the lack of detection of more complex organic species is set by the sensitivity. 
    \item Deuterated methanol, CH$_2$DOH, is firmly detected towards one source. The deuteration fraction for this source is high, $\sim$ 1\%, suggesting that this source either is cold now ($T$ < 30 K) or has been so for an extended period in the past, whereas the other nine sources have not. 
    \item There are no apparent differences between the sources in terms of chemistry as a function of distance to the OB2 association, nor location within the DR21 filament. This suggests, although based on a small number of sources, that the external environment only plays a minor role in shaping the chemistry and its associated evolution of these sources. Instead, local factors such as the thermal history and outflows, appear more important for the observed chemistry. 
\end{itemize}

With these observations, characterisation of molecular emission from intermediate- and high-mass protostellar sources in Cygnus-X is started. The observations demonstrate the usefulness and necessity for observing a large frequency range in order to properly understand the chemistry of such complex objects. Finally, the initial results presented here indicate that the external environment only plays a minor role in setting the chemistry on smaller scales ($\sim$ 2000 AU), but clearly more observations are needed to verify this statement.


\begin{acknowledgements}
We acknowledge and thank the staff of the SMA for their assistance and continued support. The authors wish to recognise and acknowledge the very significant cultural role and reverence that the summit of Mauna Kea has always had within the indigenous Hawaiian community. We are most fortunate to have had the opportunity to conduct observations from this mountain. The research of SJvdW and LEK is supported by a research grant (19127) from VILLUM FONDEN. The research of HC is supported by an OPUS research grant (2021/41/B/ST9/03958) from the Narodowe Centrum Nauki. 

\end{acknowledgements}


\bibliographystyle{aa.bst} 
\bibliography{refs.bib} %


\appendix 


\section{Source sample spectra and emission maps}
\label{App:Spectra_emission-maps}

\begin{figure*}
    \centering
    \includegraphics[width=\textwidth]{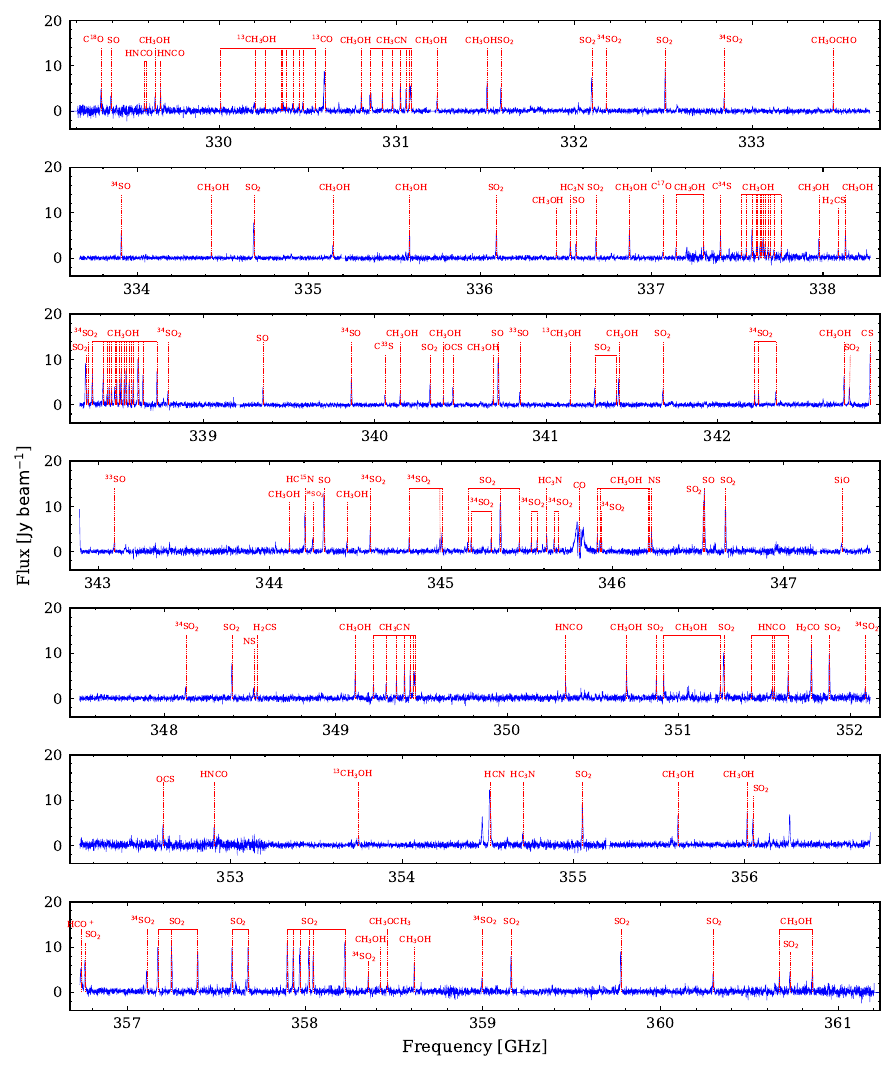}
    \caption{Spectrum towards S26, at the continuum peak. The spectrum shows many molecular lines, with the identified lines marked and labelled.}
    \label{Fig:S26_spectrum}%
\end{figure*}

\begin{figure*}
    \centering
    \includegraphics[width=\textwidth]{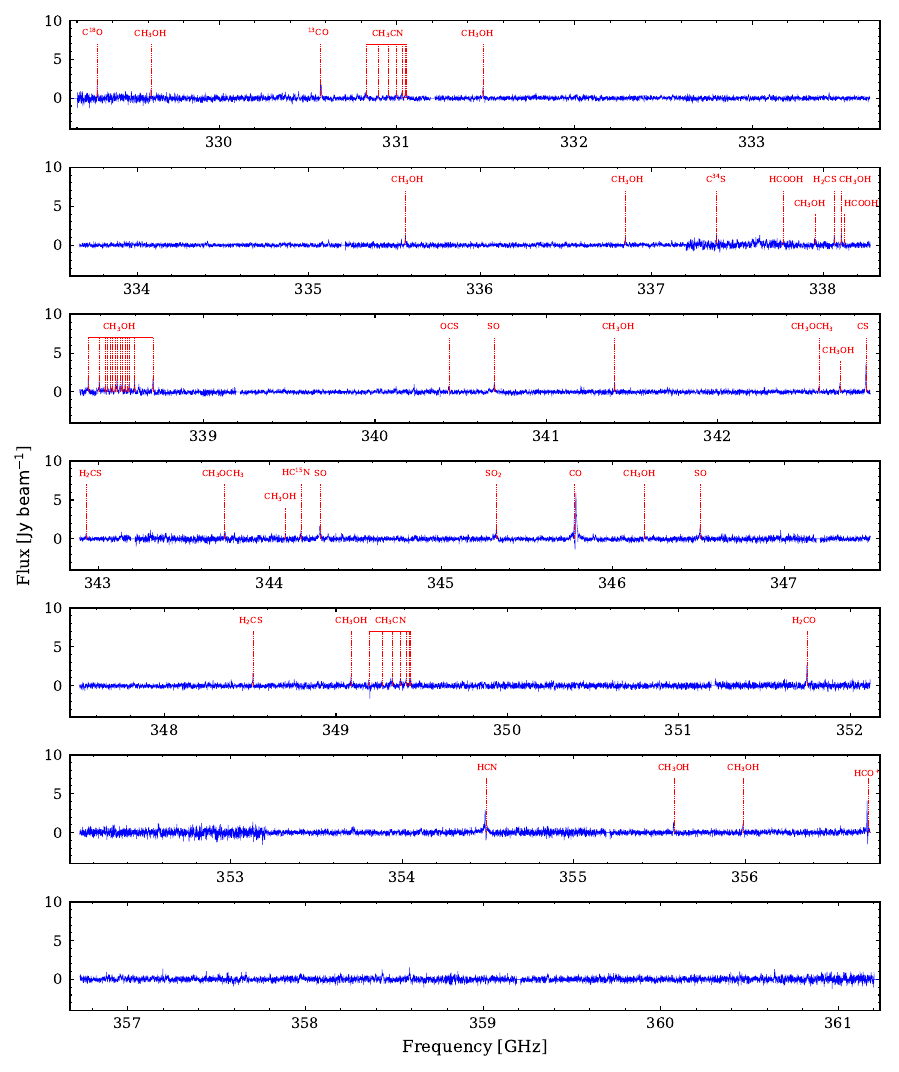}
    \caption{Spectrum towards N12, at the continuum peak of the MM2 core. The brightest lines are marked and labelled.}
    \label{Fig:N12_spectrum}%
\end{figure*}

\begin{figure*}
    \centering
    \includegraphics[width=\textwidth]{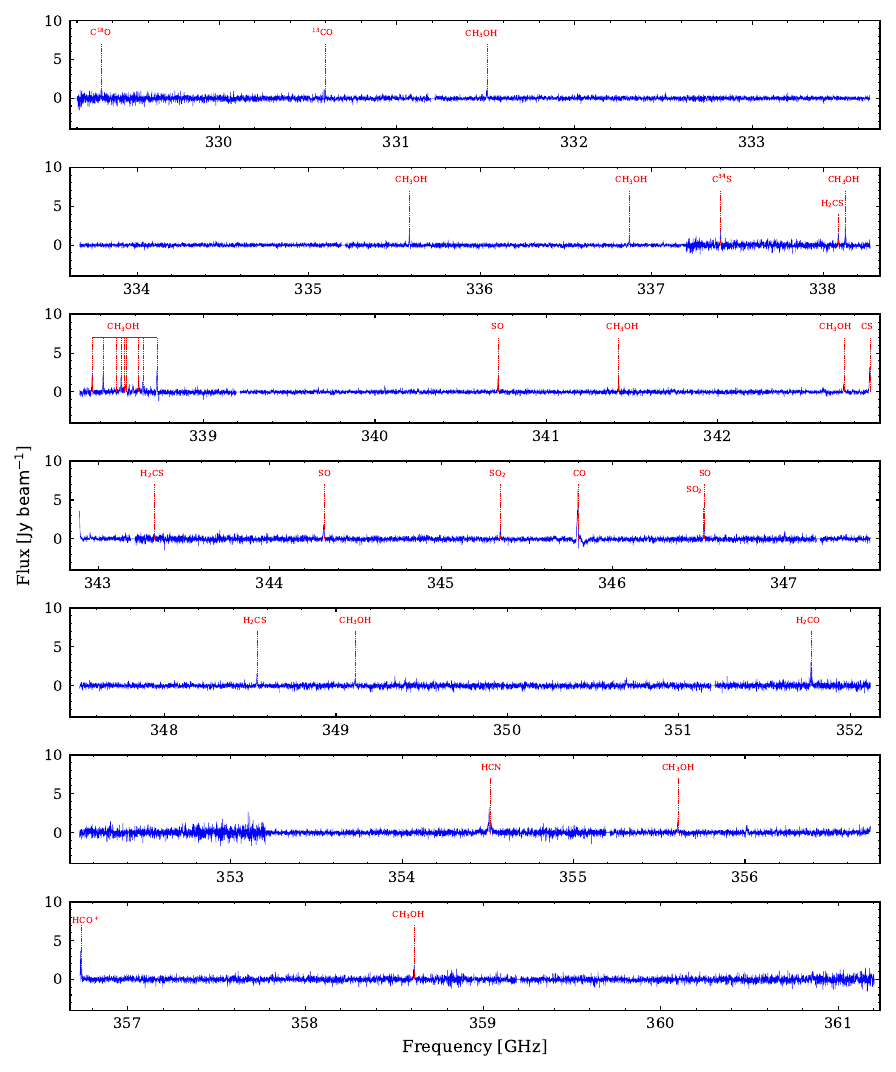}
    \caption{Spectrum towards N53, at the continuum peak of the eastern core. The brightest lines are marked and labelled.}
    \label{Fig:N53_spectrum}%
\end{figure*}

\begin{figure*}
    \centering
    \includegraphics[width=\textwidth]{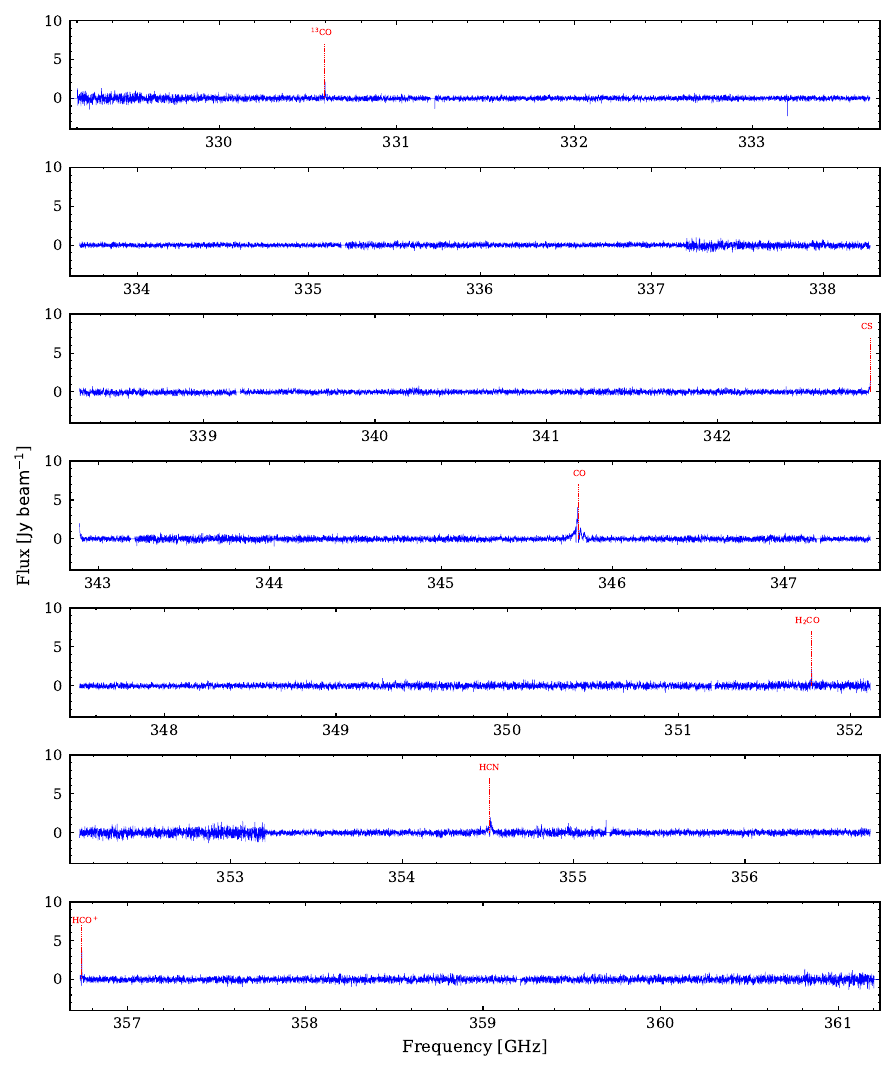}
    \caption{Spectrum towards N54, at the continuum peak of the eastern core. The brightest lines are marked and labelled.}
    \label{Fig:N54_spectrum}%
\end{figure*}

\begin{figure*}
    \centering
    \includegraphics[width=\textwidth]{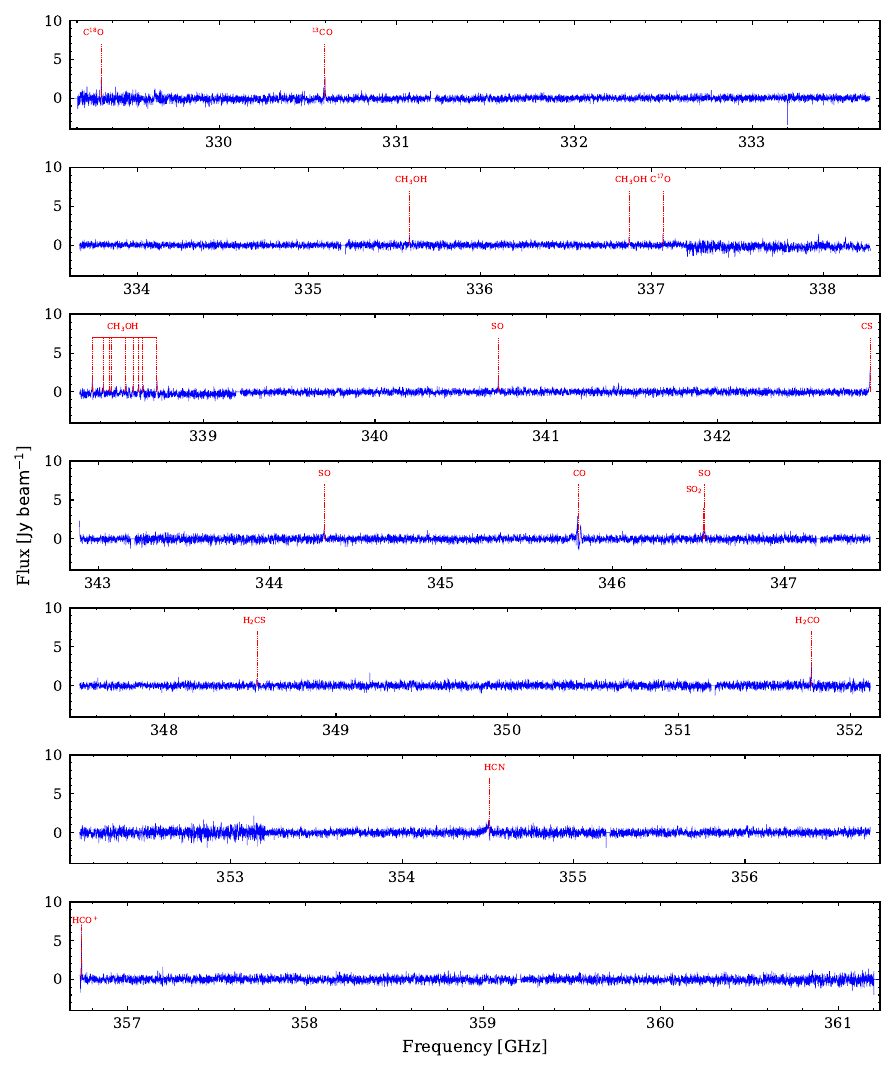}
    \caption{Spectrum towards N51, at the continuum peak. The brightest lines are marked and labelled.}
    \label{Fig:N51_spectrum}%
\end{figure*}

\begin{figure*}
    \centering
    \includegraphics[width=\textwidth]{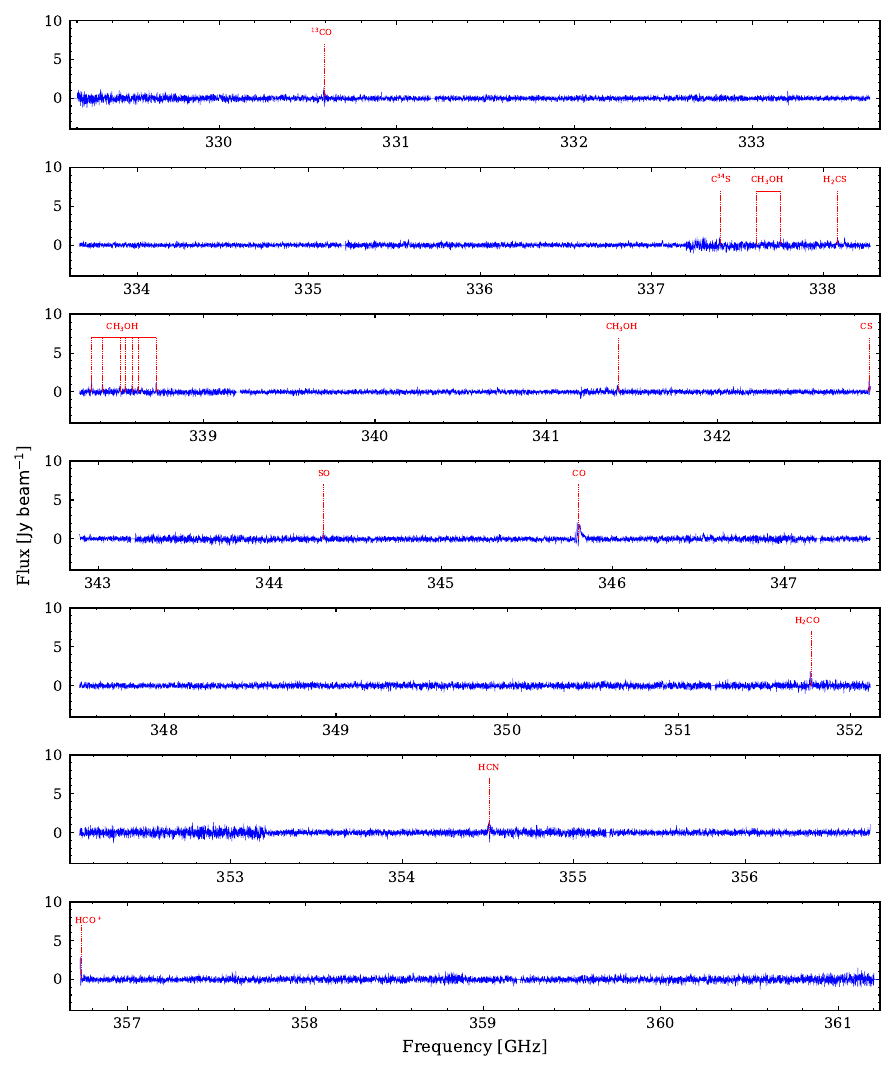}
    \caption{Spectrum towards N38, at the continuum peak. The brightest lines are marked and labelled.}
    \label{Fig:N38_spectrum}%
\end{figure*}

\begin{figure*}
    \centering
    \includegraphics[width=\textwidth]{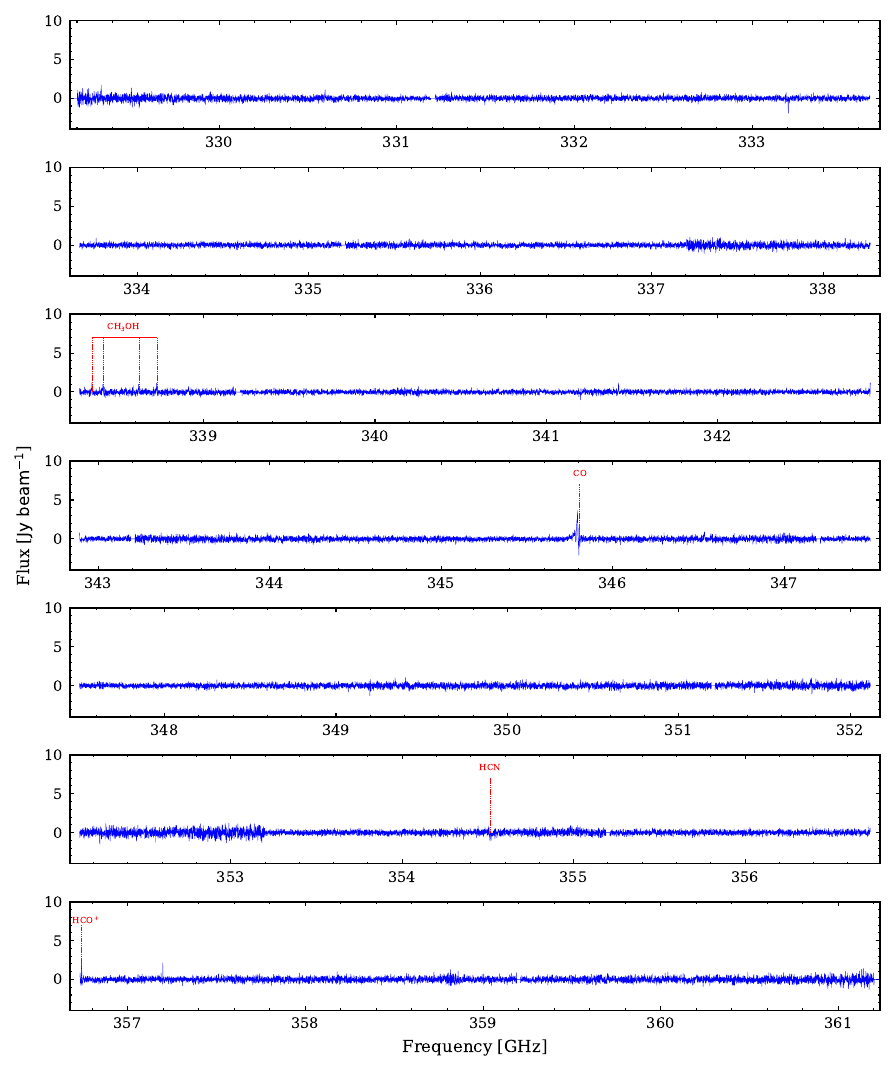}
    \caption{Spectrum towards N48, at the continuum peak. The brightest lines are marked and labelled.}
    \label{Fig:N48_spectrum}%
\end{figure*}

\begin{figure*}
    \centering
    \includegraphics[width=\textwidth]{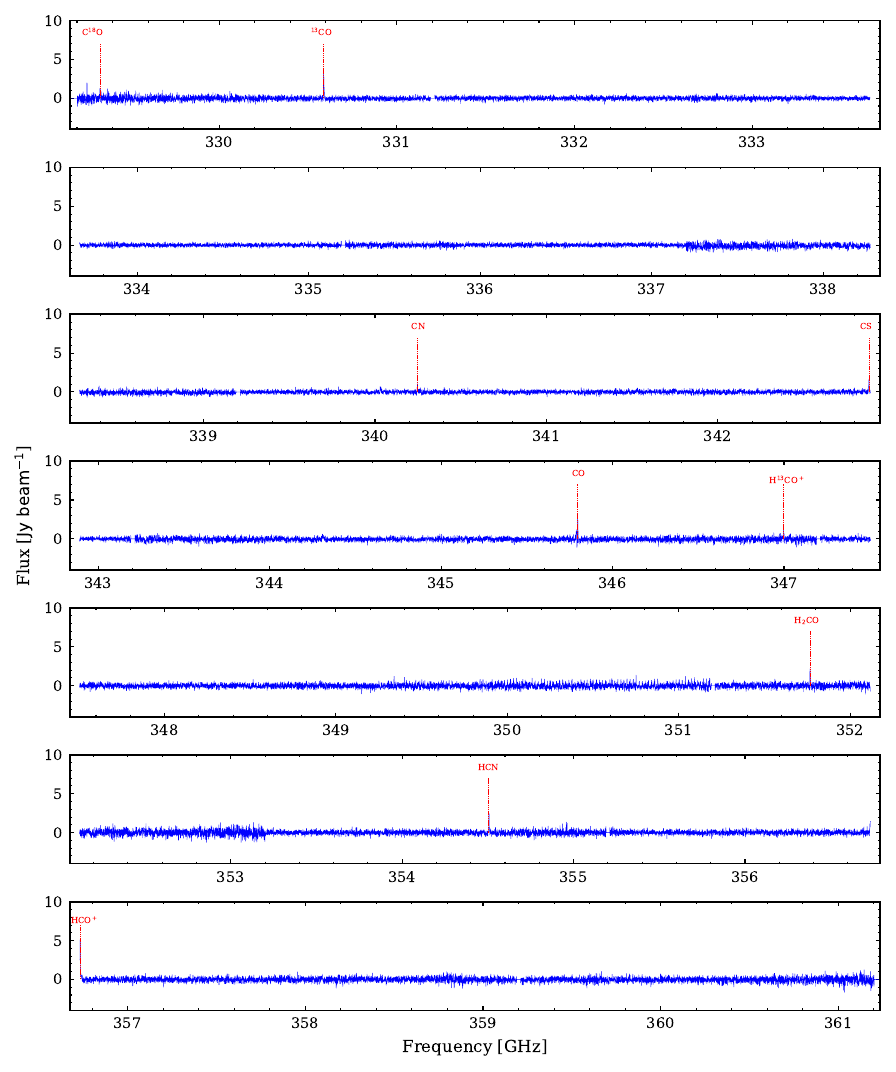}
    \caption{Spectrum towards S8, at the eastern continuum peak. The brightest lines are marked and labelled.}
    \label{Fig:S8_spectrum}%
\end{figure*}

\begin{figure*}
    \centering
    \includegraphics[width=\textwidth]{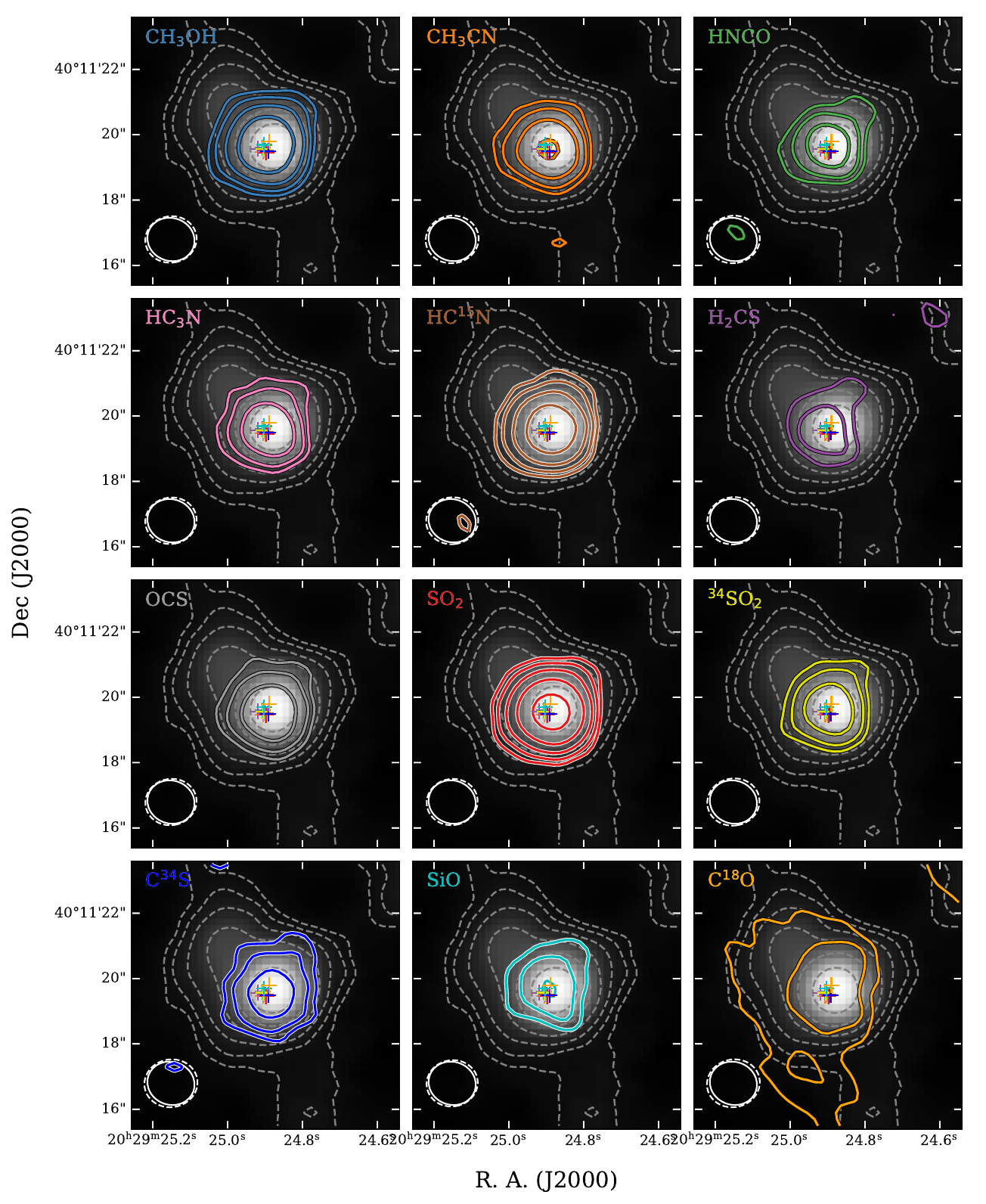}
    \caption{Molecules observed towards S26. The contour levels for both the continuum (grey dashed lines) and molecular emission (coloured) are 3, 6, 12, 24 and 48$\sigma$. The pluses are the molecular emission peak positions derived from 2D Gaussian fits.}
    \label{Fig:S26_molecules}%
\end{figure*}

\begin{figure*}
    \centering
    \includegraphics[width=\textwidth]{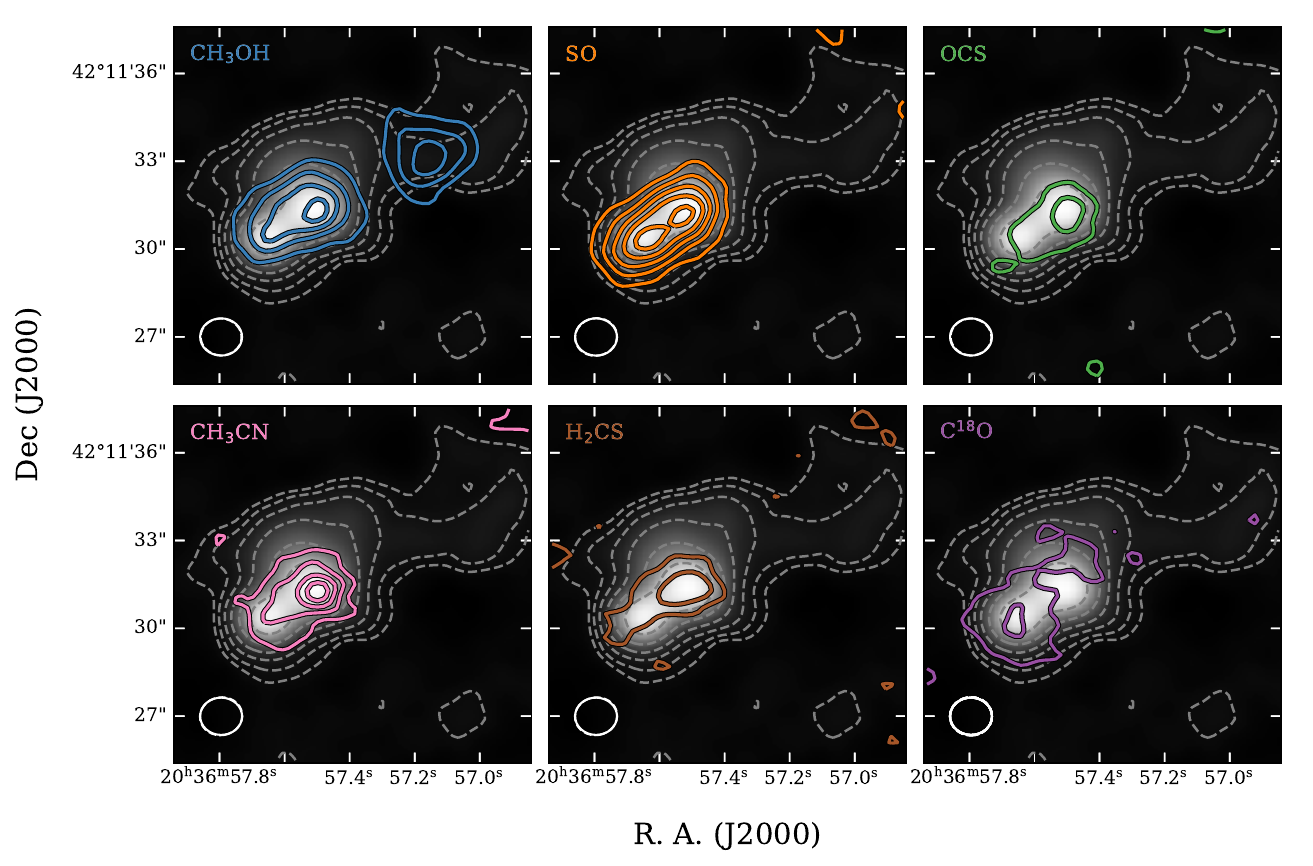}
    \caption{Molecules observed towards N12. The contour levels for the molecular emission are here 3, 6, 9, 12, and 15$\sigma$. The continuum contours are as in Fig. \ref{Fig:S26_molecules}}
    \label{Fig:N12_molecules}%
\end{figure*}

\begin{figure*}
    \centering
    \includegraphics[width=\textwidth]{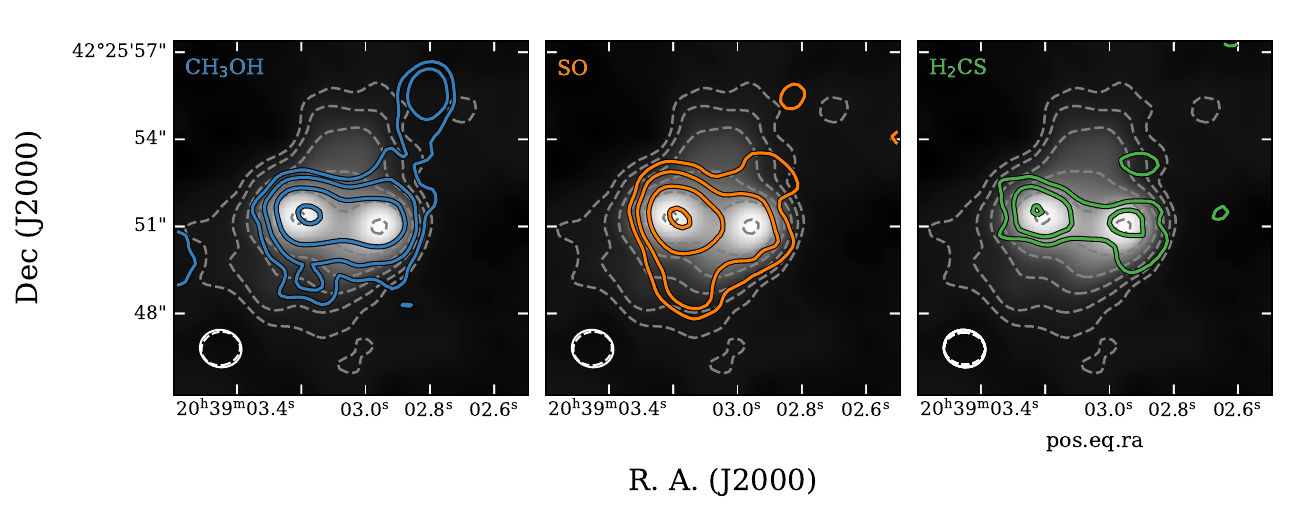}
    \caption{Molecules observed towards N53. The contour levels for the molecular emission are here 3, 6, 9 and 12$\sigma$. The continuum contours are as in Fig. \ref{Fig:S26_molecules}}
    \label{Fig:N53_molecules}%
\end{figure*}

\begin{figure*}
    \centering
    \includegraphics[width=\textwidth]{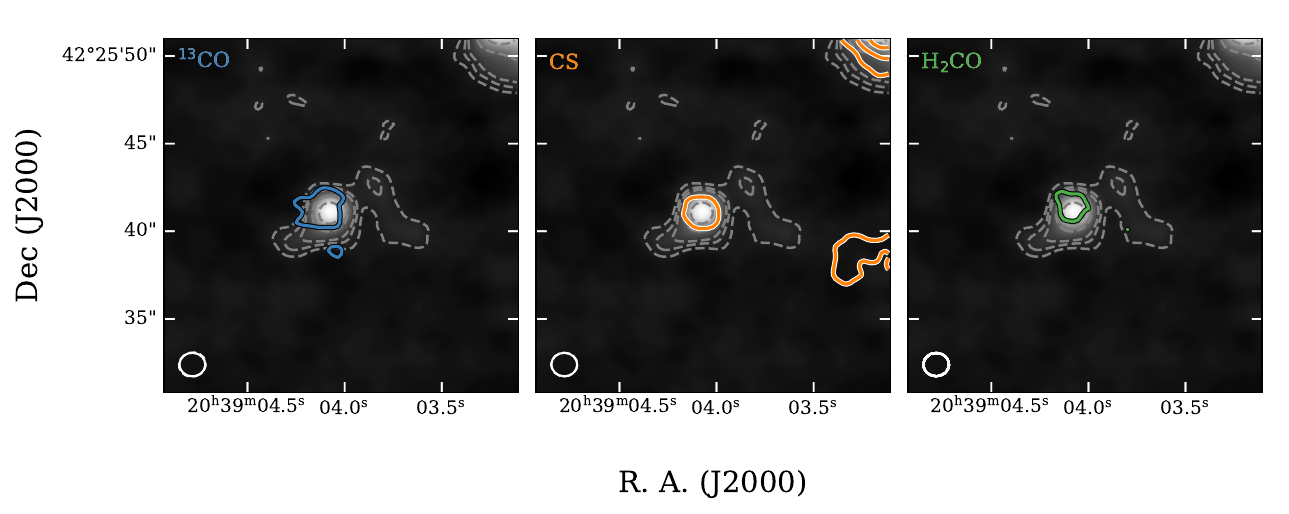}
    \caption{Molecules observed towards N54. The contour levels for the molecular emission are here 3$\sigma$. The continuum contours are as in Fig. \ref{Fig:S26_molecules}}
    \label{Fig:N54_molecules}%
\end{figure*}

\begin{figure*}
    \centering
    \includegraphics[width=\textwidth]{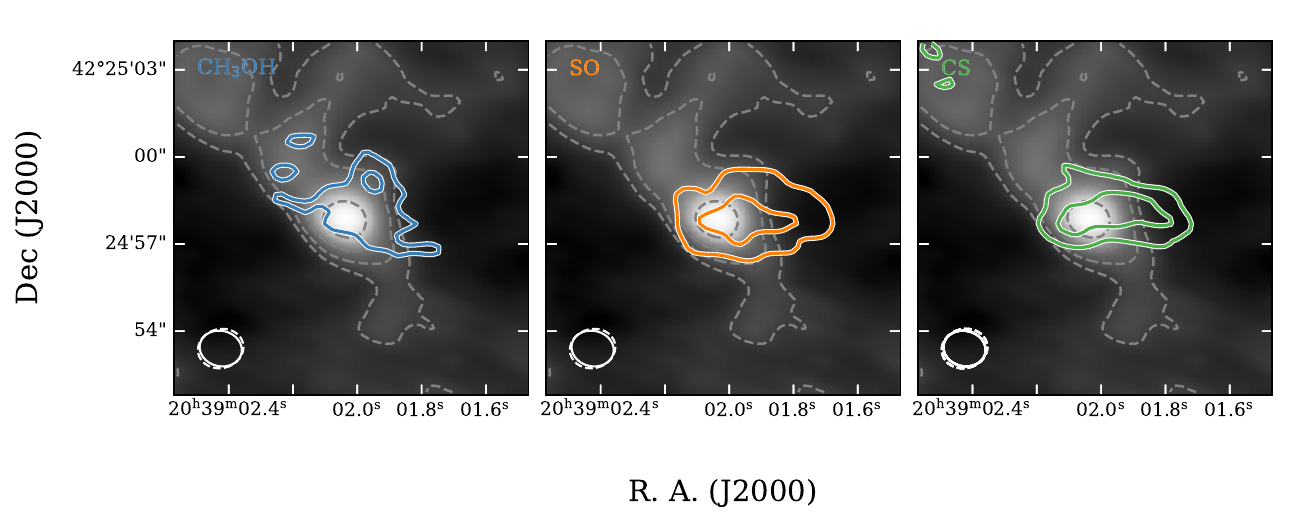}
    \caption{Molecules observed towards N51. The contour levels for the molecular emission are here 3 and 6$\sigma$. The continuum contours are as in Fig. \ref{Fig:S26_molecules}}
    \label{Fig:N51_molecules}%
\end{figure*}

\begin{figure*}
    \centering
    \includegraphics[width=\textwidth]{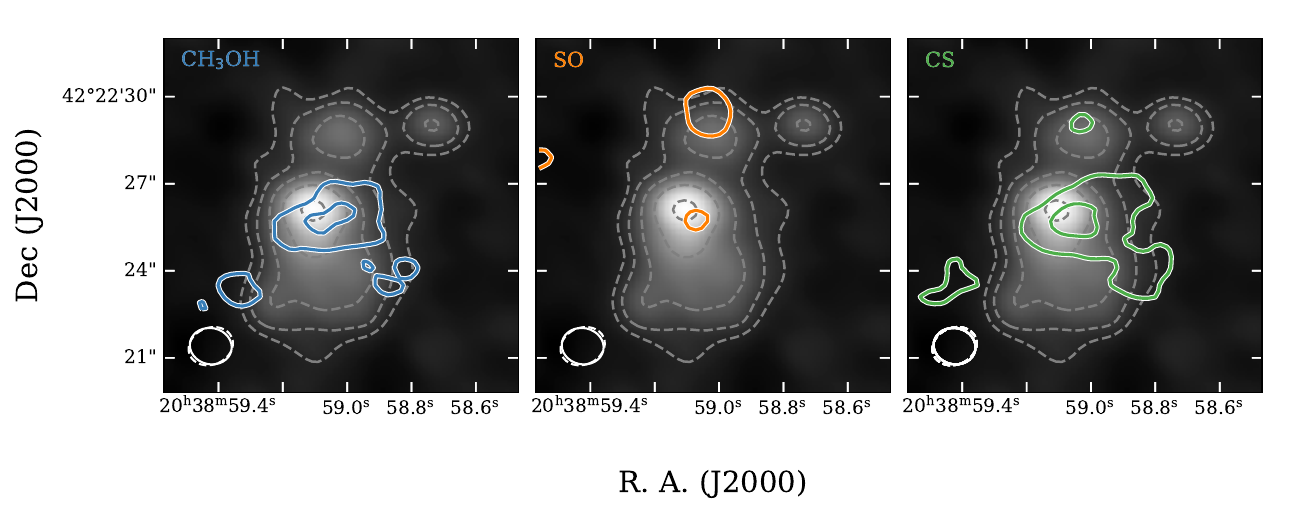}
    \caption{Molecules observed towards N38. The contour levels for the molecular emission are here 3 and 6$\sigma$. The continuum contours are as in Fig. \ref{Fig:S26_molecules}}
    \label{Fig:N38_molecules}%
\end{figure*}

\begin{figure*}
    \centering
    \includegraphics[width=\textwidth]{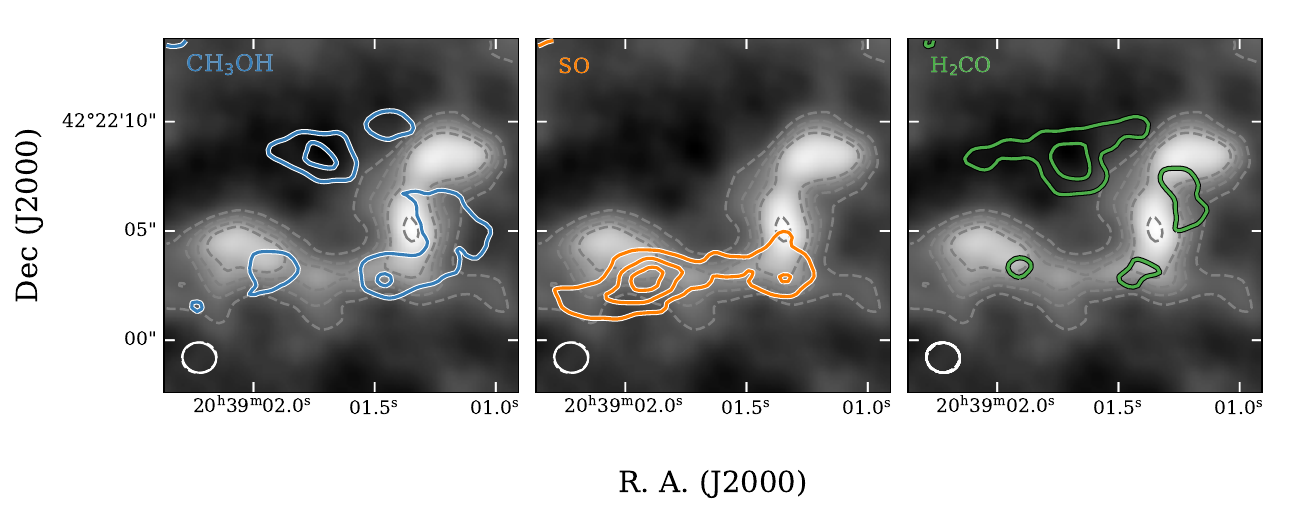}
    \caption{Molecules observed towards N48. The contour levels for the molecular emission are here 3, 6 and 9$\sigma$. The continuum contours are as in Fig. \ref{Fig:S26_molecules}}
    \label{Fig:N48_molecules}%
\end{figure*}

\begin{figure*}
    \centering
    \includegraphics[width=\textwidth]{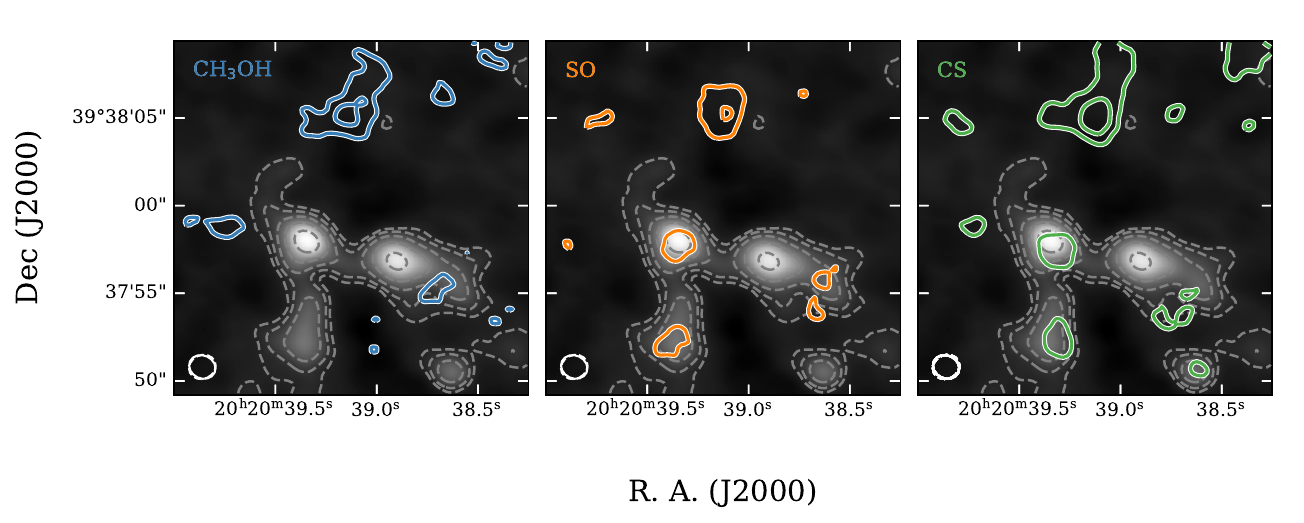}
    \caption{Molecules observed towards S8. The contour levels for the molecular emission are here 3 and 6$\sigma$. The continuum contours are as in Fig. \ref{Fig:S26_molecules}}
    \label{Fig:S8_molecules}%
\end{figure*}

In this appendix the source sample spectra (Figs. \ref{Fig:S26_spectrum} -- \ref{Fig:S8_spectrum}) and emission maps (Figs. \ref{Fig:S26_molecules} -- \ref{Fig:S8_molecules}) are shown. The molecular transitions used to produce the molecular emission maps for each source are listed in Tables \ref{table:S26_Molecules_COMs} -- \ref{table:S8_Molecules_all}. Readers can refer to the main text for the spectrum and molecular maps for N63.

\begin{table*}
\caption{Gaussian fits for molecular transitions of Complex Organic Molecules, including t-HCOOH and H$_2$CO, observed towards S26\tablefootmark{a}.}
\label{table:S26_Molecules_COMs}      
\centering          
\begin{tabular}{l c c c c c c r}     
\hline
\hline 
    Molecule & Frequency & Transition & Position & $I_{\mathrm{peak}}$\tablefootmark{b} & $I_{\mathrm{int}}$\tablefootmark{b}  & FWHM\tablefootmark{b}  & $\varv_{\text{source}}$\tablefootmark{b} \\ 
     &  [GHz] & &  [$'',''$] & [$\mathrm{Jy \ beam^{-1}}$] &  [$\mathrm{Jy \ beam^{-1} \ km \ s^{-1}}$] & [$\mathrm{km \ s^{-1}}$] & [$\mathrm{km \ s^{-1}}$]\\ 
\hline
    \multicolumn{8}{c}{S26}\\
\hline
    A-CH$_3$OH & 331.5023 & $11_{1, -0} - 11_{0, +0}$ &  $(      0.15,      0.01)$ &        6.9 &       31.1 &        4.1 &       $-$4.9 \\
      & 335.5820 & $7_{1, +0} - 6_{1, +0}$ &  $(      0.31,      0.02)$ &        5.8 &       24.3 &        4.0 &       $-$4.6 \\
      & 336.8651 & $12_{1, -0} - 12_{0, +0}$ &  $(      0.19,      0.01)$ &        6.7 &       33.4 &        4.3 &       $-$3.7 \\
      & 338.4087 & $7_{0, +0} - 6_{0, +0}$ & $(      0.15,      0.00)$ &        6.1 &       41.2 &        6.1 &       $-$4.2 \\
      & 349.1070 & $14_{1, -0} - 11_{0, +0}$ &  $(      0.20,     -0.09)$ &        6.2 &       32.0 &        4.5 &       $-$2.9 \\
      & 355.6029 & $13_{0, +0} - 11_{1, +0}$ &  $(      0.21,      0.05)$ &        5.9 &       27.6 &        4.4 &       $-$2.7 \\
      & 356.0072 & $15_{1, -0} - 15_{0, +0}$ &  $(      0.23,     -0.08)$ &        6.9 &       33.8 &        4.2 &       $-$4.0 \\
      & 360.6616 & $3_{1, +1} - 4_{2, +1}$ &  $(      0.36,      0.18)$ &        4.0 &       19.9 &        3.9 &       $-$4.4 \\
    E-CH$_3$OH & 338.1244 & $7_{0, 0} - 6_{0, 0}$ & $(      0.17,      0.04)$ &        6.2 &       29.2 &        4.1 &       $-$5.5 \\
     & 338.3446 & $7_{-1, 0} - 6_{-1, 0}$ & $(      0.18,      0.03)$ &        6.5 &       36.3 &        4.7 &       $-$5.1 \\
     & 360.8489 & $11_{0, 0} - 10_{1, 0}$ & $(      0.13,      0.12)$ &        5.5 &       26.8 &        4.2 &       $-$5.3 \\
\hline 
    H$_2$CO & 351.7686 & $5_{1, 5} - 4_{1, 4}$ & $(      0.01,     -0.16)$ &       10.6 &       64.8 &        5.7 &       $-$4.2 \\
\hline
\end{tabular}
\tablefoot{\tablefoottext{a}{The transitions listed here, as well as the transitions listed in Tables \ref{table:S26_Molecules_S_species} and \ref{table:S26_Molecules_N_SiO_HDO}, were relatively isolated and unblended, and were used to make the emission maps shown in Fig. \ref{Fig:S26_molecules}}
 \tablefoottext{b}{Typical uncertainties on the Gaussian fits are $(0\farcs01,0\farcs01)$ for the position, 0.3 $\mathrm{Jy \ beam^{-1}}$ for  $I_{\mathrm{peak}}$, 0.8 $\mathrm{Jy \ beam^{-1} \ km \ s^{-1}}$ for  $I_{\mathrm{int}}$, and 0.4 $\mathrm{km \ s^{-1}}$ for the FWHM and $\varv_{\text{source}}$.} }
\end{table*}


\begin{table*}
\caption{Gaussian fits for molecular transitions of S-bearing species, observed towards S26.} 
\label{table:S26_Molecules_S_species}      
\centering          
\begin{tabular}{l c c c c c c r}     
\hline
\hline 
    Molecule & Frequency & Transition & Position & $I_{\mathrm{peak}}$ & $I_{\mathrm{int}}$  & FWHM  & $\varv_{\text{source}}$ \\ 
     &  [GHz] & &  [$'',''$] & [$\mathrm{Jy \ beam^{-1}}$] &  [$\mathrm{Jy \ beam^{-1} \ km \ s^{-1}}$] & [$\mathrm{km \ s^{-1}}$] & [$\mathrm{km \ s^{-1}}$]\\ 
\hline
\hline
    \multicolumn{8}{c}{S26}\\
\hline
    CS & 342.8828 & $7_{0} - 6_{0}$ & $(      0.01,      0.07)$ &  11.0  &       59.9 &   5.2 & $-$6.4 \\
\hline
    C$^{34}$S & 337.3965 & $7_{0} - 6_{0}$ &  $(      0.02,     -0.10)$ &        5.4 &       25.7 &        4.5 &       $-$4.9 \\
\hline
   SO & 340.7141 & $8_{7} - 6_{ 6} $ & $(      0.04,     -0.10)$ &       10.7 &       73.1 &        6.4 &       $-$4.5 \\
    & 334.4311 & $8_{8} - 7_{7} $ & $(      0.03,     -0.11)$ &       13.0 &       89.1 &        6.5 &       $-$5.1 \\
    & 346.5281 & $8_{9} - 7_{8} $ & $(      0.08,     -0.16)$ &       12.8 &      120.3 &        9.5 &       $-$4.8 \\
\hline
   $^{34}$SO & 333.9010 & $8_{7} - 7_{ 6} $ & $(      0.10,     -0.09)$ &        5.4 &       28.3 &        4.6 &       $-$5.3 \\
    & 337.5801 & $8_{8} - 7_{7} $ & $(      0.04,     -0.22)$ &        6.6 &       39.2 &        5.3 &       $-$5.0 \\
    & 339.8573 & $8_{9} - 7_{8} $ & $(      0.10,     -0.05)$ &        5.6 &       31.0 &        4.8 &       $-$4.3 \\
\hline
   $^{33}$SO & 340.8379 & $8_{8,7.5} - 7_{8,6.5} $ & $(      0.35,     -0.07)$ &        2.1 &       11.3 &        4.8 &       $-$5.4 \\
    & 343.0881 & $8_{9,9.5} - 7_{8,8.5} $ & $(      0.18,      0.03)$ &        2.6 &       18.0 &        5.5 &       $-$5.1 \\
\hline
   SO$_2$ & 331.5802 & $11_{6, 6} - 12_{ 5, 7} $ & $(      0.16,     -0.17)$ &        4.7 &       23.3 &        4.1 &       $-$5.6 \\
    & 332.0914 & $21_{2, 20} - 21_{ 1, 21} $ & $(      0.10,     -0.17)$ &        7.4 &       39.4 &        4.9 &       $-$4.2 \\
    & 332.5052 & $4_{3, 1} -  3_{ 2, 2} $ & $(      0.07,     -0.19)$ &        9.0 &       51.2 &        5.3 &       $-$3.6 \\
    & 336.0892 & $23_{ 3, 21 } -  23_{ 2, 22}$ & $(      0.09,     -0.14)$ &        6.1 &       35.5 &        5.2 &       $-$4.5 \\
    & 336.6695 & $16_{ 7, 9 } -  17_{ 6, 12  }$ & $(      0.13,     -0.08)$ &        4.1 &       20.9 &        4.3 &       $-$5.2 \\
    & 338.3060 & $18_{ 4, 14 } - 18_{ 3, 15}$ & $(      0.11,     -0.09)$ &        8.6 &       50.9 &        5.3 &       $-$5.7 \\
    & 340.3164 & $28_{ 2, 26 } - 28_{ 1, 27}$ & $(      0.14,     -0.06)$ &        4.6 &       27.4 &        5.3 &       $-$4.4 \\
    & 341.2755 & $21_{ 8, 14 } -  22_{ 7, 15}  $ & $(      0.21,     -0.11)$ &        3.4 &       16.9 &        4.0 &       $-$6.0 \\
    & 345.4490 & $26_{ 9, 17 } - 27_{ 8, 20}  $ & $(      0.28,     -0.17)$ &        1.8 &       12.7 &        5.7 &       $-$7.6 \\
    & 346.6522 & $19_{ 1, 19 } -  18_{ 0, 18}  $ & $(      0.08,     -0.17)$ &       10.4 &       67.0 &        5.9 &       $-$6.3 \\
    & 350.8628 & $10_{ 6, 4 } -  11_{ 5, 7  } $ & $(      0.12,     -0.17)$ &        4.7 &       24.0 &        4.6 &       $-$6.7 \\
    & 355.0455 & $12_{ 4, 8 } - 12_{ 3, 9  }$ &  $(      0.04,     -0.12)$ &        9.2 &       54.1 &        5.5 &       $-$7.4 \\
    & 356.0406 & $15_{ 7, 9 } - 16_{ 6, 10 }$ &  $(      0.05,     -0.10)$ &        5.5 &       31.3 &        4.8 &       $-$5.4 \\
    & 357.1654 & $13_{ 4, 10 } -  13_{ 3, 11} $ & $(      0.04,     -0.12)$ &       10.0 &       62.4 &        5.7 &       $-$3.5 \\
    & 357.2412 & $15_{ 4, 12 } -  15_{ 3, 13}  $ & $(      0.05,     -0.12)$ &        9.8 &       63.3 &        5.9 &       $-$8.3 \\
    & 357.3876 & $11_{ 4, 8 } -  11_{ 3, 9}$  & $(      0.02,     -0.18)$ &       10.0 &       59.0 &        5.5 &       $-$5.6 \\
    & 357.5814 & $8_{ 4 ,4 } -  8_{ 3, 5 }$ & $(      0.04,     -0.18)$ &        9.2 &       60.9 &        6.0 &       $-$6.6 \\
    & 357.8924 & $7_{ 4, 4 } - 7_{ 3, 5  }$ & $(      0.01,     -0.14)$ &       10.7 &       66.6 &        5.8 &       $-$6.5 \\
    & 357.9258 & $6_{ 4, 2 } -  6_{ 3, 3 }$ & $(      0.05,     -0.16)$ &        9.7 &       60.5 &        5.5 &       $-$7.5 \\
    & 357.9629 & $17_{ 4, 14 } -  17_{ 3, 15} $ & $(      0.04,     -0.14)$ &        9.5 &       60.1 &        5.7 &       $-$7.1 \\
    & 358.0131 & $5_{ 4, 2 } -  5_{ 3, 3}$ & $(      0.09,     -0.12)$ &        9.8 &       61.4 &        5.7 &       $-$6.6 \\
    & 358.2156 & $20_{ 0, 20 } -  19_{ 1, 19} $ & $(      0.08,     -0.16)$ &       11.5 &       75.3 &        6.0 &       $-$7.0 \\
    & 359.7707 & $19_{ 4, 16 } -  19_{ 3, 17}$ & $(      0.07,     -0.10)$ &        9.1 &       58.8 &        5.6 &       $-$3.3 \\
    & 360.2904 & $34_{ 5, 29} -  34_{ 4,30}$ & $(      0.08,     -0.04)$ &        4.4 &       26.2 &        5.6 &       $-$7.3 \\
    & 360.7218  & $20_{ 8, 12} - 21_{ 7,15}$ & $(      0.17,     -0.15)$ &        3.6 &       19.8 &        4.7 &       $-$4.3 \\
\hline
   $^{34}$SO$_2$ & 330.6676 & $21_{2, 20} - 21_{1, 21}$  & $(      0.32,     -0.10)$ &        1.8 &        7.8 &        3.7 &       $-$7.0 \\
    & 332.8362 & $16_{4, 12} - 16_{3, 13}$ & $(      0.09,      0.05)$ &        2.3 &       12.8 &        4.8 &       $-$4.8 \\
    & 342.2089 & $5_{3, 3} - 4_{2, 2}$ & $(      0.12,     -0.00)$ &        2.5 &       11.3 &        3.9 &       $-$7.1 \\
    & 342.3320 & $12_{4, 8} - 12_{3, 9}$ & $(      0.31,      0.09)$ &        2.2 &       13.5 &        5.0 &       $-$6.4 \\
    & 344.2453 & $10_{4, 6} - 10_{3, 7}$ & $(      0.16,     -0.03)$ &        2.9 &       16.3 &        4.3 &       $-$6.8 \\
    & 344.5810 & $19_{1, 19} - 18_{0, 18}$ &  $(      0.15,     -0.08)$ &        4.5 &       24.6 &        4.7 &       $-$4.5 \\
    & 357.1022 & $20_{0, 20} - 19_{1, 19}$ &  $(      0.10,     -0.10)$ &        2.9 &       18.7 &        5.6 &       $-$7.1 \\
\hline
   OCS & 340.4493 & $ 28 - 27 $ & $(      0.08,     -0.12)$ &        3.6 &       19.4 &        4.7 &       $-$3.5 \\
    & 352.5996 & $ 29 - 28 $ & $(      0.16,     -0.14)$ &        4.5 &       25.6 &        5.1 &       $-$5.9 \\
\hline
   H$_2$CS & 338.0832 & $ 10_{ 1, 10} - 9_{ 1, 9} $  & $(      0.34,     -0.06)$ &        2.1 &       10.4 &        4.1 &        $-$6.2 \\
    & 342.9464 & $ 10_{ 0, 10} -  9_{ 0, 9} $ & $(      0.34,      0.01)$ &        1.8 &        7.8 &        3.7 &       $-$5.1 \\
\hline    
\end{tabular}
\tablefoot{Uncertainties are as in Table \ref{table:S26_Molecules_COMs}.}
\end{table*}


\begin{table*}
\caption{Gaussian fits for molecular transitions of N-bearing species, CO, HCO$^+$, SiO and HDO, observed towards S26.}             
\label{table:S26_Molecules_N_SiO_HDO}      
\centering          
\begin{tabular}{l c c c c c c r}     
\hline
\hline 
    Molecule & Frequency & Transition & Position & $I_{\mathrm{peak}}$ & $I_{\mathrm{int}}$  & FWHM  & $\varv_{\text{source}}$ \\ 
     &  [GHz] & &  [$'',''$] & [$\mathrm{Jy \ beam^{-1}}$] &  [$\mathrm{Jy \ beam^{-1} \ km \ s^{-1}}$] & [$\mathrm{km \ s^{-1}}$] & [$\mathrm{km \ s^{-1}}$]\\ 

\hline
   \multicolumn{8}{c}{S26} \\
\hline
   \multicolumn{8}{c}{N-bearing species} \\
\hline 
   CH$_3$CN & 330.9126 &  $18_{5} - 17_{5}$   &  $(      0.31,     -0.13)$ &        2.8 &       15.2 &        4.6 &       $-$6.3 \\
    & 330.9698 &  $18_{4} - 17_{4}$   &  $(      0.25,     -0.22)$ &        3.6 &       18.4 &        4.6 &       $-$5.9 \\
    & 331.0143 & $18_{3} - 17_{3}$    &  $(      0.15,     -0.08)$ &        5.9 &       31.1 &        4.3 &       $-$6.0 \\
    & 331.0461 &  $18_{2} - 17_{2}$   &  $(      0.18,     -0.15)$ &        5.2 &       25.4 &        4.4 &       $-$6.3 \\
    & 349.3933 & $19_{3} - 18_{3}$    &  $(      0.15,     -0.04)$ &        6.1 &       32.7 &        4.8 &       $-$6.9 \\
    & 349.4268 &  $19_{2} - 18_{2}$   &  $(      0.16,     -0.11)$ &        5.9 &       30.8 &        4.4 &       $-$6.2 \\
\hline
   HNCO  & 329.6644 & $15_{0,15} - 14_{0,14} $ &  $(      0.29,     -0.04)$ &        3.9 &       20.2 &        4.7 &       $-$6.4 \\
    &  350.3330 & $16_{1,16} - 15_{1,15} $ &  $(      0.18,      0.03)$ &        3.4 &       18.4 &        4.7 &       $-$8.5 \\
\hline
    HCN & 354.5055 & 4 - 3 & $(      0.00,      0.03)$ &       11.9 &      101.8 &        9.6 &       $-$5.3 \\
\hline
    HC$^{15}$N & 344.2001 & $4 - 3$ & $(      0.10,     -0.08)$ &        8.3 &       50.5 &        5.4 &       $-$4.0 \\
\hline
   HC$_3$N  & 336.5201 & $37 - 36 $ &  $(      0.15,     -0.04)$ &        3.0 &       17.3 &        5.1 &       $-$4.4 \\
    & 345.6090  & $38 - 37 $ &  $(      0.19,     -0.04)$ &        3.5 &       21.3 &        5.3 &       $-$7.1 \\
    & 354.6975 & $39 - 38 $ &  $(      0.05,     -0.27)$ &        2.7 &       16.6 &        5.5 &       $-$8.7 \\
\hline    
   SiO & 347.3306 & $8_{0} - 7_{0}$ &  $(      0.15,      0.10)$ &        1.5 &       13.8 &        9.6 &       $-$4.7 \\
\hline 
    C$^{18}$O & 329.33050 & $3 - 2$ & $(      0.54,     -0.45)$ &        4.0 &       18.7 &        5.5 &       $-$6.6 \\

\hline  
\end{tabular}
\tablefoot{Uncertainties are as in Table \ref{table:S26_Molecules_COMs}.}
\end{table*}


\begin{table*}
\caption{Molecular transitions of CH$_3$OH, SO and CH$_3$CN observed towards N12, used to produce the contour maps shown in Fig. \ref{Fig:N12_molecules}.}
\label{table:N12_Molecules_all}      
\centering          
\begin{tabular}{l c c c c c r}     
\hline
\hline 
    Molecule & Frequency & Transition  & $I_{\mathrm{peak}}$$^a$ & $I_{\mathrm{int}}$$^a$  & FWHM$^a$  & $\varv_{\text{source}}$$^a$ \\ 
     &  [GHz] & &  [$\mathrm{Jy \ beam^{-1}}$] &  [$\mathrm{Jy \ beam^{-1} \ km \ s^{-1}}$] & [$\mathrm{km \ s^{-1}}$] & [$\mathrm{km \ s^{-1}}$]\\ 
\hline
    A-CH$_3$OH & 331.5023 & $11_{1, -0} - 11_{0, +0}$ &    1.3 &        5.8 &        3.9 &       16.5 \\
     & 336.8651 & $12_{1, -0} - 12_{0, +0}$ &      1.1 &        4.7 &        4.2 &       18.1 \\
     & 341.4156 & $7_{1, -0} - 6_{1, -0}$ &       1.1 &        3.8 &        3.9 &       14.3 \\
     & 342.7298 & $13_{1, -0} - 13_{0, +0}$ &        0.9 &        3.0 &        4.6 &       15.1 \\
    E-CH$_3$OH & 338.1244 & $7_{0, 0} - 6_{0, 0}$ &       1.5 &        4.7 &        3.4 &       16.3 \\
     & 338.3446 & $7_{-1, 0} - 6_{-1, 0}$ &       1.6 &        5.7 &        3.9 &       17.1 \\
     & 338.6149  & $7_{1, 0} - 6_{1, 0}$ &      1.3 &        3.3 &        3.6 &       16.9 \\
     & 358.6058 & $4_{1, 0} - 3_{0, 0}$ &      1.0 &        4.6 &        5.0 &       16.5 \\
\hline 
   SO & 340.7141 & $8_{7} - 6_{ 6} $ &       1.1 &        9.5 &        6.6 &       17.8 \\
    & 344.3106 & $8_{8} - 7_{7} $ &      1.4 &       11.0 &        6.5 &       18.7 \\
    & 346.5281 & $8_{9} - 7_{8} $ &      1.6 &       13.4 &        6.7 &       15.4 \\
\hline     
   OCS & 340.4493 & $ 28 - 27 $  &   0.7 &        2.3 &        5.0 &       18.3 \\
    & 352.5996 & $ 29 - 28 $  &   0.8 &        4.5 &        7.9 &       15.5 \\
\hline  
   H$_2$CS & 338.0832 & $ 10_{ 1, 10} - 9_{ 1, 9} $   &   1.2 &    4.9 &        4.3 &       17.1 \\
    & 342.9464 & $ 10_{ 0, 10} -  9_{ 0, 9} $  &   0.6 &  2.1 &    3.9 &       15.3 \\
\hline 
   CH$_3$CN  & 331.0143 & $18_{3} - 17_{3}$    &        0.8 &        3.3 &        4.0 &       15.9 \\
    & 331.0461 &  $18_{2} - 17_{2}$  &        0.7 &        3.1 &        5.1 &       15.7 \\
    & 331.0652\tablefootmark{a} &  $18_{1} - 17_{1}$   &  0.7 &        6.3 &        9.7 &       13.7 \\
    & 331.0715 &  $18_{0} - 17_{0}$   &   - &       - &        - &       - \\
C$^{18}$O & 329.33050 & $3 - 2$ &   1.5 &   4.3 &        3.6 &       15.1 \\
\hline
\end{tabular}
\tablefoot{\tablefoottext{a}{The two CH$_3$CN transitions at 331.0652 and 331.0715 are blended and the table entries for these lines therefore represent the values of the two lines combined.}}
\end{table*}


\begin{table*}
\caption{Molecular transitions of CH$_3$OH, SO and H$_2$CS observed towards N53, used to produce the contour maps shown in Fig. \ref{Fig:N53_molecules}.}
\label{table:N53_Molecules_all}      
\centering          
\begin{tabular}{l c c c c c r}     
\hline
\hline 
    Molecule & Frequency & Transition  & $I_{\mathrm{peak}}$$^a$ & $I_{\mathrm{int}}$$^a$  & FWHM$^a$  & $\varv_{\text{source}}$$^a$ \\ 
     &  [GHz] & &  [$\mathrm{Jy \ beam^{-1}}$] &  [$\mathrm{Jy \ beam^{-1} \ km \ s^{-1}}$] & [$\mathrm{km \ s^{-1}}$] & [$\mathrm{km \ s^{-1}}$]\\ 
\hline
    A-CH$_3$OH & 331.5023 & $11_{1, -0} - 11_{0, +0}$  & 1.1 &   5.1 &        4.9 &  $-$3.8 \\
      & 335.5820 & $7_{1, +0} - 6_{1, +0}$  &    1.4 &   5.3 &  4.3 &       $-$3.1 \\
      & 336.8651 & $12_{1, -0} - 12_{0, +0}$  &   0.7 &   5.0 &    8.3 &       $-$2.0 \\
      & 338.4087 & $7_{0, +0} - 6_{0, +0}$  &   1.6 &   10.1 &        6.9 &       $-$4.6 \\

      & 341.4156 & $7_{1, -0} - 6_{1, -0}$ &   1.8 &        8.4 &        4.4 &       $-$5.3 \\
      & 349.1070 & $14_{1, -0} - 11_{0, +0}$  &   0.5 &        3.3 &        7.8 &       $-$5.8 \\
      & 355.6029 & $13_{0, +0} - 11_{1, +0}$ &  2.0 &  6.6 &        3.3 &       $-$2.3 \\
    E-CH$_3$OH & 338.1244 & $7_{0, 0} - 6_{0, 0}$  &  1.9 &   7.7 &    4.3 &       $-$3.7 \\
      & 338.3446 & $7_{-1, 0} - 6_{-1, 0}$ &   2.5 &        7.9 &        4.0 &       $-$4.2 \\
      & 338.7217  & $7_{2, 0} - 6_{2, 0}$ & 2.7 &   8.7 &   3.9 &       $-$5.1 \\
\hline 
   SO & 340.7141 & $8_{7} - 6_{ 6} $   &        1.7 &        7.6 &        4.9 &       $-$5.5 \\
    & 334.4311 & $8_{8} - 7_{7} $   &        1.7 &        9.0 &        6.1 &       $-$3.5 \\
    & 346.5281 & $8_{9} - 7_{8} $   &        2.0 &       10.5 &        5.9 &       $-$6.4 \\
\hline
   H$_2$CS & 338.0832 & $ 10_{ 1, 10} - 9_{ 1, 9} $    &    1.3 &        3.9 &        4.2 &       $-$5.0 \\
    & 342.9464 & $ 10_{ 0, 10} -  9_{ 0, 9} $   &    0.5 &     1.7 &        5.6 &       $-$5.3 \\
    & 348.5344 & $ 10_{ 1, 9} - 9_{ 1, 8} $   &   0.7 &   3.2 &    4.9 &       $-$2.2 \\
\hline
\end{tabular}
\end{table*}


\begin{table*}
\caption{Molecular transitions of  $^{13}$CO, CS and H$_2$CO observed towards N54, used to produce the contour maps shown in Fig. \ref{Fig:N54_molecules}.}
\label{table:N54_Molecules_all}      
\centering          
\begin{tabular}{l c c c c c r}     
\hline
\hline 
    Molecule & Frequency & Transition  & $I_{\mathrm{peak}}$$^a$ & $I_{\mathrm{int}}$$^a$  & FWHM$^a$  & $\varv_{\text{source}}$$^a$ \\ 
     &  [GHz] & &  [$\mathrm{Jy \ beam^{-1}}$] &  [$\mathrm{Jy \ beam^{-1} \ km \ s^{-1}}$] & [$\mathrm{km \ s^{-1}}$] & [$\mathrm{km \ s^{-1}}$]\\ 
\hline
   $^{13}$CO & 330.5880 & $3 - 2$ &        1.8 &        4.9 &        2.2 &       $-$9.1 \\
\hline 
    CS & 342.8828 & $7_{0} - 6_{0}$ &    1.7 &  7.0 &        3.6 &       $-$5.9 \\
\hline
    H$_2$CO & 351.7686 & $5_{1, 5} - 4_{1, 4}$ &       1.7 &        6.0 &        3.0 &       $-$4.4 \\
\hline
\end{tabular}
\end{table*}


\begin{table*}
\caption{Molecular transitions of CH$_3$OH, SO and CS observed towards N51, used to produce the contour maps shown in Fig. \ref{Fig:N51_molecules}.}
\label{table:N51_Molecules_all}      
\centering          
\begin{tabular}{l c c c c c r}     
\hline
\hline 
    Molecule & Frequency & Transition  & $I_{\mathrm{peak}}$$^a$ & $I_{\mathrm{int}}$$^a$  & FWHM$^a$  & $\varv_{\text{source}}$$^a$ \\ 
     &  [GHz] & &  [$\mathrm{Jy \ beam^{-1}}$] &  [$\mathrm{Jy \ beam^{-1} \ km \ s^{-1}}$] & [$\mathrm{km \ s^{-1}}$] & [$\mathrm{km \ s^{-1}}$]\\ 
\hline
    A-CH$_3$OH & 331.5023 & $11_{1, -0} - 11_{0, +0}$ &       1.5 &        3.8 &        2.6 &       $-$2.2 \\
      & 335.5820 & $7_{1, +0} - 6_{1, +0}$ &     0.5 &        1.9 &        6.7 &       $-$3.1 \\
      & 336.8651 & $12_{1, -0} - 12_{0, +0}$ &        1.2 &        3.5 &        2.5 &       $-$2.7 \\
      & 338.5129 & $7_{2, -0} - 6_{2, -0}$ &        2.2 &        5.9 &        2.2 &       $-$3.8 \\
     & 341.4156 & $7_{1, -0} - 6_{1, -0}$ &         1.0 &        3.5 &        4.5 &       $-$5.4 \\
\hline 
   SO & 340.7141 & $8_{7} - 6_{ 6} $ &    0.8 &        6.4 &        7.2 &       $-$4.5 \\
    & 334.4311 & $8_{8} - 7_{7} $  &        1.4 &        7.1 &        4.8 &       $-$4.1 \\
    & 346.5281 & $8_{9} - 7_{8} $  &        1.1 &        9.4 &        9.1 &       $-$6.2 \\
\hline
    CS & 342.8828 & $7_{0} - 6_{0}$  &   2.4 &   12.9 &        4.4 &        1.2 \\
\hline
\end{tabular}
\end{table*}


\begin{table*}
\caption{Molecular transitions of CH$_3$OH, SO and CS observed towards N38, used to produce the contour maps shown in Fig. \ref{Fig:N38_molecules}.}
\label{table:N38_Molecules_all}      
\centering          
\begin{tabular}{l c c c c c r}     
\hline
\hline 
    Molecule & Frequency & Transition  & $I_{\mathrm{peak}}$$^a$ & $I_{\mathrm{int}}$$^a$  & FWHM$^a$  & $\varv_{\text{source}}$$^a$ \\ 
     &  [GHz] & &  [$\mathrm{Jy \ beam^{-1}}$] &  [$\mathrm{Jy \ beam^{-1} \ km \ s^{-1}}$] & [$\mathrm{km \ s^{-1}}$] & [$\mathrm{km \ s^{-1}}$]\\ 
\hline
    A-CH$_3$OH & 335.5820 & $7_{1, +0} - 6_{1, +0}$ &        0.7 &        2.2 &        4.6 &        0.2 \\
      & 336.8651 & $12_{1, -0} - 12_{0, +0}$ &          0.4 &        2.1 &        7.6 &       $-$4.3 \\
      & 338.4087 & $7_{0, +0} - 6_{0, +0}$ &         1.4 &        6.0 &        5.2 &       $-$3.5 \\
      & 341.4156 & $7_{1, -0} - 6_{1, -0}$ &        0.9 &        4.0 &        5.7 &       $-$2.7 \\
      & 342.7298 & $13_{1, -0} - 13_{0, +0}$ &          0.5 &        3.0 &        3.8 &       $-$0.8 \\
    E-CH$_3$OH & 338.1244 & $7_{0, 0} - 6_{0, 0}$ &         0.9 &        1.8 &        5.3 &        0.5 \\
      & 338.3446 & $7_{-1, 0} - 6_{-1, 0}$ &         1.2 &        5.7 &        5.6 &       $-$3.0 \\
      & 338.6149  & $7_{1, 0} - 6_{1, 0}$ &         0.7 &        3.4 &        4.4 &        0.2 \\
      & 338.7217  & $7_{2, 0} - 6_{2, 0}$ &         1.2 &        4.8 &        4.4 &       $-$2.1 \\
      & 358.6058 & $4_{1, 0} - 3_{0, 0}$ &          0.6 &        3.7 &        6.8 &       $-$0.6 \\
\hline 
   SO & 340.7141 & $8_{7} - 6_{ 6} $ &        0.3 &        1.5 &       12.6 &        1.1 \\
    & 334.4311 & $8_{8} - 7_{7} $ &       0.6 &        3.4 &       12.4 &        1.3 \\
    & 346.5281 & $8_{9} - 7_{8} $ &     0.7 &        4.0 &       13.7 &        0.5 \\
\hline
    CS & 342.8828 & $7_{0} - 6_{0}$ &  1.9 &        7.5 &        4.1 &        0.2 \\
\hline
\end{tabular}
\end{table*}


\begin{table*}
\caption{Molecular transitions of CH$_3$OH and SO observed towards N48, used to produce the contour maps shown in Fig. \ref{Fig:N48_molecules}.}
\label{table:N48_Molecules_all}      
\centering          
\begin{tabular}{l c c c c c r}     
\hline
\hline 
    Molecule & Frequency & Transition  & $I_{\mathrm{peak}}$$^a$ & $I_{\mathrm{int}}$$^a$  & FWHM$^a$  & $\varv_{\text{source}}$$^a$ \\ 
     &  [GHz] & &  [$\mathrm{Jy \ beam^{-1}}$] &  [$\mathrm{Jy \ beam^{-1} \ km \ s^{-1}}$] & [$\mathrm{km \ s^{-1}}$] & [$\mathrm{km \ s^{-1}}$]\\ 
\hline
    A-CH$_3$OH & 331.5023 & $11_{1, -0} - 11_{0, +0}$   &        0.5 &        0.5 &        6.2 &       $-$2.1 \\
      & 335.5820 & $7_{1, +0} - 6_{1, +0}$   &        0.8 &        2.6 &        4.9 &       $-$4.8 \\
      & 338.4087 & $7_{0, +0} - 6_{0, +0}$   &        2.5 &        8.7 &        5.3 &       $-$8.4 \\
      & 341.4156 & $7_{1, -0} - 6_{1, -0}$   &        1.3 &        4.1 &       -4.0 &      $-$10.3 \\
    E-CH$_3$OH & 338.1244 & $7_{0, 0} - 6_{0, 0}$  &        0.7 &        5.6 &       10.3 &       $-$7.8 \\
      & 338.3446 & $7_{-1, 0} - 6_{-1, 0}$   &        2.3 &        7.4 &        4.8 &       $-$8.1 \\
      & 338.7217  & $7_{2, 0} - 6_{2, 0}$   &        1.0 &        5.2 &        4.6 &       $-$4.4 \\
      & 358.6058 & $4_{1, 0} - 3_{0, 0}$   &        0.5 &        2.9 &       10.7 &       $-$7.9 \\
\hline 
   SO & 340.7141 & $8_{7} - 6_{ 6}$  &        0.6 &        5.4 &        6.8 &       $-$5.3 \\
    & 334.4311 & $8_{8} - 7_{7} $   &        0.6 &        4.2 &        8.0 &       $-$5.8 \\
    & 346.5281 & $8_{9} - 7_{8} $   &        0.8 &        5.9 &        8.0 &       $-$8.0 \\
\hline
\end{tabular}
\end{table*}


\begin{table*}
\caption{Molecular transitions of CH$_3$OH and SO observed towards S8, used to produce the contour maps shown in Fig. \ref{Fig:S8_molecules}.}
\label{table:S8_Molecules_all}      
\centering          
\begin{tabular}{l c c c c c r}     
\hline
\hline 
    Molecule & Frequency & Transition  & $I_{\mathrm{peak}}$$^a$ & $I_{\mathrm{int}}$$^a$  & FWHM$^a$  & $\varv_{\text{source}}$$^a$ \\ 
     &  [GHz] & &  [$\mathrm{Jy \ beam^{-1}}$] &  [$\mathrm{Jy \ beam^{-1} \ km \ s^{-1}}$] & [$\mathrm{km \ s^{-1}}$] & [$\mathrm{km \ s^{-1}}$]\\ 
\hline
    A-CH$_3$OH & 331.5023 & $11_{1, -0} - 11_{0, +0}$   &        0.5 &        0.5 &        6.2 &       $-$2.1 \\
      & 335.5820 & $7_{1, +0} - 6_{1, +0}$   &        0.8 &        2.6 &        4.9 &       $-$4.8 \\
      & 338.4087 & $7_{0, +0} - 6_{0, +0}$   &        2.5 &        8.7 &        5.3 &       $-$8.4 \\
      & 341.4156 & $7_{1, -0} - 6_{1, -0}$   &        1.3 &        4.1 &       -4.0 &      $-$10.3 \\
    E-CH$_3$OH & 338.1244 & $7_{0, 0} - 6_{0, 0}$  &        0.7 &        5.6 &       10.3 &       $-$7.8 \\
      & 338.3446 & $7_{-1, 0} - 6_{-1, 0}$   &        2.3 &        7.4 &        4.8 &       $-$8.1 \\
      & 338.7217  & $7_{2, 0} - 6_{2, 0}$   &        1.0 &        5.2 &        4.6 &       $-$4.4 \\
      & 358.6058 & $4_{1, 0} - 3_{0, 0}$   &        0.5 &        2.9 &       10.7 &       $-$7.9 \\
\hline 
   SO & 340.7141 & $8_{7} - 6_{ 6}$  &        0.6 &        5.4 &        6.8 &       $-$5.3 \\
    & 334.4311 & $8_{8} - 7_{7} $   &        0.6 &        4.2 &        8.0 &       $-$5.8 \\
    & 346.5281 & $8_{9} - 7_{8} $   &        0.8 &        5.9 &        8.0 &       $-$8.0 \\
\hline
\end{tabular}
\end{table*}


\section{Source sample column densities}

\label{App:Column_densities_tables}
\begin{figure*}
    \centering
    \includegraphics[width=17.2cm]{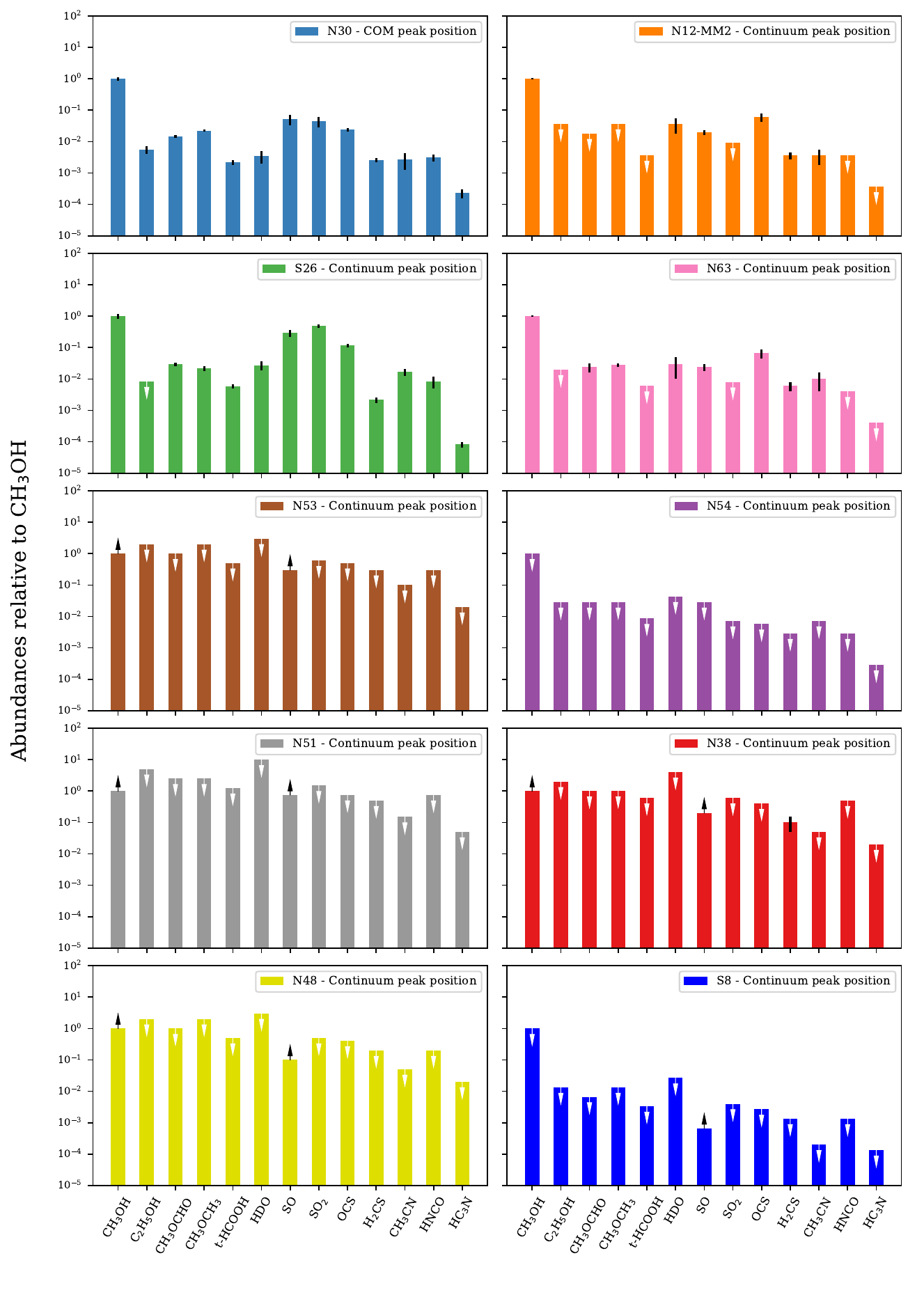}
    \caption{Relative abundances to CH$_3$OH for the source sample. White and black arrows represent upper and lower limits respectively, while black bars represent uncertainties.}
    \label{Fig:column_densities_fig}
\end{figure*}

The derived values for the source sample obtained from synthetic spectrum fitting using the software package CASSIS are presented in this appendix, in Tables \ref{table:S26_column_densities} -- \ref{table:S8_column_densities} (see Table \ref{table:N63_column_densities} for values obtained for N63). The parameters used in the fitting were the column density ($N$), excitation temperature ($T_\mathrm{ex}$), source velocity ($\varv_\mathrm{peak}$) and line width (FWHM). These column densities are shown relative to CH$_3$OH in Fig. \ref{Fig:column_densities_fig}.

%

\begin{table*}
\caption{Column densities, excitation temperatures, FWHM and $\varv_{\text{peak}}$ derived at the continuum peak position of S26\tablefootmark{a} }             
\label{table:S26_column_densities}      
\centering          
\begin{tabular}{l c c c c }    
\hline  \hline 
    Molecule & Column density & $T_\mathrm{ex}$& FWHM & $\varv_{\text{peak}}$   \\ 
     & [$\mathrm{cm^{-2} }$]& [K] & [$\mathrm{km \ s^{-1} }$] & [$\mathrm{km \ s^{-1} }$]  \\
\hline
   \multicolumn{5}{c}{COMs and O-bearing species} \\
\hline   
    CH$_3$OH\tablefootmark{b}    & 1.2 (0.2)\tablefootmark{c} $\times 10^{18}$ & \hspace{0.6cm} 100 (25)\tablefootmark{c} \hspace{0.6cm} & \hspace{1cm} 4.0 \hspace{1cm}  & \hspace{1cm} $-$5.5 \hspace{1cm}  \\
    $^{13}$CH$_3$OH   & 1.8 (0.3) $\times 10^{16}$ & 100 (25) & 4.0 & $-$5.5  \\
    CH$_3$$^{18}$OH\tablefootmark{c} & <1.0 $\times 10^{16}$ & 100 & 4.0 & $-$5.5  \\
    CH$_2$DOH\tablefootmark{c}  & <1.0 $\times 10^{16}$ & 100 & 4.0 & $-$5.5  \\
    C$_2$H$_5$OH\tablefootmark{c}      & <1.5 $\times 10^{16}$ & 100 & 4.0 & $-$5.5  \\ 
    CH$_3$OCHO        & 3.5 (0.5) $\times 10^{16}$ & 100 (25) & 4.0 & $-$5.5  \\ 
    CH$_3$OCH$_3$     & 2.6 (0.5) $\times 10^{16}$ & 100 (25) & 4.0 & $-$5.5  \\ 
    t-HCOOH            & 7.0 (1.0) $\times 10^{15}$ & 190 (60) & 4.0 & $-$5.5 \\
    c-HCOOH\tablefootmark{c}     & <1.0 $\times 10^{15}$ & 190  & 4.0 & $-$5.5  \\
    HDO\tablefootmark{d}         & 3.3 (1.0) $\times 10^{16}$ & 100 (25) & 4.0 & $-$5.5   \\
\hline 
\multicolumn{5}{c}{S-bearing species} \\
\hline 
    SO\tablefootmark{b}   & 3.5 (0.9) $\times 10^{17}$ & 120 (40) & 4.0 & $-$5.0  \\
    $^{33}$SO   & 3.6 (1.0) $\times 10^{15}$ & 120 (40) & 4.0 & -5.0  \\
    $^{34}$SO   & 1.6 (0.4) $\times 10^{16}$ & 120 (20) & 3.5 & -5.0  \\
    SO$_{2}$\tablefootmark{b}    & 5.9 (0.7) $\times 10^{17}$ & 120 (20) & 4.0 & $-$5.0  \\
    $^{33}$SO$_{2}$\tablefootmark{c}   & <7.0 $\times 10^{15}$ & 120 & 4.0 & $-$5.0  \\
    $^{34}$SO$_{2}$   & 2.6 (0.3) $\times 10^{16}$ & 120 (20) & 4.0 & $-$5.0 \\ 
    OCS\tablefootmark{b}  & 1.4 (0.2) $\times 10^{17}$ & 100 (50) & 4.0 & $-$4.5 \\
    OC$^{34}$S         & 6.4 (1.0)$\times 10^{15}$ & 100 (50) & 4.0 & $-$4.5  \\
    H$_2$CS           &  2.6 (0.5) $\times 10^{15}$ & 130 (40) & 4.0 & $-$5.0  \\
\hline 
\multicolumn{5}{c}{N-bearing species} \\
\hline 
    CH$_3$CN, v$_8$=1  &  2.0 (0.5)$\times 10^{16}$ & 140 (20) & 4.5 & $-$5.0  \\ 
    $^{13}$CH$_3$CN\tablefootmark{e}  &  < 5.0 $\times 10^{14}$ &  120  & 5.0 & $-$5.5   \\ 
    HNCO    &  1.0 (0.4)$\times 10^{16}$ & 140 (30) & 4.5 & $-$5.5  \\
    HC$_3$N    &  1.0 (0.2)$\times 10^{14}$ & 140 (20) & 4.5 & $-$5.5 \\
\hline 

\end{tabular}
\tablefoot{\tablefoottext{a}{Values are derived from synthetic spectrum fitting with CASSIS, with the assumption of LTE.}
\tablefoottext{b}{Column densities for optically thick lines derived using local ISM ratios \citep[][]{Wilson-1999,Milam-2005}}
\tablefoottext{c}{Uncertainties on $N$ and $T_\mathrm{ex}$ are in parenthesis next to the derived values.}
\tablefoottext{d}{Upper limits were determined by setting $T_\mathrm{ex}$, FWHM, and $\varv_{\text{peak}}$, equal to that derived for CH$_3$OH and SO, for the O- and S-bearing species respectively.}
\tablefoottext{e}{To estimate a column density for HDO, the excitation temperature and line width of methanol was assumed.}
}

\end{table*}

%

\begin{table*}
\caption{Column densities, excitation temperatures, FWHM and $\varv_{\text{peak}}$ derived at the continuum peak position of N12-MM2\tablefootmark{a} }             
\label{table:N12_column_densities}      
\centering          
\begin{tabular}{l c c c c }    
\hline  \hline 
    Molecule & Column density & $T_\mathrm{ex}$& FWHM & $\varv_{\text{peak}}$   \\ 
     & [$\mathrm{cm^{-2} }$]& [K] & [$\mathrm{km \ s^{-1} }$] & [$\mathrm{km \ s^{-1} }$]  \\
\hline
   \multicolumn{5}{c}{COMs and O-bearing species} \\
\hline   
    CH$_3$OH\tablefootmark{b}    & 5.5 (0.2) $\times 10^{17}$ & \hspace{0.6cm} 150 (50) \hspace{0.6cm} & \hspace{1cm} 3.5 \hspace{1cm}  & \hspace{1cm} 16.0 \hspace{1cm}  \\
    $^{13}$CH$_3$OH\tablefootmark{c}   & 8.0 (0.3) $\times 10^{15}$ & 150 (50) & 3.5 & 16.0  \\
    CH$_3$$^{18}$OH & <1.0 $\times 10^{16}$ & 150  & 3.5 & 16.0   \\
    CH$_2$DOH  & <2.0 $\times 10^{16}$ & 150  & 3.5 & 16.0  \\
    C$_2$H$_5$OH     & <2.0 $\times 10^{16}$ & 150  & 3.5 & 16.0    \\ 
    CH$_3$OCHO        & <1.0 $\times 10^{16}$ & 150  & 3.5 & 16.0   \\ 
    CH$_3$OCH$_3$     & <2.0 $\times 10^{16}$ & 150  & 3.5 & 16.0  \\ 
    t-HCOOH            & <2.0  $\times 10^{15}$ & 150  & 3.5 & 16.0 \\
    c-HCOOH            & <2.0 $\times 10^{14}$ & 150  & 3.5 & 16.0 \\
    HDO\tablefootmark{d}         & 2.0 (1.0) $\times 10^{16}$ & 150  & 3.5 & 16.0   \\
\hline 
\multicolumn{5}{c}{S-bearing species} \\
\hline 
    SO\tablefootmark{b}  & 1.1 (0.2) $\times 10^{16}$ & 250 (50) & 3.0 & 15.0 \\
    $^{33}$SO\tablefootmark{c}   & <5.0  $\times 10^{14}$ & 250  & 3.0 & 15.0 \\
    $^{34}$SO   & 5.0 (1.0) $\times 10^{14}$ & 250 (50) & 3.0 & 15.0  \\
    SO$_{2}$\tablefootmark{b}    & >5.0  $\times 10^{15}$ & 250  & 3.0 & 15.0  \\
    $^{33}$SO$_{2}$   & <5.0 $\times 10^{15}$ & 250  & 3.0 & 15.0  \\
    $^{34}$SO$_{2}$   & <5.0  $\times 10^{15}$ & 250  & 3.0 & 15.0 \\ 
    OCS\tablefootmark{b}        & 3.3 (1.0) $\times 10^{16}$ & 200 (100) & 4.0 & 16.0 \\
    OC$^{34}$S   & 1.5 (0.5) $\times 10^{15}$ & 200 (100) & 4.0 & 16.0  \\
    H$_2$CS       &  2.0 (0.5) $\times 10^{15}$ & 100 (50) & 4.0 & 16.0 \\
\hline 
\multicolumn{5}{c}{N-bearing species} \\
\hline 
    CH$_3$CN, v$_8$=1  &  2.0 (1.0)$\times 10^{15}$ & 250 (50) & 4.0 & 15.5  \\ 
    $^{13}$CH$_3$CN\tablefootmark{d}  &  < 2.0 $\times 10^{14}$ & 250 & 4.0  & 15.5 \\
    HNCO    &  <2.0 $\times 10^{15}$ &  250  & 4.0  & 15.5  \\
    HC$_3$N    &  <2.0 $\times 10^{14}$  & 250 & 4.0 & 15.5 \\
    HC$^{15}$N    &  1.0 (0.3) $\times 10^{14}$ & 250 (50) & 3.5 & 16.0 \\
\hline 

\end{tabular}
\tablefoot{\tablefoottext{a}{Values are derived from synthetic spectrum fitting with CASSIS, with the assumption of LTE.}
\tablefoottext{b}{Column densities for optically thick lines derived using local ISM ratios \citep[][]{Wilson-1999,Milam-2005}. In the case of SO$_2$ only a lower limit is given since $^{34}$SO$_2$ was not detected.}
\tablefoottext{c}{Upper limits were determined by setting $T_\mathrm{ex}$, FWHM, and $\varv_{\text{peak}}$, equal to that derived for CH$_3$OH and SO, for the O- and S-bearing species respectively.}
\tablefoottext{d}{To estimate a column density for HDO, the excitation temperature and line width of methanol was assumed.}
}

\end{table*}

%

\begin{table*}
\caption{Column densities, excitation temperatures, FWHM and $\varv_{\text{peak}}$ derived at the continuum peak position of N53\tablefootmark{a} }             
\label{table:N53_column_densities}      
\centering          
\begin{tabular}{l c c c c }    
\hline  \hline 
    Molecule & Column density & $T_\mathrm{ex}$& FWHM & $\varv_{\text{peak}}$   \\ 
     & [$\mathrm{cm^{-2} }$]& [K] & [$\mathrm{km \ s^{-1} }$] & [$\mathrm{km \ s^{-1} }$]  \\
\hline
   \multicolumn{5}{c}{COMs and O-bearing species} \\
\hline  
    CH$_3$OH  & >1.0 $\times 10^{16}$ & \hspace{0.6cm} 150  \hspace{0.6cm} & \hspace{1cm} 4.5 \hspace{1cm}  & \hspace{1cm} $-$5.0 \hspace{1cm}  \\
    $^{13}$CH$_3$OH   & <5.0  $\times 10^{15}$ & 150  & 4.5 & $-$5.0  \\
    CH$_3$$^{18}$OH & <1.0 $\times 10^{16}$ & 150  & 4.5 & $-$5.0    \\
    CH$_2$DOH  & <1.0 $\times 10^{16}$ & 100  & 4.5 & $-$5.0  \\
    C$_2$H$_5$OH     & <2.0 $\times 10^{16}$ & 150  & 4.5 & $-$5.0    \\ 
    CH$_3$OCHO        & <1.0 $\times 10^{16}$ & 150  & 4.5 & $-$5.0   \\ 
    CH$_3$OCH$_3$     & <2.0 $\times 10^{16}$ & 150  & 4.5 & $-$5.0  \\ 
    t-HCOOH            & <5.0  $\times 10^{15}$ & 150  & 4.5 & $-$5.0 \\
    c-HCOOH            & <1.0 $\times 10^{15}$ & 150  & 4.5 & $-$5.0 \\
    HDO            & <3.0 $\times 10^{16}$ & 150  & 4.5 & $-$5.0  \\
\hline 
\multicolumn{5}{c}{S-bearing species} \\
\hline 
    SO          & >3.0  $\times 10^{15}$ & 200  & 4.5 & $-$4.5  \\
    $^{33}$SO   & <5.0  $\times 10^{14}$ & 200  & 4.5 & $-$4.5  \\
    $^{34}$SO   & <5.0 $\times 10^{14}$ & 200  & 4.5 & $-$4.5  \\
    SO$_{2}$    & <6.0  $\times 10^{15}$ & 200  & 4.5 & $-$4.5 \\
    $^{33}$SO$_{2}$   & <3.0 $\times 10^{15}$ & 200  & 4.5 & $-$4.5   \\
    $^{34}$SO$_{2}$   & <3.0  $\times 10^{15}$ & 200  & 4.5 & $-$4.5 \\ 
    OCS        & >5.0 $\times 10^{15}$ & 200  & 4.5 & $-$5.5 \\
    OC$^{34}$S   & <3.0 $\times 10^{15}$ & 200  & 4.5 & $-$4.5  \\
    H$_2$CS       &  <3.0 $\times 10^{15}$ & 200  & 4.5 & $-$4.5 \\
\hline 
\multicolumn{5}{c}{N-bearing species} \\
\hline 
    CH$_3$CN   &  <1.0 $\times 10^{15}$ & 200  & 4.5 & $-$4.5 \\ 
    $^{13}$CH$_3$CN &  <2.0 $\times 10^{14}$ & 200  & 4.5 & $-$4.5 \\
    HNCO    &  <3.0 $\times 10^{15}$ &  200  & 4.5 & $-$4.5  \\
    HC$_3$N    &  <2.0 $\times 10^{14}$  & 200  & 4.5 & $-$4.5 \\
    HC$^{15}$N    &  <1.0 $\times 10^{14}$ & 200  & 4.5 & $-$4.5 \\
\hline 

\end{tabular}
\tablefoot{\tablefoottext{a}{Values are derived from synthetic spectrum fitting with CASSIS, with the assumption of LTE. Only CH$_3$OH and SO are detected, but with no isotopologues detected, only lower limits are derived. Upper limits are derived for all other molecules.}
}
\end{table*}

%

\begin{table*}
\caption{Column densities, excitation temperatures, FWHM and $\varv_{\text{peak}}$ derived at the continuum peak position of N54\tablefootmark{a} }             
\label{table:N54_column_densities}      
\centering          
\begin{tabular}{l c c c c }    
\hline  \hline 
    Molecule & Column density & $T_\mathrm{ex}$& FWHM & $\varv_{\text{peak}}$   \\ 
     & [$\mathrm{cm^{-2} }$]& [K] & [$\mathrm{km \ s^{-1} }$] & [$\mathrm{km \ s^{-1} }$]  \\
\hline
   \multicolumn{5}{c}{COMs and O-bearing species} \\
\hline   
    CH$_3$OH\tablefootmark{b}  & <7.0  $\times 10^{17}$ & \hspace{0.6cm} 150  \hspace{0.6cm} & \hspace{1cm} 4.0 \hspace{1cm}  & \hspace{1cm} $-$5.5 \hspace{1cm}  \\
    $^{13}$CH$_3$OH\tablefootmark{b}   & <1.0  $\times 10^{16}$ & 150  & 4.0 & $-$5.5  \\
    CH$_3$$^{18}$OH & <2.0 $\times 10^{16}$ & 150  & 4.0 & $-$5.5   \\
    CH$_2$DOH  & <2.0 $\times 10^{16}$ & 150  & 4.0 & $-$5.5  \\
    C$_2$H$_5$OH     & <2.0 $\times 10^{16}$ & 150  & 4.0 & $-$5.5  \\ 
    CH$_3$OCHO        & <2.0 $\times 10^{16}$ & 150  & 4.0 & $-$5.5  \\ 
    CH$_3$OCH$_3$     & <2.0 $\times 10^{16}$ & 150  & 4.0 & $-$5.5 \\ 
    t-HCOOH            & <6.0  $\times 10^{15}$ & 150  & 4.0 & $-$5.5 \\
    c-HCOOH            & <2.0 $\times 10^{15}$ & 150  & 4.0 & $-$5.5\\
    HDO            & <3.0 $\times 10^{16}$ & 150  & 4.0 & $-$5.5  \\
\hline 
\multicolumn{5}{c}{S-bearing species} \\
\hline 
    SO         & <2.0  $\times 10^{16}$ & 200  & 4.5 & $-$5.5  \\
    $^{33}$SO   & <5.0  $\times 10^{14}$ & 200  & 4.5 & $-$5.5  \\
    $^{34}$SO   & <8.0 $\times 10^{14}$ & 200  & 4.5 & $-$5.5  \\
    SO$_{2}$    & <5.0  $\times 10^{15}$ & 200  & 4.5 & $-$5.5 \\
    $^{33}$SO$_{2}$   & <4.0 $\times 10^{15}$ & 200  & 4.5 & $-$5.5   \\
    $^{34}$SO$_{2}$   & <4.0  $\times 10^{15}$ & 200  & 4.5 & $-$5.5 \\ 
    OCS        & <4.0 $\times 10^{15}$ & 200  & 4.5 & $-$5.5 \\
    OC$^{34}$S   & <4.0 $\times 10^{15}$ & 200  & 4.5 & $-$5.5  \\
    H$_2$CS       &  <2.0 $\times 10^{15}$ & 200  & 4.5 & $-$5.5 \\
\hline 
\multicolumn{5}{c}{N-bearing species} \\
\hline 
    CH$_3$CN   &  <5.0 $\times 10^{15}$ & 200  & 4.5 & $-$5.5   \\ 
    $^{13}$CH$_3$CN &  <3.0 $\times 10^{14}$ & 200  & 4.5 & $-$5.5  \\
    HNCO    &  <2.0 $\times 10^{15}$ & 200  & 4.5 & $-$5.5  \\
    HC$_3$N    &  <2.0 $\times 10^{14}$  & 200  & 4.5 & $-$5.5   \\
    HC$^{15}$N    &  <1.0 $\times 10^{14}$ & 200  & 4.5 & $-$5.5 \\
\hline 
\end{tabular}
\tablefoot{\tablefoottext{a}{Values are derived from synthetic spectrum fitting with CASSIS, with the assumption of LTE.}
\tablefoottext{b}{No COMs were detected, and only upper limits are given. In the case of CH$_3$OH, and SO, the upper limits given are derived from the $^{13}$C and ${34}$S isotopologues, respectively. We assumed T$_\mathrm{ex}$ of 150 K for the O-bearing species, and 200 K for S- and N-bearing species, which is an average of those derived for N30, N12, N63 and S26.}
}

\end{table*}

%

\begin{table*}
\caption{Column densities, excitation temperatures, FWHM and $\varv_{\text{peak}}$ derived at the continuum peak position of N51\tablefootmark{a} }             
\label{table:N51_column_densities}      
\centering          
\begin{tabular}{l c c c c }    
\hline  \hline 
    Molecule & Column density & $T_\mathrm{ex}$& FWHM & $\varv_{\text{peak}}$   \\ 
     & [$\mathrm{cm^{-2} }$]& [K] & [$\mathrm{km \ s^{-1} }$] & [$\mathrm{km \ s^{-1} }$]  \\
\hline
   \multicolumn{5}{c}{COMs and O-bearing species} \\
\hline   
    CH$_3$OH  & >4.0  $\times 10^{15}$ & \hspace{0.6cm} 150  \hspace{0.6cm} & \hspace{1cm} 4.0 \hspace{1cm}  & \hspace{1cm} $-$4.5 \hspace{1cm}  \\
    $^{13}$CH$_3$OH   & <5.0  $\times 10^{15}$ & 150  & 4.0 & $-$4.5  \\
    CH$_3$$^{18}$OH & <1.0 $\times 10^{16}$ & 150  & 4.0 & $-$4.5   \\
    CH$_2$DOH  & <1.0 $\times 10^{16}$ & 150  & 4.0 & $-$4.5 \\
    C$_2$H$_5$OH     & <2.0 $\times 10^{16}$ & 150  & 4.0 & $-$4.5   \\ 
    CH$_3$OCHO        & <1.0 $\times 10^{16}$ & 150  & 4.0 & $-$4.5  \\ 
    CH$_3$OCH$_3$     & <1.0 $\times 10^{16}$ & 150  & 4.0 & $-$4.5 \\ 
    t-HCOOH            & <5.0  $\times 10^{15}$ & 150  & 4.0 & $-$4.5\\
    c-HCOOH            & <1.0 $\times 10^{15}$ & 150  & 4.0 & $-$4.5 \\
    HDO            & <4.0 $\times 10^{16}$ & 150  & 4.0 & $-$4.5  \\
\hline 
\multicolumn{5}{c}{S-bearing species} \\
\hline 
    SO          & >3.0  $\times 10^{15}$ & 200  & 4.0 & $-$4.5 \\
    $^{33}$SO   & <5.0  $\times 10^{14}$ & 200  & 4.0 & $-$4.5  \\
    $^{34}$SO   & <6.0 $\times 10^{14}$ & 200  & 4.0 & $-$4.5  \\
    SO$_{2}$    & <6.0  $\times 10^{15}$ & 200  & 4.0 & $-$4.5 \\
    $^{33}$SO$_{2}$   & <3.0 $\times 10^{15}$ & 200  & 4.0 & $-$4.5   \\
    $^{34}$SO$_{2}$   & <3.0  $\times 10^{15}$ & 200  & 4.0 & $-$4.5 \\ 
    OCS        & <3.0 $\times 10^{15}$ & 200  & 4.0 & $-$4.5 \\
    OC$^{34}$S   & <2.0 $\times 10^{15}$ & 200  & 4.0 & $-$4.5  \\
    H$_2$CS       &  <2.0 $\times 10^{15}$ & 200  & 4.0 & $-$4.5 \\
\hline 
\multicolumn{5}{c}{N-bearing species} \\
\hline 
    CH$_3$CN   &  <6.0 $\times 10^{14}$ & 200  & 4.5 & $-$4.5 \\ 
    $^{13}$CH$_3$CN &  <3.0 $\times 10^{14}$ & 200  & 4.5 & $-$4.5\\
    HNCO    &  <3.0 $\times 10^{15}$ &  200  & 4.5 & $-$4.5 \\
    HC$_3$N    &  <2.0 $\times 10^{14}$  & 200  & 4.5 & $-$4.5 \\
    HC$^{15}$N    &  <1.0 $\times 10^{14}$ & 200  & 4.5 & $-$4.5 \\
\hline 

\end{tabular}
\tablefoot{\tablefoottext{a}{Values are derived from synthetic spectrum fitting with CASSIS, with the assumption of LTE. Only CH$_3$OH and SO are detected, but with no isotopologues detected, only lower limits are derived. Upper limits are derived for all other molecules, using average values for T$_\mathrm{ex}$ derived from the sources where these molecules are detected.}
}

\end{table*}

%

\begin{table*}
\caption{Column densities, excitation temperatures, FWHM and $\varv_{\text{peak}}$ derived at the continuum peak position of N38\tablefootmark{a} }             
\label{table:N38_column_densities}      
\centering          
\begin{tabular}{l c c c c }    
\hline  \hline 
    Molecule & Column density & $T_\mathrm{ex}$& FWHM & $\varv_{\text{peak}}$   \\ 
     & [$\mathrm{cm^{-2} }$]& [K] & [$\mathrm{km \ s^{-1} }$] & [$\mathrm{km \ s^{-1} }$]  \\
\hline
   \multicolumn{5}{c}{COMs and O-bearing species} \\
\hline   
    CH$_3$OH  & >1.0  $\times 10^{16}$ & \hspace{0.6cm} 150  \hspace{0.6cm} & \hspace{1cm} 4.0 \hspace{1cm}  & \hspace{1cm} $-$1.0 \hspace{1cm}  \\
    $^{13}$CH$_3$OH   & <1.0  $\times 10^{16}$ & 150  & 4.0 & $-$1.0  \\
    CH$_3$$^{18}$OH & <1.0 $\times 10^{16}$ & 150  & 4.0 & $-$1.0    \\
    CH$_2$DOH  & <1.0 $\times 10^{16}$ & 150  & 4.0 & $-$1.0  \\
    C$_2$H$_5$OH     & <2.0 $\times 10^{16}$ & 150  & 4.0 & $-$1.0    \\ 
    CH$_3$OCHO        & <1.0 $\times 10^{16}$ & 150  & 4.0 & $-$1.0  \\ 
    CH$_3$OCH$_3$     & <1.0 $\times 10^{16}$ & 150  & 4.0 & $-$1.0 \\ 
    t-HCOOH            & <6.0  $\times 10^{15}$ & 150  & 4.0 & $-$1.0 \\
    c-HCOOH            & <1.0 $\times 10^{15}$ & 150  & 4.0 & $-$1.0 \\
    HDO            & <4.0 $\times 10^{16}$ & 150  & 4.0 & $-$1.0  \\
\hline 
\multicolumn{5}{c}{S-bearing species} \\
\hline 
    SO          & >2.0  $\times 10^{15}$ & 200  & 4.0 & $-$1.0  \\
    $^{33}$SO   & <5.0  $\times 10^{14}$ & 200  &  4.0 & $-$1.0 \\
    $^{34}$SO   & <6.0 $\times 10^{14}$ & 200  &  4.0 & $-$1.0  \\
    SO$_{2}$    & <6.0  $\times 10^{15}$ & 200  & 4.0 & $-$1.0 \\
    $^{33}$SO$_{2}$   & <3.0 $\times 10^{15}$ & 200  &  4.0 & $-$1.0   \\
    $^{34}$SO$_{2}$   & <3.0  $\times 10^{15}$ & 200  &  4.0 & $-$1.0 \\ 
    OCS        & <4.0 $\times 10^{15}$ & 200  &  4.0 & $-$1.0 \\
    OC$^{34}$S   & <3.0 $\times 10^{15}$ & 200  & 4.0 & $-$1.0 \\
    H$_2$CS       &  1.0 (0.5) $\times 10^{15}$ & 100 (50)  &  3.5 & $-$1.0 \\
\hline 
\multicolumn{5}{c}{N-bearing species} \\
\hline 
    CH$_3$CN   &  <5.0 $\times 10^{14}$ & 200  & 4.0 & $-$1.0 \\ 
    $^{13}$CH$_3$CN &  <2.0 $\times 10^{14}$ & 200 & 4.0 & $-$1.0\\
    HNCO    &  <5.0 $\times 10^{14}$ &  200  & 4.0 & $-$1.0 \\
    HC$_3$N    &  <2.0 $\times 10^{14}$  & 200 & 4.0 & $-$1.0 \\
    HC$^{15}$N    &  <1.0 $\times 10^{14}$ & 200  & 4.0 & $-$1.0 \\
\hline 

\end{tabular}
\tablefoot{\tablefoottext{a}{Values are derived from synthetic spectrum fitting with CASSIS, with the assumption of LTE. Only CH$_3$OH and SO are detected, but with no isotopologues detected, only lower limits are derived. Upper limits are derived for all other molecules, using average values for T$_\mathrm{ex}$ derived from the sources where these molecules are detected.}
}

\end{table*}

%

\begin{table*}
\caption{Column densities, excitation temperatures, FWHM and $\varv_{\text{peak}}$ derived at the continuum peak position of N48\tablefootmark{a} }             
\label{table:N48_column_densities}      
\centering          
\begin{tabular}{l c c c c }    
\hline  \hline 
    Molecule & Column density & $T_\mathrm{ex}$& FWHM & $\varv_{\text{peak}}$   \\ 
     & [$\mathrm{cm^{-2} }$]& [K] & [$\mathrm{km \ s^{-1} }$] & [$\mathrm{km \ s^{-1} }$]  \\
\hline
   \multicolumn{5}{c}{COMs and O-bearing species} \\
\hline   
    CH$_3$OH  & >1.0  $\times 10^{16}$ & \hspace{0.6cm} 150  \hspace{0.6cm} & \hspace{1cm} 4.0 \hspace{1cm}  & \hspace{1cm} $-$5.0 \hspace{1cm}  \\
    $^{13}$CH$_3$OH   & <1.0  $\times 10^{16}$ & 150  & 4.0 & $-$5.0  \\
    CH$_3$$^{18}$OH & <1.0 $\times 10^{16}$ & 150  & 4.0 & $-$5.0    \\
    CH$_2$DOH  & <2.0 $\times 10^{16}$ & 150  & 4.0 & $-$5.0\\
    C$_2$H$_5$OH     & <2.0 $\times 10^{16}$ & 150  & 4.0 & $-$5.0  \\ 
    CH$_3$OCHO        & <1.0 $\times 10^{16}$ & 150  & 4.0 & $-$5.0  \\ 
    CH$_3$OCH$_3$     & <2.0 $\times 10^{16}$ & 150  & 4.0 & $-$5.0 \\ 
    t-HCOOH            & <5.0  $\times 10^{15}$ & 150  & 4.0 & $-$5.0 \\
    c-HCOOH            & <1.0 $\times 10^{15}$ & 150  & 4.0 & $-$5.0\\
    HDO            & <3.0 $\times 10^{16}$ & 150  & 4.0 & $-$5.0 \\
\hline 
\multicolumn{5}{c}{S-bearing species} \\
\hline 
    SO          & >1.0  $\times 10^{15}$ & 200  & 4.0 & $-$5.0 \\
    $^{33}$SO   & <5.0  $\times 10^{14}$ & 200  & 4.0 & $-$5.0 \\
    $^{34}$SO   & <5.0 $\times 10^{14}$ & 200  &4.0 & $-$5.0  \\
    SO$_{2}$    & <5.0  $\times 10^{15}$ & 200  & 4.0 & $-$5.0 \\
    $^{33}$SO$_{2}$   & <3.0 $\times 10^{15}$ & 200  & 4.0 & $-$5.0   \\
    $^{34}$SO$_{2}$   & <3.0  $\times 10^{15}$ & 200  & 4.0 & $-$5.0 \\ 
    OCS        & <4.0 $\times 10^{15}$ & 200  & 4.0 & $-$5.0 \\
    OC$^{34}$S   & <2.0 $\times 10^{15}$ & 200  & 4.0 & $-$5.0  \\
    H$_2$CS       &  <2.0 $\times 10^{15}$ & 200  & 4.0 & $-$5.0 \\
\hline 
\multicolumn{5}{c}{N-bearing species} \\
\hline 
    CH$_3$CN   &  <5.0 $\times 10^{14}$ & 200 & 4.0 & $-$5.0 \\ 
    $^{13}$CH$_3$CN &  <3.0 $\times 10^{14}$ & 200 & 4.0 & $-$5.0 \\
    HNCO    &  <2.0 $\times 10^{15}$ &  200  & 4.0 & $-$5.0  \\
    HC$_3$N    &  <2.0 $\times 10^{14}$  & 200 & 4.0 & $-$5.0 \\
    HC$^{15}$N  &  <2.0 $\times 10^{14}$ & 200  & 4.0 & $-$5.0 \\
\hline 

\end{tabular}
\tablefoot{\tablefoottext{a}{Values are derived from synthetic spectrum fitting with CASSIS, with the assumption of LTE. Only CH$_3$OH and SO are detected, but with no isotopologues detected, only lower limits are derived. Upper limits are derived for all other molecules.}

}

\end{table*}

%

\begin{table*}
\caption{Column densities, excitation temperatures, FWHM and $\varv_{\text{peak}}$ derived at the continuum peak position of S8\tablefootmark{a} }             
\label{table:S8_column_densities}      
\centering          
\begin{tabular}{l c c c c }    
\hline  \hline 
    Molecule & Column density & $T_\mathrm{ex}$& FWHM & $\varv_{\text{peak}}$   \\ 
     & [$\mathrm{cm^{-2} }$]& [K] & [$\mathrm{km \ s^{-1} }$] & [$\mathrm{km \ s^{-1} }$]  \\
\hline
   \multicolumn{5}{c}{COMs and O-bearing species} \\
\hline   
    CH$_3$OH  & <8.0  $\times 10^{15}$ & \hspace{0.6cm} 150  \hspace{0.6cm} & \hspace{1cm} 4.0 \hspace{1cm}  & \hspace{1cm} 1.0 \hspace{1cm}  \\
    $^{13}$CH$_3$OH   & <6.0  $\times 10^{15}$ & 150  & 4.0 & 1.0  \\
    CH$_3$$^{18}$OH & <1.0 $\times 10^{16}$ & 150  & 4.0 & 1.0     \\
    CH$_2$DOH  & <2.0 $\times 10^{16}$ & 150  & 4.0 & 1.0   \\
    C$_2$H$_5$OH     & <2.0 $\times 10^{16}$ & 150  & 4.0 & 1.0    \\ 
    CH$_3$OCHO        & <1.0 $\times 10^{16}$ & 150  & 4.0 & 1.0   \\ 
    CH$_3$OCH$_3$     & <2.0 $\times 10^{16}$ & 150  & 4.0 & 1.0   \\ 
    H$_2$CO           & >2.0 $\times 10^{15}$ & 150  & 4.0 & 1.0  \\
    H$_{2}$$^{13}$CO  & <1.0 $\times 10^{15} $ & 150  & 4.0 & 1.0  \\
    t-HCOOH            & <5.0  $\times 10^{15}$ & 150  & 4.0 & 1.0  \\
    c-HCOOH            & <1.0 $\times 10^{15}$ & 150  & 4.0 & 1.0  \\
    HDO            & <4.0 $\times 10^{16}$ & 150  & 4.0 & 1.0   \\
\hline 
\multicolumn{5}{c}{S-bearing species} \\
\hline 
    SO          & >1.0  $\times 10^{15}$ & 200  & 4.0 & 1.0  \\
    $^{33}$SO   & <5.0  $\times 10^{14}$ & 200  & 4.0 & 1.0  \\
    $^{34}$SO   & <5.0 $\times 10^{14}$ & 200  & 4.0 & 1.0  \\
    SO$_{2}$    & <6.0  $\times 10^{15}$ & 200  & 4.0 & 1.0 \\
    $^{33}$SO$_{2}$   & <4.0 $\times 10^{15}$ & 200  & 4.0 & 1.0   \\
    $^{34}$SO$_{2}$   & <4.0  $\times 10^{15}$ & 200  & 4.0 & 1.0 \\ 
    OCS           & <4.0 $\times 10^{15}$ & 200  & 4.0 & 1.0 \\
    OC$^{34}$S    & <3.0 $\times 10^{15}$ & 200  & 4.0 & 1.0  \\
    H$_2$CS       &  <2.0 $\times 10^{15}$ & 200  & 4.0 & 1.0 \\
\hline 
\multicolumn{5}{c}{N-bearing species} \\
\hline 
    CH$_3$CN   &  <3.0 $\times 10^{14}$ & 200  & 4.0 & 1.0  \\ 
    $^{13}$CH$_3$CN &  <2.0 $\times 10^{14}$ & 200  & 4.0 & 1.0 \\
    HNCO    &  <2.0 $\times 10^{15}$ &  200  & 4.0 & 1.0  \\
    HC$_3$N    &  <2.0 $\times 10^{14}$  & 200  & 4.0 & 1.0 \\
    HC$^{15}$N  &  <1.0 $\times 10^{14}$ & 200  & 4.0 & 1.0 \\
\hline 

\end{tabular}
\tablefoot{\tablefoottext{a}{Values are derived from synthetic spectrum fitting with CASSIS, with the assumption of LTE. Only CH$_3$OH and SO are detected, but with no isotopologues detected, only lower limits are derived. Upper limits are derived for all other molecules.}
}

\end{table*}


\section{Molecular line plots}
\label{App:Line_plots}
In this Appendix the plots of all the lines for each of the molecules listed in Tables \ref{table:S26_Molecules_COMs} -- \ref{table:S8_Molecules_all} are presented. In all the plots, the model is overplotted in blue over the observed spectrum in black. Only lines detected above $3\sigma$ $\sim$10 K intensity are shown.

\begin{figure*}
    \centering
    \includegraphics[width=16.8cm]{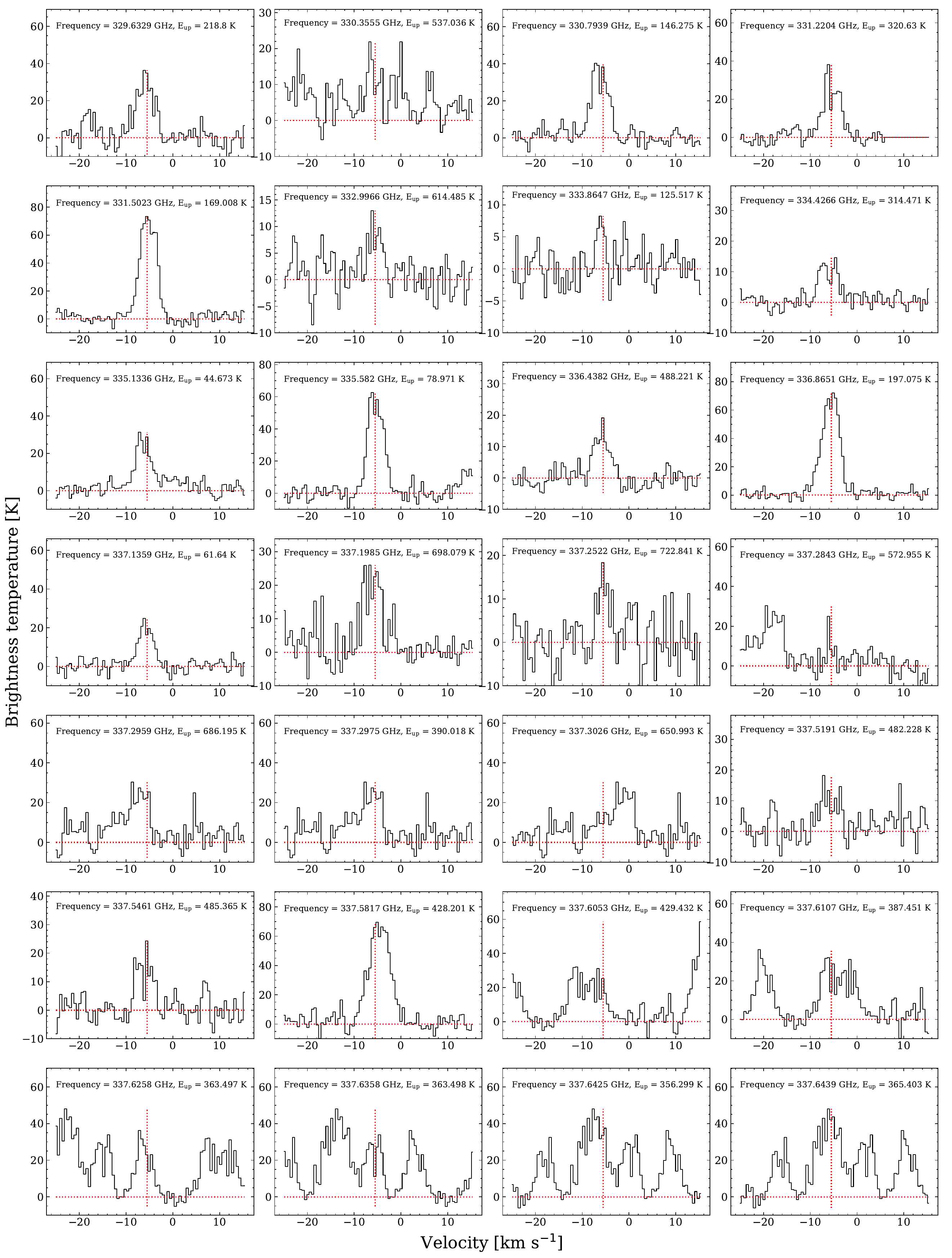}
    \caption{Detected CH$_3$OH lines towards S26. Since the lines are optically thick, we used $^{13}$CH$_3$OH (multiplied by the local ISM value of $^{12}$C/$^{13}$C = 68) to derive a column density of 1.2 (0.2)$\times 10^{18} \mathrm{cm^{-2}}$.}
    \label{Fig:S26_CH3OH_1}%
\end{figure*}

\begin{figure*}
    \centering
    \includegraphics[width=16.8cm]{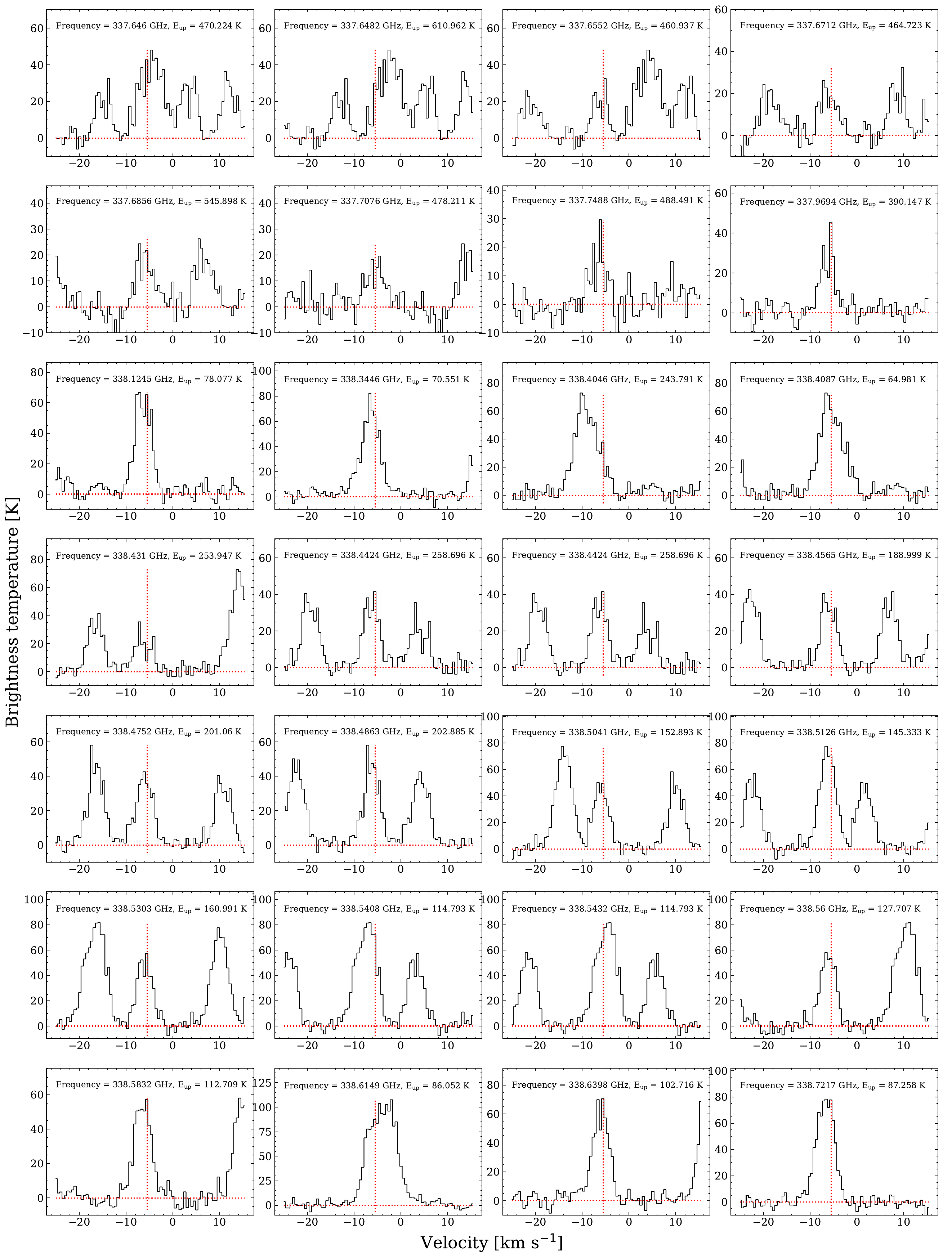}
    \caption{Detected CH$_3$OH lines towards S26. Since the lines are optically thick, we used $^{13}$CH$_3$OH (multiplied by the local ISM value of $^{12}$C/$^{13}$C = 68) to derive a column density of 1.2 (0.2)$\times 10^{18} \mathrm{cm^{-2}}$.}
    \label{Fig:S26_CH3OH_2}%
\end{figure*} 

\begin{figure*}
    \centering
    \includegraphics[width=16.8cm]{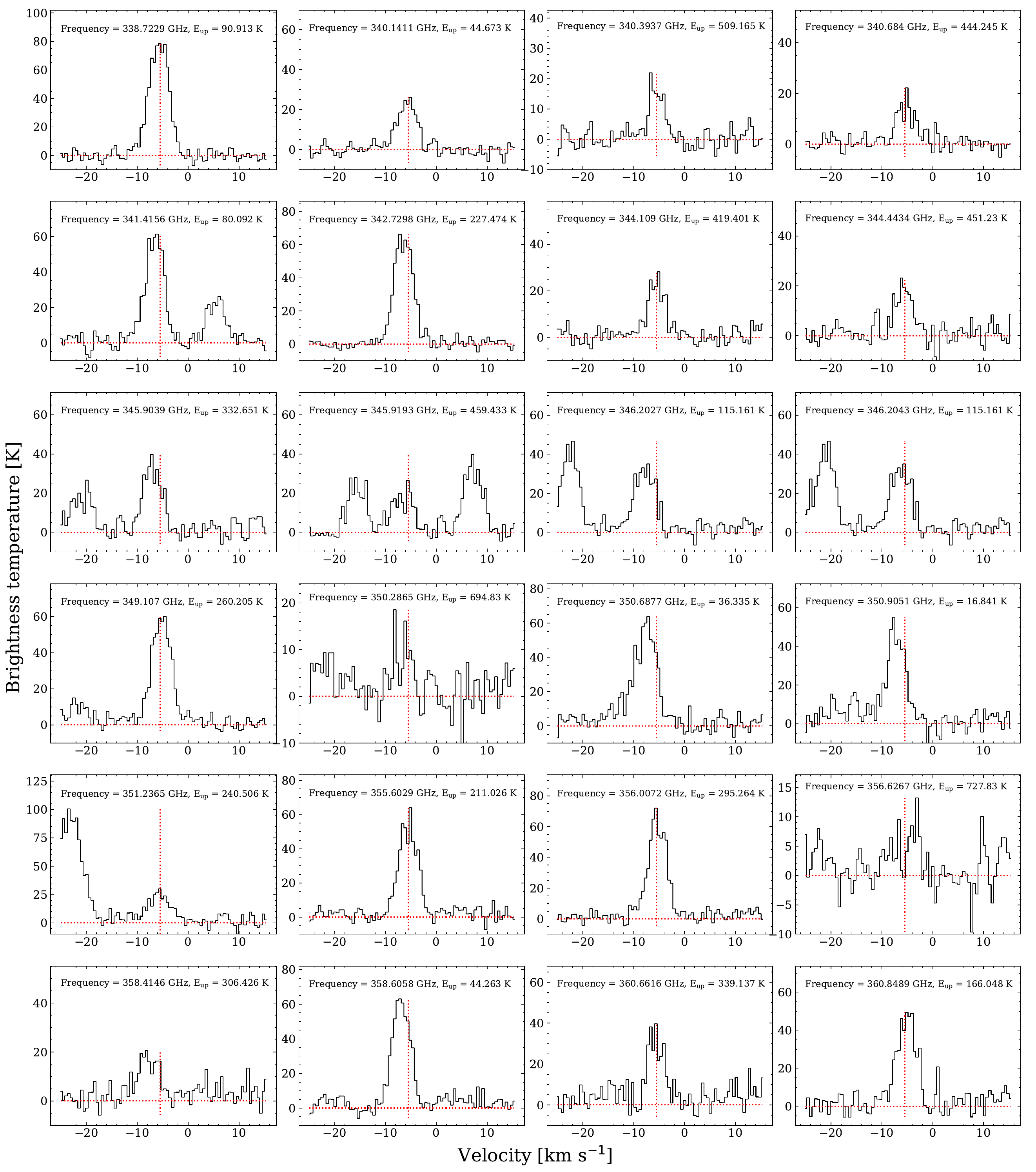}
    \caption{Detected CH$_3$OH lines towards S26. Since the lines are optically thick, we used $^{13}$CH$_3$OH (multiplied by the local ISM value of $^{12}$C/$^{13}$C = 68) to derive a column density of 1.2 (0.2)$\times 10^{18} \mathrm{cm^{-2}}$.}
    \label{Fig:S26_CH3OH_3}%
\end{figure*} 


\begin{figure*}
    \centering
    \includegraphics[width=16.8cm]{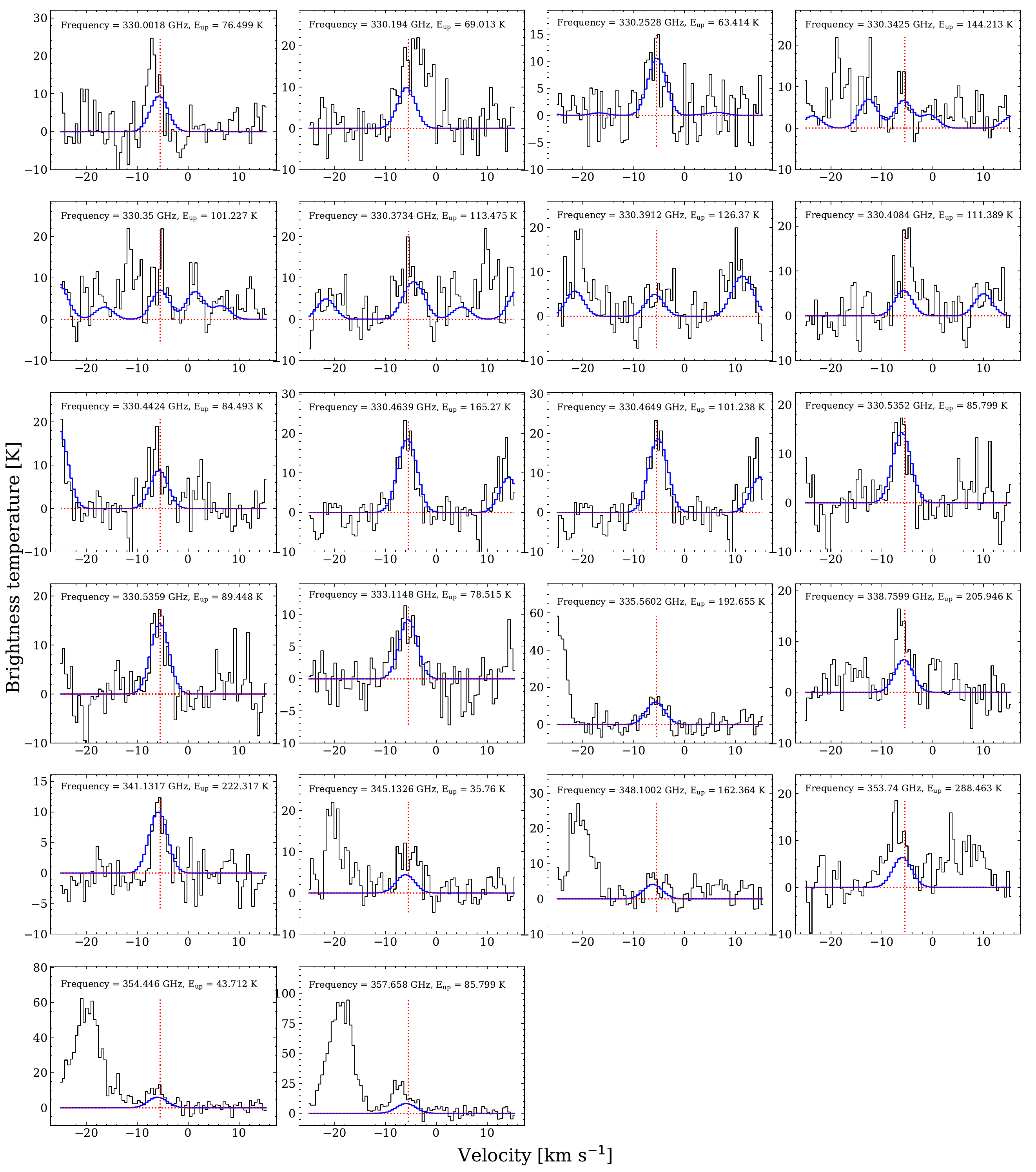}
    \caption{Detected $^{13}$CH$_3$OH lines towards S26. The values obtained from model fitting (model in blue) overplotted on the observed spectrum (in black) for this molecule were column density of 1.8 (0.3)$\times 10^{16} \mathrm{cm^{-2}}$, with excitation temperature 100 (25)K.}
    \label{Fig:S26_13CH3OH_1}%
\end{figure*}

  
\begin{figure*}
    \centering
    \includegraphics[width=16.8cm]{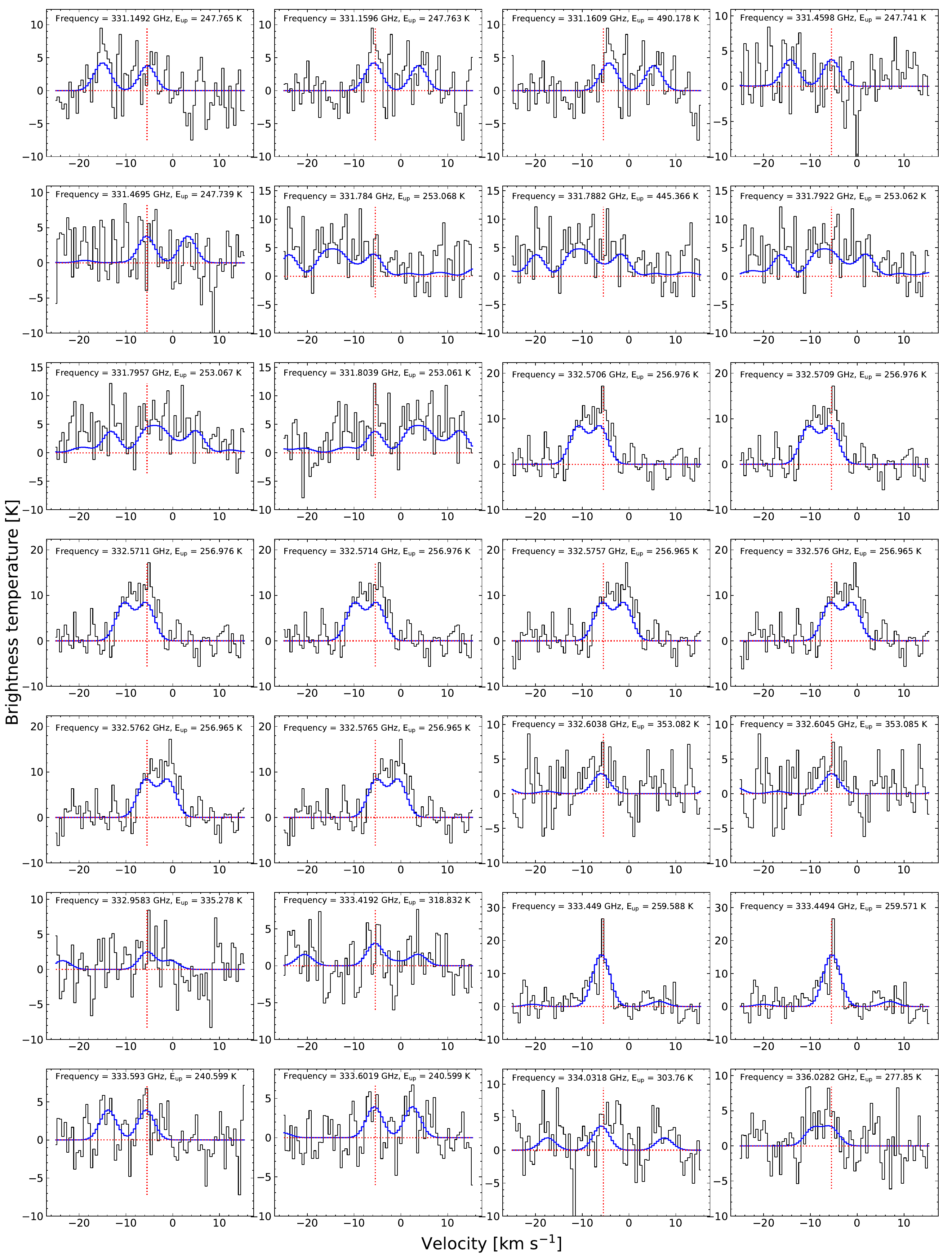}
    \caption{Detected CH$_3$OCHO lines towards S26. The values obtained from model fitting (model in blue) overplotted on the observed spectrum (in black) for this molecule were column density of 3.5 (0.5)$\times 10^{16} \mathrm{cm^{-2}}$, with excitation temperature 100 (25)K.}
    \label{Fig:S26_CH3OCHO_1}%
\end{figure*}

\begin{figure*}
    \centering
    \includegraphics[width=16.8cm]{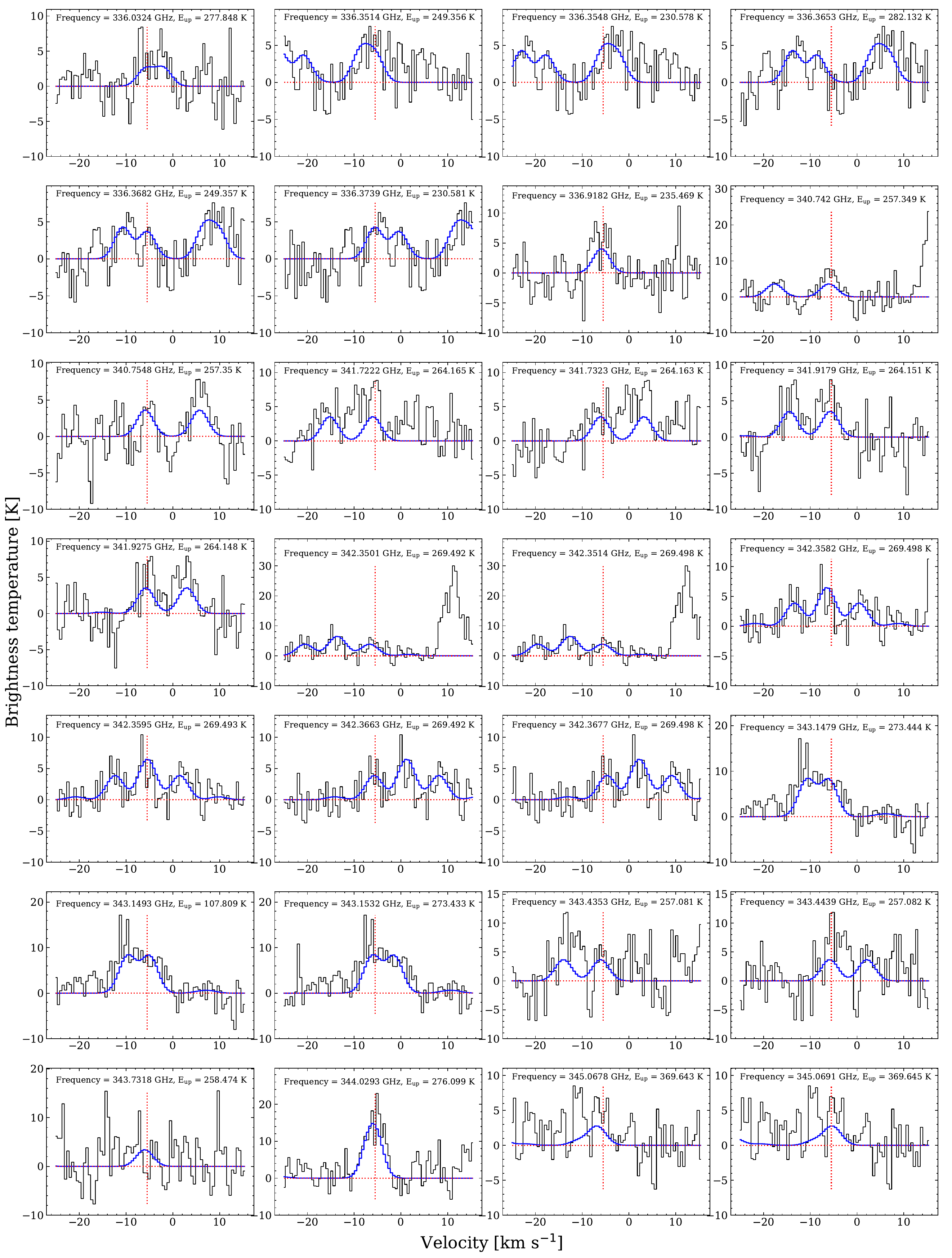}
    \caption{Detected CH$_3$OCHO lines towards S26. The values obtained from model fitting (model in blue) overplotted on the observed spectrum (in black) for this molecule were column density of 3.5 (0.5)$\times 10^{16} \mathrm{cm^{-2}}$, with excitation temperature 100 (25)K.}
    \label{Fig:S26_CH3OCHO_2}%
\end{figure*} 

\begin{figure*}
    \centering
    \includegraphics[width=16.8cm]{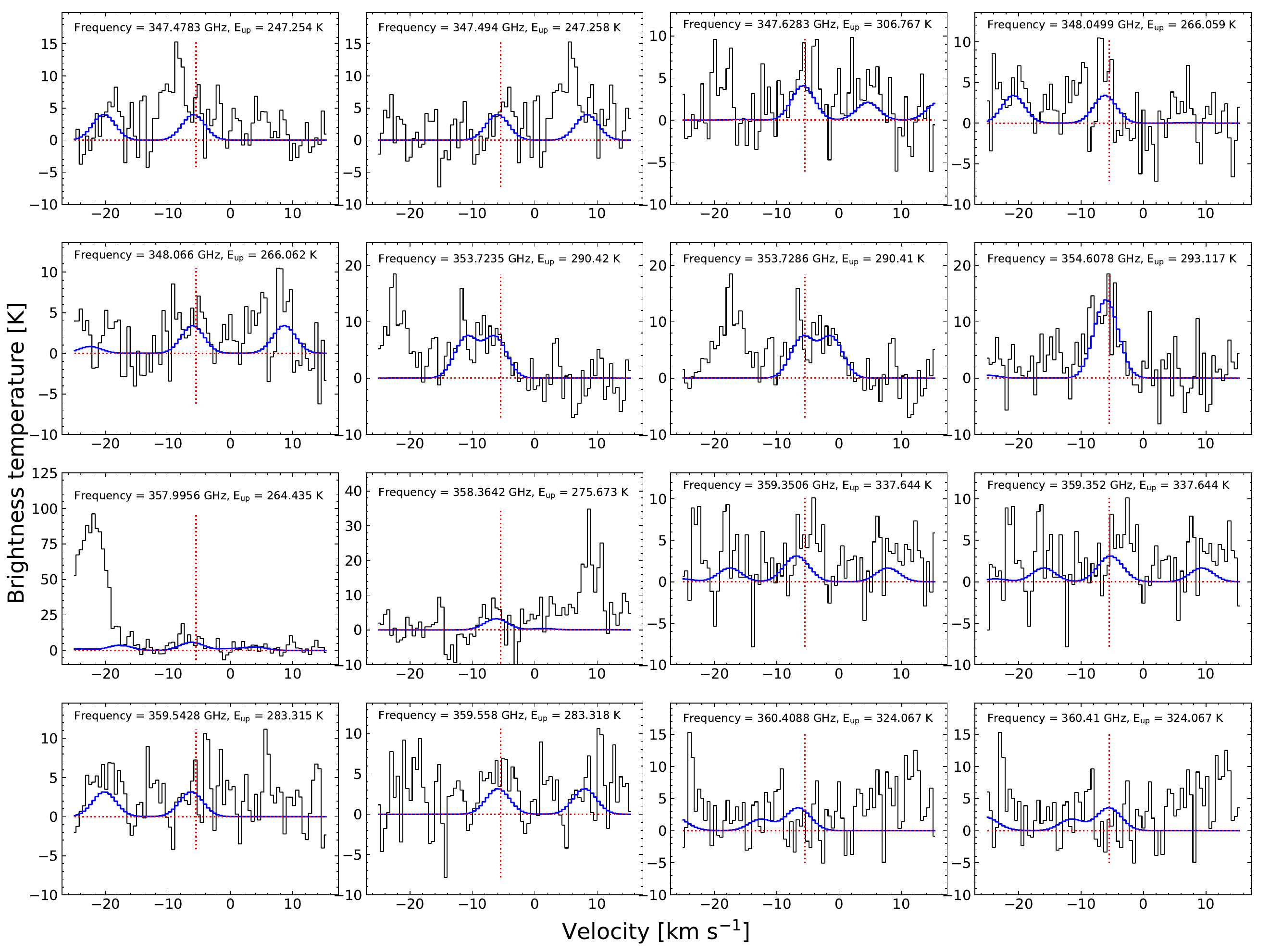}
    \caption{Detected CH$_3$OCHO lines towards S26. The values obtained from model fitting (model in blue) overplotted on the observed spectrum (in black) for this molecule were column density of 3.5 (0.5)$\times 10^{16} \mathrm{cm^{-2}}$, with excitation temperature 100 (25)K.}
    \label{Fig:S26_CH3OCHO_3}%
\end{figure*} 

 
 \begin{figure*}
    \centering
    \includegraphics[width=16.8cm]{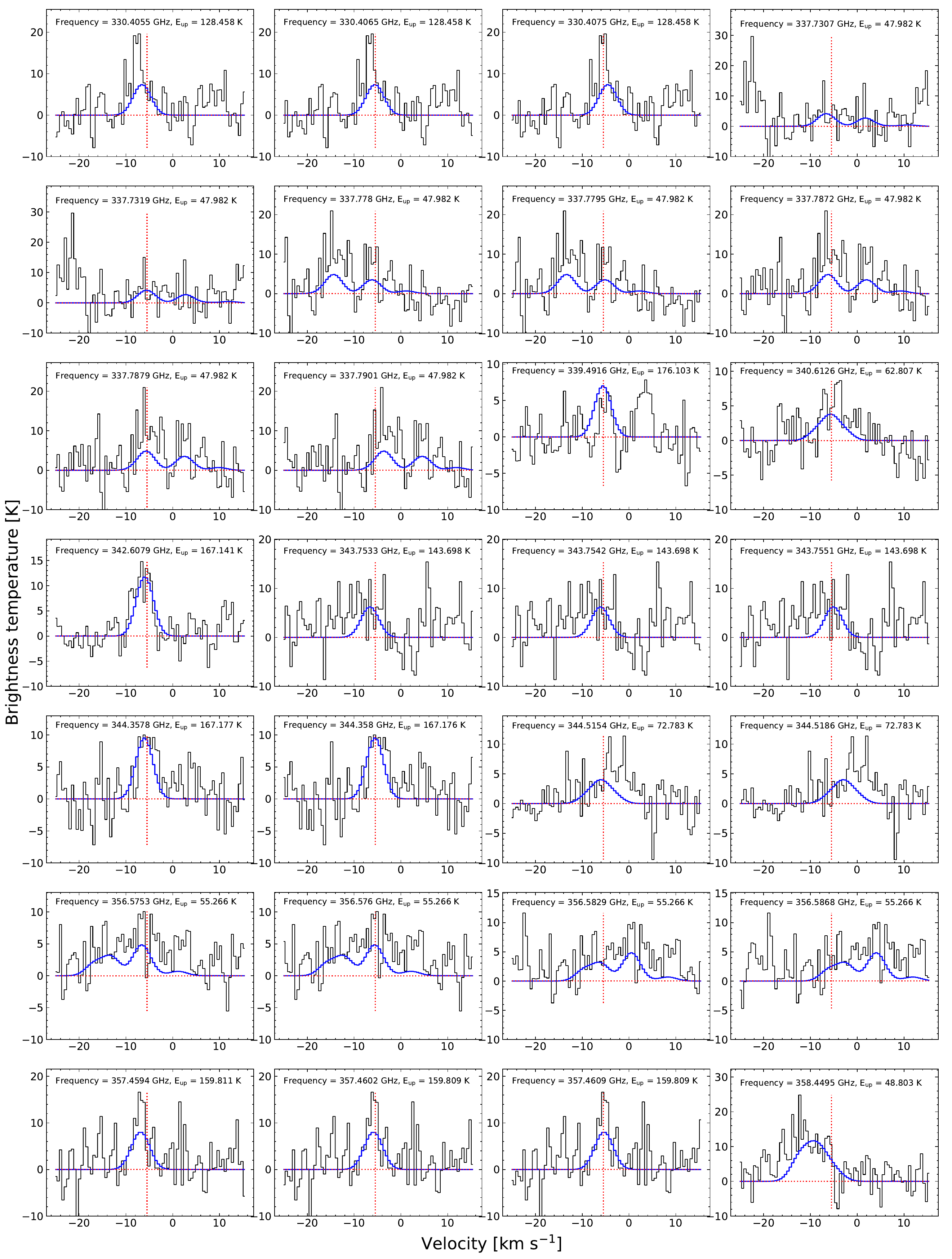}
    \caption{Detected CH$_3$OCHO lines towards S26. The values obtained from model fitting (model in blue) overplotted on the observed spectrum (in black) for this molecule were column density of 2.6 (0.5)$\times 10^{16} \mathrm{cm^{-2}}$, with excitation temperature 100 (25)K.}
    \label{Fig:S26_CH3OCH3_1}%
\end{figure*} 

\begin{figure*}
    \centering
    \includegraphics[width=16.8cm]{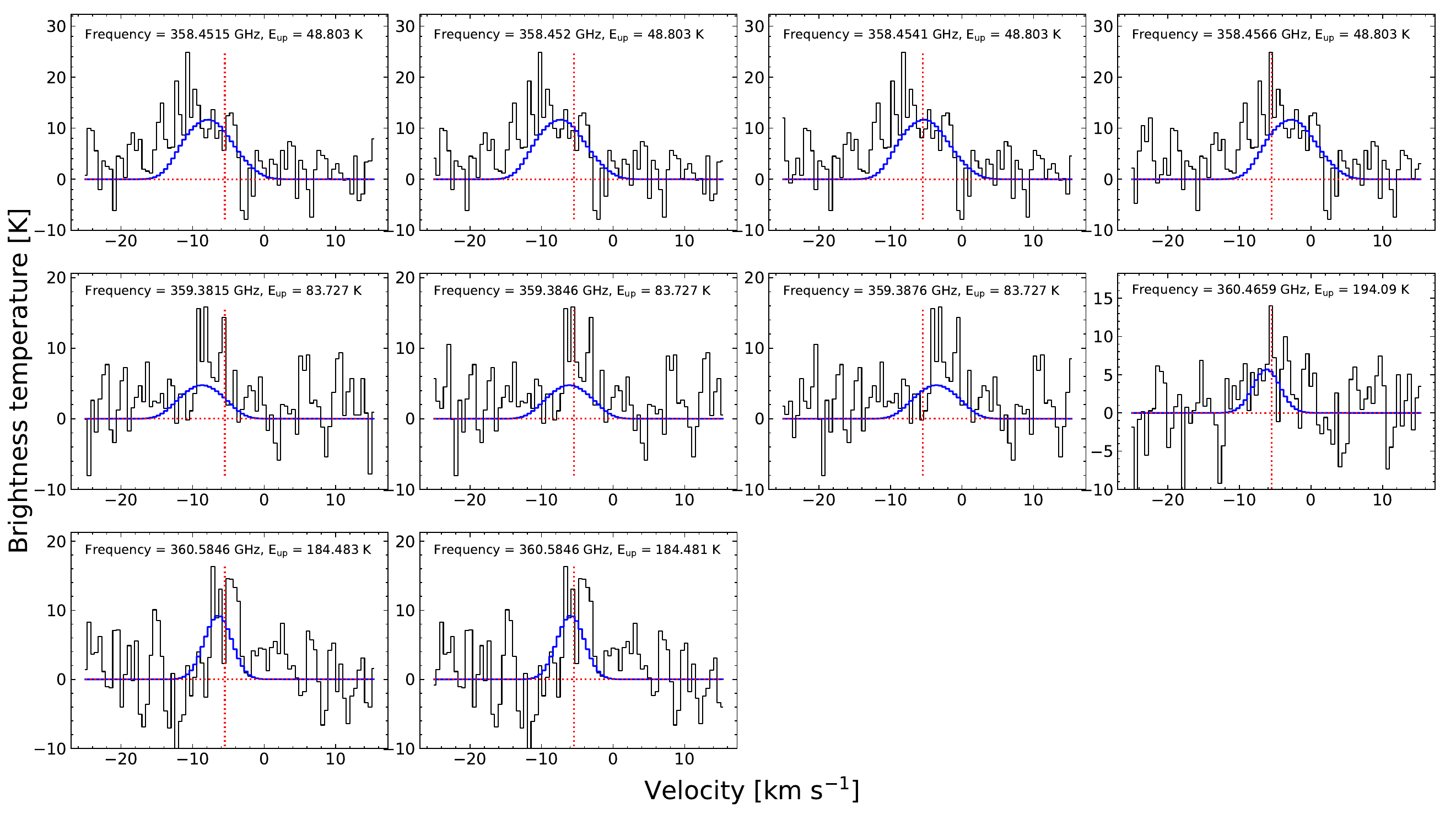}
    \caption{Detected CH$_3$OCH$_3$ lines towards S26. The values obtained from model fitting (model in blue) overplotted on the observed spectrum (in black) for this molecule were column density of 2.6 (0.5)$\times 10^{16} \mathrm{cm^{-2}}$, with excitation temperature 100 (25)K.}
    \label{Fig:S26_CH3OCH3_2}%
\end{figure*} 


\begin{figure*}
    \centering
    \includegraphics[width=16.8cm]{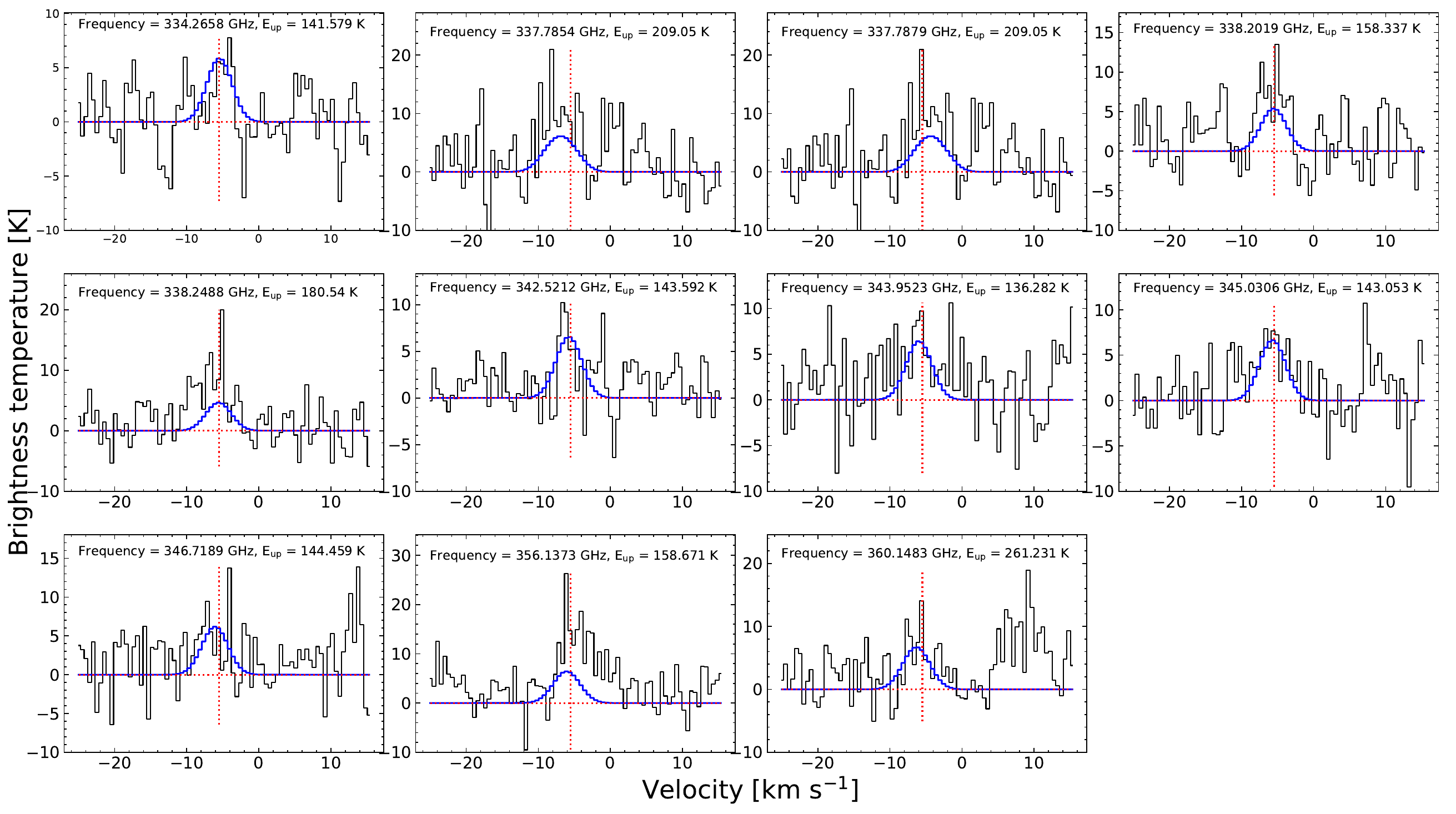}
    \caption{Detected t-HCOOH lines towards S26. The values obtained from model fitting (model in blue) overplotted on the observed spectrum (in black) for this molecule were column density of 7.0 (1.0)$\times 10^{15} \mathrm{cm^{-2}}$, with excitation temperature 190 (60)K.}
    \label{Fig:S26_t-HCOOH_1}%
\end{figure*} 


\begin{figure*}
    \centering
    \includegraphics[width=16.8cm]{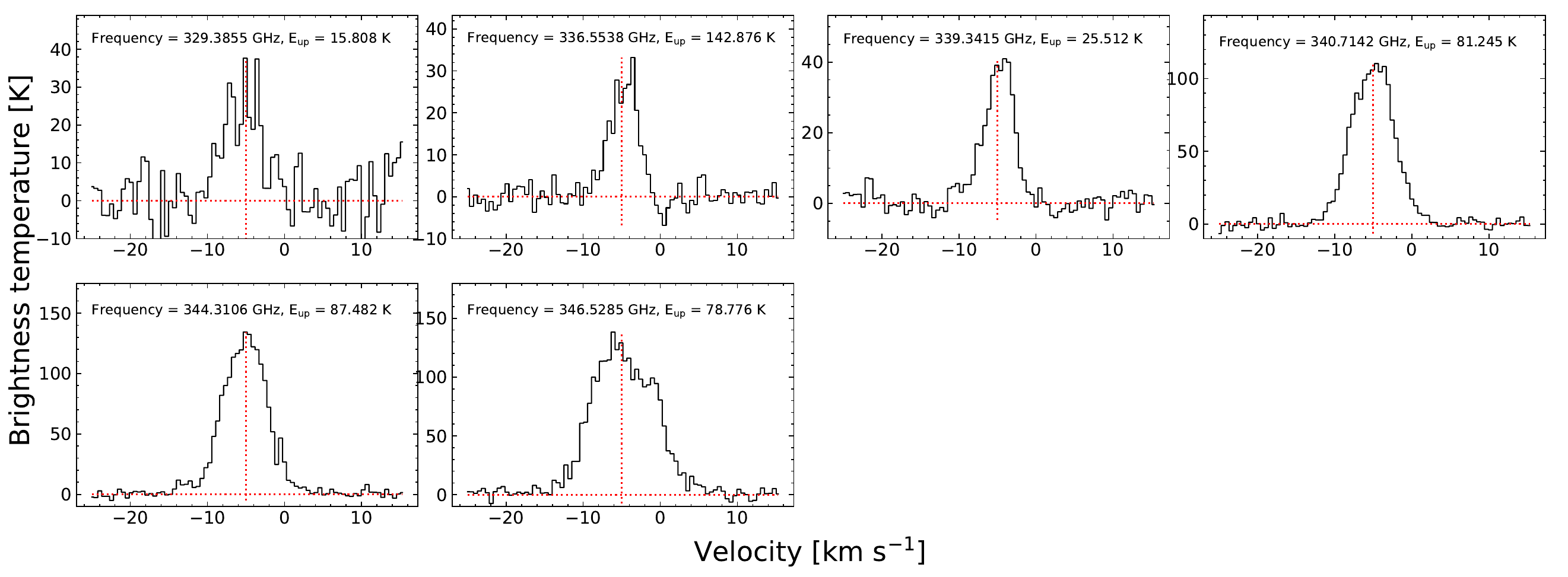}
    \caption{Since SO lines are optically thick, we used $^{34}$SO (multiplied by the local ISM value of $^{32}$S/$^{34}$S = 22) to derive a column density of 3.5 (0.9)$\times 10^{17} \mathrm{cm^{-2}}$.}
    \label{Fig:S26_SO}%
\end{figure*} 


\begin{figure*}
    \centering
    \includegraphics[width=16.8cm]{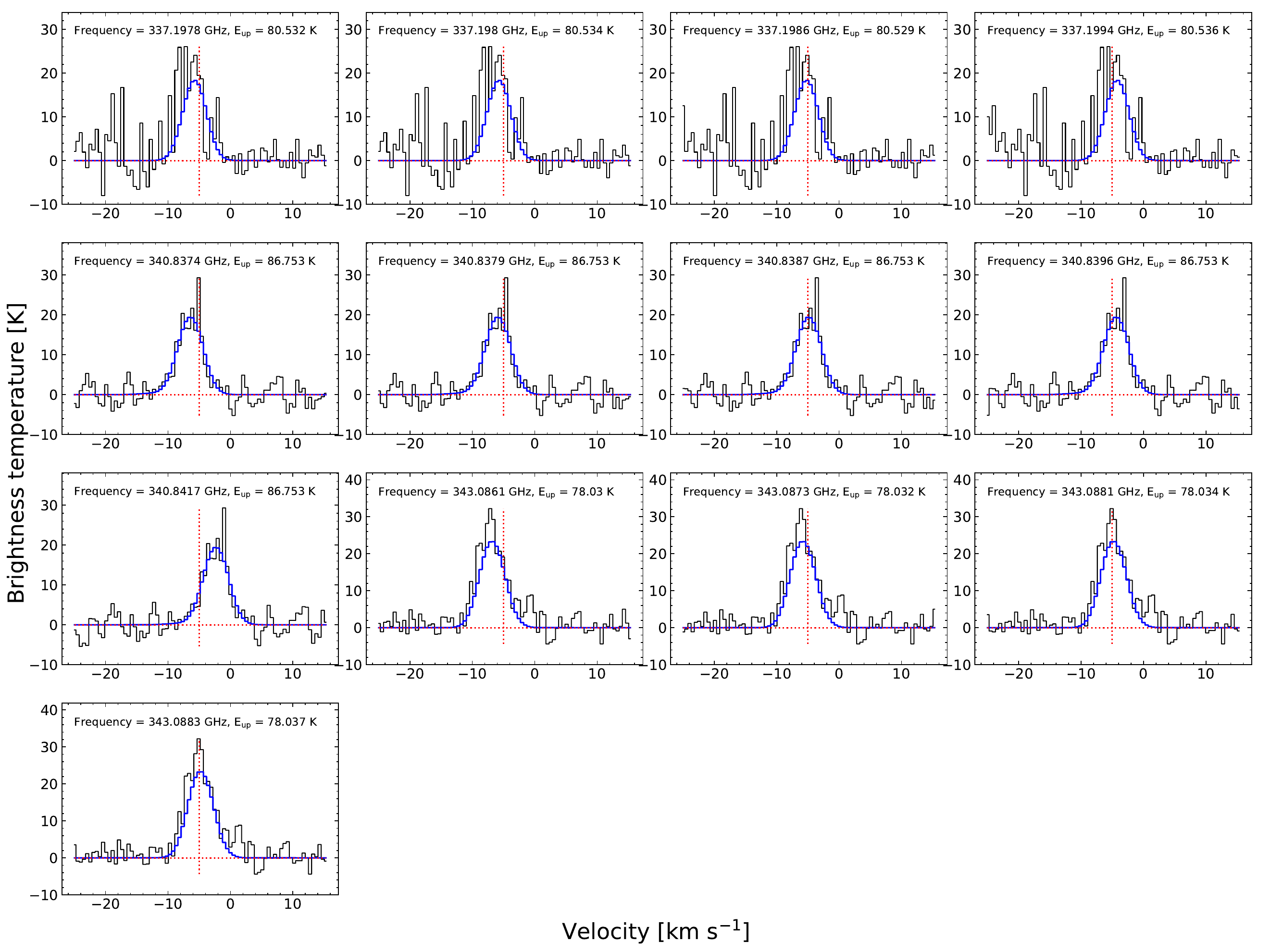}
    \caption{Detected $^{33}$SO lines towards S26. The values obtained from model fitting (model in blue) overplotted on the observed spectrum (in black) for this molecule were column density of 3.6 (1.0)$\times 10^{15} \mathrm{cm^{-2}}$, with excitation temperature 120 (40)K.}
    \label{Fig:S26_33SO}%
\end{figure*} 


\begin{figure*}
    \centering
    \includegraphics[width=16.8cm]{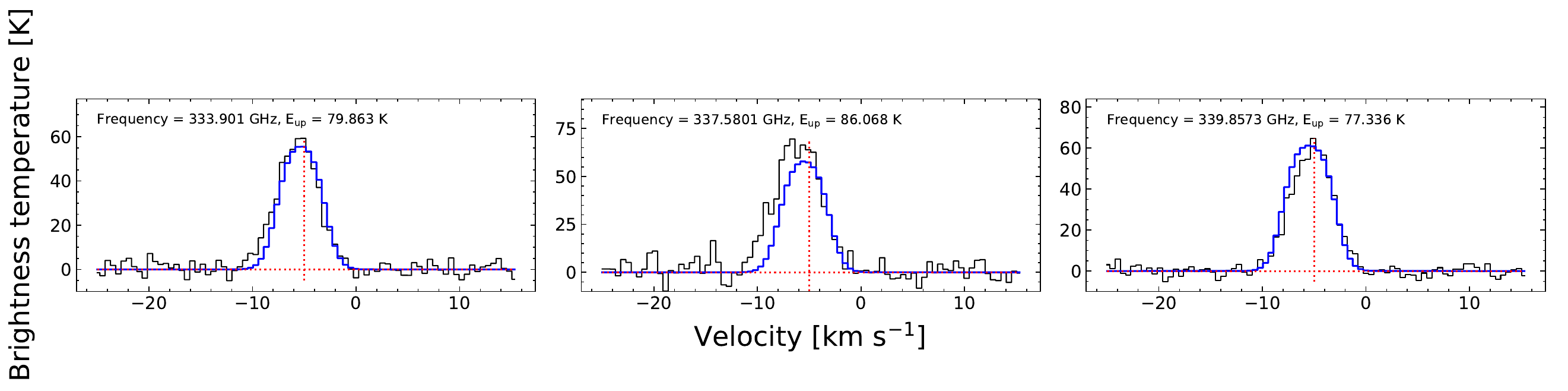}
    \caption{Detected  $^{34}$SO lines towards S26. The values obtained from model fitting (model in blue) overplotted on the observed spectrum (in black) for this molecule were column density of 1.6 (0.4)$\times 10^{16} \mathrm{cm^{-2}}$, with excitation temperature 120 (20)K.}
    \label{Fig:S26_34SO}%
\end{figure*} 


\begin{figure*}
    \centering
    \includegraphics[width=16.8cm]{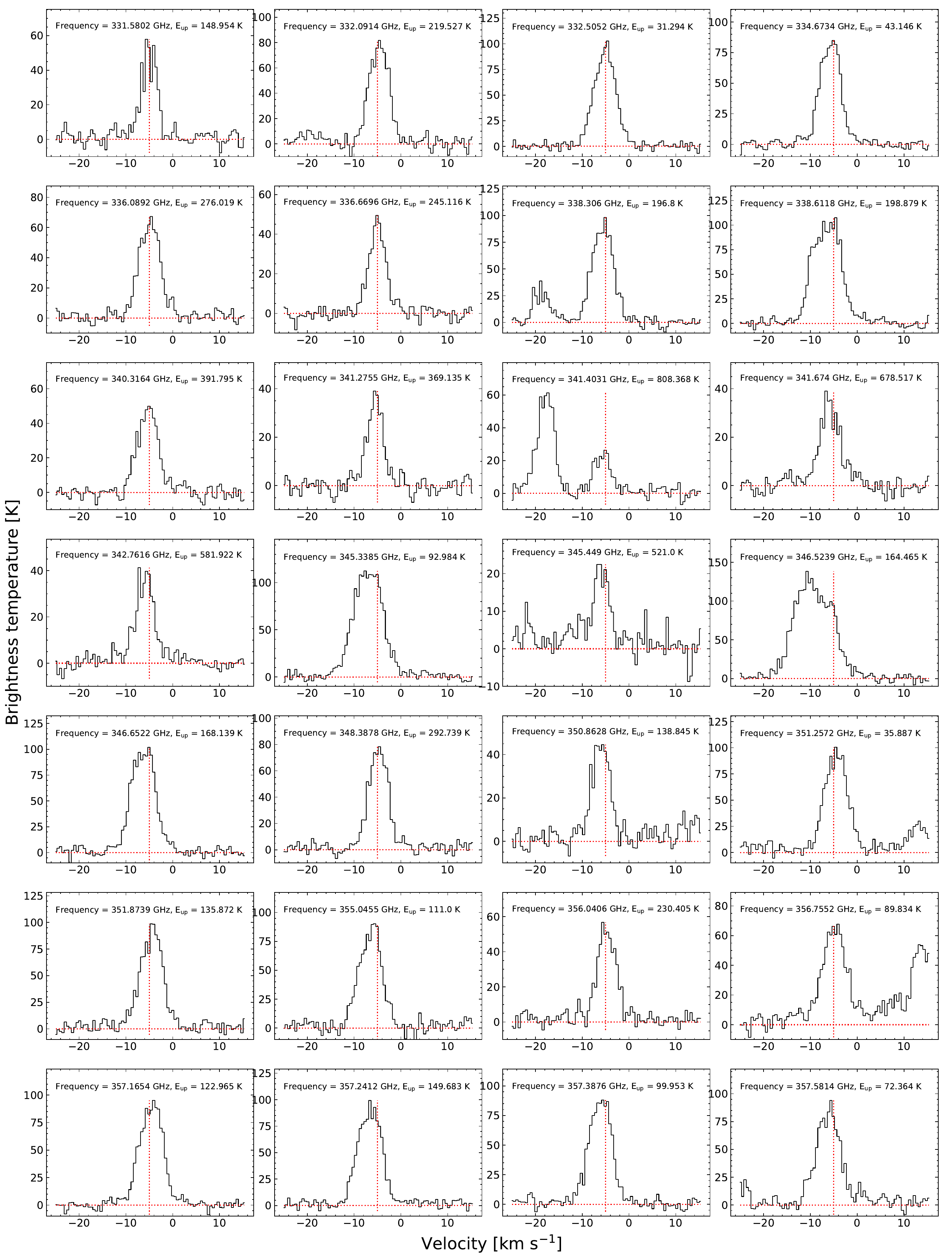}
    \caption{Since SO$_2$ lines are optically thick, we used $^{34}$SO$_2$ (multiplied by the local ISM value of $^{32}$S/$^{34}$S = 22) to derive a column density of 5.9 (0.7)$\times 10^{17} \mathrm{cm^{-2}}$.}
    \label{Fig:S26_SO2_1}%
\end{figure*} 

\begin{figure*}
    \centering
    \includegraphics[width=16.8cm]{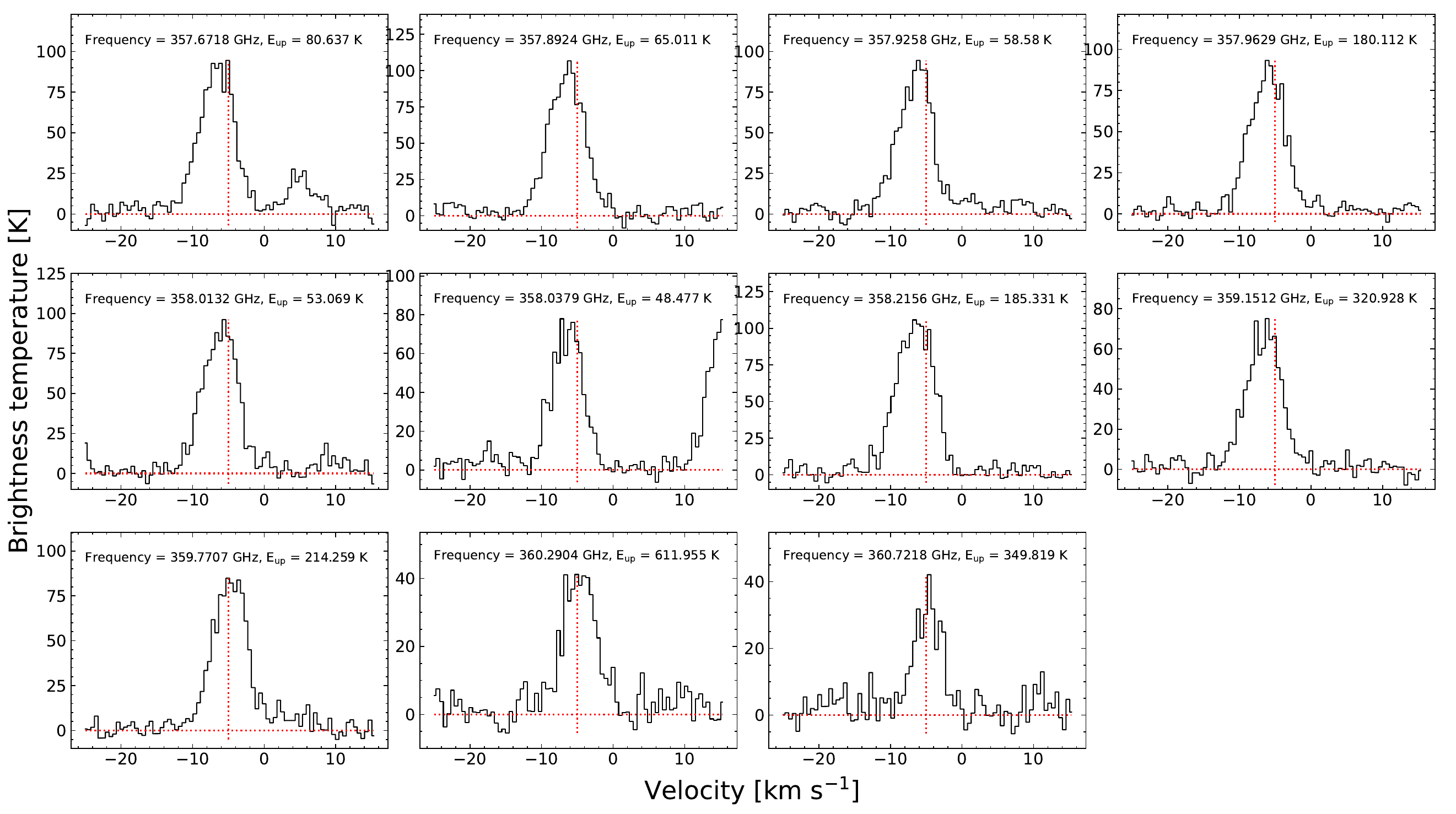}
    \caption{Since SO$_2$ lines are optically thick, we used $^{34}$SO$_2$ (multiplied by the local ISM value of $^{32}$S/$^{34}$S = 22) to derive a column density of 5.9 (0.7)$\times 10^{17} \mathrm{cm^{-2}}$.}
    \label{Fig:S26_SO2_2}%
\end{figure*} 


\begin{figure*}
    \centering
    \includegraphics[width=16.8cm]{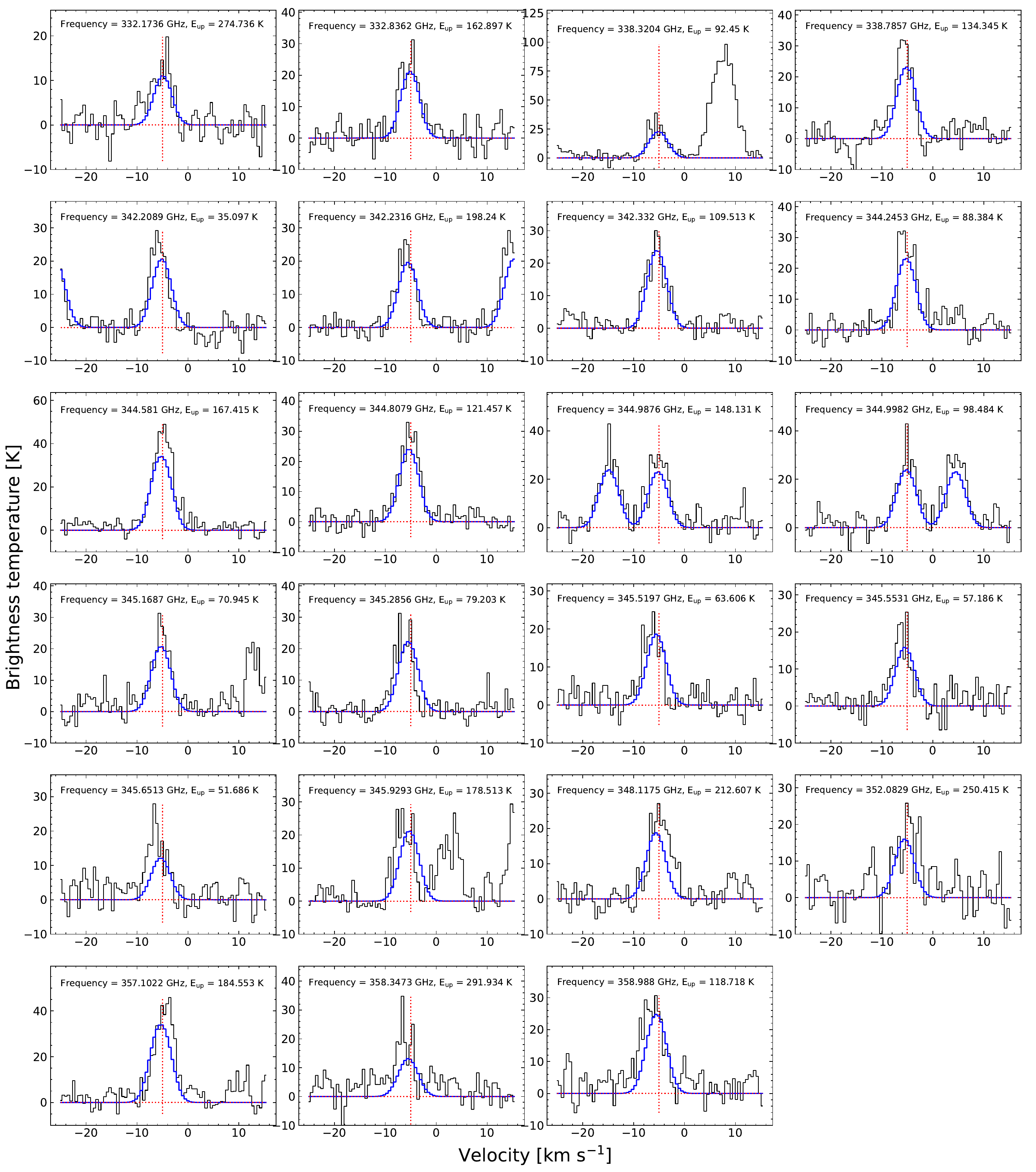}
    \caption{Detected  $^{34}$SO2 lines towards S26. The values obtained from model fitting (model in blue) overplotted on the observed spectrum (in black) for this molecule were column density of 2.6 (0.3)$\times 10^{16} \mathrm{cm^{-2}}$, with excitation temperature 120 (20)K.}
    \label{Fig:S26_34SO2_1}%
\end{figure*}


\begin{figure*}
    \centering
    \includegraphics[width=16.8cm]{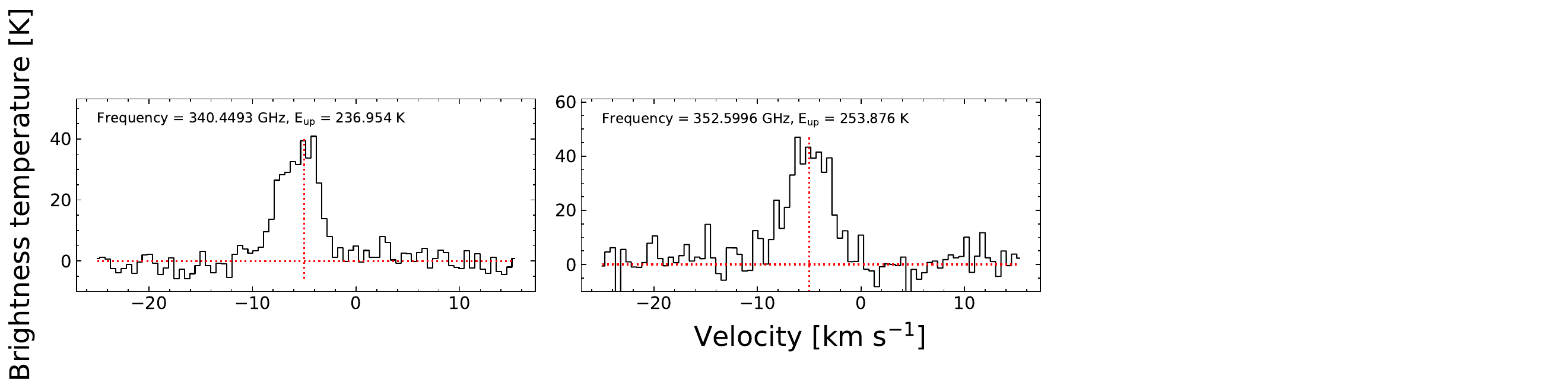}
    \caption{Since OCS lines are optically thick, we used OC$^{34}$S (multiplied by the local ISM value of $^{32}$S/$^{34}$S = 22) to derive a column density of 1.4 (0.2)$\times 10^{17} \mathrm{cm^{-2}}$.}
    \label{Fig:S26_OCS_1}%
\end{figure*} 


\begin{figure*}
    \centering
    \includegraphics[width=16.8cm]{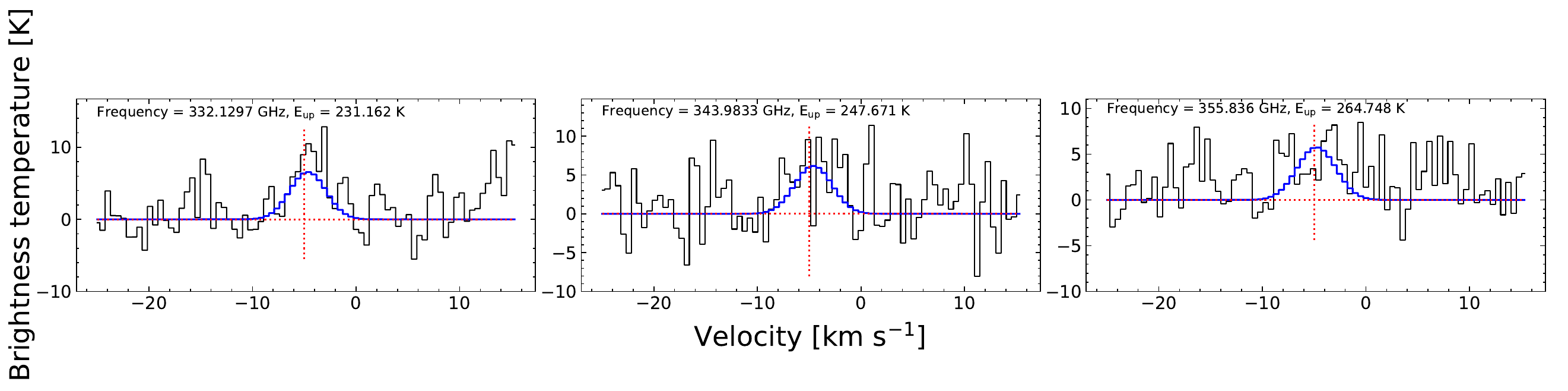}
    \caption{Detected OC$^{34}$S lines towards S26. The values obtained from model fitting (model in blue) overplotted on the observed spectrum (in black) for this molecule were column density of 6.4 (1.0)$\times 10^{15} \mathrm{cm^{-2}}$, with excitation temperature 100 (50)K.}
    \label{Fig:S26_OC34S_1}%
\end{figure*} 


\begin{figure*}
    \centering
    \includegraphics[width=16.8cm]{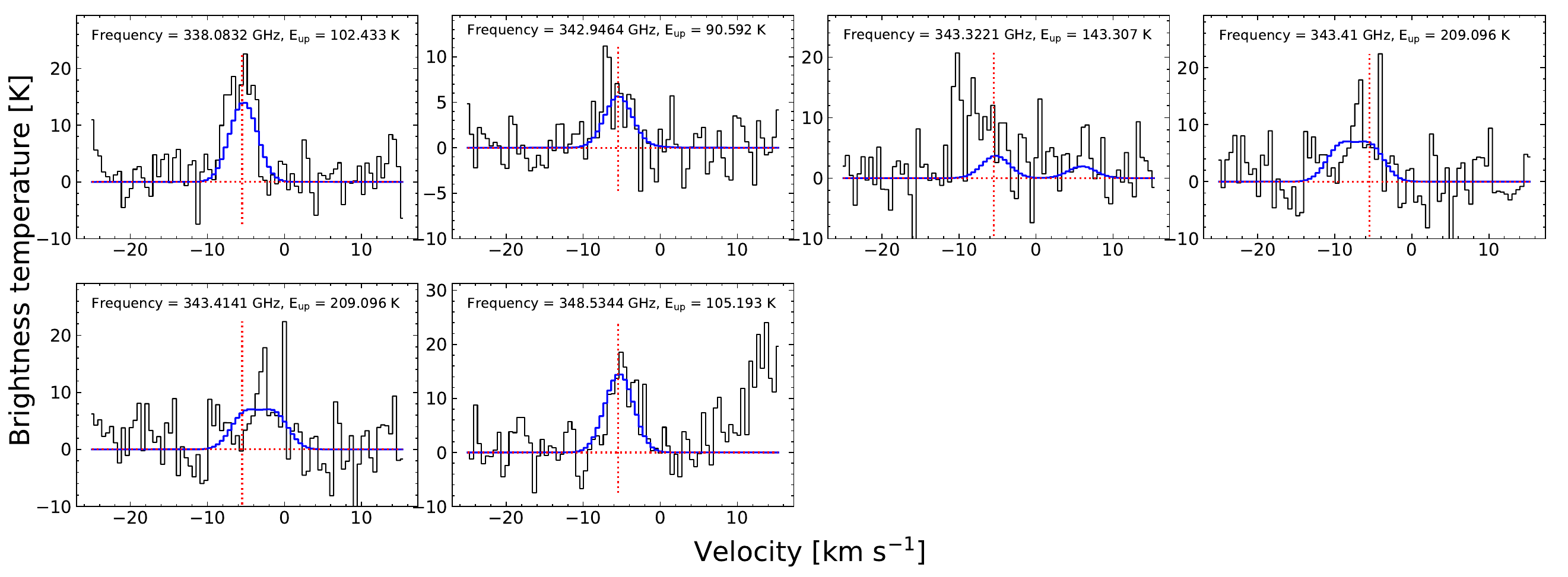}
    \caption{Detected H$_{2}$CS lines towards S26. The values obtained from model fitting (model in blue) overplotted on the observed spectrum (in black) for this molecule were column density of 2.6 (0.2)$\times 10^{15} \mathrm{cm^{-2}}$, with excitation temperature 130 (40)K.}
    \label{Fig:S26_H2CS_1}%
\end{figure*} 


\begin{figure*}
    \centering
    \includegraphics[width=16.8cm]{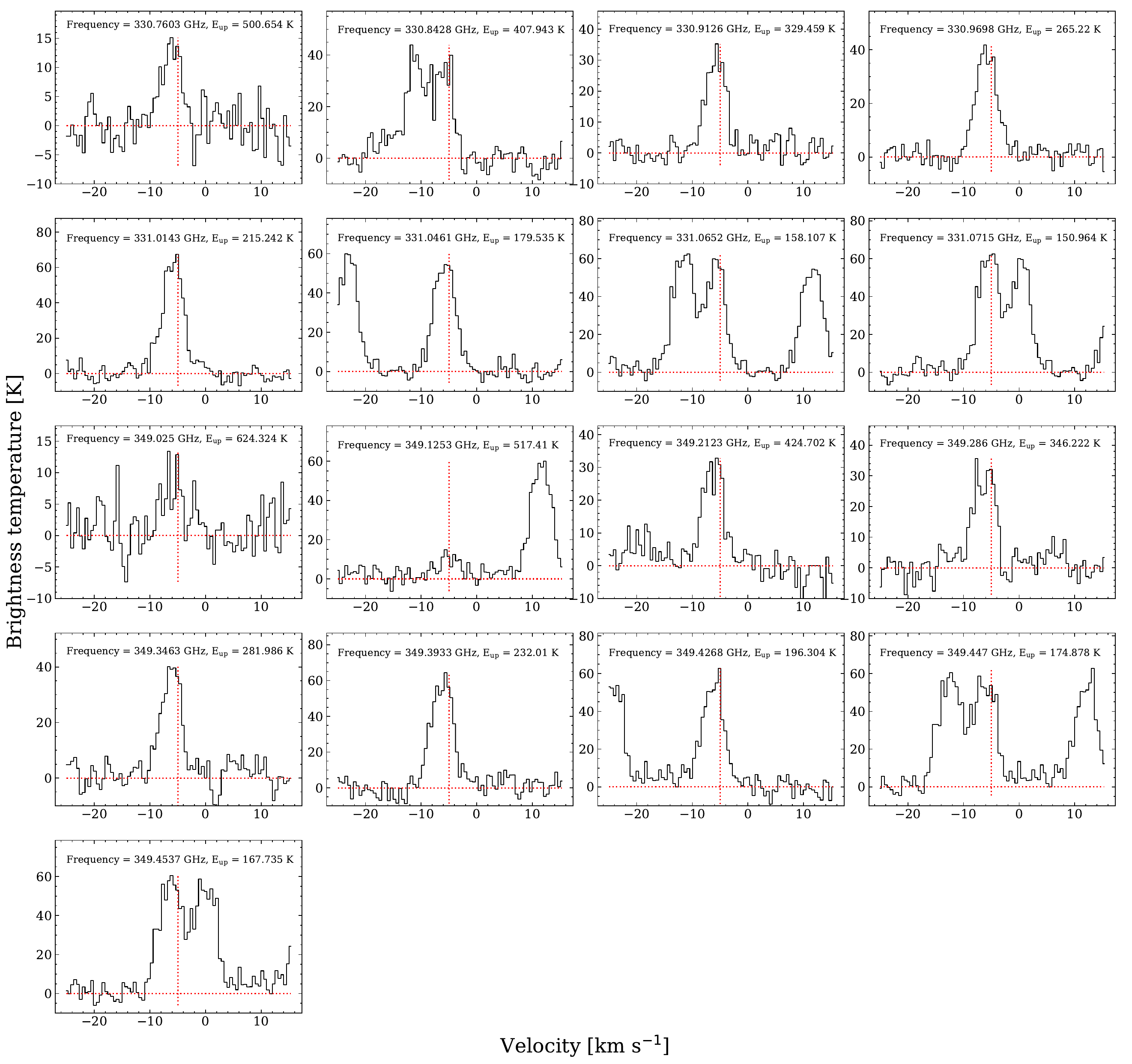}
    \caption{Detected CH$_{3}$CN,v=0 lines towards S26. Since the lines are optically thick, we used CH$_{3}$CN,v8=1 to derive a column density of2.0 (0.5)$\times 10^{16} \mathrm{cm^{-2}}$, with excitation temperature 140 (20)K.}
    \label{Fig:S26_CH3CN,v=0_1}%
\end{figure*} 


\begin{figure*}
    \centering
    \includegraphics[width=16.8cm]{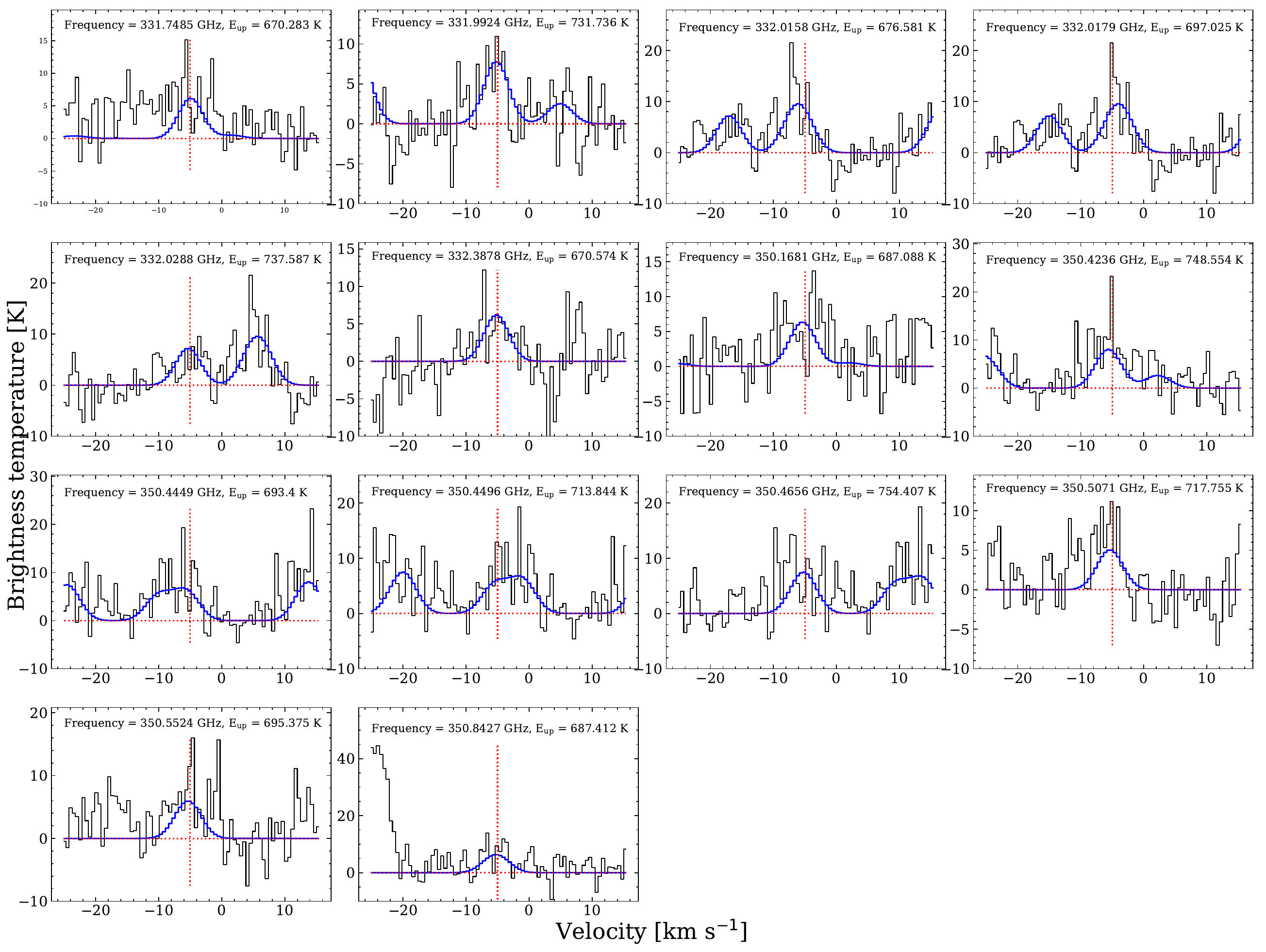}
    \caption{Detected CH$_{3}$CN,v8=1 lines towards S26. The values obtained from model fitting (model in blue) overplotted on the observed spectrum (in black) for this molecule were column density of 2.0 (0.5)$\times 10^{16} \mathrm{cm^{-2}}$, with excitation temperature 140 (20)K.}
    \label{Fig:S26_CH3CN,v8=1_1}%
\end{figure*}


\begin{figure*}
    \centering
    \includegraphics[width=16.8cm]{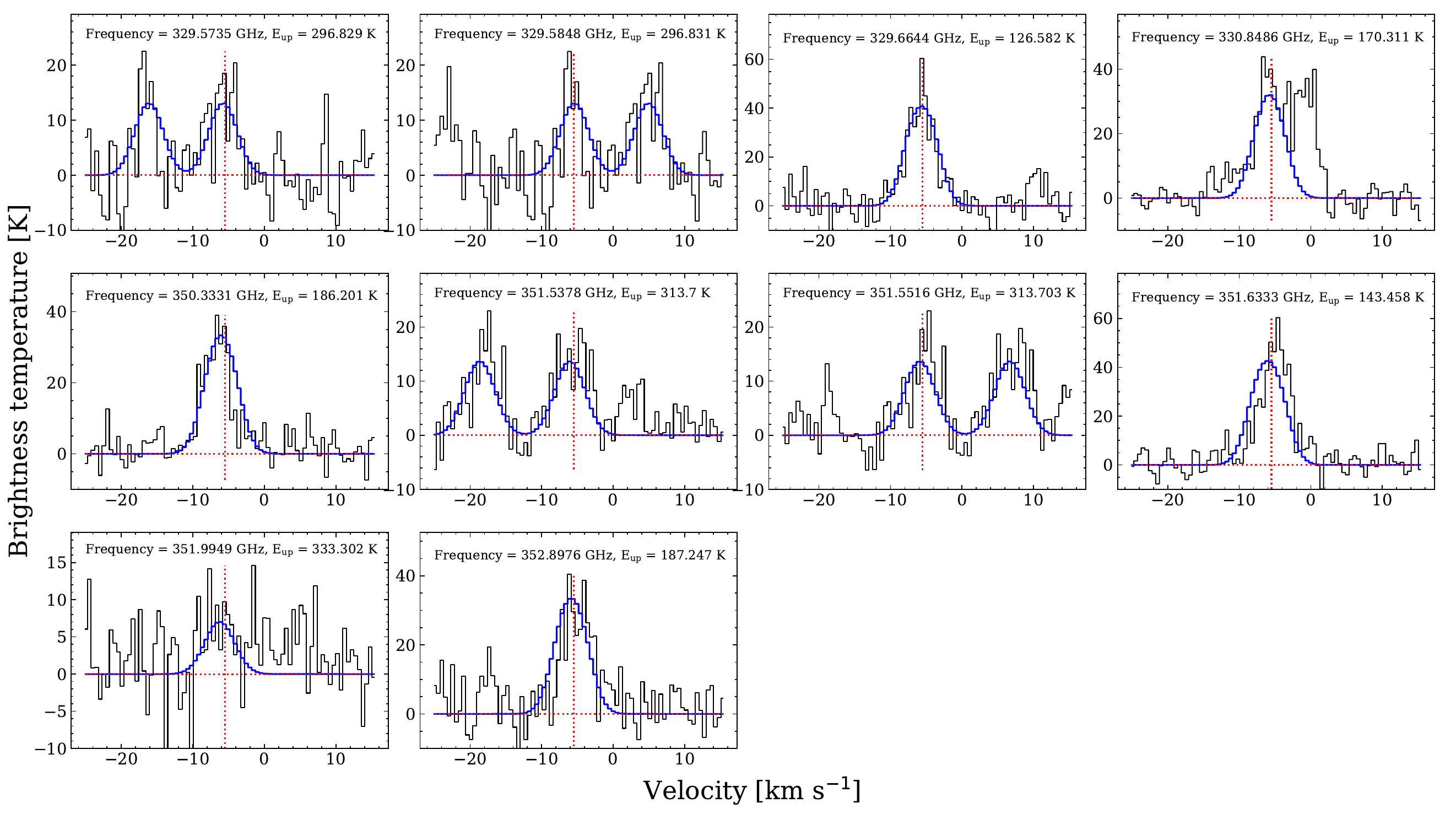}
    \caption{Detected HNCO lines towards S26. The values obtained from model fitting (model in blue) overplotted on the observed spectrum (in black) for this molecule were column density of 1.0 (0.4)$\times 10^{16} \mathrm{cm^{-2}}$, with excitation temperature 140 (30)K.}
    \label{Fig:S26_HNCO_1}%
\end{figure*} 


\begin{figure*}
    \centering
    \includegraphics[width=16.8cm]{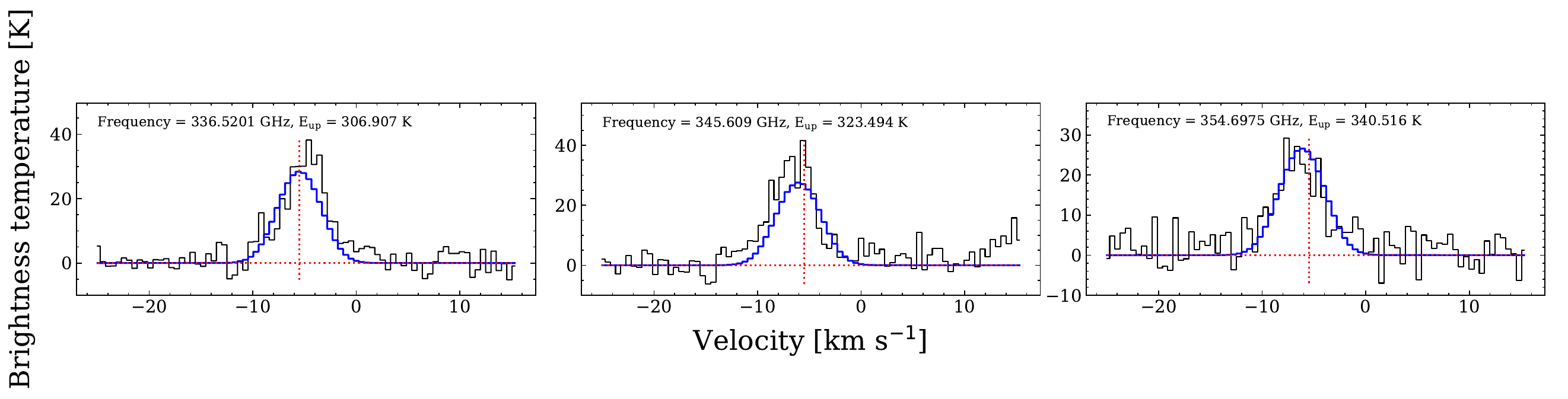}
    \caption{Detected HC$_3$N lines towards S26. The values obtained from model fitting (model in blue) overplotted on the observed spectrum (in black) for this molecule were column density of 1.0 (0.2)$\times 10^{16} \mathrm{cm^{-2}}$, with excitation temperature 140 (20)K.}
    \label{Fig:S26_HC3N_1}%
\end{figure*}



\begin{figure*}
    \centering
    \includegraphics[width=16.8cm]{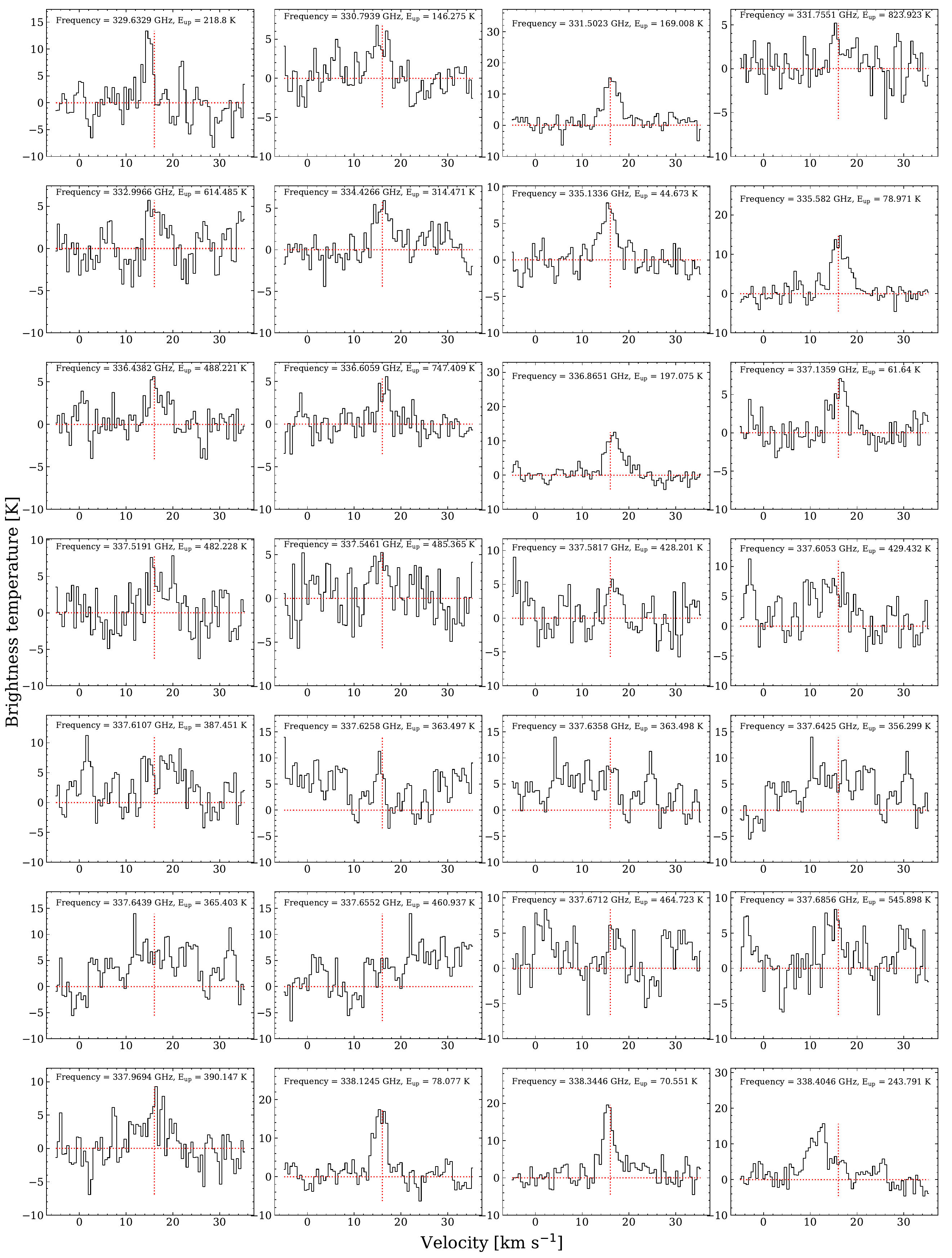}
    \caption{Detected CH$_3$OH lines towards N12. Since the lines are optically thick, we used $^{13}$CH$_3$OH (multiplied by the local ISM value of $^{12}$C/$^{13}$C = 68) to derive a column density of 5.5 (0.2)$\times 10^{17} \mathrm{cm^{-2}}$.}
    \label{Fig:N12_CH3OH_1}%
\end{figure*}

\begin{figure*}
    \centering
    \includegraphics[width=16.8cm]{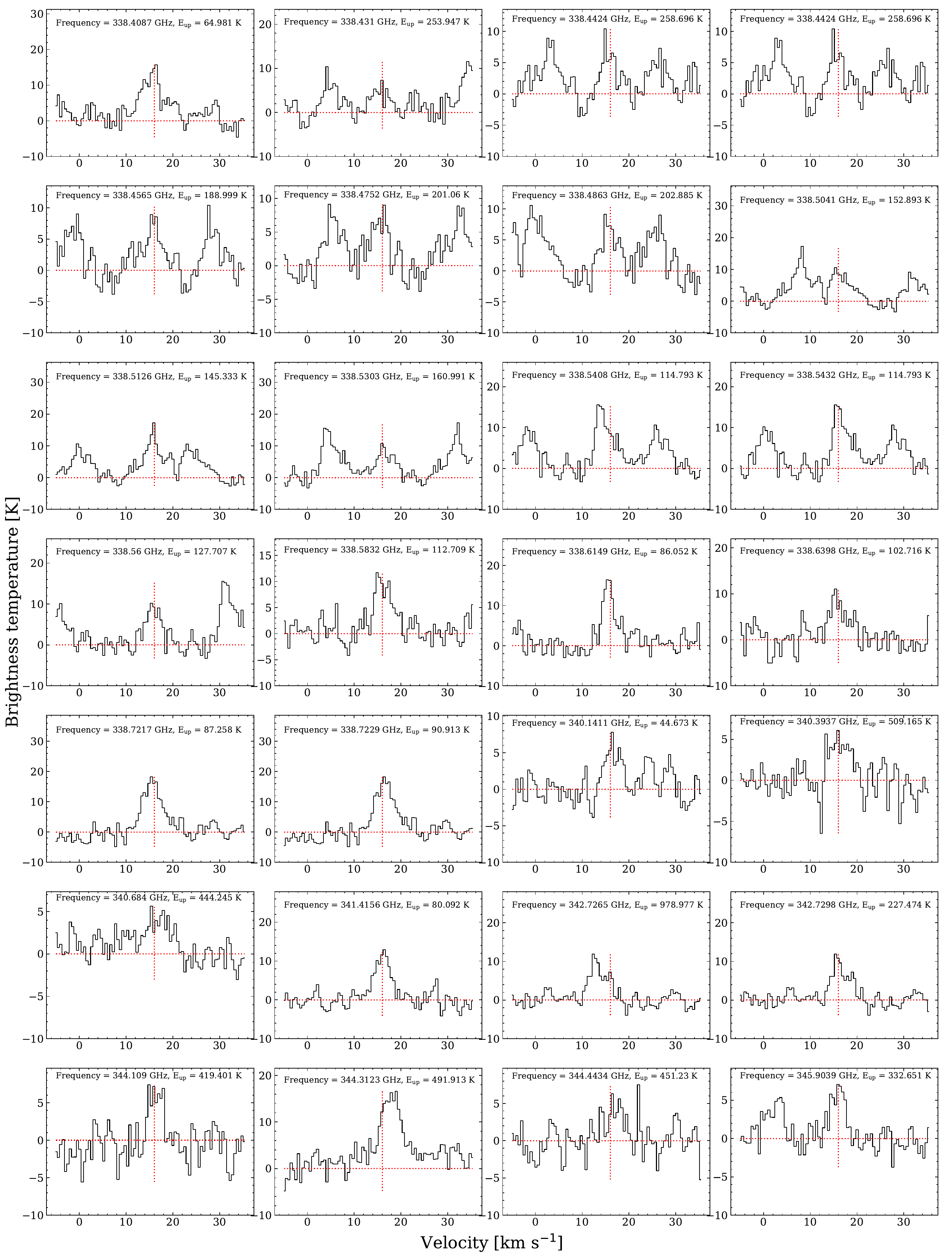}
    \caption{Detected CH$_3$OH lines towards N12. Since the lines are optically thick, we used $^{13}$CH$_3$OH (multiplied by the local ISM value of $^{12}$C/$^{13}$C = 68) to derive a column density of 5.5 (0.2)$\times 10^{17} \mathrm{cm^{-2}}$.}
    \label{Fig:N12_CH3OH_2}%
\end{figure*} 

\begin{figure*}
    \centering
    \includegraphics[width=16.8cm]{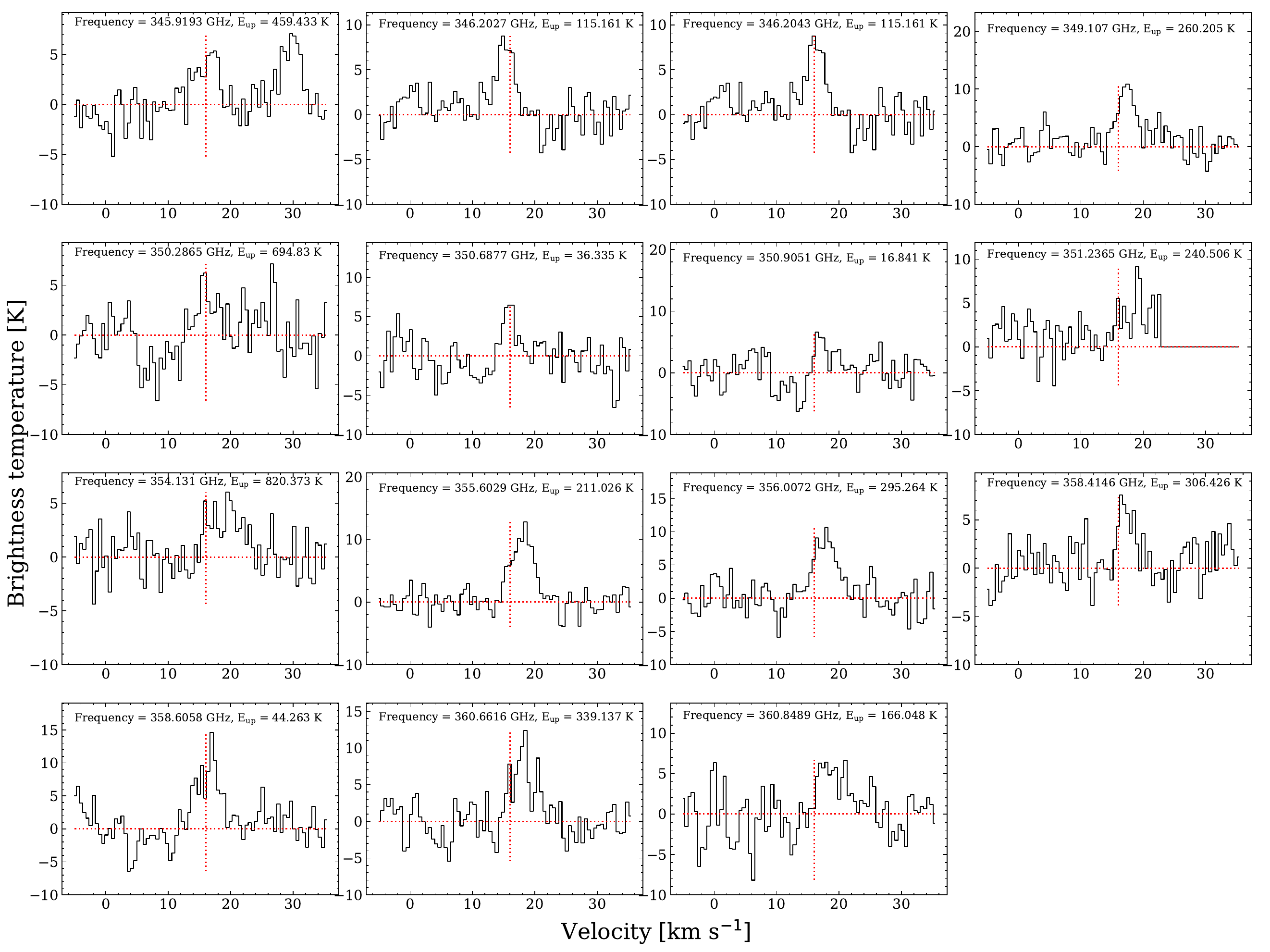}
    \caption{Detected CH$_3$OH lines towards N12. Since the lines are optically thick, we used $^{13}$CH$_3$OH (multiplied by the local ISM value of $^{12}$C/$^{13}$C = 68) to derive a column density of 5.5 (0.2)$\times 10^{17} \mathrm{cm^{-2}}$.}
    \label{Fig:N12_CH3OH_3}%
\end{figure*} 


\begin{figure*}
    \centering
    \includegraphics[width=16.8cm]{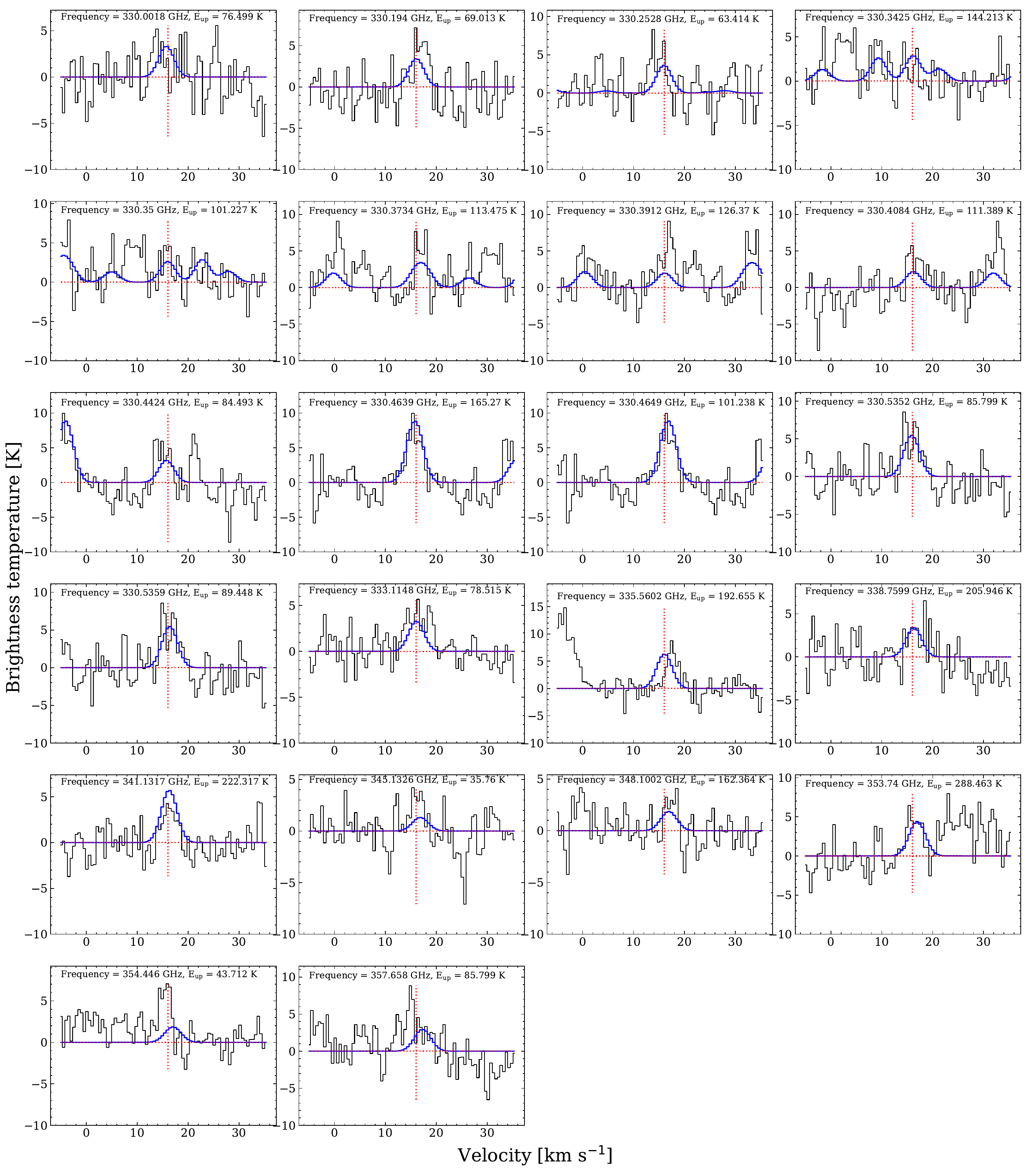}
    \caption{Detected $^{13}$CH$_3$OH lines towards N12. The values obtained from model fitting (model in blue) overplotted on the observed spectrum (in black) for this molecule were column density of 8.0 (0.3)$\times 10^{15} \mathrm{cm^{-2}}$, with excitation temperature 150 (25)K.}
    \label{Fig:N12_13CH3OH_1}%
\end{figure*}

  
\begin{figure*}
    \centering
    \includegraphics[width=16.8cm]{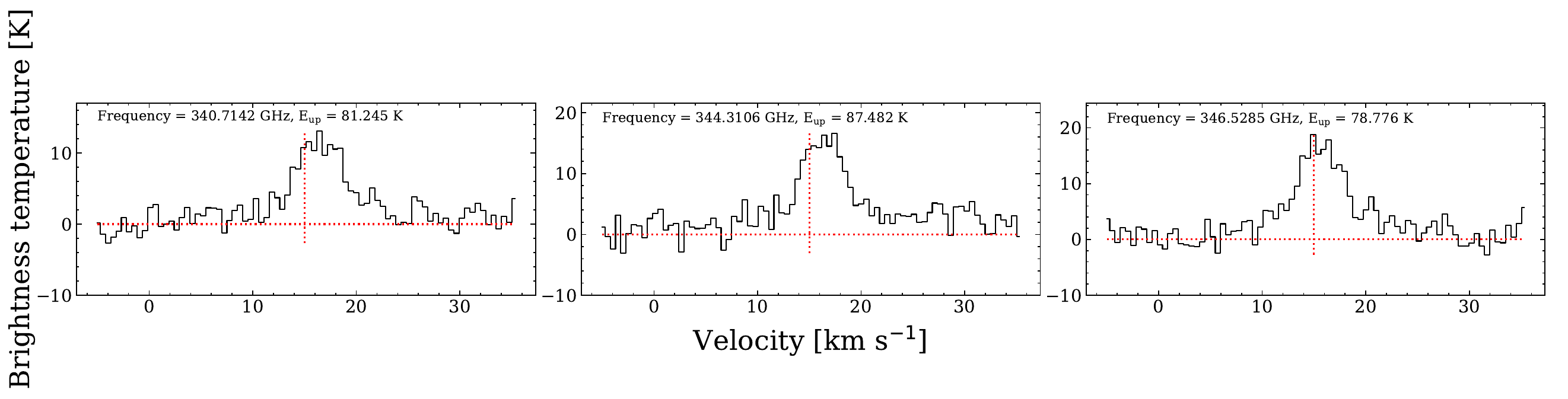}
    \caption{Detected SO lines towards N12. Since the lines are optically thick, we used $^{34}$SO (multiplied by the local ISM value of $^{32}$S/$^{34}$S = 22) to derive a column density of 1.1 (0.2)$\times 10^{16} \mathrm{cm^{-2}}$.}
    \label{Fig:N12_SO_1}%
\end{figure*} 
 
  
\begin{figure*}
    \centering
    \includegraphics[width=16.8cm]{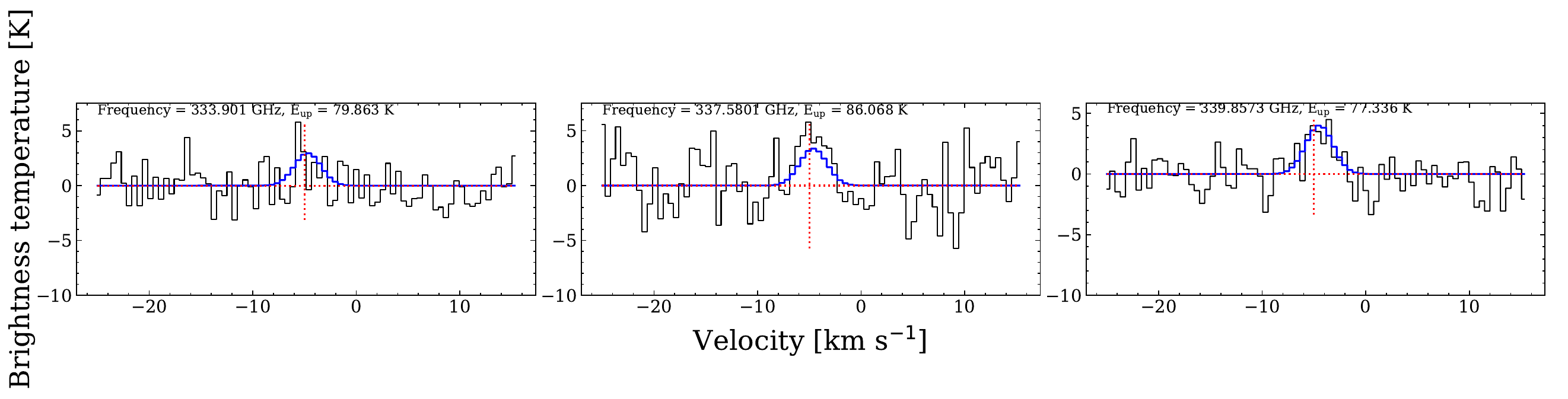}
    \caption{Detected $^{34}$SO lines towards N12. The values obtained from model fitting (model in blue) overplotted on the observed spectrum (in black) for this molecule were column density of 3.5 (0.5)$\times 10^{16} \mathrm{cm^{-2}}$, with excitation temperature 250 (50)K.}
    \label{Fig:N12_34SO_1}%
\end{figure*}

  
\begin{figure*}
    \centering
    \includegraphics[width=16.8cm]{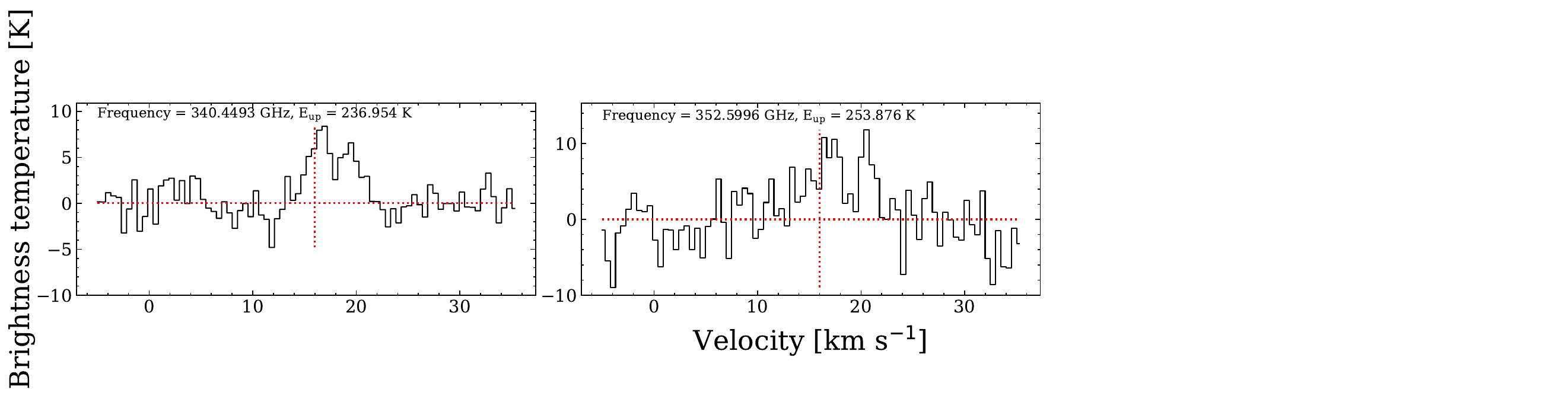}
    \caption{Detected OCS lines towards N12. Since the lines are optically thick, we used OC$^{34}$S (multiplied by the local ISM value of $^{32}$S/$^{34}$S = 22) to derive a column density of 3.3 (1.0)$\times 10^{16} \mathrm{cm^{-2}}$.}
    \label{Fig:N12_OCS_1}%
\end{figure*} 

  
\begin{figure*}
    \centering
    \includegraphics[width=16.8cm]{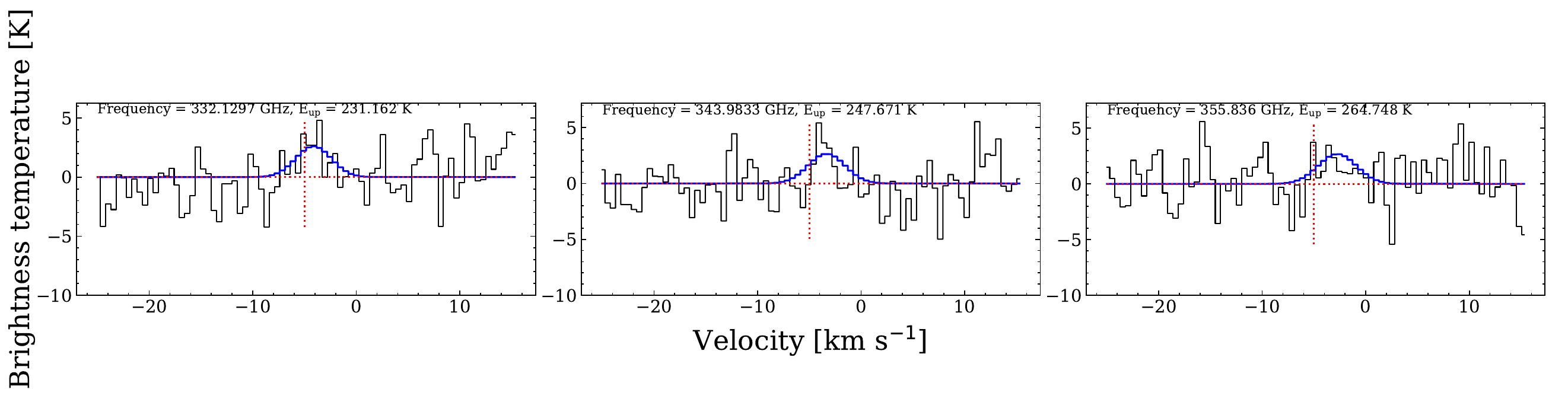}
    \caption{Detected OC$^{34}$S lines towards N12. The values obtained from model fitting (model in blue) overplotted on the observed spectrum (in black) for this molecule were column density of 1.5 (0.5)$\times 10^{15} \mathrm{cm^{-2}}$, with excitation temperature 200 (100)K.}
    \label{Fig:N12_OC34S_1}%
\end{figure*} 

  
\begin{figure*}
    \centering
    \includegraphics[width=16.8cm]{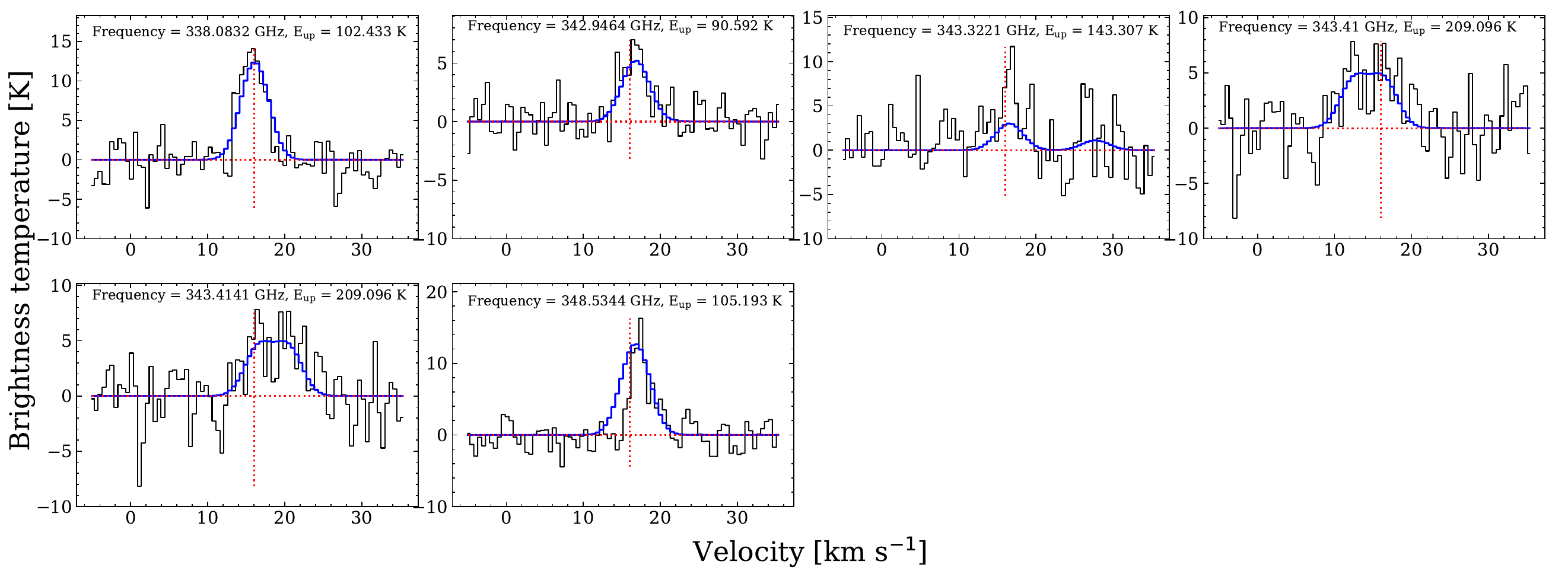}
    \caption{Detected H2CS lines towards N12. The values obtained from model fitting (model in blue) overplotted on the observed spectrum (in black) for this molecule were column density of 2.0 (0.5)$\times 10^{15} \mathrm{cm^{-2}}$, with excitation temperature 100 (50)K.}
    \label{Fig:N12_H2CS_1}%
\end{figure*} 

  
\begin{figure*}
    \centering
    \includegraphics[width=16.8cm]{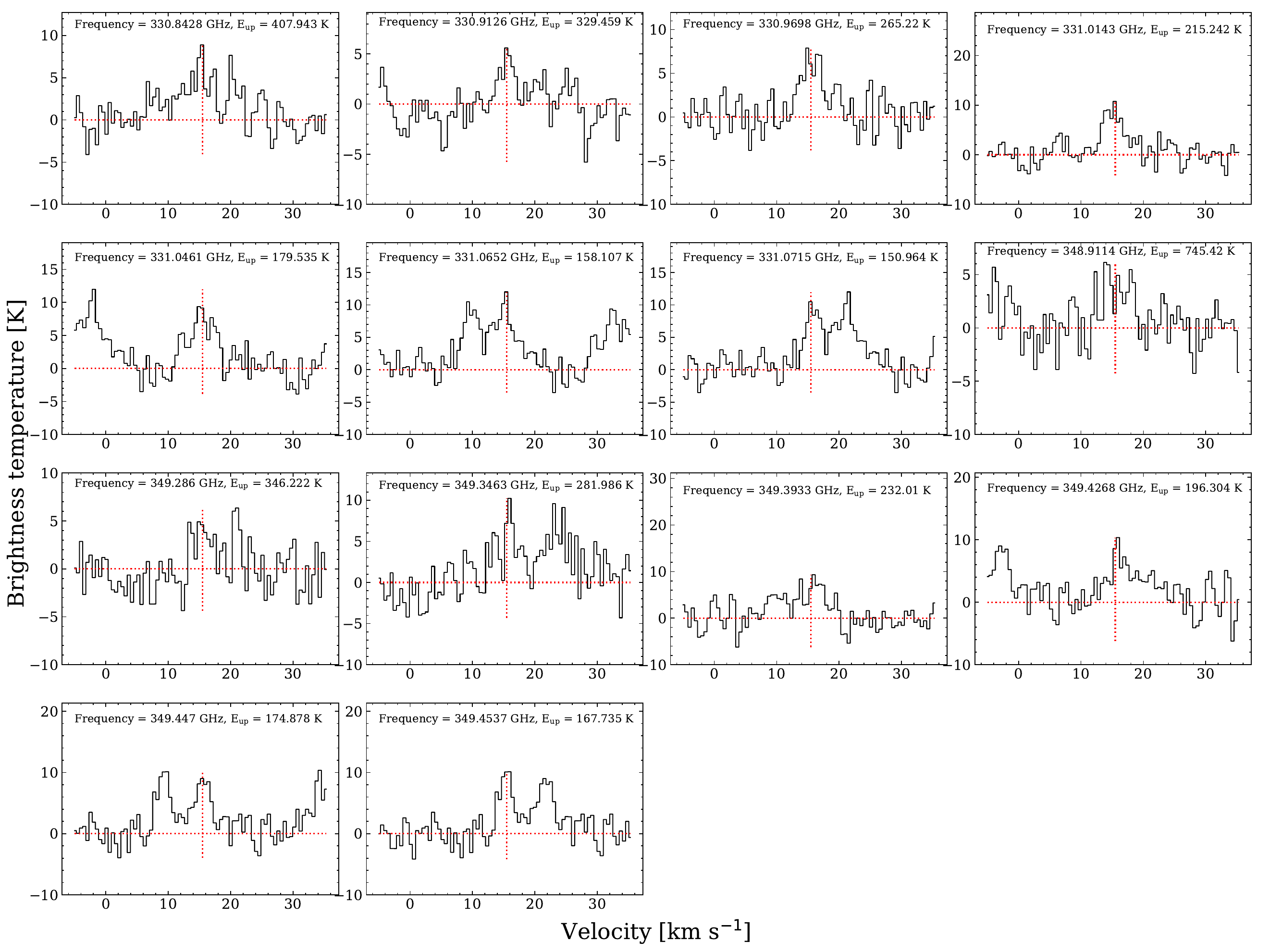}
    \caption{Detected CH$_3$CN lines towards N12. Since the lines are optically thick, we used the higher vibration transitions CH$_3$CN,v$_8=1$ to derive a column density of 2.0 (1.0)$\times 10^{16} \mathrm{cm^{-2}}$.}
    \label{Fig:N12_CH3CN_1}%
\end{figure*} 

  
\begin{figure*}
    \centering
    \includegraphics[width=16.8cm]{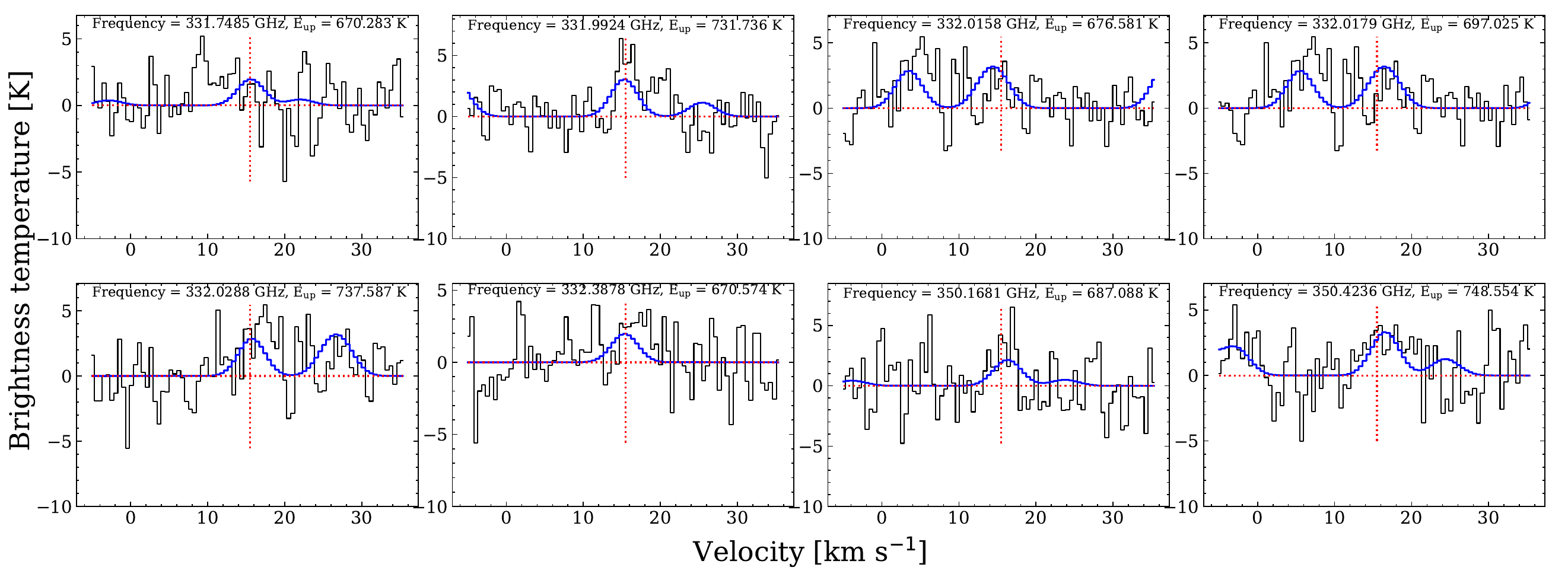}
    \caption{Detected CH$_3$CN,v$_8=1$ lines towards N12. The values obtained from model fitting (model in blue) overplotted on the observed spectrum (in black) for this molecule were column density of 2.0 (1.0)$\times 10^{15} \mathrm{cm^{-2}}$, with excitation temperature 250 (50)K.}
    \label{Fig:N12_CH3CNv8_1}%
\end{figure*}



\begin{figure*}
    \centering
    \includegraphics[width=16.8cm]{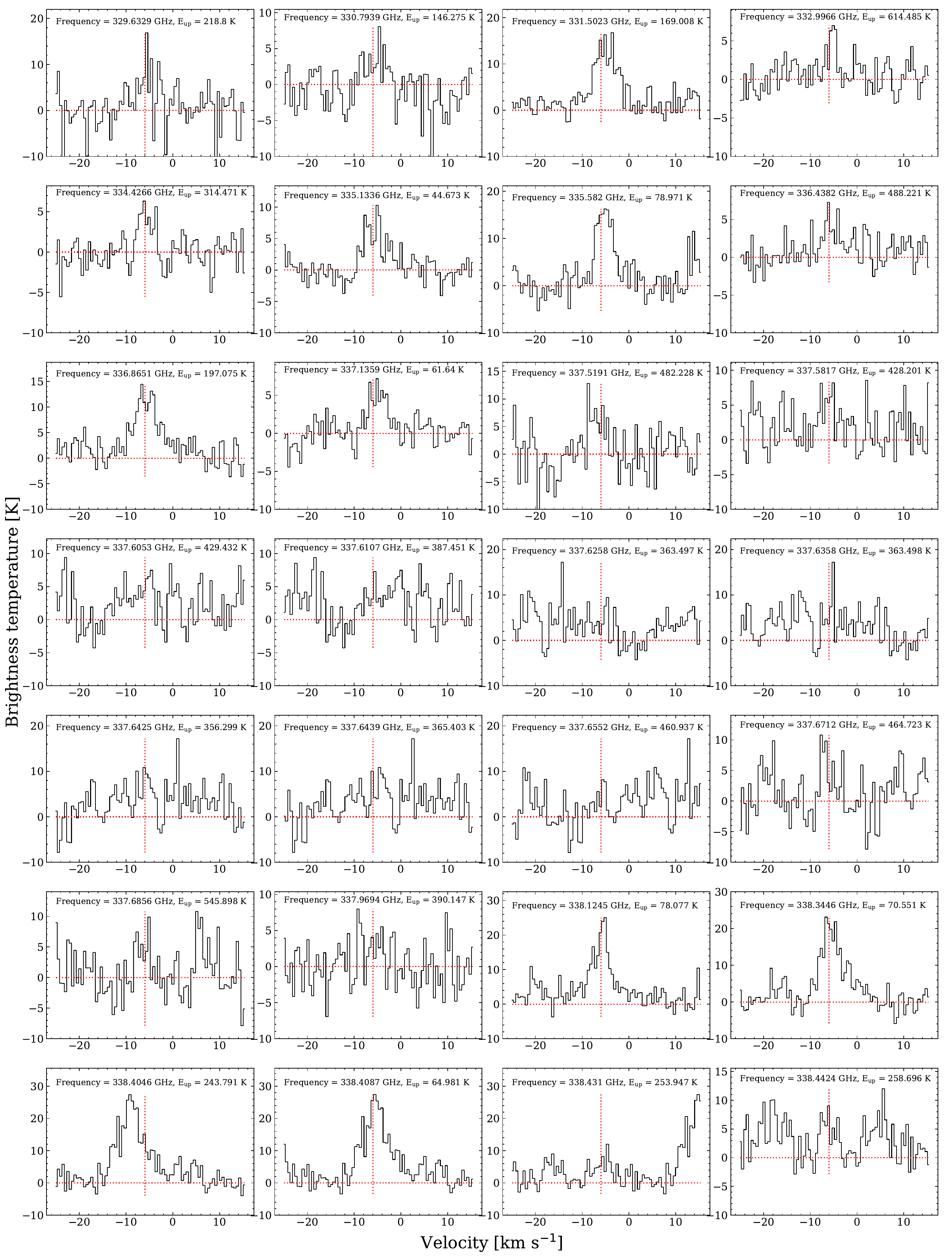}
    \caption{Detected CH$_3$OH lines towards N63. Since the lines are optically thick, we used $^{13}$CH$_3$OH (multiplied by the local ISM value of $^{12}$C/$^{13}$C = 68) to derive a column density of 5.0 (0.2)$\times 10^{17} \mathrm{cm^{-2}}$.}
    \label{Fig:N63_CH3OH_1}%
\end{figure*}

\begin{figure*}
    \centering
    \includegraphics[width=16.8cm]{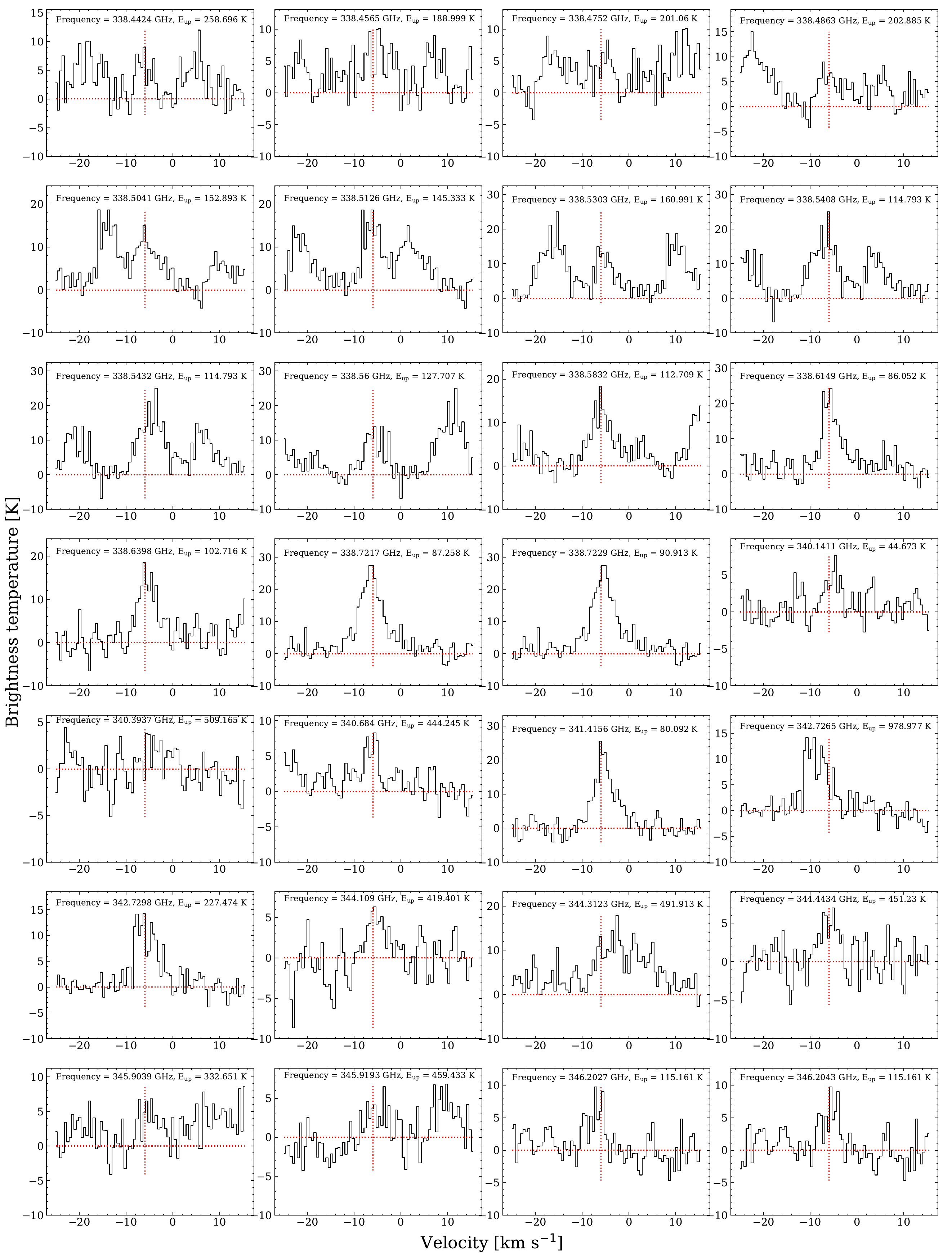}
    \caption{Detected CH$_3$OH lines towards N63. Since the lines are optically thick, we used $^{13}$CH$_3$OH (multiplied by the local ISM value of $^{12}$C/$^{13}$C = 68) to derive a column density of 5.0 (0.2)$\times 10^{17} \mathrm{cm^{-2}}$.}
    \label{Fig:N63_CH3OH_2}%
\end{figure*} 

\begin{figure*}
    \centering
    \includegraphics[width=16.8cm]{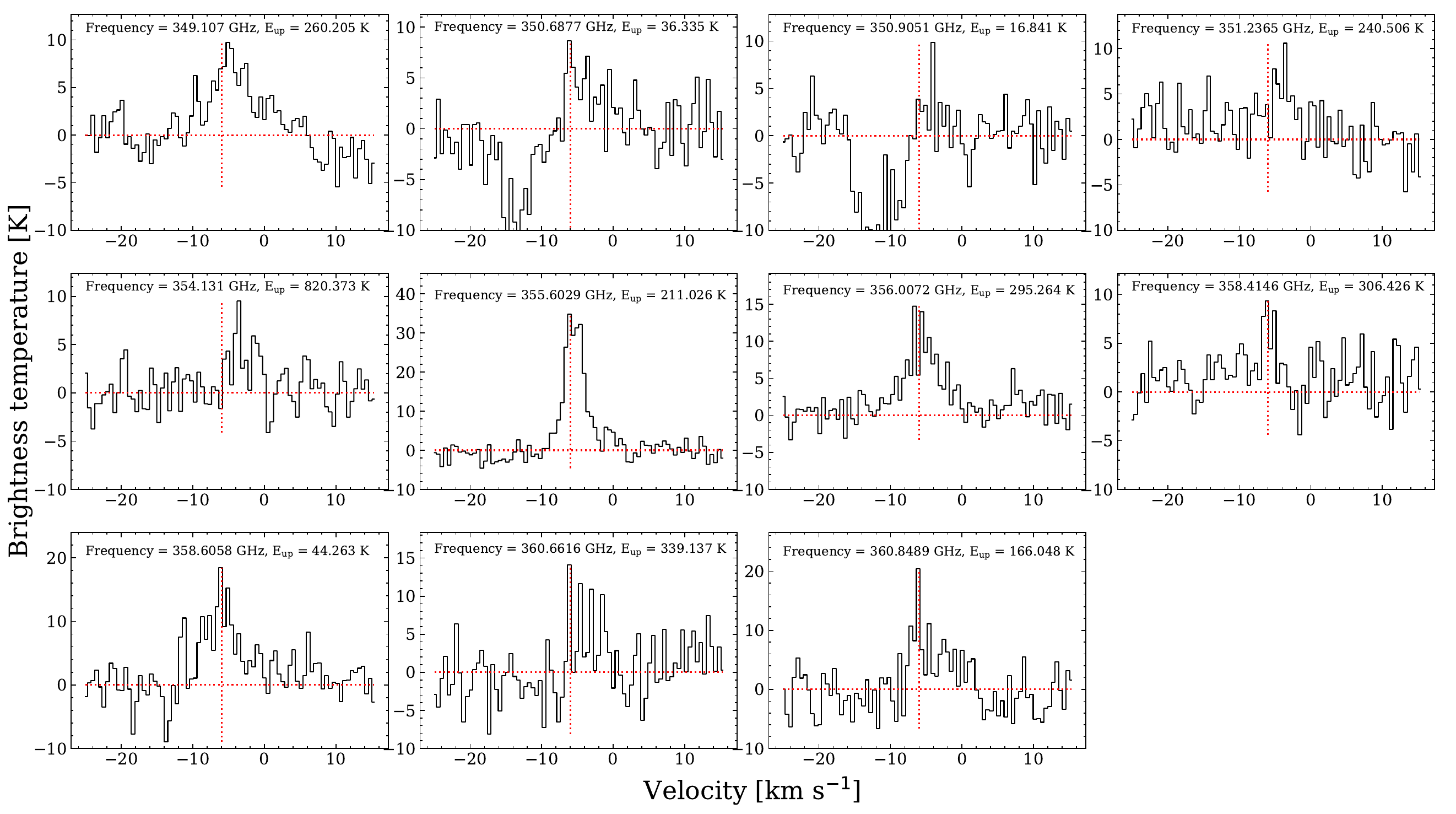}
    \caption{Detected CH$_3$OH lines towards N63. Since the lines are optically thick, we used $^{13}$CH$_3$OH (multiplied by the local ISM value of $^{12}$C/$^{13}$C = 68) to derive a column density of 5.0 (0.2)$\times 10^{17} \mathrm{cm^{-2}}$.}
    \label{Fig:N63_CH3OH_3}%
\end{figure*} 


\begin{figure*}
    \centering
    \includegraphics[width=16.8cm]{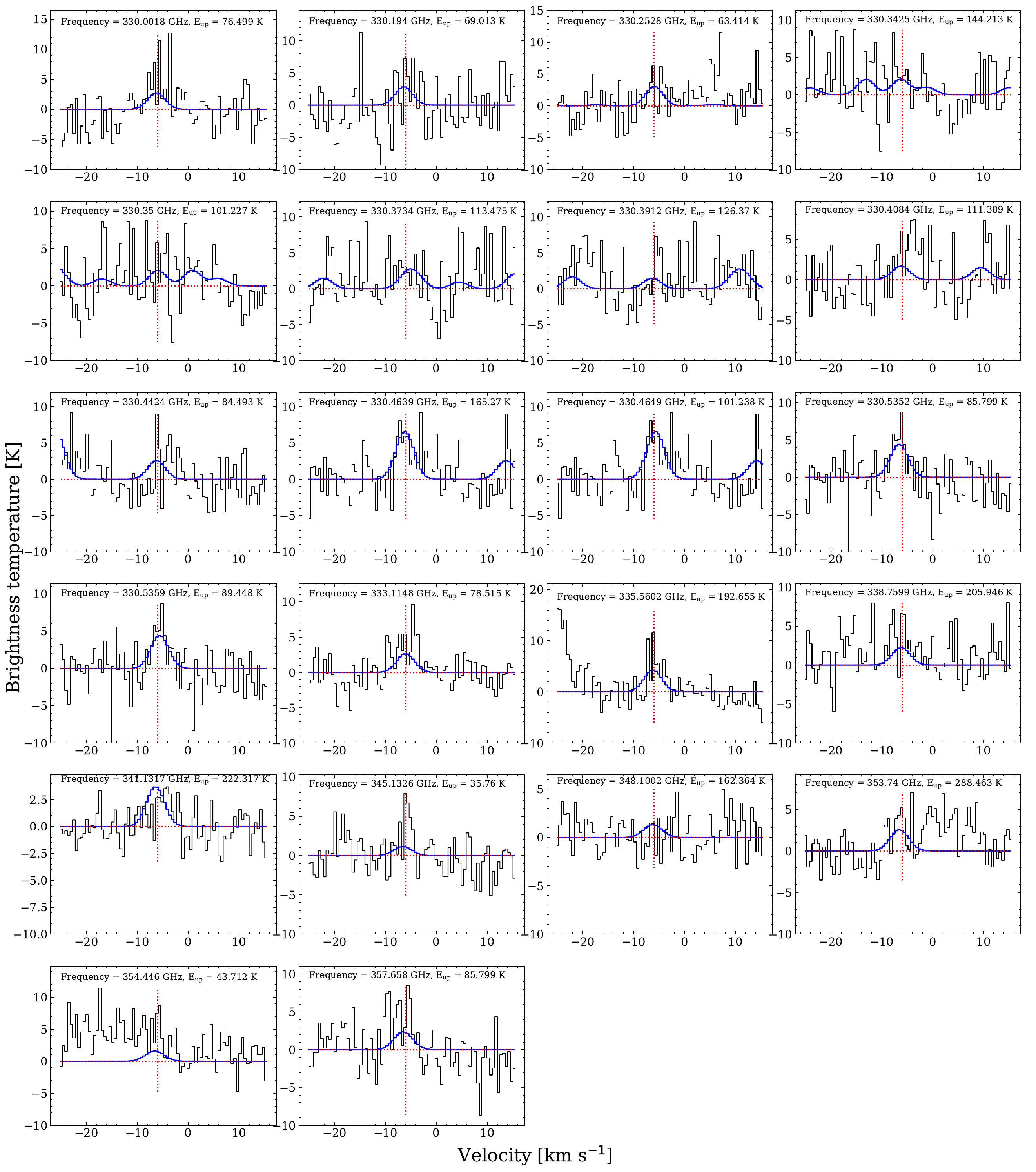}
    \caption{Detected $^{13}$CH$_3$OH lines towards N63. The values obtained from model fitting (model in blue) overplotted on the observed spectrum (in black) for this molecule were column density of 5.8 (0.3)$\times 10^{15} \mathrm{cm^{-2}}$, with excitation temperature 120 (50)K.}
    \label{Fig:N63_13CH3OH_1}%
\end{figure*} 


\begin{figure*}
    \centering
    \includegraphics[width=16.8cm]{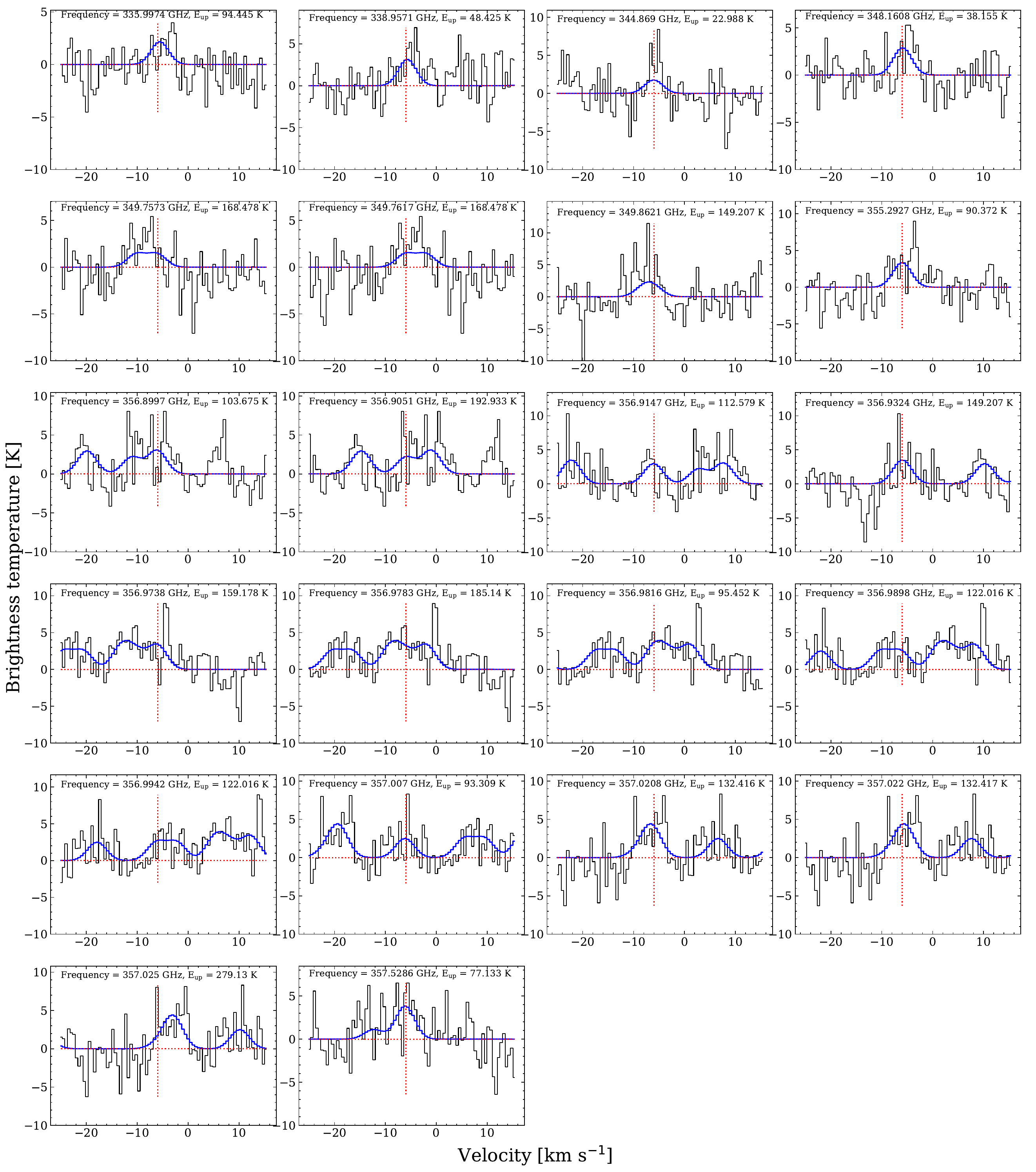}
    \caption{Detected CH$_2$DOH lines towards N63. The values obtained from model fitting (model in blue) overplotted on the observed spectrum (in black) for this molecule were column density of 1.5 (0.4)$\times 10^{16} \mathrm{cm^{-2}}$, with excitation temperature 150 (40)K.}
    \label{Fig:N63_CH2DOH_1}%
\end{figure*} 

  
\begin{figure*}
    \centering
    \includegraphics[width=16.8cm]{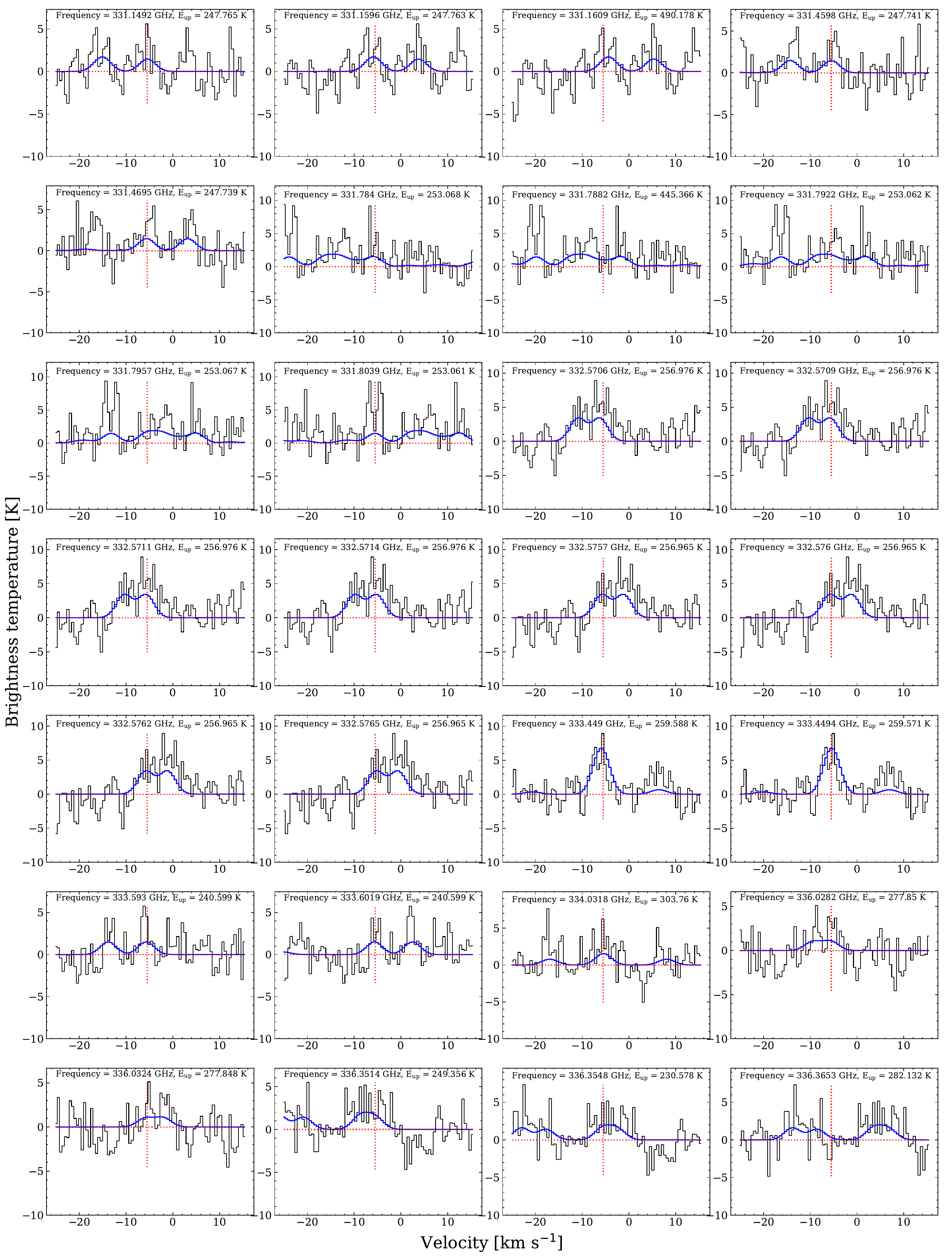}
    \caption{Detected CH$_3$OCHO lines towards N63. The values obtained from model fitting (model in blue) overplotted on the observed spectrum (in black) for this molecule were column density of 1.2 (0.4)$\times 10^{16} \mathrm{cm^{-2}}$, with excitation temperature 120 (40)K.}
    \label{Fig:N63_CH3OCHO_1}%
\end{figure*}

\begin{figure*}
    \centering
    \includegraphics[width=16.8cm]{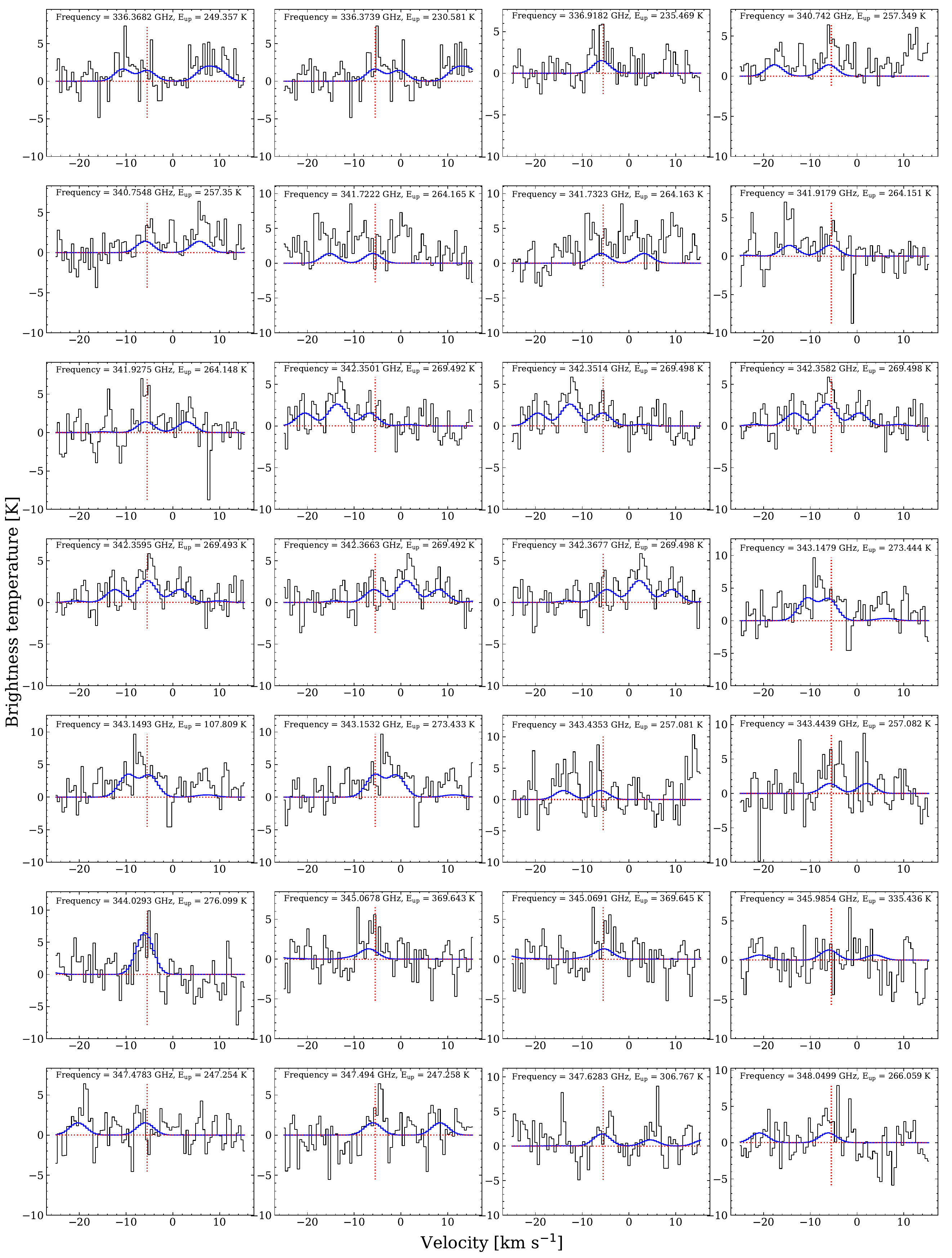}
    \caption{Detected CH$_3$OCHO lines towards N63. The values obtained from model fitting (model in blue) overplotted on the observed spectrum (in black) for this molecule were column density of 1.2 (0.4)$\times 10^{16} \mathrm{cm^{-2}}$, with excitation temperature 120 (40)K.}
    \label{Fig:N63_CH3OCHO_2}%
\end{figure*} 

\begin{figure*}
    \centering
    \includegraphics[width=16.8cm]{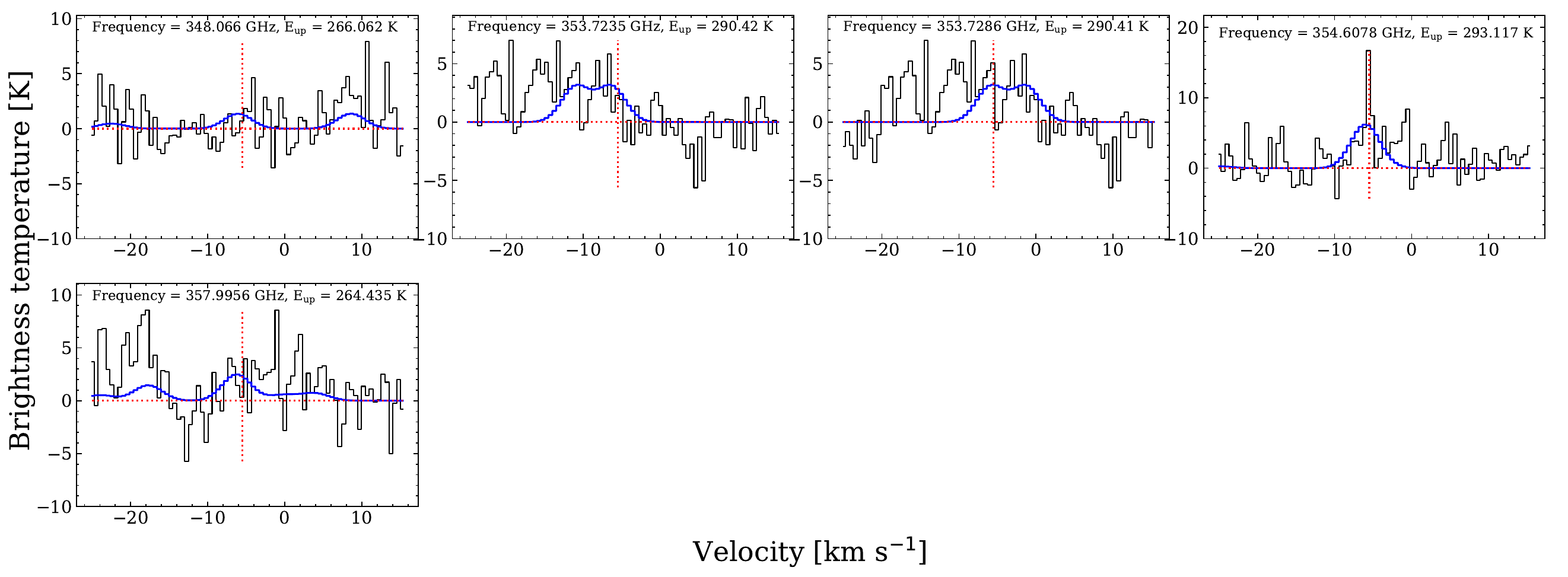}
    \caption{Detected CH$_3$OCHO lines towards N63. The values obtained from model fitting (model in blue) overplotted on the observed spectrum (in black) for this molecule were column density of 1.2 (0.4)$\times 10^{16} \mathrm{cm^{-2}}$, with excitation temperature 120 (40)K.}
    \label{Fig:N63_CH3OCHO_3}%
\end{figure*} 

 
 \begin{figure*}
    \centering
    \includegraphics[width=16.8cm]{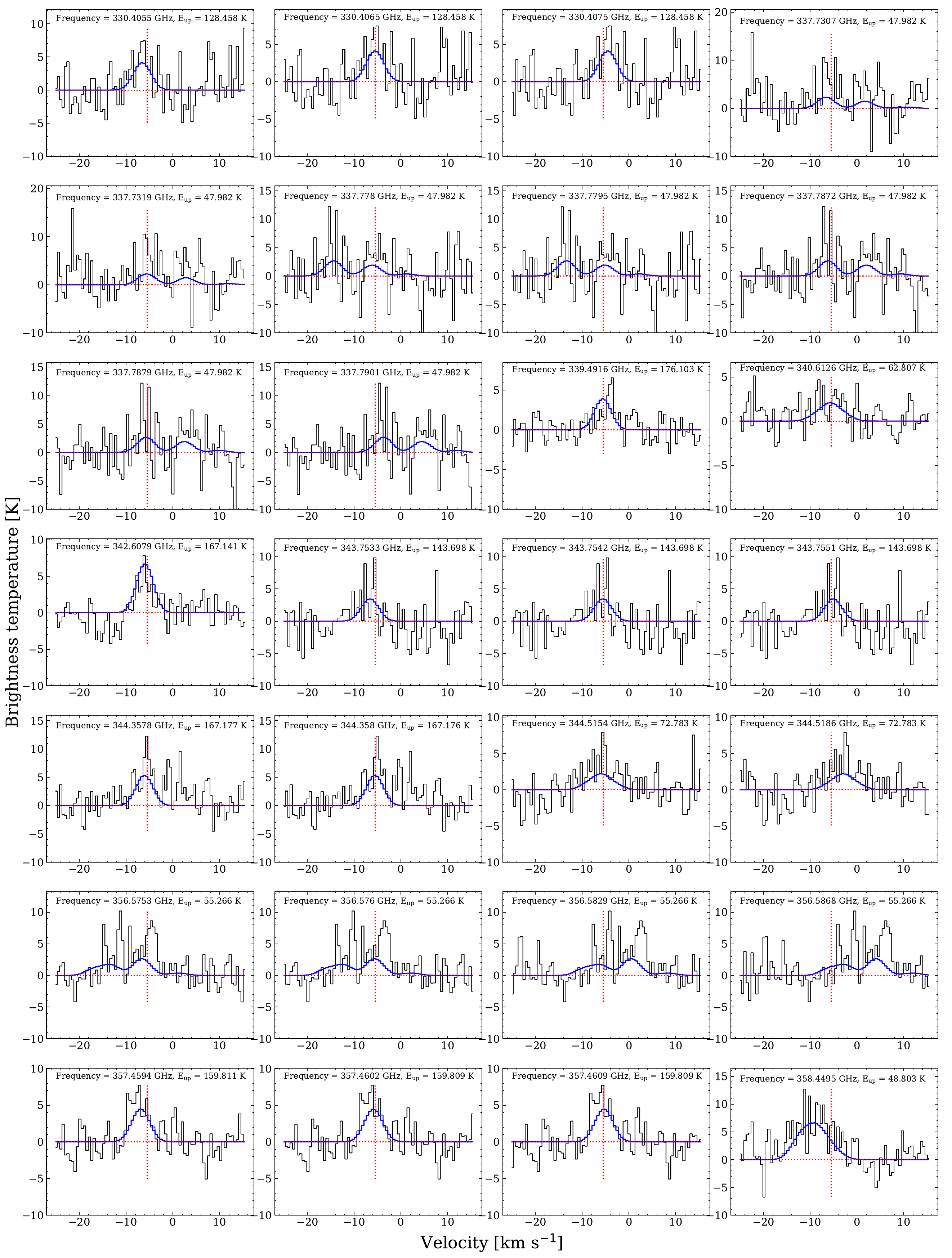}
    \caption{Detected CH$_3$OCHO lines towards N63. The values obtained from model fitting (model in blue) overplotted on the observed spectrum (in black) for this molecule were column density of 1.4 (0.4)$\times 10^{16} \mathrm{cm^{-2}}$, with excitation temperature 120 (40)K.}
    \label{Fig:N63_CH3OCH3_1}%
\end{figure*} 

\begin{figure*}
    \centering
    \includegraphics[width=16.8cm]{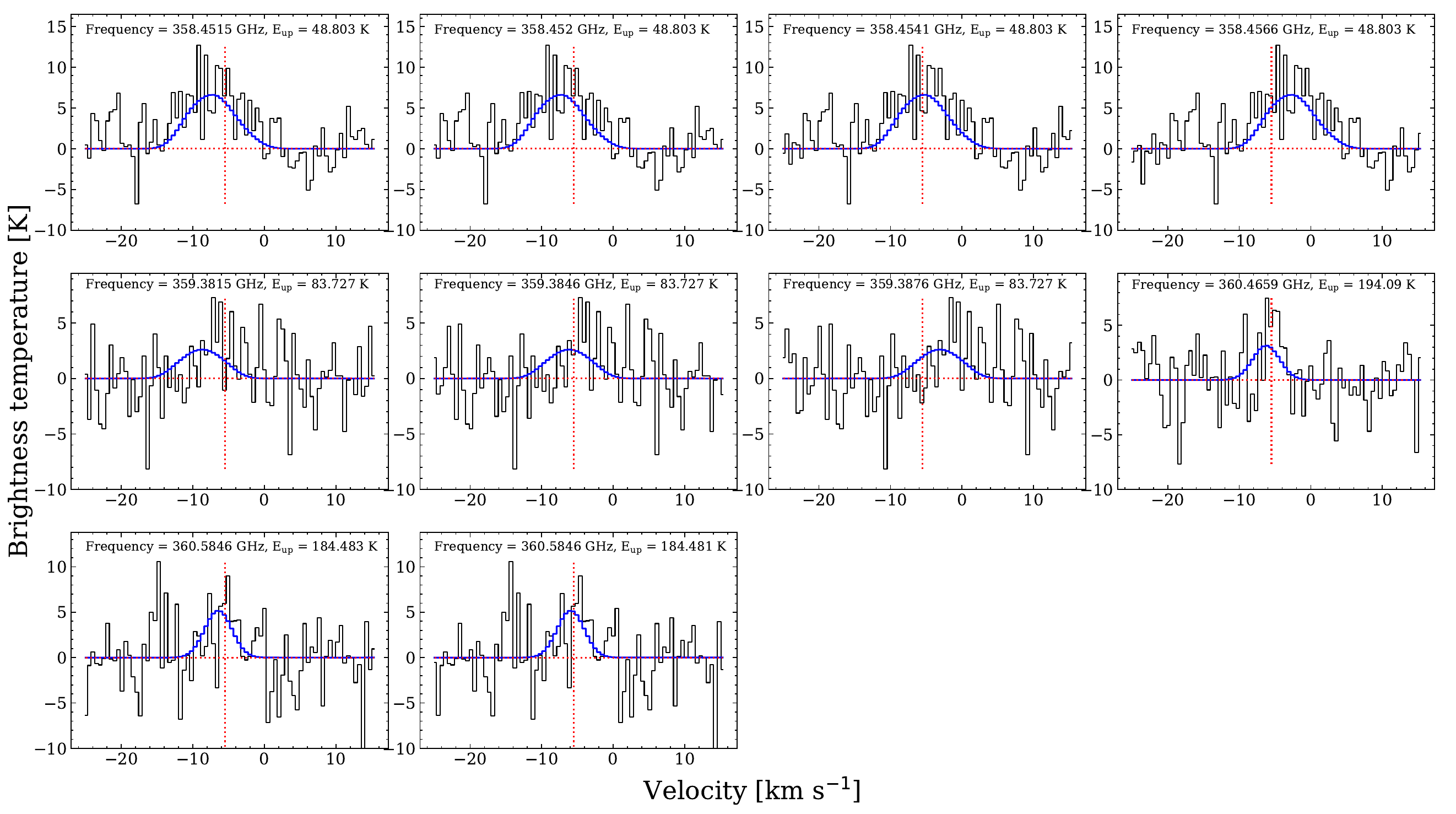}
    \caption{Detected CH$_3$OCH$_3$ lines towards N63. The values obtained from model fitting (model in blue) overplotted on the observed spectrum (in black) for this molecule were column density of 1.4 (0.4)$\times 10^{16} \mathrm{cm^{-2}}$, with excitation temperature 120 (40)K.}
    \label{Fig:N63_CH3OCH3_2}%
\end{figure*} 


\begin{figure*}
    \centering
    \includegraphics[width=16.8cm]{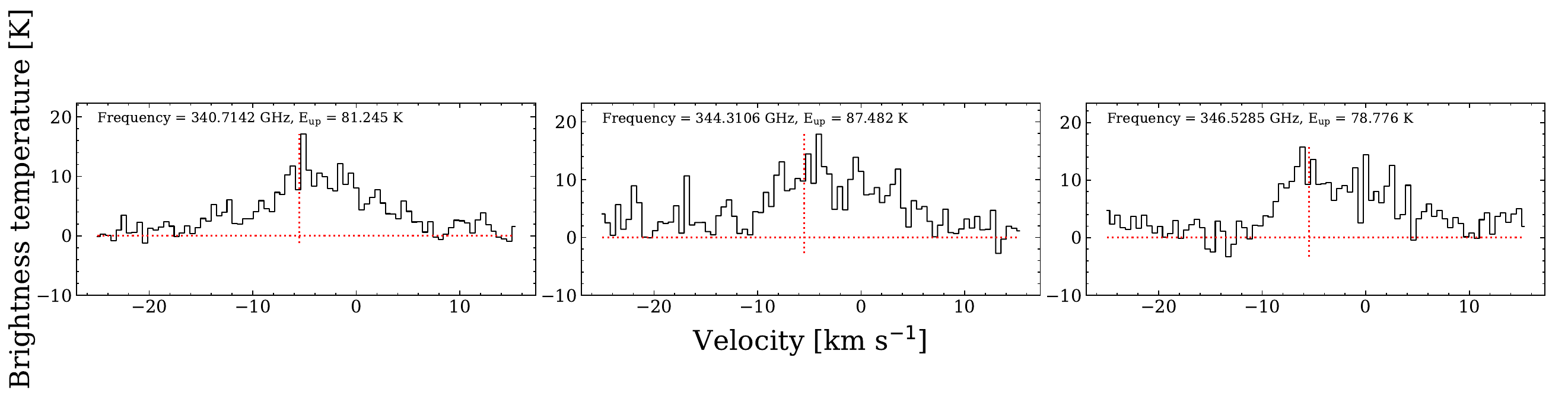}
    \caption{Since SO lines are optically thick, we used $^{34}$SO (multiplied by the local ISM value of $^{32}$S/$^{34}$S = 22) to derive a column density of 1.2 (0.3)$\times 10^{16} \mathrm{cm^{-2}}$.}
    \label{Fig:N63_SO}%
\end{figure*}


\begin{figure*}
    \centering
    \includegraphics[width=16.8cm]{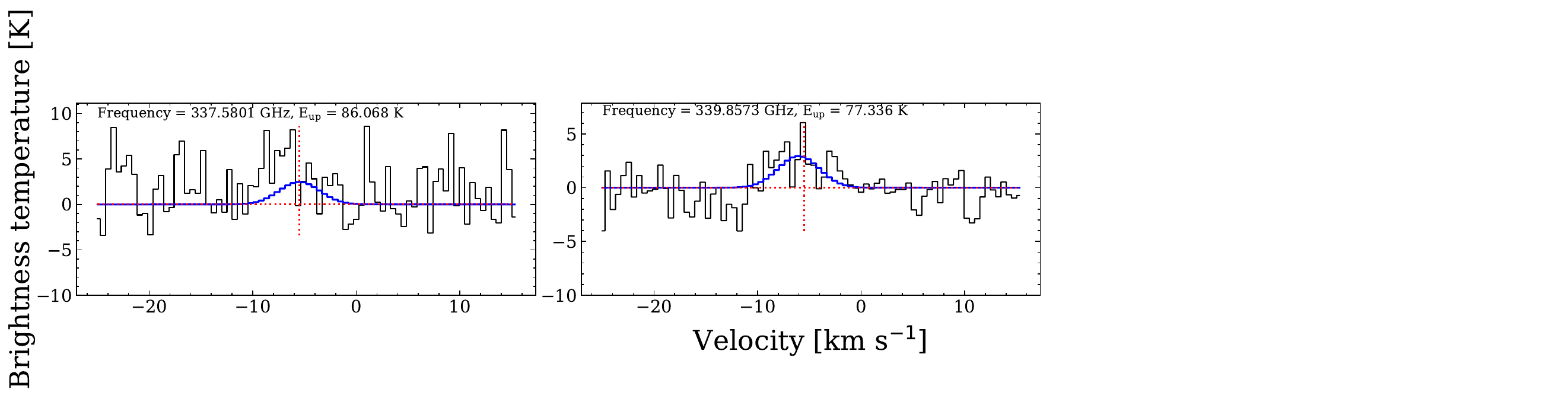}
    \caption{Detected  $^{34}$SO lines towards N63. The values obtained from model fitting (model in blue) overplotted on the observed spectrum (in black) for this molecule were column density of 5.5 (1.0)$\times 10^{14} \mathrm{cm^{-2}}$, with excitation temperature 250 (50)K.}
    \label{Fig:N63_34SO}%
\end{figure*} 


\begin{figure*}
    \centering
    \includegraphics[width=16.8cm]{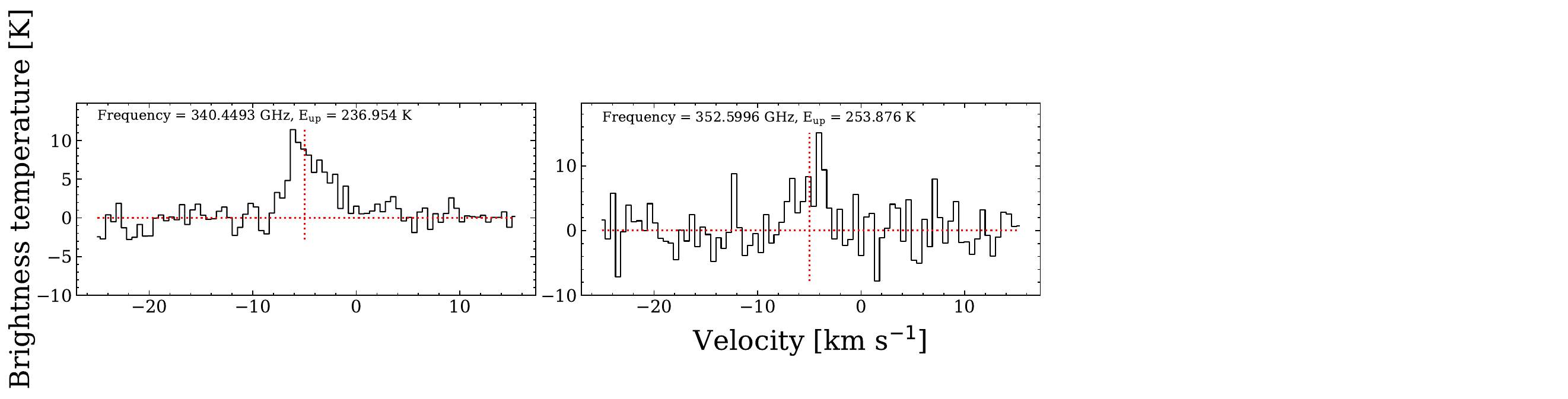}
    \caption{Since OCS lines are optically thick, we used OC$^{34}$S (multiplied by the local ISM value of $^{32}$S/$^{34}$S = 22) to derive a column density of 3.3 (1.1)$\times 10^{16} \mathrm{cm^{-2}}$.}
    \label{Fig:N63_OCS_1}%
\end{figure*} 


\begin{figure*}
    \centering
    \includegraphics[width=16.8cm]{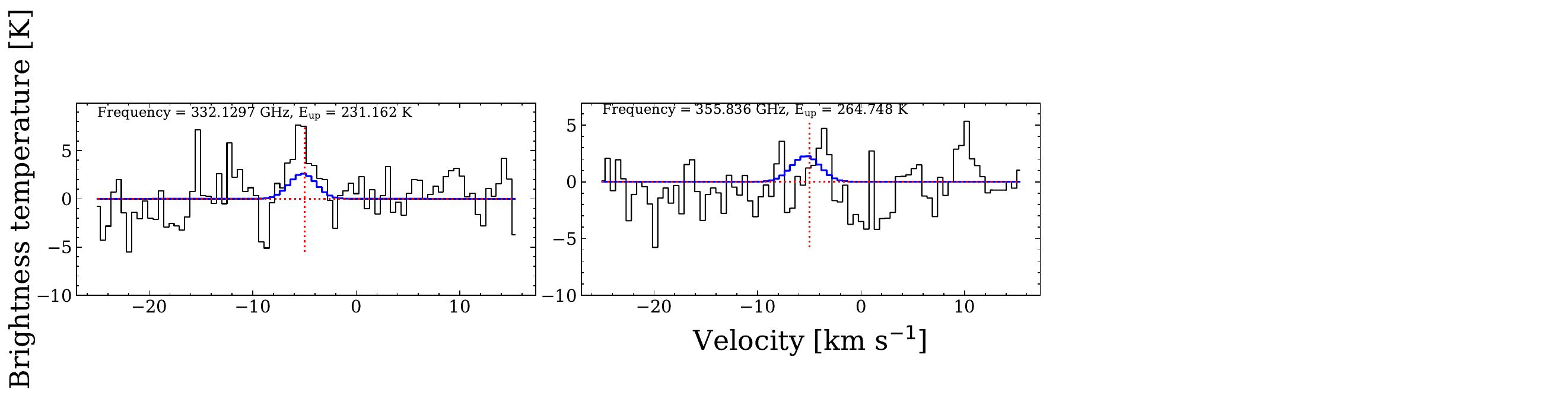}
    \caption{Detected OC$^{34}$S lines towards N63. The values obtained from model fitting (model in blue) overplotted on the observed spectrum (in black) for this molecule were column density of 1.5 (0.5)$\times 10^{15} \mathrm{cm^{-2}}$, with excitation temperature 100 (30)K.}
    \label{Fig:N63_OC34S_1}%
\end{figure*} 


\begin{figure*}
    \centering
    \includegraphics[width=16.8cm]{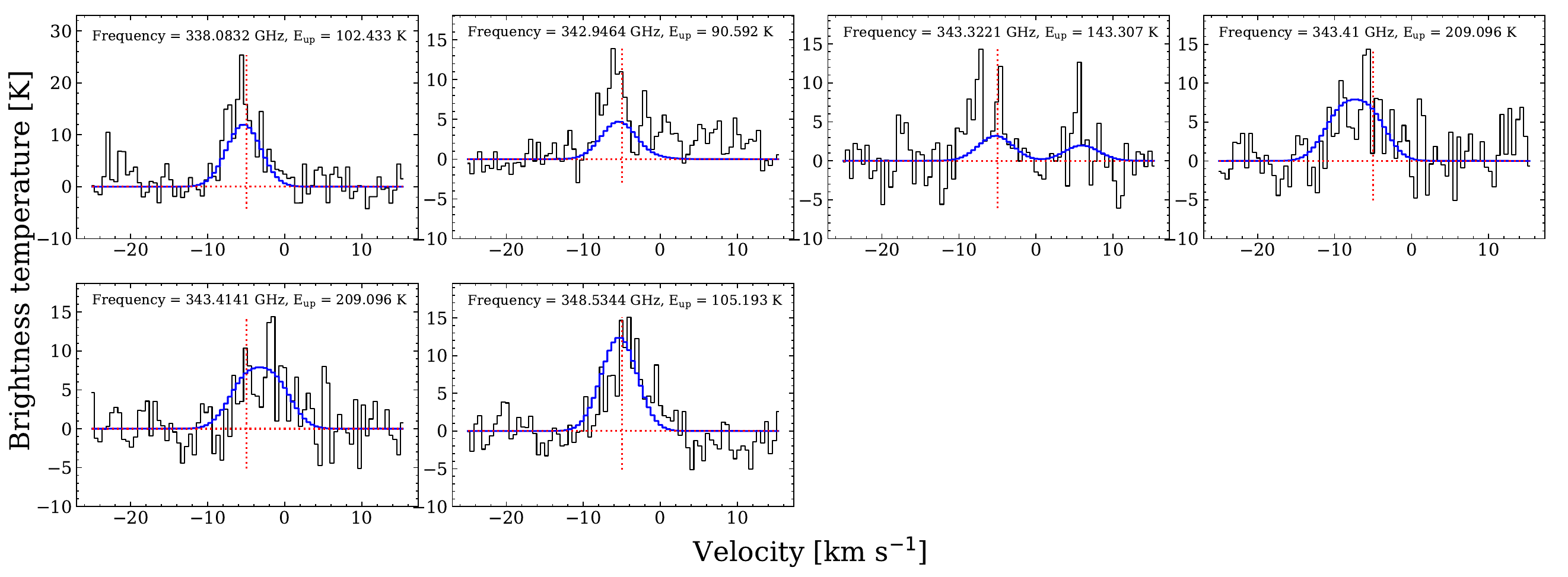}
    \caption{Detected H$_{2}$CS lines towards N63. The values obtained from model fitting (model in blue) overplotted on the observed spectrum (in black) for this molecule were column density of 3.0 (1.0)$\times 10^{15} \mathrm{cm^{-2}}$, with excitation temperature 150 (40)K.}
    \label{Fig:N63_H2CS_1}%
\end{figure*} 


\begin{figure*}
    \centering
    \includegraphics[width=16.8cm]{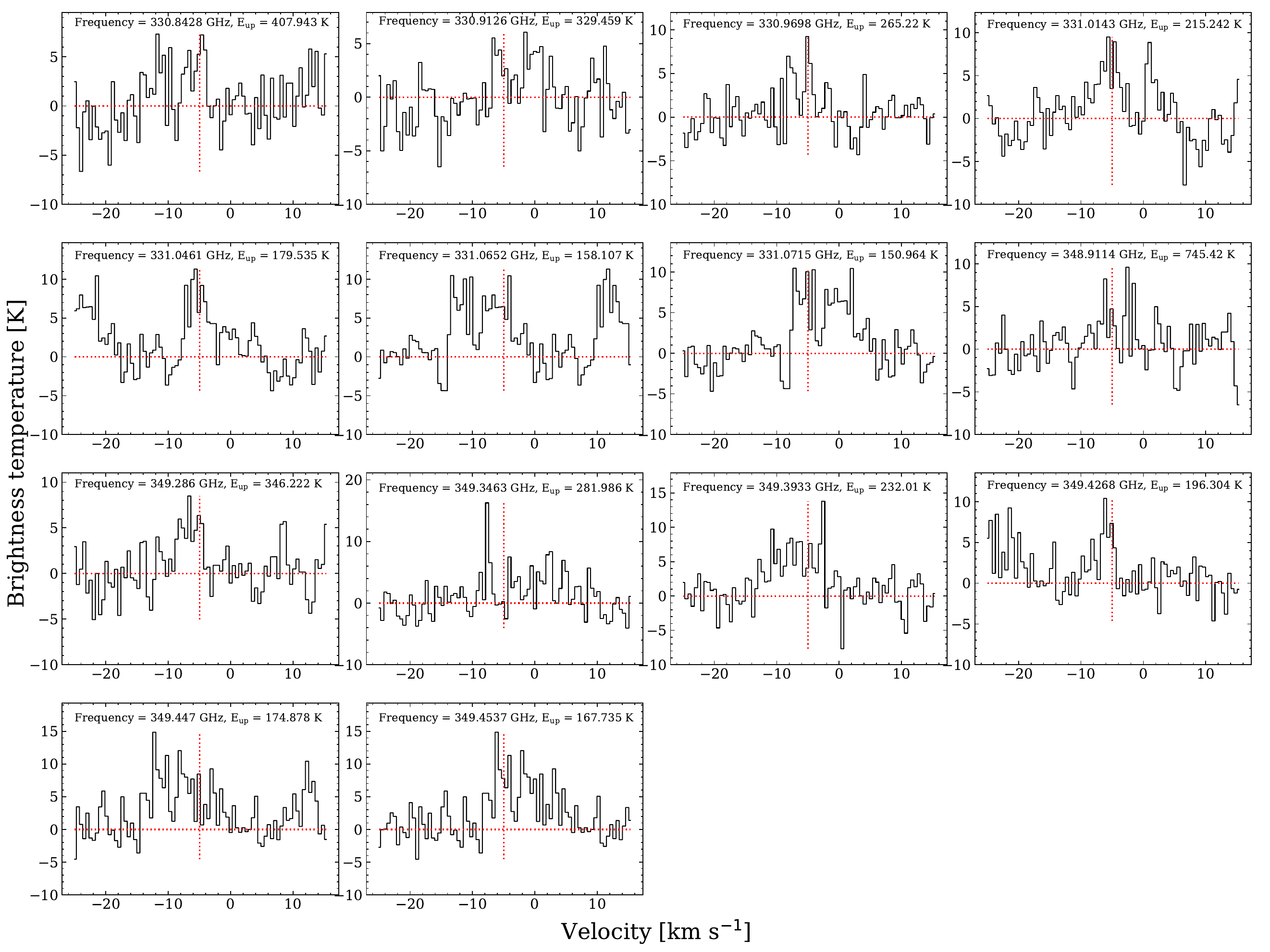}
    \caption{Detected CH$_{3}$CN,v=0 lines towards N64. Since the lines are optically thick, we used CH$_{3}$CN,v8=1 to derive a column density of 5.0 (3.0)$\times 10^{15} \mathrm{cm^{-2}}$, with excitation temperature 120 (20)K.}
    \label{Fig:N64_CH3CN,v=0_1}%
\end{figure*} 


\begin{figure*}
    \centering
    \includegraphics[width=16.8cm]{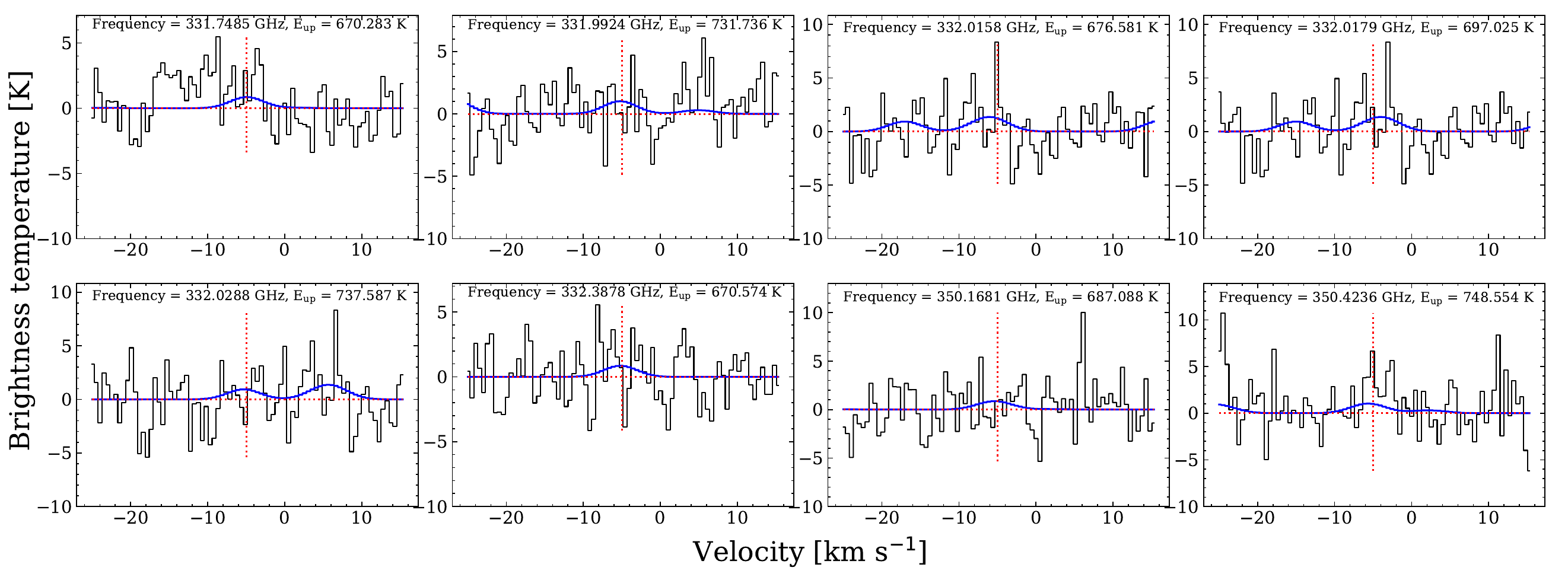}
    \caption{Detected CH$_{3}$CN,v8=1 lines towards N64. The values obtained from model fitting (model in blue) overplotted on the observed spectrum (in black) for this molecule were column density of 5.0 (3.0)$\times 10^{15} \mathrm{cm^{-2}}$, with excitation temperature 120 (20)K.}
    \label{Fig:N64_CH3CN,v8=1_1}%
\end{figure*}



\begin{figure*}
    \centering
    \includegraphics[width=16.8cm]{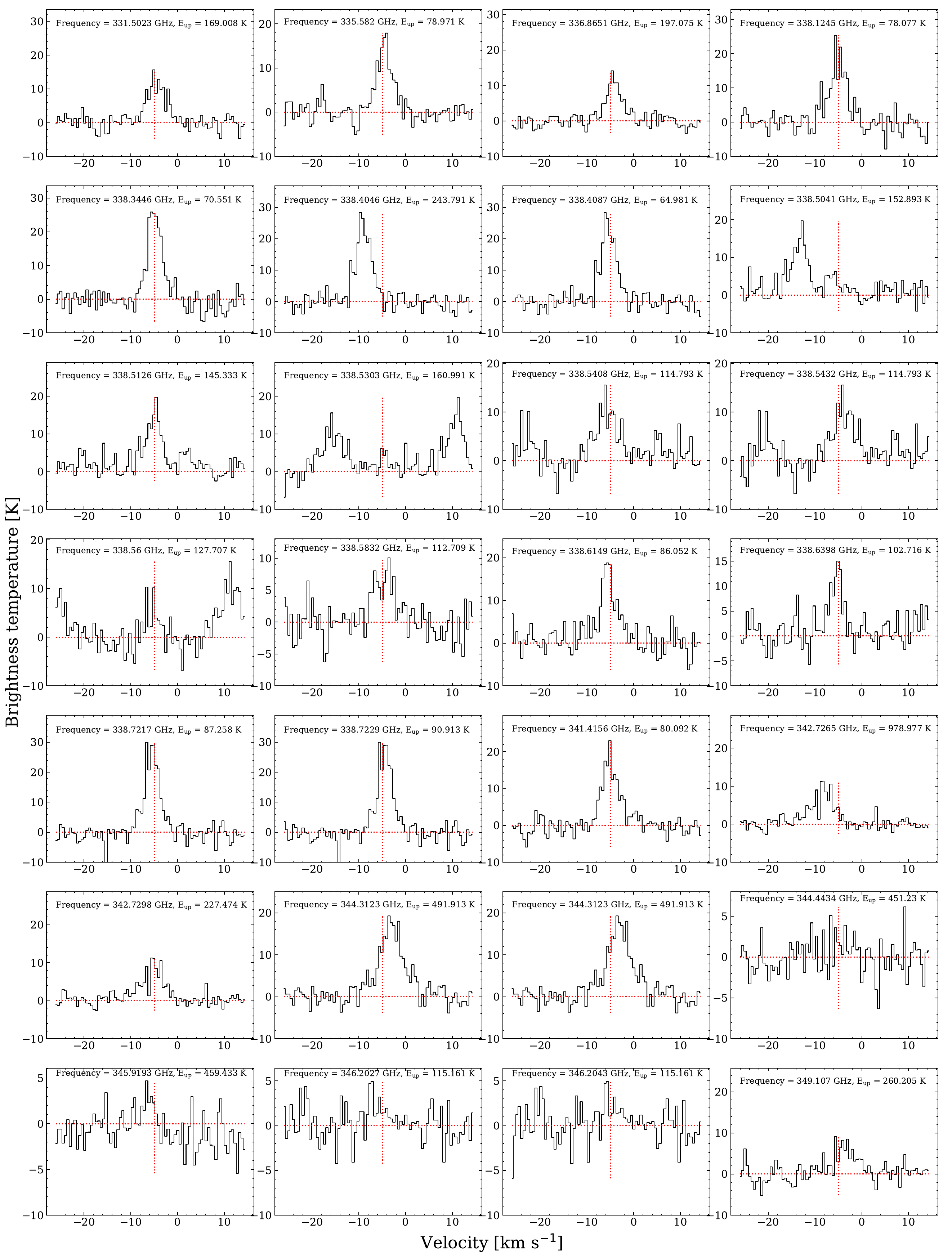}
    \caption{Detected CH$_3$OH lines towards N53. Based on results from the source sample we conclude that the detected lines are optically thick, and since no isotopologues were detected we were not able to derive a column density for this molecule.}
    \label{Fig:N53_CH3OH_1}%
\end{figure*}

\begin{figure*}
    \centering
    \includegraphics[width=16.8cm]{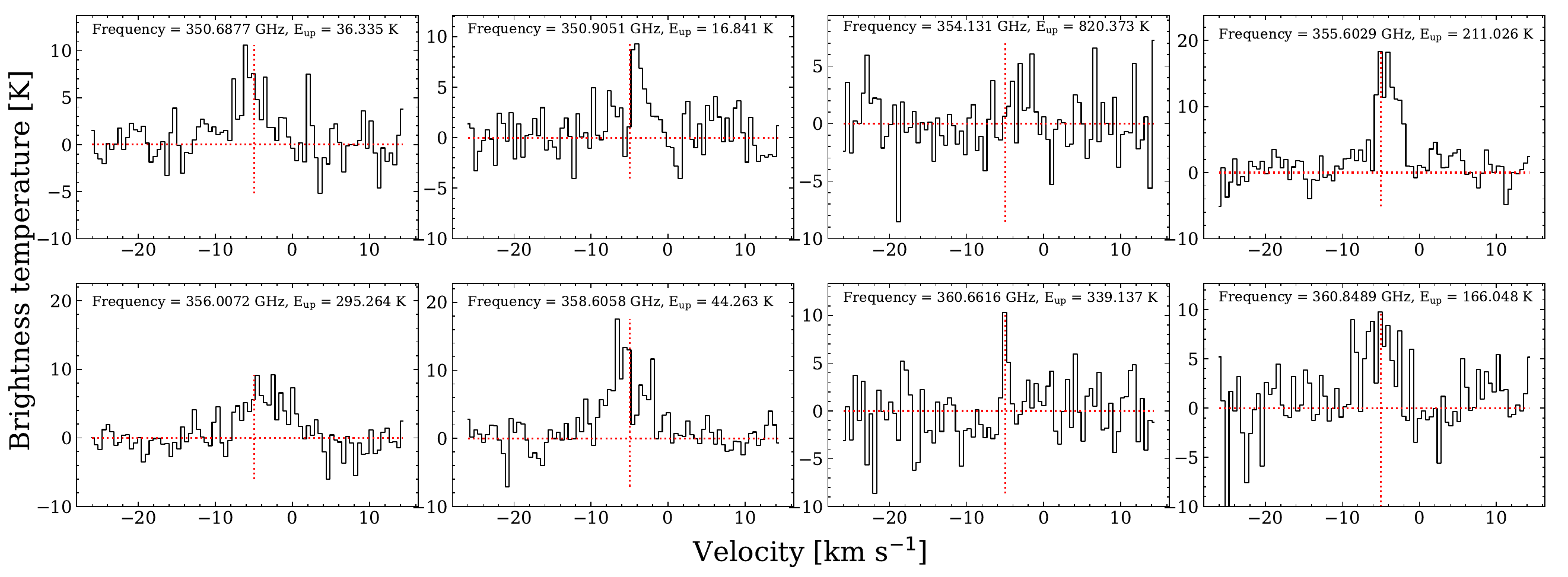}
    \caption{Detected CH$_3$OH lines towards N53. Based on results from the source sample we conclude that the detected lines are optically thick, and since no isotopologues were detected we were not able to derive a column density for this molecule.}
    \label{Fig:N53_CH3OH_2}%
\end{figure*} 


\begin{figure*}
    \centering
    \includegraphics[width=16.8cm]{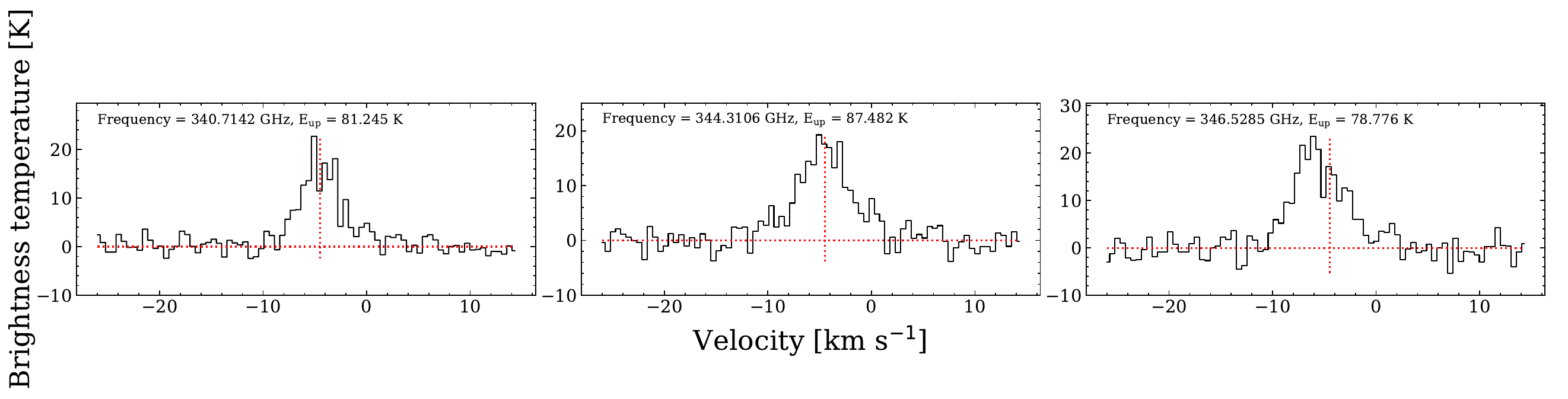}
    \caption{Detected SO lines towards N53. Based on results from the source sample we conclude that the detected lines are optically thick, and since no isotopologues were detected we were not able to derive a column density for this molecule.}
    \label{Fig:N53_SO_1}%
\end{figure*} 


\begin{figure*}
    \centering
    \includegraphics[width=16.8cm]{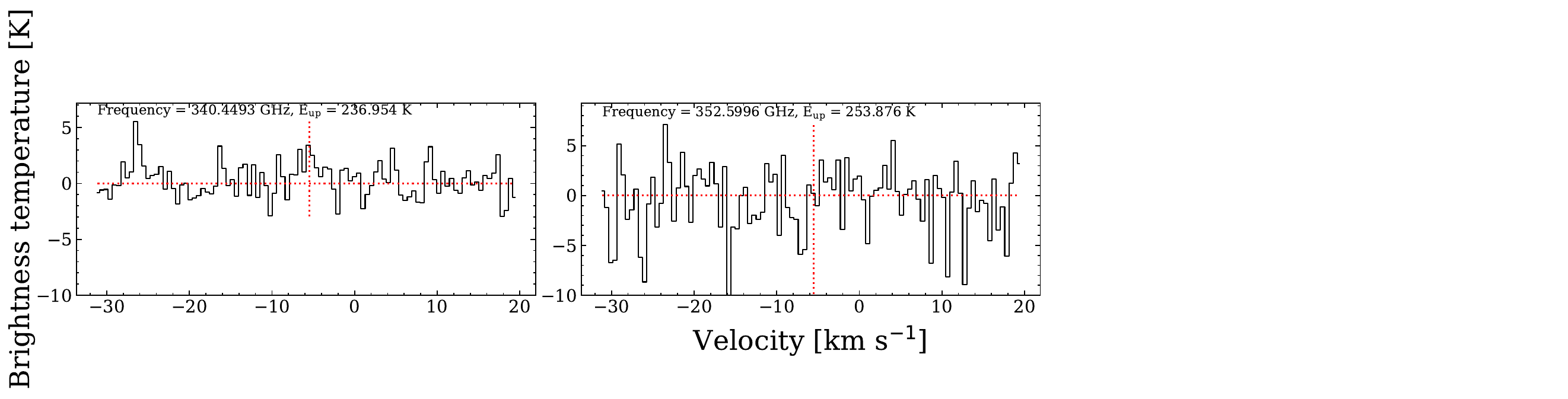}
    \caption{Detected OCS lines towards N53. Based on results from the source sample we conclude that the detected lines are optically thick, and since no isotopologues were detected we were not able to derive a column density for this molecule.}
    \label{Fig:N53_OCS_1}%
\end{figure*}

 
\begin{figure*}
    \centering
    \includegraphics[width=16.8cm]{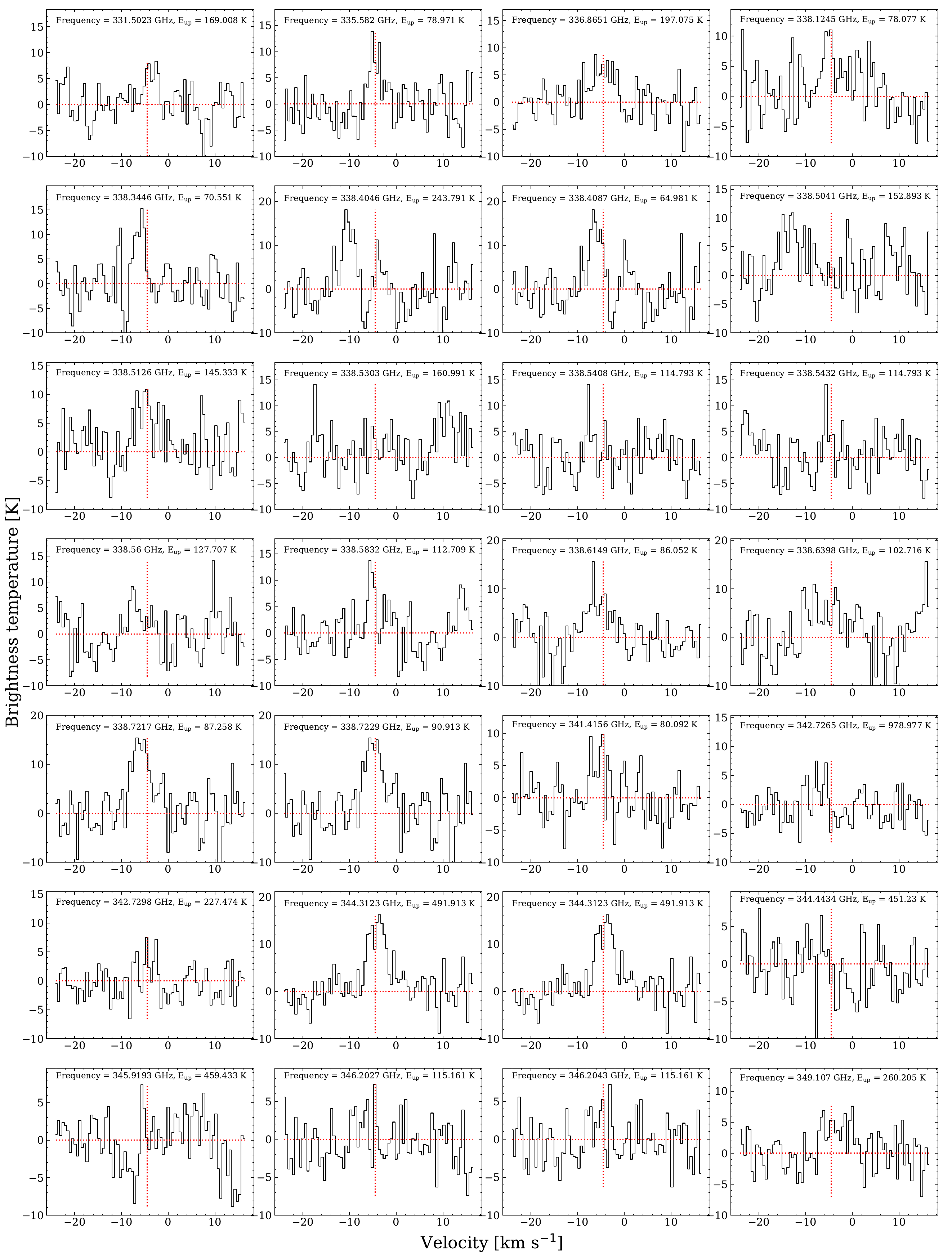}
    \caption{Detected CH$_3$OH lines towards N51. Based on results from the source sample we conclude that the detected lines are optically thick, and since no isotopologues were detected we were not able to derive a column density for this molecule.}
    \label{Fig:N51_CH3OH_1}%
\end{figure*}

\begin{figure*}
    \centering
    \includegraphics[width=16.8cm]{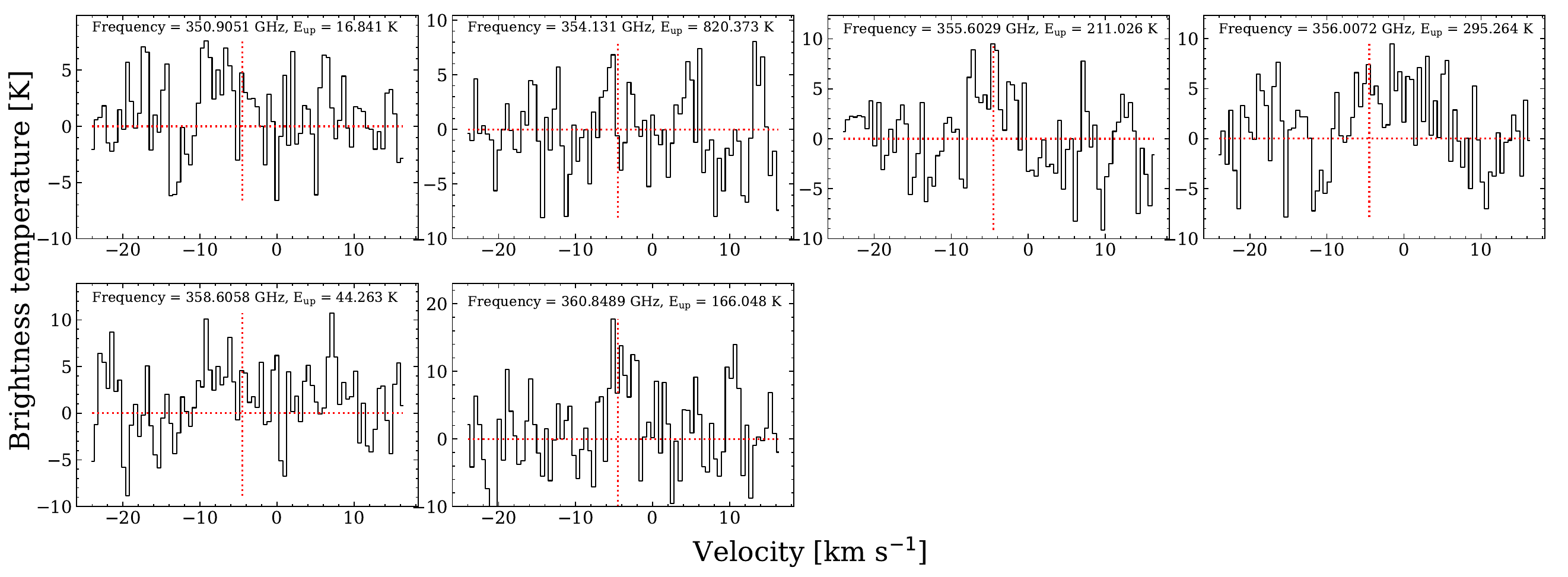}
    \caption{Detected CH$_3$OH lines towards N51. Based on results from the source sample we conclude that the detected lines are optically thick, and since no isotopologues were detected we were not able to derive a column density for this molecule.}
    \label{Fig:N51_CH3OH_2}%
\end{figure*} 


\begin{figure*}
    \centering
    \includegraphics[width=16.8cm]{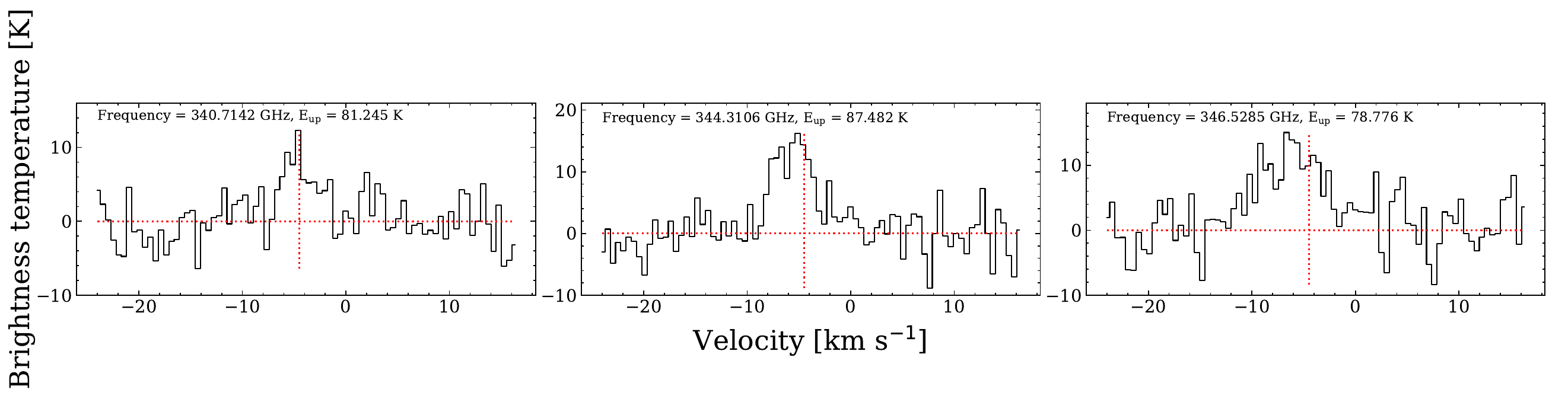}
    \caption{Detected SO lines towards N51. Based on results from the source sample we conclude that the detected lines are optically thick, and since no isotopologues were detected we were not able to derive a column density for this molecule.}
    \label{Fig:N51_SO_1}%
\end{figure*}

 
\begin{figure*}
    \centering
    \includegraphics[width=16.8cm]{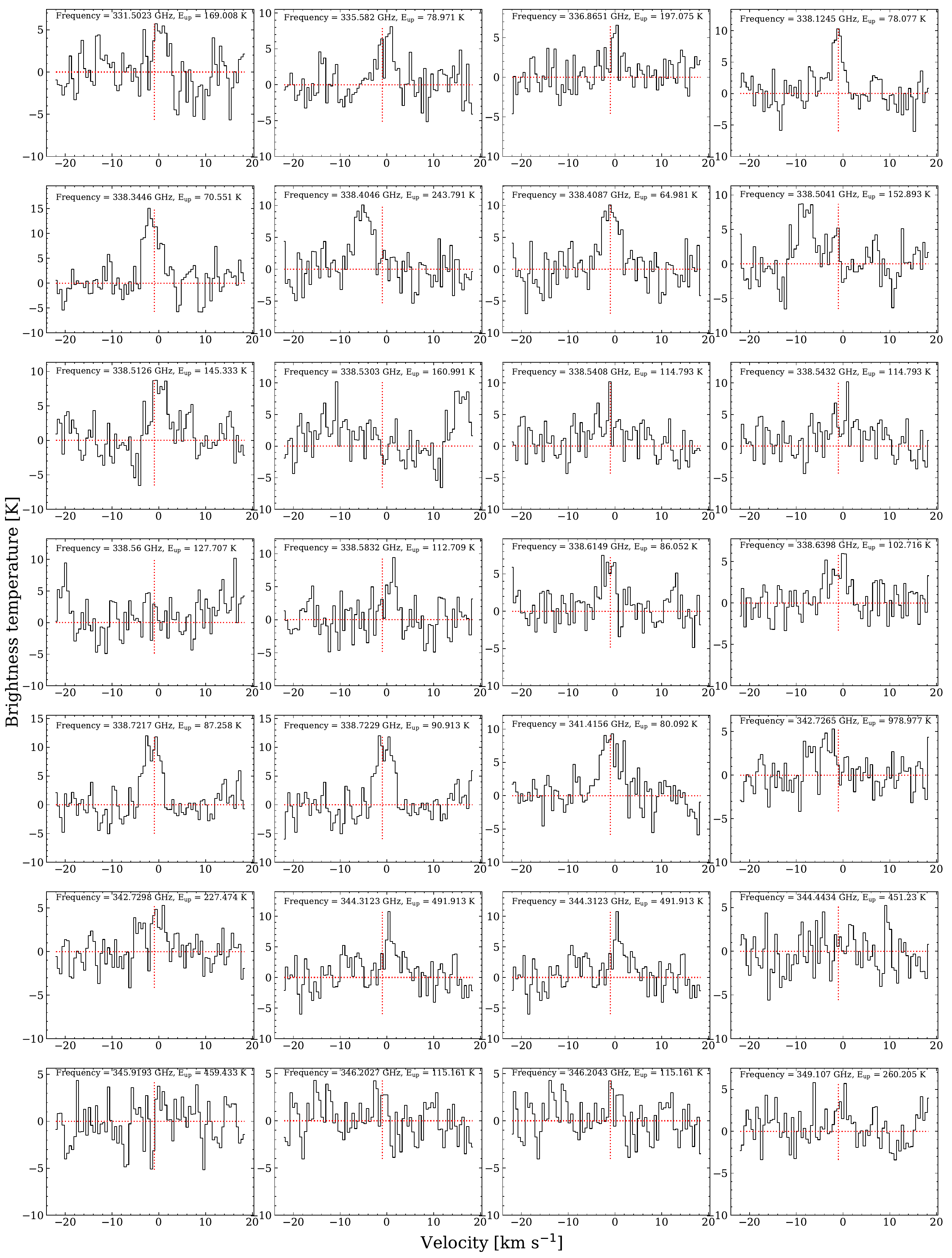}
    \caption{Detected CH$_3$OH lines towards N38. Based on results from the source sample we conclude that the detected lines are optically thick, and since no isotopologues were detected we were not able to derive a column density for this molecule.}
    \label{Fig:N38_CH3OH_1}%
\end{figure*}

\begin{figure*}
    \centering
    \includegraphics[width=16.8cm]{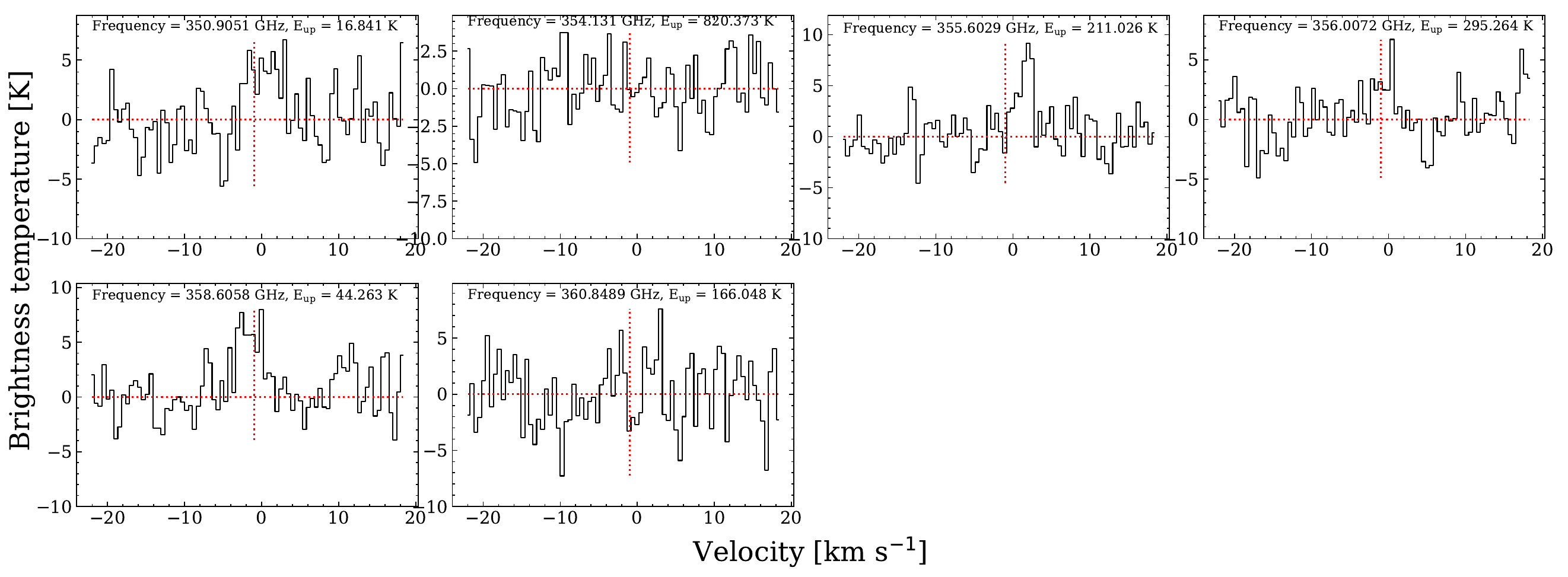}
    \caption{Detected CH$_3$OH lines towards N38. Based on results from the source sample we conclude that the detected lines are optically thick, and since no isotopologues were detected we were not able to derive a column density for this molecule.}
    \label{Fig:N38_CH3OH_2}%
\end{figure*} 


\begin{figure*}
    \centering
    \includegraphics[width=16.8cm]{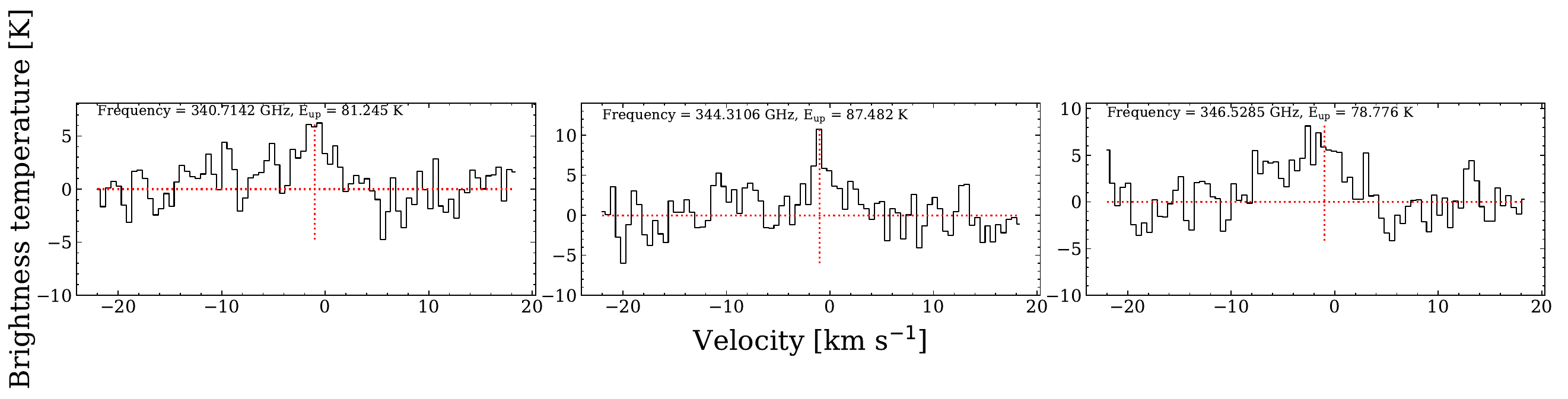}
    \caption{Detected SO lines towards N38. Based on results from the source sample we conclude that the detected lines are optically thick, and since no isotopologues were detected we were not able to derive a column density for this molecule.}
    \label{Fig:N38_SO_1}%
\end{figure*} 


\begin{figure*}
    \centering
    \includegraphics[width=16.8cm]{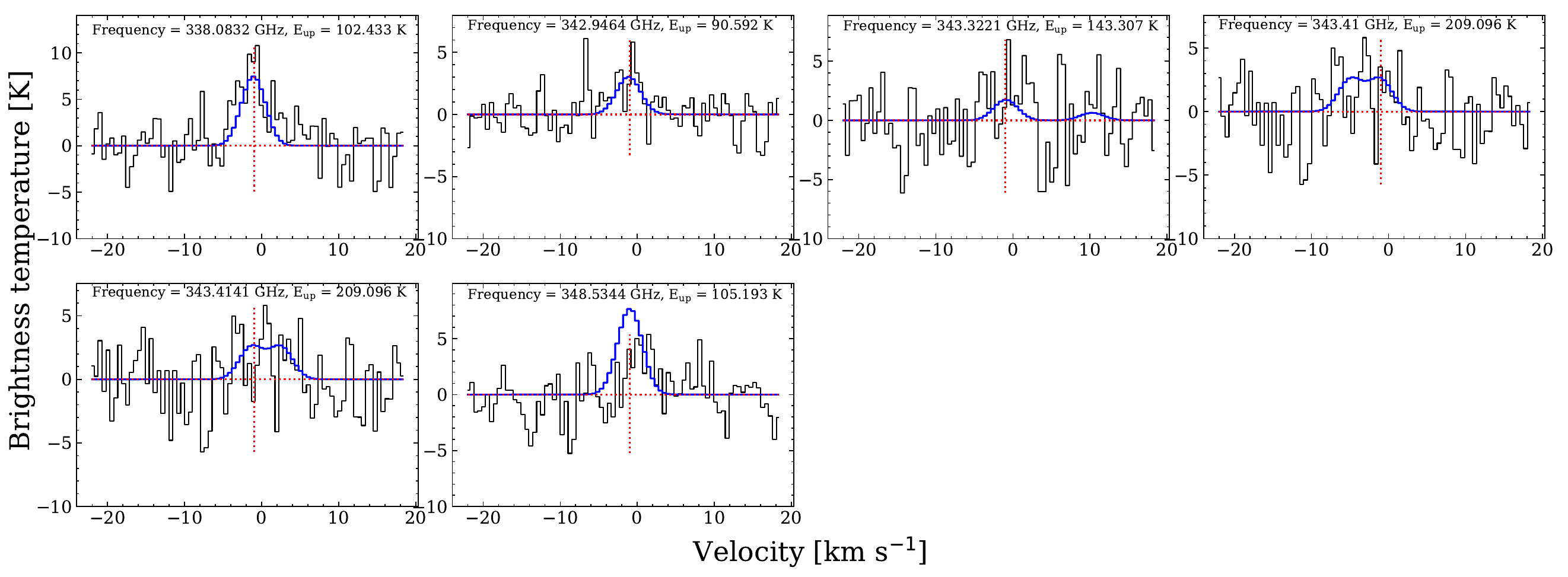}
    \caption{Detected H2CS lines towards N38. The values obtained from model fitting (model in blue) overplotted on the observed spectrum (in black) for this molecule were column density of 1.0 (0.5)$\times 10^{15} \mathrm{cm^{-2}}$, with excitation temperature 100 (50)K.}
    \label{Fig:N38_H2CS_1}%
\end{figure*}

 
\begin{figure*}
    \centering
    \includegraphics[width=16.8cm]{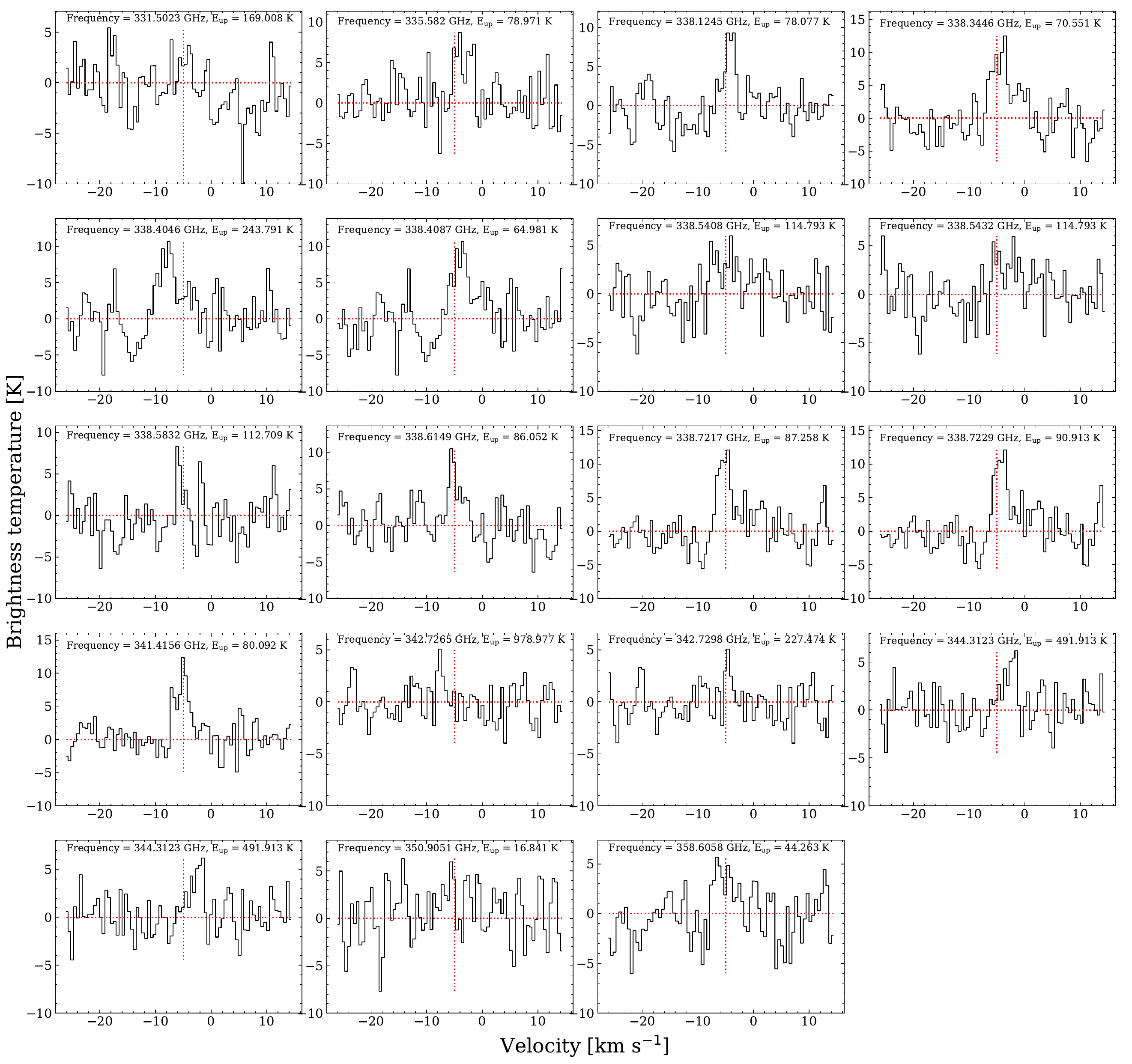}
    \caption{Detected CH$_3$OH lines towards N48. Based on results from the source sample we conclude that the detected lines are optically thick, and since no isotopologues were detected we were not able to derive a column density for this molecule.}
    \label{Fig:N48_CH3OH_1}%
\end{figure*} 


\begin{figure*}
    \centering
    \includegraphics[width=16.8cm]{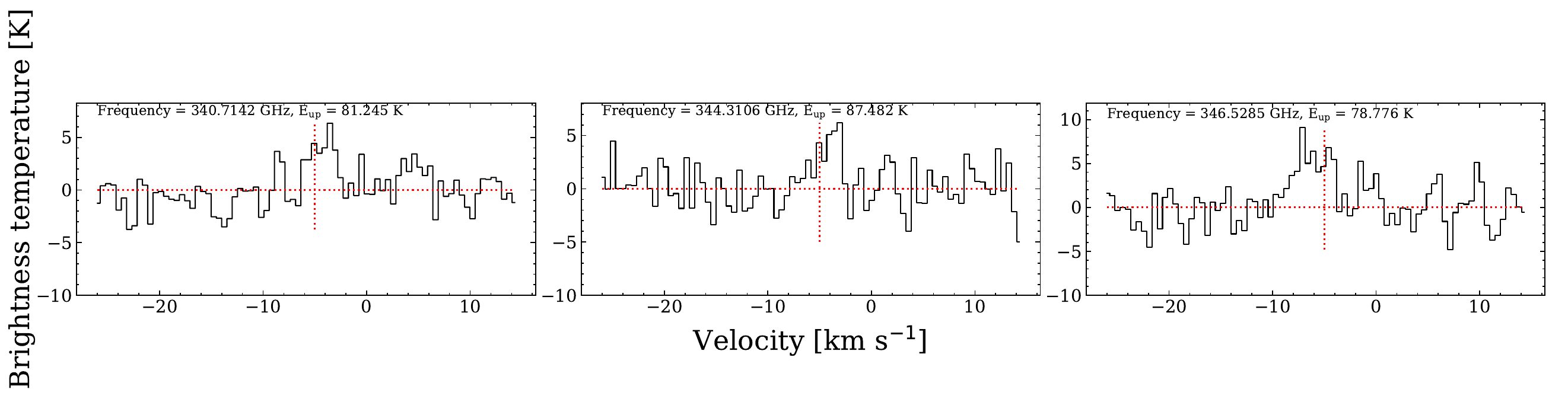}
    \caption{Detected SO lines towards N48. Based on results from the source sample we conclude that the detected lines are optically thick, and since no isotopologues were detected we were not able to derive a column density for this molecule.}
    \label{Fig:N48_SO_1}%
\end{figure*}


\begin{figure*}
    \centering
    \includegraphics[width=16.8cm]{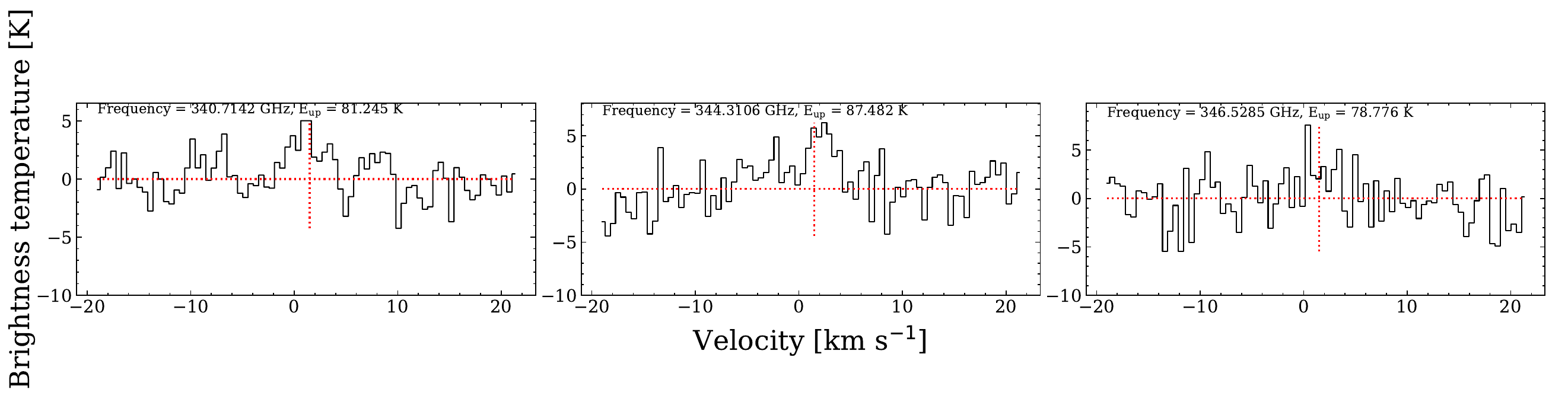}
    \caption{Detected SO lines towards S8. Based on results from the source sample we conclude that the detected lines are optically thick, and since no isotopologues were detected we were not able to derive a column density for this molecule.}
    \label{Fig:S8_SO_1}%
\end{figure*}


\section{Molecule lists in spectra}
\label{App:line_tables}
This appendix is available in electronic form only at Zenedo.org, url: https://zenodo.org/deposit/8200687. The tables contain the detected molecular lines for the source sample. 

\longtab[1]{
\begin{longtable}{l c c c c r}     
\caption{\label{table:Molecules_S26} Detected molecular transitions towards S26. The listed line transitions are from best fit line models obtained using the software package CASSIS, on the spectrum taken at the position of peak continuum emission. The model parameters are listed in Table \ref{table:S26_column_densities}. Only transitions above 6 K are listed ($\sim 3 \sigma$), corresponding to about 0.6 $\mathrm{Jy \ beam^{-1}}$.} \\
\hline\hline 
     Frequency & Molecule & Transition & $E_{\mathrm{up}}$/$k_{\mathrm{B}}$ & $A_{\mathrm{ij}}$  & Opacity ($\tau$)  \\ 
      $[\mathrm{GHz}] $  &  & &  $[\mathrm{K}]$ & $\times 10^{-4} [\mathrm{s^{-1}}]$  &   \\ 
\hline
\endfirsthead
\caption{continued.}\\
\hline\hline
     Frequency & Molecule & Transition & $E_{\mathrm{up}}$/$k_{\mathrm{B}}$ & $A_{\mathrm{ij}}$  & Opacity ($\tau$)    \\ 
      $[\mathrm{GHz}] $  &  & &  $[\mathrm{K}]$ & $\times 10^{-4} [\mathrm{s^{-1}}]$  &  \\ 
\hline
\endhead
\hline
\endfoot
  329.330552 & C$^{18}$O & $(3 - 2)$ &      31.61 &       0.02 &       -- \\
  329.385477 & SO & $(2_{1} - 1_{0)}$ &      15.81 &       0.14 &       0.04 \\
  329.573452 & HNCO & $(15_{2,14} - 14_{2,13)}$ &     296.83 &       4.72 &       0.23 \\
  329.584800 & HNCO & $(15_{2,13} - 14_{2,12)}$ &     296.83 &       4.72 &       0.23 \\
  329.628168 & CH$_3$OH & $(24_{7,18,3} - 24_{4,21,6)}$ &    1323.36 &       0.00 &       0.00 \\
  329.632881 & CH$_3$OH & $(12_{2,10,0} - 11_{3,9,0)}$ &     218.80 &       0.60 &       8.25 \\
  329.664367 & HNCO & $(15_{0,15} - 14_{0,14)}$ &     126.58 &       5.04 &       0.99 \\
  330.001752 & CH$_3$OH & $(7_{0,7,0} - 6_{0,6,0)}$ &      76.50 &       1.58 &       0.19 \\
  330.191103 & $^{34}$SO$_{2}$ & $(8_{2,6} - 7_{1,7)}$ &      42.77 &       1.19 &       0.22 \\
  330.194042 & CH$_3$OH & $(7_{-1,7,0} - 6_{-1,6,0)}$ &      69.01 &       1.55 &       0.21 \\
  330.252798 & CH$_3$OH & $(7_{0,7,+0} - 6_{0,6,+0)}$ &      63.41 &       1.58 &       0.22 \\
  330.342534 & CH$_3$OH & $(7_{4,3,+0} - 6_{4,2,+0)}$ &     144.21 &       1.07 &       0.07 \\
  330.342534 & CH$_3$OH & $(7_{4,4,-0} - 6_{4,3,-0)}$ &     144.21 &       1.07 &       0.07 \\
  330.349987 & CH$_3$OH & $(7_{2,6,-0} - 6_{2,5,-0)}$ &     101.23 &       1.46 &       0.14 \\
  330.355512 & CH$_3$OH & $(20_{3,17,0} - 19_{4,16,0)}$ &     537.04 &       0.64 &       0.60 \\
  330.371293 & CH$_3$OH & $(7_{3,5,+0} - 6_{3,4,+0)}$ &     113.47 &       1.29 &       0.11 \\
  330.373443 & CH$_3$OH & $(7_{3,4,-0} - 6_{3,3,-0)}$ &     113.47 &       1.29 &       0.11 \\
  330.405456 & CH$_3$OCH$_3$ & $(16_{2,15,3} - 15_{1,14,3)}$ &     128.46 &       1.65 &       0.02 \\
  330.405456 & CH$_3$OCH$_3$ & $(16_{2,15,5} - 15_{1,14,5)}$ &     128.46 &       1.65 &       0.03 \\
  330.406503 & CH$_3$OCH$_3$ & $(16_{2,15,1} - 15_{1,14,1)}$ &     128.46 &       1.65 &       0.07 \\
  330.407010 & CH$_3$OCH$_3$ & $(49_{10,40,3} - 48_{11,37,3)}$ &    1256.39 &       0.20 &       0.00 \\
  330.407549 & CH$_3$OCH$_3$ & $(16_{2,15,0} - 15_{1,14,0)}$ &     128.46 &       1.65 &       0.04 \\
  330.442421 & CH$_3$OH & $(7_{1,6,0} - 6_{1,5,0)}$ &      84.49 &       1.59 &       0.18 \\
  330.463873 & CH$_3$OH & $(11_{1,10,-0} - 11_{0,11,+0)}$ &     165.27 &       3.90 &       0.30 \\
  330.464947 & CH$_3$OH & $(7_{2,5,+0} - 6_{2,4,+0)}$ &     101.24 &       1.46 &       0.14 \\
  330.535222 & CH$_3$OH & $(7_{2,5,0} - 6_{2,4,0)}$ &      85.80 &       1.44 &       0.16 \\
  330.535890 & CH$_3$OH & $(7_{-2,6,0} - 6_{-2,5,0)}$ &      89.45 &       1.46 &       0.16 \\
  330.557569 & CH$_3$CN, v=0 & $(18_{-9,0} - 17_{9,0)}$ &     728.67 &      23.60 &       0.03 \\
  330.557569 & CH$_3$CN, v=0 & $(18_{9,0} - 17_{-9,0)}$ &     728.67 &      23.60 &       0.03 \\
  330.587965 & $^{13}$CO & $(3 - 2)$ &      31.73 &       0.02 &       -- \\
  330.665207 & CH$_3$CN, v=0 & $(18_{8,0} - 17_{8,0)}$ &     607.57 &      25.30 &       0.05 \\
  330.667565 & $^{34}$SO$_{2}$ & $(21_{2,20} - 21_{1,21)}$ &     218.58 &       1.45 &       0.16 \\
  330.760284 & CH$_3$CN, v=0 & $(18_{7,0} - 17_{7,0)}$ &     500.65 &      26.80 &       0.08 \\
  330.793887 & CH$_3$OH & $(8_{3,5,2} - 9_{2,7,2)}$ &     146.28 &       0.54 &      10.28 \\
  330.842762 & CH$_3$CN, v=0 & $(18_{-6,0} - 17_{6,0)}$ &     407.94 &      28.00 &       0.13 \\
  330.842762 & CH$_3$CN, v=0 & $(18_{6,0} - 17_{-6,0)}$ &     407.94 &      28.00 &       0.13 \\
  330.848569 & HNCO & $(15_{1,14} - 14_{1,13)}$ &     170.31 &       5.01 &       0.68 \\
  330.912608 & CH$_3$CN, v=0 & $(18_{5,0} - 17_{5,0)}$ &     329.46 &      29.10 &       0.18 \\
  330.969794 & CH$_3$CN, v=0 & $(18_{4,0} - 17_{4,0)}$ &     265.22 &      30.00 &       0.25 \\
  331.014296 & CH$_3$CN, v=0 & $(18_{-3,0} - 17_{3,0)}$ &     215.24 &      30.70 &       0.31 \\
  331.014296 & CH$_3$CN, v=0 & $(18_{3,0} - 17_{-3,0)}$ &     215.24 &      30.70 &       0.31 \\
  331.046096 & CH$_3$CN, v=0 & $(18_{2,0} - 17_{2,0)}$ &     179.53 &      31.20 &       0.37 \\
  331.065182 & CH$_3$CN, v=0 & $(18_{1,0} - 17_{1,0)}$ &     158.11 &      31.50 &       0.40 \\
  331.071544 & CH$_3$CN, v=0 & $(18_{0,0} - 17_{0,0)}$ &     150.96 &      31.60 &       0.42 \\
  331.220371 & CH$_3$OH & $(16_{1,16,2} - 15_{2,13,2)}$ &     320.63 &       0.52 &       3.38 \\
  331.502319 & CH$_3$OH & $(11_{1,10,0} - 11_{0,11,0)}$ &     169.01 &       3.93 &      80.45 \\
  331.580244 & SO$_{2}$ & $(11_{6,6} - 12_{5,7)}$ &     148.95 &       0.43 &       1.11 \\
  331.748518 & CH$_3$CN, v$_8$=1 & $(18_{-1,3} - 17_{1,3)}$ &     670.28 &      31.40 &       0.08 \\
  331.992354 & CH$_3$CN, v$_8$=1 & $(18_{-2,2} - 17_{2,2)}$ &     731.74 &      31.20 &       0.05 \\
  331.992354 & CH$_3$CN, v$_8$=1 & $(18_{2,2} - 17_{-2,2)}$ &     731.74 &      31.20 &       0.05 \\
  332.015818 & CH$_3$CN, v$_8$=1 & $(18_{0,2} - 17_{0,2)}$ &     676.58 &      31.60 &       0.08 \\
  332.017856 & CH$_3$CN, v$_8$=1 & $(18_{1,2} - 17_{1,2)}$ &     697.02 &      31.50 &       0.07 \\
  332.028758 & CH$_3$CN, v$_8$=1 & $(18_{-4,3} - 17_{4,3)}$ &     737.59 &      30.00 &       0.05 \\
  332.028758 & CH$_3$CN, v$_8$=1 & $(18_{4,3} - 17_{-4,3)}$ &     737.59 &      30.00 &       0.05 \\
  332.091431 & SO$_{2}$ & $(21_{2,20} - 21_{1,21)}$ &     219.53 &       1.51 &       3.99 \\
  332.129700 & OC$^{34}$S & $(28 - 27)$ &     231.16 &       1.07 &       0.13 \\
  332.173573 & $^{34}$SO$_{2}$ & $(23_{3,21} - 23_{2,22)}$ &     274.74 &       2.54 &       0.19 \\
  332.387753 & CH$_3$CN, v$_8$=1 & $(18_{1,3} - 17_{-1,3)}$ &     670.57 &      31.60 &       0.08 \\
  332.505242 & SO$_{2}$ & $(4_{3,1} - 3_{2,2)}$ &      31.29 &       3.29 &       8.73 \\
  332.570617 & CH$_3$OCHO & $(30_{1,29,2} - 29_{2,28,1)}$ &     256.98 &       0.80 &       0.01 \\
  332.570888 & CH$_3$OCHO & $(30_{2,29,1} - 29_{2,28,1)}$ &     256.98 &       5.53 &       0.07 \\
  332.571107 & CH$_3$OCHO & $(30_{1,29,2} - 29_{1,28,2)}$ &     256.98 &       5.53 &       0.07 \\
  332.571378 & CH$_3$OCHO & $(30_{2,29,1} - 29_{1,28,2)}$ &     256.98 &       0.80 &       0.01 \\
  332.575715 & CH$_3$OCHO & $(30_{1,29,0} - 29_{2,28,0)}$ &     256.96 &       0.80 &       0.01 \\
  332.575985 & CH$_3$OCHO & $(30_{2,29,0} - 29_{2,28,0)}$ &     256.96 &       5.53 &       0.07 \\
  332.576202 & CH$_3$OCHO & $(30_{1,29,0} - 29_{1,28,0)}$ &     256.96 &       5.53 &       0.07 \\
  332.576472 & CH$_3$OCHO & $(30_{2,29,0} - 29_{1,28,0)}$ &     256.96 &       0.80 &       0.01 \\
  332.836225 & $^{34}$SO$_{2}$ & $(16_{4,12} - 16_{3,13)}$ &     162.90 &       3.02 &       0.39 \\
  332.996563 & CH$_3$OH & $(22_{2,20,2} - 21_{3,18,2)}$ &     614.49 &       0.63 &       0.29 \\
  333.114779 & CH$_3$OH & $(7_{1,6,-0} - 6_{1,5,-0)}$ &      78.52 &       1.59 &       0.19 \\
  333.448974 & CH$_3$OCHO & $(31_{0,31,2} - 30_{1,30,1)}$ &     259.59 &       0.82 &       0.01 \\
  333.448976 & CH$_3$OCHO & $(31_{1,31,1} - 30_{1,30,1)}$ &     259.59 &       5.73 &       0.08 \\
  333.448977 & CH$_3$OCHO & $(31_{0,31,2} - 30_{0,30,2)}$ &     259.59 &       5.73 &       0.08 \\
  333.448979 & CH$_3$OCHO & $(31_{1,31,1} - 30_{0,30,2)}$ &     259.59 &       0.82 &       0.01 \\
  333.449417 & CH$_3$OCHO & $(31_{0,31,0} - 30_{1,30,0)}$ &     259.57 &       0.91 &       0.01 \\
  333.449419 & CH$_3$OCHO & $(31_{1,31,0} - 30_{1,30,0)}$ &     259.57 &       5.64 &       0.08 \\
  333.449421 & CH$_3$OCHO & $(31_{0,31,0} - 30_{0,30,0)}$ &     259.57 &       5.64 &       0.08 \\
  333.449422 & CH$_3$OCHO & $(31_{1,31,0} - 30_{0,30,0)}$ &     259.57 &       0.91 &       0.01 \\
  333.449750 & CH$_3$OCHO & $(40_{13,27,5} - 40_{12,28,5)}$ &     782.89 &       0.53 &       0.00 \\
  333.864722 & CH$_3$OH & $(9_{1,8,1} - 8_{2,6,2)}$ &     125.52 &       0.01 &       0.21 \\
  333.900983 & $^{34}$SO & $(8_{7} - 7_{6)}$ &      79.86 &       4.69 &       1.92 \\
  334.426571 & CH$_3$OH & $(3_{0,3,4} - 2_{1,2,4)}$ &     314.47 &       0.56 &       0.79 \\
  334.673353 & SO$_{2}$ & $(8_{2,6} - 7_{1,7)}$ &      43.15 &       1.27 &       5.69 \\
  335.133570 & CH$_3$OH & $(2_{2,1,0} - 3_{1,2,0)}$ &      44.67 &       0.27 &       4.07 \\
  335.395500 & HDO & $(3_{3,1} - 4_{2,2)}$ &     335.27 &       0.26 &       0.09 \\
  335.560207 & CH$_3$OH & $(12_{1,11,-0} - 12_{0,12,+0)}$ &     192.66 &       4.04 &       0.25 \\
  335.582017 & CH$_3$OH & $(7_{1,7,0} - 6_{1,6,0)}$ &      78.97 &       1.63 &      52.22 \\
  336.089228 & SO$_{2}$ & $(23_{3,21} - 23_{2,22)}$ &     276.02 &       2.67 &       4.71 \\
  336.438224 & CH$_3$OH & $(14_{7,7,0} - 15_{6,9,0)}$ &     488.22 &       0.36 &       0.37 \\
  336.438224 & CH$_3$OH & $(14_{7,8,0} - 15_{6,10,0)}$ &     488.22 &       0.36 &       0.37 \\
  336.520084 & HC$_3$N & $(37 - 36)$ &     306.91 &      30.50 &       0.46 \\
  336.553811 & SO & $(11_{10} - 10_{10)}$ &     142.88 &       0.06 &       0.04 \\
  336.669581 & SO$_{2}$ & $(16_{7,9} - 17_{6,12)}$ &     245.12 &       0.58 &       0.93 \\
  336.865149 & CH$_3$OH & $(12_{1,11,0} - 12_{0,12,0)}$ &     197.07 &       4.07 &      66.37 \\
  337.060513 & C$^{17}$O & $(3_{1.5} - 2_{2.5)}$ &      32.35 &       0.00 &       0.00 \\
  337.060709 & C$^{17}$O & $(3_{0.5} - 2_{1.5)}$ &      32.35 &       0.01 &       0.00 \\
  337.060831 & C$^{17}$O & $(3_{2.5} - 2_{3.5)}$ &      32.35 &       0.00 &       0.00 \\
  337.060936 & C$^{17}$O & $(3_{2.5} - 2_{2.5)}$ &      32.35 &       0.01 &       0.03 \\
  337.060988 & C$^{17}$O & $(3_{4.5} - 2_{3.5)}$ &      32.35 &       0.02 &       0.08 \\
  337.060988 & C$^{17}$O & $(3_{5.5} - 2_{4.5)}$ &      32.35 &       0.02 &       0.12 \\
  337.061050 & C$^{17}$O & $(3_{1.5} - 2_{1.5)}$ &      32.35 &       0.01 &       0.02 \\
  337.061123 & C$^{17}$O & $(3_{3.5} - 2_{3.5)}$ &      32.35 &       0.01 &       0.03 \\
  337.061214 & C$^{17}$O & $(3_{0.5} - 2_{0.5)}$ &      32.35 &       0.02 &       0.02 \\
  337.061226 & C$^{17}$O & $(3_{3.5} - 2_{2.5)}$ &      32.35 &       0.01 &       0.05 \\
  337.061471 & C$^{17}$O & $(3_{2.5} - 2_{1.5)}$ &      32.35 &       0.01 &       0.03 \\
  337.061553 & C$^{17}$O & $(3_{1.5} - 2_{0.5)}$ &      32.35 &       0.01 &       0.01 \\
  337.061951 & C$^{17}$O & $(3_{4.5} - 2_{4.5)}$ &      32.35 &       0.00 &       0.02 \\
  337.062093 & C$^{17}$O & $(3_{3.5} - 2_{4.5)}$ &      32.35 &       0.00 &       0.00 \\
  337.135853 & CH$_3$OH & $(3_{3,1,1} - 4_{2,3,1)}$ &      61.64 &       0.16 &       2.79 \\
  337.197845 & $^{33}$SO & $(8_{7,7.5} - 7_{6,6.5)}$ &      80.53 &       4.70 &       0.06 \\
  337.198022 & $^{33}$SO & $(8_{7,6.5} - 7_{6,5.5)}$ &      80.53 &       4.62 &       0.05 \\
  337.198620 & $^{33}$SO & $(8_{7,8.5} - 7_{6,7.5)}$ &      80.53 &       4.83 &       0.07 \\
  337.199371 & $^{33}$SO & $(8_{7,5.5} - 7_{6,4.5)}$ &      80.54 &       4.65 &       0.05 \\
  337.252172 & CH$_3$OH & $(7_{3,5,6} - 6_{3,4,6)}$ &     722.84 &       1.39 &       0.07 \\
  337.252173 & CH$_3$OH & $(7_{3,4,6} - 6_{3,3,6)}$ &     722.84 &       1.39 &       0.07 \\
  337.273561 & CH$_3$OH & $(7_{4,3,6} - 6_{4,2,6)}$ &     679.25 &       1.13 &       0.09 \\
  337.273561 & CH$_3$OH & $(7_{4,4,6} - 6_{4,3,6)}$ &     679.25 &       1.13 &       0.09 \\
  337.284320 & CH$_3$OH & $(7_{0,7,6} - 6_{0,6,6)}$ &     572.96 &       1.68 &       0.38 \\
  337.295913 & CH$_3$OH & $(7_{3,5,7} - 6_{3,4,7)}$ &     686.20 &       1.37 &       0.10 \\
  337.297484 & CH$_3$OH & $(7_{1,7,3} - 6_{1,6,3)}$ &     390.02 &       1.65 &       2.35 \\
  337.302644 & CH$_3$OH & $(7_{2,6,7} - 6_{2,5,7)}$ &     650.99 &       1.55 &       0.16 \\
  337.312360 & CH$_3$OH & $(7_{1,7,8} - 6_{1,6,8)}$ &     596.79 &       1.65 &       0.30 \\
  337.396459 & C$^{34}$S & $(7_{0} - 6_{0)}$ &      64.77 &       8.00 &       1.21 \\
  337.463703 & CH$_3$OH & $(7_{6,1,3} - 6_{6,0,3)}$ &     533.02 &       0.45 &       0.15 \\
  337.463703 & CH$_3$OH & $(7_{6,2,3} - 6_{6,1,3)}$ &     533.02 &       0.45 &       0.15 \\
  337.519138 & CH$_3$OH & $(7_{3,4,4} - 6_{3,3,4)}$ &     482.23 &       1.38 &       0.78 \\
  337.546116 & CH$_3$OH & $(7_{5,2,3} - 6_{5,1,3)}$ &     485.37 &       0.82 &       0.45 \\
  337.546116 & CH$_3$OH & $(7_{5,3,3} - 6_{5,2,3)}$ &     485.37 &       0.82 &       0.45 \\
  337.580147 & $^{34}$SO & $(8_{8} - 7_{7)}$ &      86.07 &       4.89 &       2.11 \\
  337.581680 & CH$_3$OH & $(7_{4,4,4} - 6_{4,3,4)}$ &     428.20 &       1.13 &       1.09 \\
  337.590319 & t-HCOOH & $(15_{6,10} - 14_{6,9)}$ &     243.95 &       3.68 &       0.03 \\
  337.590319 & t-HCOOH & $(15_{6,9} - 14_{6,8)}$ &     243.95 &       3.68 &       0.03 \\
  337.605288 & CH$_3$OH & $(7_{2,6,5} - 6_{2,5,5)}$ &     429.43 &       1.56 &       1.49 \\
  337.610661 & CH$_3$OH & $(7_{3,5,5} - 6_{3,4,5)}$ &     387.45 &       1.37 &       2.00 \\
  337.610680 & CH$_3$OH & $(7_{6,1,4} - 6_{6,0,4)}$ &     657.12 &       0.43 &       0.04 \\
  337.625753 & CH$_3$OH & $(7_{2,6,3} - 6_{2,5,3)}$ &     363.50 &       1.55 &       2.85 \\
  337.635754 & CH$_3$OH & $(7_{2,5,3} - 6_{2,4,3)}$ &     363.50 &       1.55 &       2.85 \\
  337.642478 & CH$_3$OH & $(7_{1,7,4} - 6_{1,6,4)}$ &     356.30 &       1.65 &       3.28 \\
  337.643915 & CH$_3$OH & $(7_{0,7,4} - 6_{0,6,4)}$ &     365.40 &       1.69 &       3.06 \\
  337.646042 & CH$_3$OH & $(7_{4,3,5} - 6_{4,2,5)}$ &     470.22 &       1.14 &       0.72 \\
  337.648209 & CH$_3$OH & $(7_{5,2,5} - 6_{5,1,5)}$ &     610.96 &       0.83 &       0.13 \\
  337.655199 & CH$_3$OH & $(7_{3,5,3} - 6_{3,4,3)}$ &     460.94 &       1.38 &       0.96 \\
  337.655236 & CH$_3$OH & $(7_{3,4,3} - 6_{3,3,3)}$ &     460.94 &       1.38 &       0.96 \\
  337.671238 & CH$_3$OH & $(7_{2,5,4} - 6_{2,4,4)}$ &     464.72 &       1.55 &       1.04 \\
  337.685248 & CH$_3$OH & $(7_{5,3,4} - 6_{5,2,4)}$ &     493.95 &       0.83 &       0.41 \\
  337.685614 & CH$_3$OH & $(7_{4,3,3} - 6_{4,2,3)}$ &     545.90 &       1.15 &       0.34 \\
  337.685614 & CH$_3$OH & $(7_{4,4,3} - 6_{4,3,3)}$ &     545.90 &       1.15 &       0.34 \\
  337.707568 & CH$_3$OH & $(7_{1,6,5} - 6_{1,5,5)}$ &     478.21 &       1.65 &       0.97 \\
  337.748830 & CH$_3$OH & $(7_{0,7,3} - 6_{0,6,3)}$ &     488.49 &       1.69 &       0.89 \\
  337.837801 & CH$_3$OH & $(20_{6,14,2} - 21_{5,17,2)}$ &     675.95 &       0.60 &       0.13 \\
  337.969438 & CH$_3$OH & $(7_{1,6,3} - 6_{1,5,3)}$ &     390.15 &       1.66 &       2.35 \\
  338.083195 & H$_2$CS & $(10_{1,10} - 9_{1,9)}$ &     102.43 &       5.77 &       0.22 \\
  338.124488 & CH$_3$OH & $(7_{0,7,1} - 6_{0,6,1)}$ &      78.08 &       1.70 &      54.13 \\
  338.305993 & SO$_{2}$ & $(18_{4,14} - 18_{3,15)}$ &     196.80 &       3.27 &       8.67 \\
  338.320356 & $^{34}$SO$_{2}$ & $(13_{2,12} - 12_{1,11)}$ &      92.45 &       2.27 &       0.42 \\
  338.344588 & CH$_3$OH & $(7_{1,7,2} - 6_{1,6,2)}$ &      70.55 &       1.67 &      57.29 \\
  338.404610 & CH$_3$OH & $(7_{6,2,1} - 6_{6,1,1)}$ &     243.79 &       0.45 &       2.75 \\
  338.408698 & CH$_3$OH & $(7_{0,7,0} - 6_{0,6,0)}$ &      64.98 &       1.70 &      61.82 \\
  338.430975 & CH$_3$OH & $(7_{6,1,2} - 6_{6,0,2)}$ &     253.95 &       0.45 &       2.49 \\
  338.442367 & CH$_3$OH & $(7_{6,1,0} - 6_{6,0,0)}$ &     258.70 &       0.45 &       2.37 \\
  338.442367 & CH$_3$OH & $(7_{6,2,0} - 6_{6,1,0)}$ &     258.70 &       0.45 &       2.37 \\
  338.456536 & CH$_3$OH & $(7_{5,3,2} - 6_{5,2,2)}$ &     189.00 &       0.83 &       8.77 \\
  338.475226 & CH$_3$OH & $(7_{5,2,1} - 6_{5,1,1)}$ &     201.06 &       0.83 &       7.78 \\
  338.486322 & CH$_3$OH & $(7_{5,2,0} - 6_{5,1,0)}$ &     202.88 &       0.84 &       7.67 \\
  338.486322 & CH$_3$OH & $(7_{5,3,0} - 6_{5,2,0)}$ &     202.88 &       0.84 &       7.67 \\
  338.504065 & CH$_3$OH & $(7_{4,4,2} - 6_{4,3,2)}$ &     152.89 &       1.15 &      17.29 \\
  338.512632 & CH$_3$OH & $(7_{4,4,0} - 6_{4,3,0)}$ &     145.33 &       1.15 &      18.68 \\
  338.512644 & CH$_3$OH & $(7_{4,3,0} - 6_{4,2,0)}$ &     145.33 &       1.15 &      18.68 \\
  338.512853 & CH$_3$OH & $(7_{2,6,0} - 6_{2,5,0)}$ &     102.70 &       1.57 &      39.20 \\
  338.530257 & CH$_3$OH & $(7_{4,3,1} - 6_{4,2,1)}$ &     160.99 &       1.15 &      16.04 \\
  338.540826 & CH$_3$OH & $(7_{3,5,0} - 6_{3,4,0)}$ &     114.79 &       1.39 &      30.64 \\
  338.543152 & CH$_3$OH & $(7_{3,4,0} - 6_{3,3,0)}$ &     114.79 &       1.39 &      30.64 \\
  338.559963 & CH$_3$OH & $(7_{3,4,2} - 6_{3,3,2)}$ &     127.71 &       1.40 &      27.12 \\
  338.583216 & CH$_3$OH & $(7_{3,5,1} - 6_{3,4,1)}$ &     112.71 &       1.39 &      31.38 \\
  338.611810 & SO$_{2}$ & $(20_{1,19} - 19_{2,18)}$ &     198.88 &       2.87 &       8.28 \\
  338.614936 & CH$_3$OH & $(7_{1,6,1} - 6_{1,5,1)}$ &      86.05 &       1.71 &      50.30 \\
  338.639802 & CH$_3$OH & $(7_{2,5,0} - 6_{2,4,0)}$ &     102.72 &       1.58 &      39.21 \\
  338.721693 & CH$_3$OH & $(7_{2,6,1} - 6_{2,5,1)}$ &      87.26 &       1.55 &      45.00 \\
  338.722898 & CH$_3$OH & $(7_{2,5,2} - 6_{2,4,2)}$ &      90.91 &       1.57 &      43.90 \\
  338.759948 & CH$_3$OH & $(13_{0,13,+0} - 12_{1,12,+0)}$ &     205.95 &       2.18 &       0.13 \\
  338.785687 & $^{34}$SO$_{2}$ & $(14_{4,10} - 14_{3,11)}$ &     134.34 &       3.08 &       0.43 \\
  339.341459 & SO & $(3_{3} - 2_{3)}$ &      25.51 &       0.14 &       0.08 \\
  339.491473 & CH$_3$OCH$_3$ & $(19_{1,18,0} - 18_{2,17,0)}$ &     176.10 &       2.03 &       0.04 \\
  339.491529 & CH$_3$OCH$_3$ & $(19_{1,18,1} - 18_{2,17,1)}$ &     176.10 &       2.03 &       0.06 \\
  339.491586 & CH$_3$OCH$_3$ & $(19_{1,18,3} - 18_{2,17,3)}$ &     176.10 &       2.03 &       0.02 \\
  339.491586 & CH$_3$OCH$_3$ & $(19_{1,18,5} - 18_{2,17,5)}$ &     176.10 &       2.03 &       0.02 \\
  339.857269 & $^{34}$SO & $(8_{9} - 7_{8)}$ &      77.34 &       5.08 &       2.60 \\
  340.052575 & C$^{33}$S & $(7_{0} - 6_{0)}$ &      65.28 &       8.19 &       0.25 \\
  340.141143 & CH$_3$OH & $(2_{2,0,0} - 3_{1,3,0)}$ &      44.67 &       0.28 &       4.08 \\
  340.316406 & SO$_{2}$ & $(28_{2,26} - 28_{1,27)}$ &     391.80 &       2.58 &       2.05 \\
  340.393659 & CH$_3$OH & $(16_{6,11,0} - 17_{5,12,0)}$ &     509.17 &       0.50 &       0.46 \\
  340.393745 & CH$_3$OH & $(16_{6,10,0} - 17_{5,13,0)}$ &     509.17 &       0.50 &       0.46 \\
  340.449273 & OCS & $(28 - 27)$ &     236.95 &       1.15 &       1.22 \\
  340.683969 & CH$_3$OH & $(11_{1,11,4} - 10_{0,10,4)}$ &     444.25 &       1.10 &       1.36 \\
  340.714155 & SO & $(8_{7} - 7_{6)}$ &      81.25 &       4.99 &       3.79 \\
  340.837357 & $^{33}$SO & $(8_{8,6.5} - 7_{7,5.5)}$ &      86.75 &       4.89 &       0.05 \\
  340.837862 & $^{33}$SO & $(8_{8,7.5} - 7_{7,6.5)}$ &      86.75 &       4.87 &       0.06 \\
  340.838679 & $^{33}$SO & $(8_{8,8.5} - 7_{7,7.5)}$ &      86.75 &       4.92 &       0.07 \\
  340.839639 & $^{33}$SO & $(8_{8,9.5} - 7_{7,8.5)}$ &      86.75 &       5.03 &       0.08 \\
  340.841735 & $^{33}$SO & $(8_{8,8.5} - 7_{7,8.5)}$ &      86.75 &       0.11 &       0.00 \\
  341.131665 & CH$_3$OH & $(13_{1,12,-0} - 13_{0,13,+0)}$ &     222.32 &       4.19 &       0.20 \\
  341.275524 & SO$_{2}$ & $(21_{8,14} - 22_{7,15)}$ &     369.13 &       0.69 &       0.49 \\
  341.403068 & SO$_{2}$ & $(40_{4,36} - 40_{3,37)}$ &     808.37 &       4.11 &       0.14 \\
  341.415615 & CH$_3$OH & $(7_{1,6,0} - 6_{1,5,0)}$ &      80.09 &       1.71 &      52.63 \\
  341.673961 & SO$_{2}$ & $(36_{5,31} - 36_{4,32)}$ &     678.52 &       4.34 &       0.40 \\
  342.208857 & $^{34}$SO$_{2}$ & $(5_{3,3} - 4_{2,2)}$ &      35.10 &       3.10 &       0.37 \\
  342.231633 & $^{34}$SO$_{2}$ & $(20_{1,19} - 19_{2,18)}$ &     198.24 &       3.06 &       0.35 \\
  342.332013 & $^{34}$SO$_{2}$ & $(12_{4,8} - 12_{3,9)}$ &     109.51 &       3.06 &       0.45 \\
  342.358225 & CH$_3$OCHO & $(30_{2,28,2} - 29_{2,27,2)}$ &     269.50 &       5.98 &       0.07 \\
  342.359508 & CH$_3$OCHO & $(30_{3,28,0} - 29_{3,27,0)}$ &     269.49 &       5.98 &       0.07 \\
  342.360148 & CH$_3$OCHO & $(57_{11,47,1} - 57_{10,48,1)}$ &    1069.40 &       0.56 &       0.00 \\
  342.521194 & t-HCOOH & $(16_{1,16} - 15_{1,15)}$ &     143.59 &       4.56 &       0.06 \\
  342.607898 & CH$_3$OCH$_3$ & $(19_{0,19,3} - 18_{1,18,3)}$ &     167.14 &       3.30 &       0.03 \\
  342.607898 & CH$_3$OCH$_3$ & $(19_{0,19,5} - 18_{1,18,5)}$ &     167.14 &       3.30 &       0.04 \\
  342.607971 & CH$_3$OCH$_3$ & $(19_{0,19,1} - 18_{1,18,1)}$ &     167.14 &       3.30 &       0.11 \\
  342.608044 & CH$_3$OCH$_3$ & $(19_{0,19,0} - 18_{1,18,0)}$ &     167.14 &       3.30 &       0.07 \\
  342.726474 & CH$_3$OH & $(20_{6,15,5} - 19_{7,13,5)}$ &     978.98 &       0.68 &       0.01 \\
  342.729796 & CH$_3$OH & $(13_{1,12,0} - 13_{0,13,0)}$ &     227.47 &       4.23 &      53.17 \\
  342.761625 & SO$_{2}$ & $(34_{3,31} - 34_{2,32)}$ &     581.92 &       3.45 &       0.67 \\
  342.882850 & CS & $(7_{0} - 6_{0)}$ &      65.83 &       8.40 &     -- \\
  343.086102 & $^{33}$SO & $(8_{9,7.5} - 7_{8,6.5)}$ &      78.03 &       5.11 &       0.07 \\
  343.087298 & $^{33}$SO & $(8_{9,8.5} - 7_{8,7.5)}$ &      78.03 &       5.09 &       0.08 \\
  343.088078 & $^{33}$SO & $(8_{9,9.5} - 7_{8,8.5)}$ &      78.03 &       5.13 &       0.09 \\
  343.088295 & $^{33}$SO & $(8_{9,10.5} - 7_{8,9.5)}$ &      78.04 &       5.23 &       0.10 \\
  343.147898 & CH$_3$OCHO & $(31_{1,30,2} - 30_{2,29,1)}$ &     273.44 &       0.89 &       0.01 \\
  343.148047 & CH$_3$OCHO & $(31_{2,30,1} - 30_{2,29,1)}$ &     273.44 &       6.08 &       0.07 \\
  343.148169 & CH$_3$OCHO & $(31_{1,30,2} - 30_{1,29,2)}$ &     273.44 &       6.08 &       0.07 \\
  343.148318 & CH$_3$OCHO & $(31_{2,30,1} - 30_{1,29,2)}$ &     273.44 &       0.89 &       0.01 \\
  343.149303 & CH$_3$OCHO & $(17_{5,12,0} - 16_{4,13,0)}$ &     107.81 &       0.25 &       0.01 \\
  343.152958 & CH$_3$OCHO & $(31_{1,30,0} - 30_{2,29,0)}$ &     273.43 &       0.89 &       0.01 \\
  343.153106 & CH$_3$OCHO & $(31_{2,30,0} - 30_{2,29,0)}$ &     273.43 &       6.08 &       0.07 \\
  343.153227 & CH$_3$OCHO & $(31_{1,30,0} - 30_{1,29,0)}$ &     273.43 &       6.08 &       0.07 \\
  343.153376 & CH$_3$OCHO & $(31_{2,30,0} - 30_{1,29,0)}$ &     273.43 &       0.89 &       0.01 \\
  343.153762 & CH$_3$OCHO & $(26_{13,13,0} - 26_{12,14,0)}$ &     319.30 &       0.46 &       0.00 \\
  343.153770 & CH$_3$OCHO & $(26_{13,14,0} - 26_{12,15,0)}$ &     319.30 &       0.46 &       0.00 \\
  343.325713 & H$_{2}$$^{13}$CO & $(5_{1,5} - 4_{1,4)}$ &      61.28 &      11.20 &       0.06 \\
  343.409963 & H$_2$CS & $(10_{3,8} - 9_{3,7)}$ &     209.10 &       5.56 &       0.09 \\
  343.414146 & H$_2$CS & $(10_{3,7} - 9_{3,6)}$ &     209.10 &       5.56 &       0.09 \\
  343.753320 & CH$_3$OCH$_3$ & $(17_{2,16,3} - 16_{1,15,3)}$ &     143.70 &       1.96 &       0.02 \\
  343.753320 & CH$_3$OCH$_3$ & $(17_{2,16,5} - 16_{1,15,5)}$ &     143.70 &       1.96 &       0.01 \\
  343.754216 & CH$_3$OCH$_3$ & $(17_{2,16,1} - 16_{1,15,1)}$ &     143.70 &       1.96 &       0.07 \\
  343.952343 & t-HCOOH & $(15_{1,14} - 14_{1,13)}$ &     136.28 &       4.60 &       0.06 \\
  343.983268 & OC$^{34}$S & $(29 - 28)$ &     247.67 &       1.19 &       0.12 \\
  344.029259 & CH$_3$OCHO & $(32_{0,32,2} - 31_{1,31,1)}$ &     276.10 &       0.77 &       0.01 \\
  344.029260 & CH$_3$OCHO & $(32_{1,32,1} - 31_{1,31,1)}$ &     276.10 &       6.43 &       0.07 \\
  344.029261 & CH$_3$OCHO & $(32_{0,32,2} - 31_{0,31,2)}$ &     276.10 &       6.43 &       0.07 \\
  344.029262 & CH$_3$OCHO & $(32_{1,32,1} - 31_{0,31,2)}$ &     276.10 &       0.77 &       0.01 \\
  344.029644 & CH$_3$OCHO & $(32_{0,32,0} - 31_{1,31,0)}$ &     276.08 &       1.00 &       0.01 \\
  344.029645 & CH$_3$OCHO & $(32_{1,32,0} - 31_{1,31,0)}$ &     276.08 &       6.20 &       0.07 \\
  344.029646 & CH$_3$OCHO & $(32_{0,32,0} - 31_{0,31,0)}$ &     276.08 &       6.20 &       0.07 \\
  344.029647 & CH$_3$OCHO & $(32_{1,32,0} - 31_{0,31,0)}$ &     276.08 &       1.00 &       0.01 \\
  344.109039 & CH$_3$OH & $(18_{2,17,1} - 17_{3,15,1)}$ &     419.40 &       0.68 &       1.71 \\
  344.200109 & HC$^{15}$N & $(4 - 3)$ &      41.30 &      18.80 &       3.51 \\
  344.245346 & $^{34}$SO$_{2}$ & $(10_{4,6} - 10_{3,7)}$ &      88.38 &       2.96 &       0.43 \\
  344.310612 & SO & $(8_{8} - 7_{7)}$ &      87.48 &       5.19 &       4.15 \\
  344.312267 & CH$_3$OH & $(10_{2,9,5} - 11_{3,9,5)}$ &     491.91 &       1.77 &       1.22 \\
  344.357816 & CH$_3$OCH$_3$ & $(19_{1,19,3} - 18_{0,18,3)}$ &     167.18 &       3.36 &       0.03 \\
  344.357816 & CH$_3$OCH$_3$ & $(19_{1,19,5} - 18_{0,18,5)}$ &     167.18 &       3.36 &       0.01 \\
  344.357826 & CH$_3$OCH$_3$ & $(38_{4,35,3} - 38_{3,36,3)}$ &     697.07 &       1.59 &       0.00 \\
  344.357826 & CH$_3$OCH$_3$ & $(38_{4,35,5} - 38_{3,36,5)}$ &     697.07 &       1.59 &       0.00 \\
  344.357929 & CH$_3$OCH$_3$ & $(19_{1,19,1} - 18_{0,18,1)}$ &     167.18 &       3.36 &       0.11 \\
  344.358041 & CH$_3$OCH$_3$ & $(19_{1,19,0} - 18_{0,18,0)}$ &     167.18 &       3.36 &       0.04 \\
  344.358087 & CH$_3$OCH$_3$ & $(38_{4,35,1} - 38_{3,36,1)}$ &     697.07 &       1.59 &       0.00 \\
  344.358347 & CH$_3$OCH$_3$ & $(38_{4,35,0} - 38_{3,36,0)}$ &     697.07 &       1.59 &       0.00 \\
  344.443433 & CH$_3$OH & $(19_{1,19,0} - 18_{2,16,0)}$ &     451.23 &       0.73 &       1.40 \\
  344.581045 & $^{34}$SO$_{2}$ & $(19_{1,19} - 18_{0,18)}$ &     167.41 &       5.16 &       0.72 \\
  344.807915 & $^{34}$SO$_{2}$ & $(13_{4,10} - 13_{3,11)}$ &     121.46 &       3.17 &       0.45 \\
  344.987585 & $^{34}$SO$_{2}$ & $(15_{4,12} - 15_{3,13)}$ &     148.13 &       3.27 &       0.42 \\
  344.998160 & $^{34}$SO$_{2}$ & $(11_{4,8} - 11_{3,9)}$ &      98.48 &       3.05 &       0.44 \\
  345.030561 & t-HCOOH & $(16_{0,16} - 15_{0,15)}$ &     143.05 &       4.66 &       0.06 \\
  345.148971 & SO$_{2}$ & $(5_{5,1} - 6_{4,2)}$ &      75.14 &       0.10 &       0.21 \\
  345.168664 & $^{34}$SO$_{2}$ & $(8_{4,4} - 8_{3,5)}$ &      70.94 &       2.75 &       0.37 \\
  345.285620 & $^{34}$SO$_{2}$ & $(9_{4,6} - 9_{3,7)}$ &      79.20 &       2.88 &       0.40 \\
  345.338538 & SO$_{2}$ & $(13_{2,12} - 12_{1,11)}$ &      92.98 &       2.38 &      10.53 \\
  345.448984 & SO$_{2}$ & $(26_{9,17} - 27_{8,20)}$ &     521.00 &       0.76 &       0.19 \\
  345.519656 & $^{34}$SO$_{2}$ & $(7_{4,4} - 7_{3,5)}$ &      63.61 &       2.59 &       0.33 \\
  345.553093 & $^{34}$SO$_{2}$ & $(6_{4,2} - 6_{3,3)}$ &      57.19 &       2.35 &       0.27 \\
  345.609010 & HC$_3$N & $(38 - 37)$ &     323.49 &      33.00 &       0.43 \\
  345.651293 & $^{34}$SO$_{2}$ & $(5_{4,2} - 5_{3,3)}$ &      51.69 &       1.96 &       0.20 \\
  345.678787 & $^{34}$SO$_{2}$ & $(4_{4,0} - 4_{3,1)}$ &      47.10 &       1.31 &       0.11 \\
  345.795990 & CO & $(3 - 2)$ &      33.19 &       0.03 &       -- \\
  345.903916 & CH$_3$OH & $(16_{1,15,0} - 15_{2,14,0)}$ &     332.65 &       1.04 &       5.49 \\
  345.919260 & CH$_3$OH & $(18_{3,15,2} - 17_{4,14,2)}$ &     459.43 &       0.73 &       1.21 \\
  345.929349 & $^{34}$SO$_{2}$ & $(17_{4,14} - 17_{3,15)}$ &     178.51 &       3.37 &       0.38 \\
  346.202719 & CH$_3$OH & $(5_{4,2,0} - 6_{3,3,0)}$ &     115.16 &       0.22 &       3.36 \\
  346.204271 & CH$_3$OH & $(5_{4,1,0} - 6_{3,4,0)}$ &     115.16 &       0.22 &       3.36 \\
  346.523878 & SO$_{2}$ & $(16_{4,12} - 16_{3,13)}$ &     164.47 &       3.39 &      10.03 \\
  346.528481 & SO & $(8_{9} - 7_{8)}$ &      78.78 &       5.38 &       5.11 \\
  346.652169 & SO$_{2}$ & $(19_{1,19} - 18_{0,18)}$ &     168.14 &       5.22 &      17.70 \\
  346.718858 & t-HCOOH & $(15_{2,13} - 14_{2,12)}$ &     144.46 &       4.66 &       0.06 \\
  347.188283 & CH$_3$OH & $(14_{1,13,-0} - 14_{0,14,+0)}$ &     254.25 &       4.36 &       0.16 \\
  347.330581 & SiO & $(8_{0} - 7_{0)}$ &      75.02 &      22.00 &       0.34 \\
  348.117469 & $^{34}$SO$_{2}$ & $(19_{4,16} - 19_{3,17)}$ &     212.61 &       3.50 &       0.33 \\
  348.387800 & SO$_{2}$ & $(24_{2,22} - 23_{3,21)}$ &     292.74 &       1.91 &       2.85 \\
  348.534365 & H$_2$CS & $(10_{1,9} - 9_{1,8)}$ &     105.19 &       6.32 &       0.22 \\
  348.911401 & CH$_3$CN, v=0 & $(19_{-9,0} - 18_{9,0)}$ &     745.42 &      28.70 &       0.03 \\
  348.911401 & CH$_3$CN, v=0 & $(19_{9,0} - 18_{-9,0)}$ &     745.42 &      28.70 &       0.03 \\
  349.024971 & CH$_3$CN, v=0 & $(19_{8,0} - 18_{8,0)}$ &     624.32 &      30.50 &       0.05 \\
  349.106997 & CH$_3$OH & $(14_{1,13,0} - 14_{0,14,0)}$ &     260.20 &       4.41 &      41.38 \\
  349.125287 & CH$_3$CN, v=0 & $(19_{7,0} - 18_{7,0)}$ &     517.41 &      32.10 &       0.09 \\
  349.212311 & CH$_3$CN, v=0 & $(19_{-6,0} - 18_{6,0)}$ &     424.70 &      33.40 &       0.13 \\
  349.212311 & CH$_3$CN, v=0 & $(19_{6,0} - 18_{-6,0)}$ &     424.70 &      33.40 &       0.13 \\
  349.286006 & CH$_3$CN, v=0 & $(19_{5,0} - 18_{5,0)}$ &     346.22 &      34.60 &       0.19 \\
  349.346343 & CH$_3$CN, v=0 & $(19_{4,0} - 18_{4,0)}$ &     281.99 &      35.50 &       0.26 \\
  349.393297 & CH$_3$CN, v=0 & $(19_{-3,0} - 18_{3,0)}$ &     232.01 &      36.30 &       0.32 \\
  349.393297 & CH$_3$CN, v=0 & $(19_{3,0} - 18_{-3,0)}$ &     232.01 &      36.30 &       0.32 \\
  349.426850 & CH$_3$CN, v=0 & $(19_{2,0} - 18_{2,0)}$ &     196.30 &      36.80 &       0.38 \\
  349.446987 & CH$_3$CN, v=0 & $(19_{1,0} - 18_{1,0)}$ &     174.88 &      37.10 &       0.42 \\
  349.453700 & CH$_3$CN, v=0 & $(19_{0,0} - 18_{0,0)}$ &     167.74 &      37.20 &       0.43 \\
  350.103118 & CH$_3$OH & $(1_{1,1,+0} - 0_{0,0,+0)}$ &      16.80 &       3.29 &       0.13 \\
  350.168100 & CH$_3$CN, v$_8$=1 & $(19_{1,3} - 18_{-1,3)}$ &     687.09 &      37.00 &       0.08 \\
  350.286493 & CH$_3$OH & $(15_{3,12,4} - 16_{4,13,4)}$ &     694.83 &       2.09 &       0.27 \\
  350.333059 & HNCO & $(16_{1,16} - 15_{1,15)}$ &     186.20 &       5.97 &       0.68 \\
  350.423619 & CH$_3$CN, v$_8$=1 & $(19_{-2,2} - 18_{2,2)}$ &     748.55 &      36.80 &       0.05 \\
  350.423619 & CH$_3$CN, v$_8$=1 & $(19_{2,2} - 18_{-2,2)}$ &     748.55 &      36.80 &       0.05 \\
  350.444906 & CH$_3$CN, v$_8$=1 & $(19_{0,2} - 18_{0,2)}$ &     693.40 &      37.20 &       0.08 \\
  350.449631 & CH$_3$CN, v$_8$=1 & $(19_{1,2} - 18_{1,2)}$ &     713.84 &      37.10 &       0.07 \\
  350.465645 & CH$_3$CN, v$_8$=1 & $(19_{-4,3} - 18_{4,3)}$ &     754.41 &      35.60 &       0.05 \\
  350.465645 & CH$_3$CN, v$_8$=1 & $(19_{4,3} - 18_{-4,3)}$ &     754.41 &      35.60 &       0.05 \\
  350.687662 & CH$_3$OH & $(4_{0,4,1} - 3_{1,3,2)}$ &      36.34 &       0.87 &      23.51 \\
  350.842670 & CH$_3$CN, v$_8$=1 & $(19_{-1,3} - 18_{1,3)}$ &     687.41 &      37.20 &       0.08 \\
  350.862756 & SO$_{2}$ & $(10_{6,4} - 11_{5,7)}$ &     138.84 &       0.44 &       1.00 \\
  350.905100 & CH$_3$OH & $(1_{1,1,0} - 0_{0,0,0)}$ &      16.84 &       3.32 &      36.36 \\
  351.236479 & CH$_3$OH & $(9_{5,4,1} - 10_{4,6,1)}$ &     240.51 &       0.36 &       2.70 \\
  351.257223 & SO$_{2}$ & $(5_{3,3} - 4_{2,2)}$ &      35.89 &       3.36 &       9.42 \\
  351.537795 & HNCO & $(16_{2,15} - 15_{2,14)}$ &     313.70 &       5.75 &       0.22 \\
  351.551573 & HNCO & $(16_{2,14} - 15_{2,13)}$ &     313.70 &       5.75 &       0.22 \\
  351.633257 & HNCO & $(16_{0,16} - 15_{0,15)}$ &     143.46 &       6.13 &       0.99 \\
  351.768645 & H$_2$CO & $(5_{1,5} - 4_{1,4)}$ &      62.45 &      12.00 &       -- \\
  351.873873 & SO$_{2}$ & $(14_{4,10} - 14_{3,11)}$ &     135.87 &       3.43 &      11.00 \\
  351.994870 & HNCO & $(23_{1,23} - 24_{0,24)}$ &     333.30 &       2.31 &       0.11 \\
  352.082921 & $^{34}$SO$_{2}$ & $(21_{4,18} - 21_{3,19)}$ &     250.41 &       3.65 &       0.27 \\
  352.599570 & OCS & $(29 - 28)$ &     253.88 &       1.28 &       1.11 \\
  352.897581 & HNCO & $(16_{1,15} - 15_{1,14)}$ &     187.25 &       6.10 &       0.68 \\
  353.723522 & CH$_3$OCHO & $(32_{1,31,2} - 31_{2,30,1)}$ &     290.42 &       0.97 &       0.01 \\
  353.728572 & CH$_3$OCHO & $(32_{1,31,0} - 31_{2,30,0)}$ &     290.41 &       0.98 &       0.01 \\
  353.739964 & CH$_3$OH & $(15_{1,14,-0} - 15_{0,15,+0)}$ &     288.46 &       4.54 &       0.12 \\
  354.127629 & CH$_3$OH & $(19_{2,17,3} - 18_{1,17,3)}$ &     738.20 &       1.43 &       0.15 \\
  354.445952 & CH$_3$OH & $(4_{1,3,0} - 3_{0,3,0)}$ &      43.71 &       1.27 &       0.11 \\
  354.505477 & HCN & $(4 - 3)$ &      42.53 &      20.50 &      -- \\
  354.604757 & CH$_3$OCHO & $(29_{15,14,3} - 28_{15,13,3)}$ &     592.88 &       4.99 &       0.00 \\
  354.604757 & CH$_3$OCHO & $(29_{15,15,3} - 28_{15,14,3)}$ &     592.88 &       4.99 &       0.00 \\
  354.607764 & CH$_3$OCHO & $(33_{0,33,2} - 32_{1,32,1)}$ &     293.12 &       0.70 &       0.01 \\
  354.607764 & CH$_3$OCHO & $(33_{1,33,1} - 32_{1,32,1)}$ &     293.12 &       7.19 &       0.06 \\
  354.607765 & CH$_3$OCHO & $(33_{0,33,2} - 32_{0,32,2)}$ &     293.12 &       7.19 &       0.06 \\
  354.607765 & CH$_3$OCHO & $(33_{1,33,1} - 32_{0,32,2)}$ &     293.12 &       0.70 &       0.01 \\
  354.608091 & CH$_3$OCHO & $(33_{0,33,0} - 32_{1,32,0)}$ &     293.10 &       1.10 &       0.01 \\
  354.608092 & CH$_3$OCHO & $(33_{0,33,0} - 32_{0,32,0)}$ &     293.10 &       6.79 &       0.06 \\
  354.608092 & CH$_3$OCHO & $(33_{1,33,0} - 32_{1,32,0)}$ &     293.10 &       6.79 &       0.06 \\
  354.608093 & CH$_3$OCHO & $(33_{1,33,0} - 32_{0,32,0)}$ &     293.10 &       1.10 &       0.01 \\
  354.697463 & HC$_3$N & $(39 - 38)$ &     340.52 &      35.70 &       0.40 \\
  355.045517 & SO$_{2}$ & $(12_{4,8} - 12_{3,9)}$ &     111.00 &       3.40 &      11.34 \\
  355.602945 & CH$_3$OH & $(13_{0,13,0} - 12_{1,12,0)}$ &     211.03 &       2.53 &      34.95 \\
  356.007235 & CH$_3$OH & $(15_{1,14,0} - 15_{0,15,0)}$ &     295.26 &       4.60 &      31.32 \\
  356.040644 & SO$_{2}$ & $(15_{7,9} - 16_{6,10)}$ &     230.41 &       0.64 &       0.98 \\
  356.137265 & t-HCOOH & $(16_{2,15} - 15_{2,14)}$ &     158.67 &       5.07 &       0.06 \\
  356.222250 & $^{34}$SO$_{2}$ & $(25_{3,23} - 25_{2,24)}$ &     319.52 &       2.97 &       0.14 \\
  356.734223 & HCO$^{+}$ & $(4 - 3)$ &      42.80 &      35.70 &     -- \\
  356.755190 & SO$_{2}$ & $(10_{4,6} - 10_{3,7)}$ &      89.83 &       3.28 &      10.88 \\
  357.102182 & $^{34}$SO$_{2}$ & $(20_{0,20} - 19_{1,19)}$ &     184.55 &       5.81 &       0.69 \\
  357.165390 & SO$_{2}$ & $(13_{4,10} - 13_{3,11)}$ &     122.97 &       3.51 &      11.34 \\
  357.241193 & SO$_{2}$ & $(15_{4,12} - 15_{3,13)}$ &     149.68 &       3.62 &      10.73 \\
  357.387579 & SO$_{2}$ & $(11_{4,8} - 11_{3,9)}$ &      99.95 &       3.38 &      11.26 \\
  357.459408 & CH$_3$OCH$_3$ & $(18_{2,17,3} - 17_{1,16,3)}$ &     159.81 &       2.31 &       0.02 \\
  357.459408 & CH$_3$OCH$_3$ & $(18_{2,17,5} - 17_{1,16,5)}$ &     159.81 &       2.31 &       0.03 \\
  357.460164 & CH$_3$OCH$_3$ & $(18_{2,17,1} - 17_{1,16,1)}$ &     159.81 &       2.31 &       0.07 \\
  357.460920 & CH$_3$OCH$_3$ & $(18_{2,17,0} - 17_{1,16,0)}$ &     159.81 &       2.31 &       0.04 \\
  357.581449 & SO$_{2}$ & $(8_{4,4} - 8_{3,5)}$ &      72.36 &       3.06 &       9.45 \\
  357.657954 & CH$_3$OH & $(7_{2,5,0} - 6_{1,5,0)}$ &      85.80 &       1.63 &       0.16 \\
  357.671821 & SO$_{2}$ & $(9_{4,6} - 9_{3,7)}$ &      80.64 &       3.20 &      10.30 \\
  357.892442 & SO$_{2}$ & $(7_{4,4} - 7_{3,5)}$ &      65.01 &       2.87 &       8.32 \\
  357.925848 & SO$_{2}$ & $(6_{4,2} - 6_{3,3)}$ &      58.58 &       2.60 &       6.89 \\
  357.962905 & SO$_{2}$ & $(17_{4,14} - 17_{3,15)}$ &     180.11 &       3.73 &       9.64 \\
  358.013154 & SO$_{2}$ & $(5_{4,2} - 5_{3,3)}$ &      53.07 &       2.18 &       5.11 \\
  358.037887 & SO$_{2}$ & $(4_{4,0} - 4_{3,1)}$ &      48.48 &       1.45 &       2.89 \\
  358.215633 & SO$_{2}$ & $(20_{0,20} - 19_{1,19)}$ &     185.33 &       5.83 &      16.89 \\
  358.347316 & $^{34}$SO$_{2}$ & $(23_{4,20} - 23_{3,21)}$ &     291.93 &       3.86 &       0.21 \\
  358.414648 & CH$_3$OH & $(10_{6,5,1} - 11_{5,6,1)}$ &     306.43 &       0.34 &       1.40 \\
  358.449468 & CH$_3$OCH$_3$ & $(5_{5,1,5} - 4_{4,0,5)}$ &      48.80 &       3.71 &       0.01 \\
  358.449478 & CH$_3$OCH$_3$ & $(5_{5,0,5} - 4_{4,1,5)}$ &      48.80 &       3.71 &       0.04 \\
  358.451541 & CH$_3$OCH$_3$ & $(5_{5,0,3} - 4_{4,0,3)}$ &      48.80 &       3.71 &       0.03 \\
  358.452015 & CH$_3$OCH$_3$ & $(5_{5,0,1} - 4_{4,0,1)}$ &      48.80 &       3.71 &       0.10 \\
  358.454082 & CH$_3$OCH$_3$ & $(5_{5,1,1} - 4_{4,1,1)}$ &      48.80 &       3.71 &       0.10 \\
  358.456618 & CH$_3$OCH$_3$ & $(5_{5,1,0} - 4_{4,0,0)}$ &      48.80 &       3.71 &       0.04 \\
  358.456629 & CH$_3$OCH$_3$ & $(5_{5,0,0} - 4_{4,1,0)}$ &      48.80 &       3.71 &       0.06 \\
  358.605799 & CH$_3$OH & $(4_{1,3,1} - 3_{0,3,1)}$ &      44.26 &       1.32 &      31.66 \\
  358.987971 & $^{34}$SO$_{2}$ & $(15_{2,14} - 14_{1,13)}$ &     118.72 &       2.92 &       0.45 \\
  359.151158 & SO$_{2}$ & $(25_{3,23} - 25_{2,24)}$ &     320.93 &       3.10 &       3.60 \\
  359.651730 & $^{34}$SO$_{2}$ & $(24_{2,22} - 23_{3,21)}$ &     292.00 &       2.20 &       0.12 \\
  359.770685 & SO$_{2}$ & $(19_{4,16} - 19_{3,17)}$ &     214.26 &       3.85 &       8.27 \\
  360.148273 & t-HCOOH & $(16_{6,11} - 15_{6,10)}$ &     261.23 &       4.58 &       0.03 \\
  360.148356 & t-HCOOH & $(16_{6,10} - 15_{6,9)}$ &     261.23 &       4.58 &       0.03 \\
  360.148518 & t-HCOOH & $(16_{15,1} - 15_{15,0)}$ &     858.50 &       0.64 &       0.00 \\
  360.148518 & t-HCOOH & $(16_{15,2} - 15_{15,1)}$ &     858.50 &       0.64 &       0.00 \\
  360.290404 & SO$_{2}$ & $(34_{5,29} - 34_{4,30)}$ &     611.96 &       4.80 &       0.66 \\
  360.584554 & CH$_3$OCH$_3$ & $(20_{0,20,3} - 19_{1,19,3)}$ &     184.48 &       3.89 &       0.03 \\
  360.584554 & CH$_3$OCH$_3$ & $(20_{0,20,5} - 19_{1,19,5)}$ &     184.48 &       3.89 &       0.01 \\
  360.584633 & CH$_3$OCH$_3$ & $(20_{0,20,1} - 19_{1,19,1)}$ &     184.48 &       3.89 &       0.10 \\
  360.584711 & CH$_3$OCH$_3$ & $(20_{0,20,0} - 19_{1,19,0)}$ &     184.48 &       3.89 &       0.04 \\
  360.661633 & CH$_3$OH & $(3_{1,3,3} - 4_{2,3,3)}$ &     339.14 &       2.64 &       2.56 \\
  360.721829 & SO$_{2}$ & $(20_{8,12} - 21_{7,15)}$ &     349.82 &       0.77 &       0.56 \\
  360.848946 & CH$_3$OH & $(11_{0,11,1} - 10_{1,9,1)}$ &     166.05 &       1.21 &      21.64 \\
\hline
\end{longtable}}

%
%


\longtab[2]{
\begin{longtable}{l c c c c r}    
\caption{\label{table:Molecules_N12} Detected molecular transitions towards N12. The listed line transitions are from best fit line models obtained using the software package CASSIS, on the spectrum taken at the position of peak continuum emission. The model parameters are listed in Table \ref{table:N12_column_densities}. Only transitions above 6 K are listed ($\sim 3 \sigma$), corresponding to about 0.6 $\mathrm{Jy \ beam^{-1}}$.} \\
\hline\hline 
     Frequency & Molecule & Transition & $E_{\mathrm{up}}$/$k_{\mathrm{B}}$ & $A_{\mathrm{ij}}$  & Opacity ($\tau$)  \\ 
      $[\mathrm{GHz}] $  &  & &  $[\mathrm{K}]$ & $\times 10^{-4} [\mathrm{s^{-1}}]$  &   \\ 
\hline
\endfirsthead
\caption{continued.}\\
\hline\hline
     Frequency & Molecule & Transition & $E_{\mathrm{up}}$/$k_{\mathrm{B}}$ & $A_{\mathrm{ij}}$  & Opacity ($\tau$)    \\ 
      $[\mathrm{GHz}] $  &  & &  $[\mathrm{K}]$ & $\times 10^{-4} [\mathrm{s^{-1}}]$  &  \\ 
\hline
\endhead
\hline
\endfoot
  329.330552 & C$^{18}$O & $(3 - 2)$ &      31.61 &       0.02 &       7.27 \\
  330.463873 & $^{13}$CH$_3$OH & $(11_{1,10,-0} - 11_{0,11,+0)}$ &     165.27 &       3.90 &       0.09 \\
  330.464947 & $^{13}$CH$_3$OH & $(7_{2,5,+0} - 6_{2,4,+0)}$ &     101.24 &       1.46 &       0.03 \\
  330.587965 & $^{13}$CO & $(3 - 2)$ &      31.73 &       0.02 &       -- \\
  330.842762 & CH$_3$CN, v=0 & $(18_{-6,0} - 17_{6,0)}$ &     407.94 &      28.00 &       0.04 \\
  330.842762 & CH$_3$CN, v=0 & $(18_{6,0} - 17_{-6,0)}$ &     407.94 &      28.00 &       0.04 \\
  330.912608 & CH$_3$CN, v=0 & $(18_{5,0} - 17_{5,0)}$ &     329.46 &      29.10 &       0.05 \\
  330.969794 & CH$_3$CN, v=0 & $(18_{4,0} - 17_{4,0)}$ &     265.22 &      30.00 &       0.07 \\
\hline
\end{longtable}
}


\longtab[3]{
\begin{longtable}{l c c c c r}     
\caption{\label{table:Molecules_N63} Detected molecular transitions towards N63. The listed line transitions are from best fit line models obtained using the software package CASSIS, on the spectrum taken at the position of peak continuum emission. The model parameters are listed in Table 6 of the main text.
Only transitions above 6 K are listed ($\sim 3 \sigma$), corresponding to about 0.6 $\mathrm{Jy \ beam^{-1}}$.} \\
\hline\hline 
     Frequency & Molecule & Transition & $E_{\mathrm{up}}$/$k_{\mathrm{B}}$ & $A_{\mathrm{ij}}$  & Opacity ($\tau$)  \\ 
      $[\mathrm{GHz}] $  &  & &  $[\mathrm{K}]$ & $\times 10^{-4} [\mathrm{s^{-1}}]$  &   \\ 
\hline
\endfirsthead
\caption{continued.}\\
\hline\hline
     Frequency & Molecule & Transition & $E_{\mathrm{up}}$/$k_{\mathrm{B}}$ & $A_{\mathrm{ij}}$  & Opacity ($\tau$)    \\ 
      $[\mathrm{GHz}] $  &  & &  $[\mathrm{K}]$ & $\times 10^{-4} [\mathrm{s^{-1}}]$  &  \\ 
\hline
\endhead
\hline
\endfoot
  329.330552 & C$^{18}$O & $(3 - 2)$ &      31.61 &       0.02 &       7.27 \\
  329.632881 & CH$_3$OH & $(12_{2,10,0} - 11_{3,9,0)}$ &     218.80 &       0.60 &       0.30 \\
  330.406503 & CH$_3$OCH$_3$ & $(16_{2,15,1} - 15_{1,14,1)}$ &     128.46 &       1.65 &       0.04 \\
  330.407010 & CH$_3$OCH$_3$ & $(49_{10,40,3} - 48_{11,37,3)}$ &    1256.39 &       0.20 &       0.00 \\
  330.463873 & $^{13}$CH$_3$OH & $(11_{1,10,-0} - 11_{0,11,+0)}$ &     165.27 &       3.90 &       0.08 \\
  330.464947 & $^{13}$CH$_3$OH & $(7_{2,5,+0} - 6_{2,4,+0)}$ &     101.24 &       1.46 &       0.03 \\
  330.587965 & $^{13}$CO & $(3 - 2)$ &      31.73 &       0.02 &       -- \\
  330.793887 & CH$_3$OH & $(8_{3,5,2} - 9_{2,7,2)}$ &     146.28 &       0.54 &       0.33 \\
\hline
\end{longtable}}

%
%


\longtab[4]{
\begin{longtable}{l c c c c r}    
\caption{\label{table:Molecules_N53} Detected molecular transitions towards N53. The listed line transitions are from best fit line models obtained using the software package CASSIS, on the spectrum taken at the position of peak continuum emission. The model parameters are listed in Table \ref{table:N53_column_densities}. Only transitions above 6 K are listed ($\sim 3 \sigma$), corresponding to about 0.6 $\mathrm{Jy \ beam^{-1}}$.} \\
\hline\hline 
     Frequency & Molecule & Transition & $E_{\mathrm{up}}$/$k_{\mathrm{B}}$ & $A_{\mathrm{ij}}$  & Opacity ($\tau$)  \\ 
      $[\mathrm{GHz}] $  &  & &  $[\mathrm{K}]$ & $\times 10^{-4} [\mathrm{s^{-1}}]$  &   \\ 
\hline
\endfirsthead
\caption{continued.}\\
\hline\hline
     Frequency & Molecule & Transition & $E_{\mathrm{up}}$/$k_{\mathrm{B}}$ & $A_{\mathrm{ij}}$  & Opacity ($\tau$)    \\ 
      $[\mathrm{GHz}] $  &  & &  $[\mathrm{K}]$ & $\times 10^{-4} [\mathrm{s^{-1}}]$  &  \\ 
\hline
\endhead
\hline
\endfoot
  329.330552 & C$^{18}$O & $(3 - 2)$ &      31.61 &       0.02 &       -- \\
  330.587965 & $^{13}$CO & $(3 - 2)$ &      31.73 &       0.02 &       -- \\
  331.502319 & CH$_3$OH & $(11_{1,10,0} - 11_{0,11,0)}$ &     169.01 &       3.93 &       0.38 \\
  335.582017 & CH$_3$OH & $(7_{1,7,0} - 6_{1,6,0)}$ &      78.97 &       1.63 &       0.18 \\
  336.865149 & CH$_3$OH & $(12_{1,11,0} - 12_{0,12,0)}$ &     197.07 &       4.07 &       0.34 \\
  337.060513 & C$^{17}$O & $(3_{1.5} - 2_{2.5)}$ &      32.35 &       0.00 &       0.00 \\
  337.060709 & C$^{17}$O & $(3_{0.5} - 2_{1.5)}$ &      32.35 &       0.01 &       0.00 \\
  337.060831 & C$^{17}$O & $(3_{2.5} - 2_{3.5)}$ &      32.35 &       0.00 &       0.00 \\
\hline
\end{longtable}
}


\longtab[5]{
\begin{longtable}{l c c c c r}    
\caption{\label{table:Molecules_N54} Detected molecular transitions towards N54. The listed line transitions are from best fit line models obtained using the software package CASSIS, on the spectrum taken at the position of peak continuum emission. The model parameters are listed in Table \ref{table:N54_column_densities}. Only transitions above 6 K are listed ($\sim 3 \sigma$), corresponding to about 0.6 $\mathrm{Jy \ beam^{-1}}$.} \\
\hline\hline 
     Frequency & Molecule & Transition & $E_{\mathrm{up}}$/$k_{\mathrm{B}}$ & $A_{\mathrm{ij}}$  & Opacity ($\tau$)  \\ 
      $[\mathrm{GHz}] $  &  & &  $[\mathrm{K}]$ & $\times 10^{-4} [\mathrm{s^{-1}}]$  &   \\ 
\hline
\endfirsthead
\caption{continued.}\\
\hline\hline
     Frequency & Molecule & Transition & $E_{\mathrm{up}}$/$k_{\mathrm{B}}$ & $A_{\mathrm{ij}}$  & Opacity ($\tau$)    \\ 
      $[\mathrm{GHz}] $  &  & &  $[\mathrm{K}]$ & $\times 10^{-4} [\mathrm{s^{-1}}]$  &  \\ 
\hline
\endhead
\hline
\endfoot
  330.587965 & $^{13}$CO & $(3 - 2)$ &      31.73 &       0.02 &       -- \\
  342.882850 & CS & $(7_{0} - 6_{0)}$ &      65.83 &       8.40 &       -- \\
  345.795990 & CO & $(3 - 2)$ &      33.19 &       0.03 &       -- \\
  351.768645 & H$_2$CO & $(5_{1,5} - 4_{1,4)}$ &      62.45 &      12.00 &       -- \\
  354.505477 & HCN & $(4 - 3)$ &      42.53 &      20.50 &       -- \\
  356.734223 & HCO$^{+}$ & $(4 - 3)$ &      42.80 &      35.70 &       -- \\
\hline
\end{longtable}
}


\longtab[6]{
\begin{longtable}{l c c c c r}    
\caption{\label{table:Molecules_N51} Detected molecular transitions towards N51. The listed line transitions are from best fit line models obtained using the software package CASSIS, on the spectrum taken at the position of peak continuum emission. The model parameters are listed in Table \ref{table:N51_column_densities}. Only transitions above 6 K are listed ($\sim 3 \sigma$), corresponding to about 0.6 $\mathrm{Jy \ beam^{-1}}$.} \\
\hline\hline 
     Frequency & Molecule & Transition & $E_{\mathrm{up}}$/$k_{\mathrm{B}}$ & $A_{\mathrm{ij}}$  & Opacity ($\tau$)  \\ 
      $[\mathrm{GHz}] $  &  & &  $[\mathrm{K}]$ & $\times 10^{-4} [\mathrm{s^{-1}}]$  &   \\ 
\hline
\endfirsthead
\caption{continued.}\\
\hline\hline
     Frequency & Molecule & Transition & $E_{\mathrm{up}}$/$k_{\mathrm{B}}$ & $A_{\mathrm{ij}}$  & Opacity ($\tau$)    \\ 
      $[\mathrm{GHz}] $  &  & &  $[\mathrm{K}]$ & $\times 10^{-4} [\mathrm{s^{-1}}]$  &  \\ 
\hline
\endhead
\hline
\endfoot

  330.587965 & $^{13}$CO & $(3 - 2)$ &      31.73 &       0.02 &       -- \\
  331.502319 & CH$_3$OH & $(11_{1,10,0} - 11_{0,11,0)}$ &     169.01 &       3.93 &       0.19 \\
  335.582017 & CH$_3$OH & $(7_{1,7,0} - 6_{1,6,0)}$ &      78.97 &       1.63 &       0.09 \\
  336.865149 & CH$_3$OH & $(12_{1,11,0} - 12_{0,12,0)}$ &     197.07 &       4.07 &       0.17 \\
  338.124488 & CH$_3$OH & $(7_{0,7,1} - 6_{0,6,1)}$ &      78.08 &       1.70 &       0.09 \\
  338.344588 & CH$_3$OH & $(7_{1,7,2} - 6_{1,6,2)}$ &      70.55 &       1.67 &       0.10 \\
  338.408698 & CH$_3$OH & $(7_{0,7,0} - 6_{0,6,0)}$ &      64.98 &       1.70 &       0.10 \\
  338.512632 & CH$_3$OH & $(7_{4,4,0} - 6_{4,3,0)}$ &     145.33 &       1.15 &       0.04 \\
\hline
\end{longtable}
}


\longtab[7]{
\begin{longtable}{l c c c c r}    
\caption{\label{table:Molecules_N38} Detected molecular transitions towards N38. The listed line transitions are from best fit line models obtained using the software package CASSIS, on the spectrum taken at the position of peak continuum emission. The model parameters are listed in Table \ref{table:N38_column_densities}. Only transitions above 6 K are listed ($\sim 3 \sigma$), corresponding to about 0.6 $\mathrm{Jy \ beam^{-1}}$.} \\
\hline\hline 
     Frequency & Molecule & Transition & $E_{\mathrm{up}}$/$k_{\mathrm{B}}$ & $A_{\mathrm{ij}}$  & Opacity ($\tau$)  \\ 
      $[\mathrm{GHz}] $  &  & &  $[\mathrm{K}]$ & $\times 10^{-4} [\mathrm{s^{-1}}]$  &   \\ 
\hline
\endfirsthead
\caption{continued.}\\
\hline\hline
     Frequency & Molecule & Transition & $E_{\mathrm{up}}$/$k_{\mathrm{B}}$ & $A_{\mathrm{ij}}$  & Opacity ($\tau$)    \\ 
      $[\mathrm{GHz}] $  &  & &  $[\mathrm{K}]$ & $\times 10^{-4} [\mathrm{s^{-1}}]$  &  \\ 
\hline
\endhead
\hline
\endfoot
  330.587965 & $^{13}$CO & $(3 - 2)$ &      31.73 &       0.02 &       -- \\
  331.502319 & CH$_3$OH & $(11_{1,10,0} - 11_{0,11,0)}$ &     169.01 &       3.93 &       0.38 \\
  335.582017 & CH$_3$OH & $(7_{1,7,0} - 6_{1,6,0)}$ &      78.97 &       1.63 &       0.18 \\
  336.865149 & CH$_3$OH & $(12_{1,11,0} - 12_{0,12,0)}$ &     197.07 &       4.07 &       0.34 \\
  337.642478 & CH$_3$OH & $(7_{1,7,4} - 6_{1,6,4)}$ &     356.30 &       1.65 &       0.03 \\
  337.643915 & CH$_3$OH & $(7_{0,7,4} - 6_{0,6,4)}$ &     365.40 &       1.69 &       0.03 \\
  338.083195 & H$_2$CS & $(10_{1,10} - 9_{1,9)}$ &     102.43 &       5.77 &       0.15 \\
  338.124488 & CH$_3$OH & $(7_{0,7,1} - 6_{0,6,1)}$ &      78.08 &       1.70 &       0.19 \\
\hline
\end{longtable}
}


\longtab[8]{
\begin{longtable}{l c c c c r}    
\caption{\label{table:Molecules_N48} Detected molecular transitions towards N48. The listed line transitions are from best fit line models obtained using the software package CASSIS, on the spectrum taken at the position of peak continuum emission. The model parameters are listed in Table \ref{table:N48_column_densities}. Only transitions above 6 K are listed ($\sim 3 \sigma$), corresponding to about 0.6 $\mathrm{Jy \ beam^{-1}}$.} \\
\hline\hline 
     Frequency & Molecule & Transition & $E_{\mathrm{up}}$/$k_{\mathrm{B}}$ & $A_{\mathrm{ij}}$  & Opacity ($\tau$)  \\ 
      $[\mathrm{GHz}] $  &  & &  $[\mathrm{K}]$ & $\times 10^{-4} [\mathrm{s^{-1}}]$  &   \\ 
\hline
\endfirsthead
\caption{continued.}\\
\hline\hline
     Frequency & Molecule & Transition & $E_{\mathrm{up}}$/$k_{\mathrm{B}}$ & $A_{\mathrm{ij}}$  & Opacity ($\tau$)    \\ 
      $[\mathrm{GHz}] $  &  & &  $[\mathrm{K}]$ & $\times 10^{-4} [\mathrm{s^{-1}}]$  &  \\ 
\hline
\endhead
\hline
\endfoot
  330.587965 & $^{13}$CO & $(3 - 2)$ &      31.73 &       0.02 &       -- \\
  331.502319 & CH$_3$OH & $(11_{1,10,0} - 11_{0,11,0)}$ &     169.01 &       3.93 &       0.19 \\
  335.582017 & CH$_3$OH & $(7_{1,7,0} - 6_{1,6,0)}$ &      78.97 &       1.63 &       0.09 \\
  336.865149 & CH$_3$OH & $(12_{1,11,0} - 12_{0,12,0)}$ &     197.07 &       4.07 &       0.17 \\
  338.124488 & CH$_3$OH & $(7_{0,7,1} - 6_{0,6,1)}$ &      78.08 &       1.70 &       0.09 \\
  338.344588 & CH$_3$OH & $(7_{1,7,2} - 6_{1,6,2)}$ &      70.55 &       1.67 &       0.10 \\
  338.408698 & CH$_3$OH & $(7_{0,7,0} - 6_{0,6,0)}$ &      64.98 &       1.70 &       0.10 \\
  338.512632 & CH$_3$OH & $(7_{4,4,0} - 6_{4,3,0)}$ &     145.33 &       1.15 &       0.04 \\
\hline
\end{longtable}
}


\longtab[9]{
\begin{longtable}{l c c c c r}    
\caption{\label{table:Molecules_S8} Detected molecular transitions towards S8. The listed line transitions are from best fit line models obtained using the software package CASSIS, on the spectrum taken at the position of peak continuum emission. The model parameters are listed in Table \ref{table:S8_column_densities}. Only transitions above 6 K are listed ($\sim 3 \sigma$), corresponding to about 0.6 $\mathrm{Jy \ beam^{-1}}$.} \\
\hline\hline 
     Frequency & Molecule & Transition & $E_{\mathrm{up}}$/$k_{\mathrm{B}}$ & $A_{\mathrm{ij}}$  & Opacity ($\tau$)  \\ 
      $[\mathrm{GHz}] $  &  & &  $[\mathrm{K}]$ & $\times 10^{-4} [\mathrm{s^{-1}}]$  &   \\ 
\hline
\endfirsthead
\caption{continued.}\\
\hline\hline
     Frequency & Molecule & Transition & $E_{\mathrm{up}}$/$k_{\mathrm{B}}$ & $A_{\mathrm{ij}}$  & Opacity ($\tau$)    \\ 
      $[\mathrm{GHz}] $  &  & &  $[\mathrm{K}]$ & $\times 10^{-4} [\mathrm{s^{-1}}]$  &  \\ 
\hline
\endhead
\hline
\endfoot
  330.587965 & $^{13}$CO & $(3 - 2)$ &      31.73 &       0.02 &       -- \\
  340.714155 & SO & $(8_{7} - 7_{6)}$ &      81.25 &       4.99 &       0.05 \\
  342.882850 & CS & $(7_{0} - 6_{0)}$ &      65.83 &       8.40 &       -- \\
  344.310612 & SO & $(8_{8} - 7_{7)}$ &      87.48 &       5.19 &       0.05 \\
  345.795990 & CO & $(3 - 2)$ &      33.19 &       0.03 &       -- \\
  346.528481 & SO & $(8_{9} - 7_{8)}$ &      78.78 &       5.38 &       0.06 \\
  351.768645 & H$_2$CO & $(5_{1,5} - 4_{1,4)}$ &      62.45 &      12.00 &       -- \\
  354.505477 & HCN & $(4 - 3)$ &      42.53 &      20.50 &       -- \\
  356.734223 & HCO$^{+}$ & $(4 - 3)$ &      42.80 &      35.70 &       -- \\
\hline
\end{longtable}
}


\end{document}